\documentclass[trackchanges]{aastex702}

\usepackage{amsmath}

\usepackage{natbib}

\newcommand{\OIIIab}{[O{\sc iii}]\,$\lambda\lambda$4959,5007}
\newcommand{\Hb}{H$\beta$}
\newcommand{\Ha}{H$\alpha$}

\newcommand{\lya}{Ly\,$\alpha$}

\def\OIII{[O\,{\sc iii}]}
\def\NII{[N\,{\sc ii}]}

\newcommand{\stsci}{\affil{Space Telescope Science Institute, 3700 San Martin Drive, Baltimore, MD 21218}}
\usepackage{hyperref}
\begin{document}

\title{Surveying the Universe in 4D: Beating Cosmic Variance with Wide-Field Slitless Spectroscopy from HST, JWST, Euclid, Roman, and Beyond}

\author[0000-0003-4030-3455]{Andreea Petric}
\stsci
\email{apetric@stsci.edu}

\author[]{Ga\"el Noirot}
\stsci
\email{gnoirot@stsci.edu}

\author[]{Stacey Alberts}
\stsci
\email{slaberts@stsci.edu}

\author[]{Russell Ryan}
\stsci
\email{Email}

\author[0000-0002-7756-4440]{Louis-Gregory Strolger}
\stsci
\email{strolger@stsci.edu}

\author[0000-0001-8489-2349]{Vicente Estrada-Carpenter}
\affiliation{School of Earth and Space Exploration, Arizona State University, Tempe, AZ 85287, USA}
\affiliation{Beus Center for Cosmic Foundations, Arizona State University, Tempe, AZ 85287, USA}
\email{vestrad9@asu.edu}

\author[]{Mainak Singha}
\affiliation{NASA/GSFC, Greenbelt, MD 20771}
\email{singham@cua.edu}

\author[0000-0002-3568-3900]{Adrian E. Bayer}
\affiliation{The NSF AI Institute for Artificial Intelligence and Fundamental Interactions, Cambridge, MA 02139, USA}
\affiliation{Laboratory for Nuclear Science, Massachusetts Institute of Technology, Cambridge, MA 02139, USA}
\affiliation{Center for Astrophysics $|$ Harvard \& Smithsonian, 60 Garden Street, Cambridge, MA 02138, USA}
\affiliation{Perimeter Institute for Theoretical Physics, 31 Caroline Street North, Waterloo, Ontario N2L 2Y5, Canada}
\email{abayer@mit.edu}

\author[0009-0004-2538-7237]{Olivier Gilbert}
\affiliation{Department of Astronomy, University of Michigan, 1085 S. University Ave., Ann Arbor, MI 48109, USA}
\email{ogilbert@umich.edu}

\author[0000-0002-0072-0281]{Farhanul Hasan}
\stsci
\email{fhasan@stsci.edu}

\author[0009-0002-9932-4461]{Mason Huberty}
\affiliation{Minnesota Institute for Astrophysics, University of Minnesota, 116 Church Street SE, Minneapolis, MN 55455, USA}
\email{huber458@umn.edu}

\author[0009-0003-4862-2925]{Bingcheng Jin}
\affiliation{Department of Astronomy, University of Michigan, 1085 S. University Ave., Ann Arbor, MI 48109, USA}
\email{bcjin@umich.edu}

\author[0000-0002-5768-738X]{Xiangyu Jin}
\affiliation{Department of Astronomy, University of Michigan, 1085 S. University Ave., Ann Arbor, MI 48109, USA}
\email{jxiangyu@umich.edu}  

\author[0009-0005-4226-0964]{Anika Kumar} 
\affiliation{Laboratory for Multiwavelength Astrophysics, School of Physics and Astronomy, Rochester Institute of Technology, 84 Lomb Memorial Drive, Rochester, NY 14623, USA} 
\email{ak8532@rit.edu}

\author[0009-0007-2139-1791]{Kasra Mokhtarpour}
\affiliation{Department of Physics, Florida Atlantic University,  777 Glades Rd, Boca Raton, FL 33431, USA}
\email{kmokhtarpour2019@fau,edu}

\author[0000-0001-7069-4026]{Preethi Nair}
\affiliation{Department of Physics and Astronomy, The University of Alabama}
\email{pnair@ua.edu}

\author[]{Eleni Tsiakaliari}
\affiliation{The Open University (UK),  Walton Hall, Kents Hill, Milton Keynes MK7 6AA, United Kingdom}
\email{eleni.tsiakaliari@open.ac.uk}

\author[orcid=0000-0002-1912-0024,gname=Vivian,sname=U]{Vivian U}
\affiliation{IPAC, California Institute of Technology, 1200 E.~California Blvd, Pasadena, CA 91125, USA}
\email{vivianu@ipac.caltech.edu}

\author[0000-0002-8163-0172]{Brittany N. Vanderhoof}
\stsci
\email{bvanderhoof@stsci.edu}

\author[0009-0003-7532-3197]{Arshia Akhtarkavan}
\affiliation{Steward Observatory, University of Arizona, 933 North Cherry Avenue, Tucson, AZ 85721, USA}
\email{aakhtarkavan@arizona.edu}

\author[]{Natalia Alvarez-Ibanez} 
\affiliation{Universidad de Navarra, 31009 Pamplona, Navarra, Spain} 
\email{nalvareziba@alumni.unav.es}

\author[0000-0002-7714-688X]{Román Fernández Aranda}
\affiliation{Centro de Astrobiología (CAB), CSIC-INTA, Carretera de Ajalvir km 4, Torrejón de Ardoz, E-28850, Madrid, Spain}
\email{rfernandez@cab.inta-csic.es}

\author[0000-0002-9888-2704]{Tri L. Astraatmadja}
\stsci
\email{tastraatmadja@stsci.edu}

\author[0009-0009-4635-9442]{Clive Binu}
\affiliation{Laboratory for Multiwavelength Astrophysics, School of Physics and Astronomy, Rochester Institute of Technology, 84 Lomb Memorial Drive, Rochester, NY 14623, USA} 
\email{ckb2084@rit.edu}

\author[orcid=0000-0002-6913-8580]{Claire Bolda}
\affiliation{Department of Physics and Astronomy, Texas A\&M University, College Station, TX, 77843-4242, USA}
\affiliation{George P. and Cynthia Woods Mitchell Institute for Fundamental Physics and Astronomy, Texas A\&M University, College Station, TX, 77843-4242, USA}
\email{boldaclaire@tamu.edu}

\author[0000-0002-6741-078X]{Westley Brown}
\affiliation{Department of Physics and Astronomy, York University, 4700 Keele Street, Toronto, Ontario, Canada, MJ3 1P3}
\email{westleyb@yorku.ca}

\author[0009-0008-8991-5536]{Guillermo Romero Cruz}
\affiliation{Instituto Nacional de Astrof\'isica, \'Optica y Electr\'onica, Luis Enrique Erro 1, Tonantzintla, C.P. 72840, Puebla, M\'exico}
\email{guillermo.romero@inaoep.mx}

\author[0000-0003-3595-7147]{Mohamed H. Abdullah}
\affiliation{Department of Physics, University of California Merced, 5200 North Lake Road, Merced, CA 95343, USA}
\affiliation{Department of Astronomy, National Research Institute of Astronomy and Geophysics, Cairo, 11421, Egypt}
\email{melha004@ucr.edu}

\author[0009-0005-6999-2073]{Jeremy Favaro}
\affiliation{Institute for Computational Astrophysics and Department of Astronomy and Physics, Saint Mary's University, 923 Robie Street, Halifax, NS B3H 3C3, Canada}
\email{jeremy.favaro@smu.ca}

\author[]{Mason Footh}
\affiliation{Department of Physics and Astronomy, The University of Alabama}
\email{Email}

\author[]{Helen Fraser}
\affiliation{The Open University (UK),  Walton Hall, Kents Hill, Milton Keynes MK7 6AA, United Kingdom}
\email{helen.fraser@open.ac.uk}

\author[0000-0002-0759-0504]{Yuming Fu}
\affiliation{Leiden Observatory, Leiden University, Einsteinweg 55, 2333 CC Leiden, The Netherlands}
\email{yfu@strw.leidenuniv.nl}

\author[0009-0000-2546-1645]{Annie Giman}
\affiliation{William H. Miller III Department of Physics and Astronomy, Johns Hopkins University, Baltimore, MD 21218, USA}
\email{egiman1@jhu.edu}

\author[0009-0007-8224-586X]{Nicolas Gomez}
\affiliation{Instituto Nacional de Astrof\'isica, \'Optica y Electr\'onica, Luis Enrique Erro 1, Tonantzintla, C.P. 72840, Puebla, M\'exico}
\email{ngomez@inaoep.mx}

\author[0000-0002-5060-1379]{Massimo Griggio}
\stsci
\email{mgriggio@stsci.edu}

\author[0009-0009-3547-9326]{Peter Gwartney}
\affiliation{Department of Physics and Astronomy, The University of Alabama, Tuscaloosa, AL 35487, USA}
\email{ptgwartney@crimson.ua.edu}

\author[0009-0006-3071-7143]{Aryana Haghjoo}
\affiliation{Department of Physics and Astronomy, University of California Riverside, Riverside, CA 92521, USA}
\email{aryana.haghjoo@email.ucr.edu}

\author[0009-0008-1839-2969]{Fabian Hervas-Peters}
\affiliation{Steward Observatory, University of Arizona, 933 North Cherry Avenue, Tucson, AZ 85721, USA}
\email{Email}

\author[0000-0002-5721-0709]{Jiamu Huang}
\affiliation{Department of Physics, University of California, Santa Barbara, CA 93106, USA}
\email{jiamu\_huang@ucsb.edu}

\author[0009-0006-7814-2334]{Sandra Jaison}
\affiliation{Departamento de Astronomia, Instituto de Física, Universidade Federal do Rio Grande do Sul, Porto Alegre, RS, Brazil}
\email{sandra.jaison@ufrgs.br}

\author[0000-0001-9187-3605]{Jeyhan S. Kartaltepe}
\affiliation{Laboratory for Multiwavelength Astrophysics, School of Physics and Astronomy, Rochester Institute of Technology, 84 Lomb Memorial Drive, Rochester, NY 14623, USA}
\email{Email}

\author[]{Sarah Kendrew}
\affiliation{European Space Agency, Space Telescope Science Institute, 3700 San Martin Drive, Baltimore MD 21218, USA}
\email{skendrew@stsci.edu}

\author[0000-0001-5346-6048]{Ivan Kramarenko}
\affiliation{Institute of Science and Technology Austria (ISTA), Am Campus 1, 3400 Klosterneuburg, Austria}
\email{ivan.kramarenko@ista.ac.at}

\author[0000-0001-8367-7591]{Sahana Kumar}
\affil{Department of Astronomy, University of Virginia, 530 McCormick Rd, Charlottesville, VA 22904, USA}
\email{sahanak@virginia.edu}

\author[0009-0002-8965-1303]{Zhaoran Liu}
\affiliation{MIT Kavli Institute for Astrophysics and Space Research, 70 Vassar Street, Cambridge, MA 02139, USA}
\email{zrliu@mit.edu}

\author[0000-0002-7530-8857]{Arianna S. Long}
\affiliation{Department of Astronomy, University of Washington, Physics-Astronomy Building, Box 351580, Seattle, WA 98195-1700, USA}
\email{aslong@uw.edu}

\author[]{Sangeeta Malhotra}
\affiliation{NASA/GSFC, Greenbelt, MD 20771}
\email{Email}

\author[0009-0008-4976-3216]{Faezeh Manesh}
\affiliation{Department of Physics and Astronomy, University of California Riverside, Riverside, CA 92521, USA}
\email{fakhl001@ucr.edu}

\author[0000-0002-7547-3385]{Jasleen Matharu}
\affiliation{Max-Planck-Institut f\"ur Astronomie, K\"onigstuhl 17, D-69117 Heidelberg, Germany}
\email{jamatharu@mpia.de}

\author[0000-0001-7166-6035]{Vihang Mehta}
\affiliation{IPAC, California Institute of Technology, 1200 E.~California Blvd, Pasadena, CA 91125, USA}
\email{vmehta@ipac.caltech.edu}

\author[0000-0003-3799-9033]{Edward M. Molter}
\affiliation{Space Telescope Science Institute, 3700 San Martin Drive, Baltimore, MD 21218, USA}
\email{emolter@stsci.edu}

\author[]{Ivelina Momcheva}
\affiliation{Max-Planck-Institut f\"ur Astronomie, K\"onigstuhl 17, D-69117 Heidelberg, Germany}
\email{Email}

\author[0000-0003-0230-6436]{Zhiwei Pan}
\affiliation{Department of Astronomy, University of Illinois Urbana-Champaign, Urbana, IL 61801, USA}
\email{zhiweip@illinois.edu}

\author[0009-0005-3823-9302]{Yuxuan Pang}
\affiliation{School of Astronomy and Space Science, University of Chinese Academy of Sciences (UCAS), Beijing 100049, China}
\email{jackdaw.pyx@gmail.com}

\author[0000-0002-9471-8499]{Pallavi Patil}
\affiliation{National Radio Astronomy Observatory, 520 Edgemont Road, Charlottesville, VA 22903, USA }
\email{ppatil@nrao.edu}

\author[0000-0002-2509-3878]{Rachel Plesha}
\stsci
\email{Email}

\author[0000-0002-9946-4731]{Marc Rafelski}
\stsci
\affiliation{Department of Physics and Astronomy, Johns Hopkins University, 3400 North Charles Street, Baltimore, MD 21218, USA}
\email{mrafelski@stsci.edu}

\author[0000-0002-4917-7873]{Mitchell Revalski}
\stsci
\email{mrevalski@stsci.edu}

\author[0009-0005-9470-0765]{Jyotika Roychowdhury}
\affiliation{Duke University, Department of Physics, Durham, NC 27708}
\email{jyotika.roychowdhury@duke.edu}

\author[0000-0001-5851-1856]{Gregory Rudnick}
\affiliation{University of Kansas, Department of Physics \& Astronomy, Lawrence, KS 66045}
\email{grudnick@ku.edu}

\author[]{Swetha Sankar}
\affiliation{William H. Miller III Department of Physics and Astronomy, Johns Hopkins University, Baltimore, MD 21218, USA}
\email{Email}

\author[]{Maya Seagraves}
\affiliation{California Polytechnic State University, California, United States}
\email{Email}

\author[0000-0001-9495-7759]{Lu Shen}
\affiliation{Center for Astronomy and Astrophysics and Department of Physics, Fudan University, Shanghai 200438, China}
\email{lushen@fudan.edu.cn}

\author[0009-0003-0248-2082]{Sabnam Shrestha}
\affiliation{Department of Physics and Astronomy, The University of Alabama, Tuscaloosa, AL 35487, USA }
\email{sshrestha22@crimson.ua.edu}

\author[0000-0002-0636-5698]{Manuchehr Taghizadeh-Popp}
\affiliation{William H. Miller III Department of Physics and Astronomy, Johns Hopkins University, Baltimore, MD 21218, USA}
\affiliation{Data Science and AI Institute, Johns Hopkins University, 6225 Smith Avenue, Baltimore, MD 21209, USA}
\email{mtaghiza@jhu.edu}

\author[0000-0003-4068-5545]{Jo Taylor}
\stsci
\email{Email}

\author[orcid=0000-0003-0747-1780]{Wei Leong Tee}
\affiliation{Department of Astronomy and Astrophysics, The Pennsylvania State University, 525 Davey Lab, University Park, PA 16802, USA}
\affiliation{Institute for Gravitation and the Cosmos, The Pennsylvania State University, University Park, PA 16802, USA}
\email{wmt5159@psu.edu}

\author[0000-0001-9052-9837]{Scott Tompkins}
\affiliation{International Centre for Radio Astronomy Research}
\email{satompki@asu.edu}

\author[0000-0002-7633-431X]{Feige Wang}
\affiliation{Department of Astronomy, University of Michigan, 1085 South University Avenue, Ann Arbor, MI 48109, USA}
\email{Email}

\author[0000-0002-9373-3865]{Xin Wang}
\affiliation{School of Astronomy and Space Science, University of Chinese Academy of Sciences (UCAS), Beijing 100049, China}
\affiliation{National Astronomical Observatories, Chinese Academy of Sciences, Beijing 100101, China}
\affiliation{Institute for Frontiers in Astronomy and Astrophysics, Beijing Normal University, Beijing 102206, China}
\email{xwang@ucas.ac.cn}

\author[]{Isak Wold}
\affiliation{Astrophysics Science Division, Goddard Space Flight Center, Center for Research and Exploration in Space Science and Technology, Maryland, USA}
\email{Email}

\author[0000-0001-5392-2701]{Jerry J.-Y. Zhang}
\affiliation{Department of Physics and Astronomy, The University of Western Ontario, 1151 Richmond St, London, ON N6A 3K7, Canada}
\affiliation{Institute for Earth and Space Exploration, The University of Western Ontario, 1151 Richmond St, London, ON N6A 3K7, Canada}
\email{junyan.zhang@uwo.ca}

\author[]{Qianqiao Zhou}
\affiliation{School of Astronomy and Space Science, University of Chinese Academy of Sciences (UCAS), Beijing 100049, China}
\email{zhouqianqiao24@mails.ucas.ac.cn}

\author[]{Zihao Zuo}
\affiliation{University of Michigan, Department of Astronomy, 085 S University, Ann Arbor, MI 48109, USA }
\email{Email}



\begin{abstract}
We summarize strategies, lessons learned, and future directions from the Space Telescope Science Institute workshop `\textit{Surveying the Universe in 4D: Beating Cosmic Variance with Wide-Field Slitless Spectroscopy from HST, JWST, Euclid, Roman, and Beyond}', held August 24--28, 2026. The workshop examined scientific results, observational and data analysis challenges, extraction tools, and future opportunities. Discussions highlighted (1) the transformative potential of WFSS for the study of transient phenomena, galaxy evolution --both spatially-resolved and within the broader context of the cosmic web--, and rare populations and (2) the synergies among Euclid and Roman surveys, Rubin-LSST monitoring, JWST WFSS, and high-resolution integral-field observations. Participants identified advances in forward modeling and physics-informed machine learning as essential for addressing spectral overlap, crowded fields, and upcoming, very large data volumes. Realizing WFSS's full potential will require community-wide infrastructure, science-ready data products, accessible cloud-based analysis tools, and robust benchmarking of reduction pipelines. Crucially, participants called for systemic changes to properly recognize early-career researchers who invest significant efforts in pipeline, code, and calibration developments that enable WFSS science, and stressed that progress requires collaborative, multidisciplinary practices that optimize the participation and benefits of the next generation.
\end{abstract}


\section{Introduction}
Massively multiplexed spectroscopy is revolutionizing our understanding of the Universe by enabling efficient, statistically robust surveys across large cosmic volumes. By capturing spectra for thousands of sources simultaneously, it enables us to map large-scale structures, trace galaxy growth in statistically significant samples, and quantify the physical processes driving galaxy evolution as a function of environment. Wide survey areas are also essential to mitigate cosmic variance and place rare or extreme populations within the broader context of the evolving web of matter and dark energy.

Wide-Field Slitless Spectroscopy (WFSS) is critical to this effort by mitigating the selection biases inherent in targeted spectroscopic surveys. Deep, unbiased emission-line samples from HST/ACS and WFC3 grism programs have demonstrated the power of this approach in both blank and cluster fields. JWST’s NIRISS, NIRCam, and MIRI now extend WFSS to fainter fluxes and higher redshifts, enabling integrated and spatially-resolved rest-frame optical spectroscopy into the epoch of reionization. Euclid and Roman WFSS further scale these capabilities to thousands of square degrees, delivering homogeneously selected samples for statistical studies of galaxy evolution, structure formation, and cosmology without target preselection. 

This 5-day workshop was held at the Space Telescope Science Institute August 24-28, 2026, to discuss key science results from WFSS programs, review extraction and analysis tools, address current WFSS challenges and pitfalls, and explore future considerations for the field. 
The program included science and technical talks, hands-on tutorials, and discussion sessions which primarily aimed to answer the following key questions:

\begin{itemize}
\item Current, upcoming, and future WFSS capabilities are poised to revolutionize our exploration of the universe. Where will the major paradigm shifts occur? What can we achieve now, what will soon become possible, and how should we approach the next generation of capabilities? Crucially, how does WFSS integrate into the broader landscape of astronomical tools? What collaborations with space- and ground-based observations are essential?

\item With current and upcoming facilities, WFSS will observe large statistical samples of galaxies with high spatial resolution; however, taking advantage of this capability is challenging. What are the high priority science cases for spatially-resolved WFSS and how do we address the challenge in reconstructing spatial information from slitless observations?

\item In the era of Euclid and Roman, how can statistical galaxy studies and large-scale structure science advance? What are the key priorities, expected challenges, and connections to cosmology? How do these surveys complement detailed, targeted observations with HST and JWST?

\item Time-domain astronomy is entering a new era with Rubin. What discoveries will large-scale time-domain surveys enable, from transients to AGN variability? What science is within reach now, what lies on the horizon, and how can we best connect Rubin’s capabilities with space-based WFSS observations to maximize scientific return?
\end{itemize}

This paper presents the main results and discussions of the workshop. Section~\ref{sec:execsummary} is the executive summary of the workshop with Sec.~\ref{sec:wfsssurveys} through \ref{sec:facilitiesandifu} focusing on the key science and technical themes of the program: WFSS surveys (Sec.~\ref{sec:wfsssurveys}), cosmic noon, AGN/LRDs, and quenching (Sec.~\ref{sec:cosmicnoonagnquench}), spatially-resolved WFSS (Sec.~\ref{sec:spatially_resolved}), time-domain astronomy (Sec.~\ref{sec:timedomain}), large scale structures (Sec.~\ref{sec:lss}), machine learning (Sec.~\ref{sec:machinelearning}), contamination subtraction (Sec.~\ref{sec:contam}), and current and new synergies (Sec.~\ref{sec:facilitiesandifu}). Section~\ref{sec:discussions} presents the main take-aways of the discussion sessions. Individual summaries of the various contributions (talks, tutorials/demos, discussion sessions, posters) are also included (Sec.~\ref{sec:Day1AM} through \ref{sec:GNoirotD}, presented in order of the workshop program\footnote{\url{https://www.stsci.edu/files/live/sites/www/files/home/events/event-assets/2026/_documents/wfss-conference-schedule.pdf}}).



\section{Executive Summary}\label{sec:execsummary}
\subsection{WFSS Surveys and Galaxy Evolution Across Cosmic Time}\label{sec:wfsssurveys}

Large, un-targeted spectroscopic samples obtained with the grisms and prisms of HST, JWST, Euclid, and Roman enable us to trace galaxy evolution from cosmic reionization to the present (sec \ref{sec:fwang}, \ref{sec:Bjin}) and across a wide range of spatial scales, from large-scale protoclusters and clusters to resolved star-formation in the near universe and at cosmic noon (sec \ref{sec:JasleenM}). A global census of star and black hole growth is essential to understanding how and why galaxies evolve. Complementary approaches to constructing luminosity functions (e.g., sec \ref{sec:QZhou}) while minimizing the impact of cosmic variance include contiguous mapping over large areas  (e.g., sec \ref{sec:Bjin}) and pure parallel programs (e.g., sec \ref{sec:Poppies}, \ref{sec:vmehta}, \ref{sec:aakhtarkavan}, \ref{sec:IMomcheva}). Significant progress has been already made in demonstrating that the morphology-density relation evolves over cosmic times Sec \ref{sec:WBrown} and that that star-forming galaxies grow inside out across cosmic time \ref{sec:JasleenM}. The torrent of data from {\it{Euclid}} and {\it{Roman}} will help elucidate the physical processes driving this evolution. New methods to disentangle the sources of emission from star-forming regions from those of AGN  \ref{sec:JasleenM} are particularly promising. Looking ahead, Euclid and the Roman Space Telescope will scale these methods to hundreds of thousands of sources, providing the statistical volume required to beat cosmic variance; an exciting example is a planned deep spectroscopic survey with Roman (Section \ref{sec:graceR}) which will provide the area and depth to answer essential questions about how galaxies influence their environments are are shaped by it. 


\subsection{Cosmic Noon, AGN/Little Red Dots, and Quenching}\label{sec:cosmicnoonagnquench}

\subsubsection{Cosmic Noon}
One of the major theme of the workshop was the use of WFSS to connect the growth of galaxies to the cycling of gas and metals from cosmic noon to the epoch of reionization. At cosmic noon, JWST/NIRISS WFSS can provide large samples of emission-line galaxies extending well below the stellar masses accessible with HST or from the ground. Programs such as JWST-GLASS (Sec.~\ref{sec:XWang}, \ref{sec:FManesh}), NGDEEP (Sec.~\ref{sec:XWang}, \ref{sec:cbolda}), CANUCS (Sec.~\ref{sec:QZhou}, \ref{sec:WBrown}, \ref{VEstradaC}, \ref{sec:JFavaro}), and PASSAGE (Sec.~\ref{sec:vmehta}, \ref{sec:fhasan}) are pushing measurements of the mass-metallicity relation (MZR), star-forming main sequence, and H$\alpha$ luminosity function to those new regimes, reaching up to 1 dex lower luminosities and masses than before (Sec.~\ref{sec:QZhou}, \ref{sec:cbolda}, \ref{sec:vmehta}, \ref{sec:fhasan}, \ref{sec:JFavaro}, \ref{sec:FManesh}). In particular, the scatter in H$\alpha$ SFRs ($\sim10$ Myr timescale) exceeds that of UV and SED-based SFRs ($\sim100$ Myr timescale), especially at low masses and higher redshifts, providing support for bursty or stochastic star-formation at early times, not yet well reproduced by current feedback models (Sec.~\ref{sec:cbolda}).
Measurements of the MZR do not find a flattening of the relation at the low-mass end, but rather a steep decrease, with systematic differences associated with morphology and SFR-level (Sec.~\ref{sec:fhasan}).
Stacked Balmer decrement maps also reveal that dust-corrected star formation in massive galaxies can be more extended than the stellar continuum, consistent with inside-out growth (Sec.~\ref{sec:XWang}). Some results also show negative metallicity gradients with mass at cosmic noon consistent with a stronger role for feedback-regulated star formation at these epochs compared to lower redshifts (Sec.~\ref{sec:XWang}), while others find no clear trend with mass as predicted by simulation and flat metallicity gradients within the error bars (Sec.~\ref{sec:vmehta}).

The role of the environment at cosmic noon was also discussed, with some results showing no strong influence of the environment in establishing the morphology-density relation in three clusters at $z\sim1.6$ (Sec.~\ref{sec:WBrown}), while others find accelerated evolution of massive members across dozens of clusters at these epochs (Sec.~\ref{sec:GNoirot}), or shallower MZR than in the field highlighting the role of the environment in gas accretion and recycling even at early times (Sec.~\ref{sec:fhasan}).

MIRI WFSS observations of PAH and gas at $1~<~z~<~3$ also provide a new method to study the evolution of the ISM at a time when most galaxies appear to reach the peak of their star-formation history (Sec.~\ref{sec:apetric}, \ref{sec:RFernandez}, \ref{sec:skendrew}).

Spatially-resolved WFSS, at cosmic noon and beyond, was also one of the major theme of the workshop and is discussed in Sec~\ref{sec:spatially_resolved}.

\subsubsection{AGN/LRDs}
WFSS surveys are key to battling cosmic variance and making progress on the puzzle posed by the correlation between the masses of SMBH and the bulges of their host galaxies, a puzzle that is almost 30 years old (sec \ref{sec:fwang}, \ref{sec:fhasan}, \ref{section:XJin}, \ref{sec:ogilbert}). Several talks aim to pin point what triggers the growth of SMBH (sec \ref{sec:pnair}, \ref{sec:Zliu}, \ref{sec:KMokhtarpour}) including the study of dual AGN, a key population to the growth pathways for SMBH (sec \ref{sec:SShreshta}, \ref{sec:mfooth}) and look for evidence for AGN driven molecular shocks with MIRI WFSS (sec \ref{sec:apetric}).

Nearby benchmark studies for AGN observations with Roman and Euclid were presented in sec \ref{sec:Sjaison}, \ref{sec:agiman}, \ref{sec:ssankar}) while an exciting method to use low-resolution ionization diagnostics to find AGN with Euclid and Roman prisms was introduced in \ref{sec:JasleenM}. Sec \ref{sec:YFu} describes how how spectroscopic observations with Euclid, combined with quasar identification pipelines that leverage Gaia and WISE data, enable the discovery of thousands of quasars.

WFSS surveys also brought astronomers a new puzzle: the ubiquitous Little Red Dots. Presentations showed significant progress in studying their properties (sec \ref{sec:fwang},\ref{sec:JHuang},\ref{sec:NAlvarez}) and the possible connections between AGN and LRDs (sec \ref{sec:YPang}, \ref{sec:zpan}).

\subsubsection{Quenching}
One major open question in galaxy evolution is how galaxies stop forming stars. A handful of presentations discussed this topic by addressing the timescales, pathways, and physical mechanisms of galaxy quenching. At cosmic noon, deep HST grism spectroscopy reveals that only galaxies with rapidly declining SFHs can fully quench and reach the red-sequence with green-valley crossing timescales of $\sim1$Gyr, and that tentative results of smaller H$\alpha$ sizes relative to the stellar continuum as quenching progresses may indicate outside-in quenching mechanisms at play for this population (Sec.~\ref{sec:gnoirot}). On the other hand, JWST/NIRISS WFSS results show the power of WFSS by directly probing where quenching happens: UV-bright clumps in star-forming cosmic-noon galaxies with no associated H$\alpha$ emission suggest that these galaxies undergo bursts of star-formation followed by rapid quenching within the clumps (Sec.~\ref{VEstradaC}).
Other surveys with JWST and Euclid show differences between mass and environmental quenching (Sec.~\ref{sec:JasleenM}) while others show how using confirmed clusters with HST grism spectroscopy at cosmic noon can help identify the environmental mechanisms responsible for both galaxy growth and quenching (Sec.~\ref{sec:GNoirot}).

\subsection{Spatially-Resolved WFSS}\label{sec:spatially_resolved}
A key theme of the workshop was the necessity to exploit the spatially-resolved capabilities offered by WFSS, including the science it enables and the technical challenges it represents. Several presentations showed approaches to extracting and analyzing slitless data while conserving (or enhancing) the spectral and spatial information it provides. Some methods combine multiple PAs/orientations to recover IFU-like spectral cubes from the WFSS data without spectral template assumptions (e.g., Sec.~\ref{sec:rryanD}, \ref{sec:rryanT}, \ref{sec:FHervas}, \ref{sec:tastraatmadja}, \ref{sec:MGriggio}, \ref{sec:JZhang}), while other approaches focus on spatially-resolved SED modeling to build clean emission-line maps or contamination-subtracted spectra (Sec.~\ref{VEstradaC}, \ref{sec:VinceDemo}).
These approaches are particularly important for extended sources with spatially-varying physical conditions. For instance, some presentations showed that spatially-resolved emission line maps have systematically lower [O~III] emission-line fluxes and higher metallicities compared to single-region models (Sec.\ref{sec:JFavaro}). Other works have showed the power of WFSS to map molecular ices in star-forming clouds with orders-of-magnitude more sightlines than previous observations, but also highlighted the challenges of deriving spatially-resolved maps of the ice species within the high-obscuration molecular clouds (Sec.~\ref{sec:HFraser}, \ref{sec:ETsiakaliari}), for which spatially-resolved forward-modeling techniques are not adapted.
Results combining NIRISS and NIRCam WFSS to probe spatially-resolved H$\alpha$ and Pa-$\alpha$ emission showed an inside-out growth of a sample of cosmic noon galaxies despite dust attenuation, as well as radial variations in their ionization conditions (Sec.~\ref{sec:JasleenM}).
Other NIRISS WFSS works are probing the role of clumps in star-forming galaxies at cosmic noon and find lower metallicities in the clumps compared to their surrounding disks, suggesting an inflow of metal poor gas triggering star-formation (Sec.~\ref{VEstradaC}).

Overall, multiple presentations showed that WFSS is a powerful observing mode to study spatially-resolved star formation, ionization, and chemical enrichment across large samples of galaxies at cosmic noon and beyond. Spatially-resolved studies pioneered by HST grism surveys are now extended to much fainter galaxies and higher redshifts with JWST, and sample sizes will dramatically increase with Euclid and Roman, enabling studies of galaxy physics together with their large scale environment. Discussions during the workshop highlighted that the key challenge in spatially-resolved WFSS is the careful treatment of spectral overlap, PSF variations, background subtraction, and spatially varying continuum and physical conditions, including host-galaxy subtraction critical for robust supernova science (see Sec.~\ref{sec:timedomain}). Tools under development, combined with machine learning algorithms (see Sec.~\ref{sec:machinelearning}) offer promising avenues to robustly exploit the upcoming wealth of spatially-resolved WFSS data. During discussions, the homogenization of benchmarks to validate tools and results across different missions, instruments, and datasets appeared as highly critical to enable the science.

\subsection{Time Domain}\label{sec:timedomain}

Presentations on WFSS's role in time-domain astronomy centered on Roman's High-Latitude Time-Domain Survey. Simulations using the OpenUniverse2024 suite were developed to validate a spectral-extraction pipeline for Type~Ia supernovae, recovering spectra to 25th magnitude at $z=1.1$ in isolated cases, while contaminated (host+neighbor) simulations showed that accurate host-galaxy subtraction will be essential for real survey performance (Section \ref{sec:JRoychowdhury}); a dedicated pipeline for decontaminating SN spectra from their host light is under development for public release (Section \ref{sec:rryanT}). More broadly, WFSS was framed as a way to scale up spectroscopic follow-up of supernovae and explosive transients beyond what is feasible amid the nightly alert volumes from Rubin-LSST. Sec \ref{sec:SKumar} presents an innovative and thorough approach to the study of Type~Ia progenitor specifically on using nickel and nebular iron lines to constrain mass and explosion mechanisms. Implications from observations of newly identified Calcium Strong Transient class were also discussed (Section \ref{sec:SKumar}). 

Underpinning these efforts, a forward-modeling framework that projects 3D data cubes into simulated Roman spectroscopic images (and can be inverted to recover cubes from multi-roll-angle data) provides realistic simulated inputs for pipeline development and testing (Section \ref{sec:tastraatmadja}). A broader look at scaling slitless pipelines from HST/JWST to Euclid/Roman data volumes emphasized the need for faster numerical methods and survey-tailored strategies (Section \ref{sec:IMomcheva}). Preparing for Roman's data volumes specifically, a pipeline extracting more than 50,000 HST stellar spectra is being developed as a testbed for the vastly larger stellar spectroscopic datasets Roman's WFI will produce (Section \ref{sec:MSeagraves}).

Discussion of these efforts also highlighted synergies with Rubin for transient and variable-AGN monitoring and the value of joint imaging-plus-spectroscopy for rare-object statistics as the community heads into the large-survey era.

\subsection{Large Scale Structure}\label{sec:lss}

Several talks addressed large-scale-structure science enabled by WFSS surveys. Euclid's Q1/DR1 quasar-identification strategy combines Gaia/WISE pre-selection, ML photometric selection, and hybrid redshift estimation to $H_E\sim21.3$. Roman will break degeneracies left by Euclid surveys (Section \ref{sec:YFu}, \ref{sec:NGomez}). Cluster work included the GalWCat spectroscopic cluster catalog, supporting galaxy-evolution and cosmological studies with planned extensions to DESI, Euclid, and Roman (Section \ref{sec:MAbdullah}); and CARLA's HST grism-confirmed showed that cosmic-noon cluster galaxies follow different growth histories than field galaxies (Section \ref{sec:GNoirot}). An intriguing application was using Roman and Euclid's wide-area imaging plus tens of millions of WFSS redshifts to enable precise measurement of the cosmic spectral energy distribution (Section \ref{sec:STompkins}).  

\subsection{Machine Learning}\label{sec:machinelearning}

Machine learning can address bottlenecks across WFSS, improving accuracy and throughput. Because every source is dispersed, overlapping traces are a central challenge, particularly in crowded fields. Neural deblending can use scene context and multiple observations to estimate individual contributions rapidly, complementing forward-modeling extraction and improving completeness and spectral recovery (sec \ref{sec:ZZuo})

Machine learning can also extract richer information from spectra. Physics-informed super-resolution using paired JWST/NIRSpec prism and grating spectra has demonstrated reconstruction from \(R\sim100\) to \(R\sim1000\), separating blended features and suggesting applications to Euclid and Roman grism spectroscopy (sec \ref{sec:AHaghjoo}). Transformer foundation models such as SpecPT complement this approach: pretraining on DESI can support spectral reconstruction and redshift estimation before adaptation to the wavelength coverage, resolution, and source populations of HST, JWST, Euclid, and Roman (sec \ref{sec:cbinu}). Combining machine learning methods with canonical template fitting can significantly improve the redshift accuracy and precision of slitless spectra (sec \ref{sec:YFu}).

End-to-end detector simulations, dispersed-image-to-spectral-time-series pipelines, three-dimensional flux-cube reconstruction, and pipeline-systematics studies provide forward models and validation data for reliable learned methods. They create opportunities to mitigate detector artifacts and accelerate demanding inverse problems within physically motivated pipelines (sections: \ref{sec:rryanD} \ref{sec:rryanT}, \ref{sec:tastraatmadja}, \ref{sec:MGriggio}, \ref{sec:NGomez}) 
Machine-learning-based field-level inference can extract information from the galaxy distribution, reconstructing latent density fields while constraining cosmological and astrophysical parameters and survey systematics. DESI-like BAO applications provide an initial demonstration, while Roman and Euclid offer opportunities for joint inference across spectroscopy, imaging, weak lensing, and external tracers (sec \ref{sec:ABayer}).

Realizing this potential requires representative data, realistic scene-level simulations, calibrated uncertainties, and cross-instrument validation. Deblending and super-resolution should be treated as conditional inference; downstream analyses should propagate probabilistic uncertainties rather than only point estimates. Shared benchmarks and data challenges using ESpRESSO, with blinded tests of spectral overlap, detector effects, source morphology, and domain shifts, would enable reproducible comparisons, expose failure modes, and build trust in ML-derived products (sec \ref{sec:msinghaD}). Physics-informed models, transfer learning, interpretability, and robustness tests will be essential for preventing simulation-specific features or survey systematics from contaminating scientific conclusions.

\subsection{Contamination Subtraction}\label{sec:contam}

Contamination subtraction is a central challenge in WFSS analysis. There were also tutorials offered on calibration tools for JWST-instrument WFSS modes, and forefront analysis tools (e.g., slitlessutils, Sleuth, Allegro, and the CASTOR simulator). 


Forward modeling is a widely used approach to contamination subtraction that combines the morphology of a galaxy with an assumed spectral shape to construct a model of its dispersed emission. The JWST/NIRISS pipeline (sec \ref{sec:JTaylor}  has recently adopted this approach, taking a similar strategy to Grizli by using polynomial templates and an iterative procedure to model field contamination. In this approach, models are generated for all objects in the field above a specified magnitude limit. Sleuth uses a similar forward-modeling approach, but is designed for targeted analyses and therefore does not generate contamination models for every object in the field. Instead, it models only the sources whose spectra contaminate the object of interest. Sleuth (sec \ref{VEstradaC}, \ref{sec:VinceDemo}) can also leverage its multi-regional approach to construct more accurate models of contaminating sources.



\subsection{New Facilities and IFU-slitless Synergies}\label{sec:facilitiesandifu}

JWST/NIRSpec can be converted into an effective slitless spectrograph by opening nearly all MSA microshutters (closing only those over irrelevant bright sources), achieving the previously unavailable combination of $1$--$2\,\mu$m coverage at $R\gtrsim1000$ (Section \ref{sec:MHuberty}). 

The MUSE Ultra Deep Field combines 142 hours of VLT/MUSE IFU data with the deepest single HST grism field ever obtained, using "Net Significance" metrics to benchmark WFSS depth and information content across surveys and reaching 5$\sigma$ continuum detections to $m_{\rm AB}\simeq27$ (Section \ref{sec:Mrevalski}). 

A conceptually new rotational slitless spectrograph design, ROSSINI, reconstructs full IFS data cubes via tomographic inversion from multiple telescope/dispersion rotation angles without an IFU, demonstrating percent-level accuracy in numerical experiments at low computational cost (Section \ref{sec:JZhang}); and a Grism ETC/Simulator for the proposed CASTOR UV/optical mission supports full-scene simulation, forward-modeled redshift fitting, and spatially resolved SED fitting ahead of any real data (Section \ref{sec:GNoirotD}). 

\subsection{Discussions}\label{sec:discussions}

 The Discussions sessions centered on the opportunities and challenges to fully leverage WFSS survey in the era of Euclid and Roman. Participants envisioned paradigm shifts in key areas from transient science (SNe Type 1A and changing look AGN) to tracing galaxy growth from high-redshift protoclusters down to present day structures. Participants were also energized by statistically significant samples across large areas un-affected by cosmic variance to study rare subpopulations like Little Red Dots (LRDs), dual AGNs, and dark stars while mapping the broader cosmic web. Achieving these goals requires deep multi-facility synergy, combining wide-area space surveys with Rubin-LSST optical monitoring, mid-infrared SED coverage from JWST/MIRI, and high-resolution IFU dynamics. 
 
 Participants emphasized that massive survey volumes need to be met by a community infrastructure, the wide availability of tools, and science ready data. The transition from JWST to Roman requires shifts from legacy tools toward modern standards, cloud-based analysis. Tackling overlapping spectral traces, including mapping ice absorption features, in crowded fields demands sophisticated forward-modeling, multi-orientation 3D flux cube reconstructions, and physics-informed machine learning deblending. 

 Discussion about community readiness and benchmarking also addressed the aspirations and concerns of early careers astronomers who examined how the community can fairily evaluate custom reduction pipelines without regularly duplicating efforts.

Crucially, early-career participants called for systemic changes to properly recognize researchers who invest substantial time in pipeline development, cross-code benchmarking, and creating well-documented public legacy data products. More broadly, many saw an opportunity to build an astronomy culture that values careful, lasting work; encourages researchers to share ideas freely and collaborate across boundaries; and creates environments where curiosity, creativity, and a sense of collective purpose can flourish alongside scientific ambition. Such a culture would strengthen not only the scientific foundations of WFSS, but also the communities and shared resources that sustain discovery over the long term, helping the next generation of WFSS astronomers build on what came before and ensure that the benefits of our work extend well beyond any single project, institution, or generation.

\section{Day 1: WFSS Surveys and General Galaxy Evolution} \label{sec:Day1AM}
\subsection{Feige Wang: Review of Wide-Field Slitless Spectroscopy from JWST to Roman, Mapping AGN, Galaxies, and Large-Scale Structures Across Cosmic Time\label{sec:fwang}}

\begin{figure}
    \centering
    \includegraphics[width=0.9\linewidth]{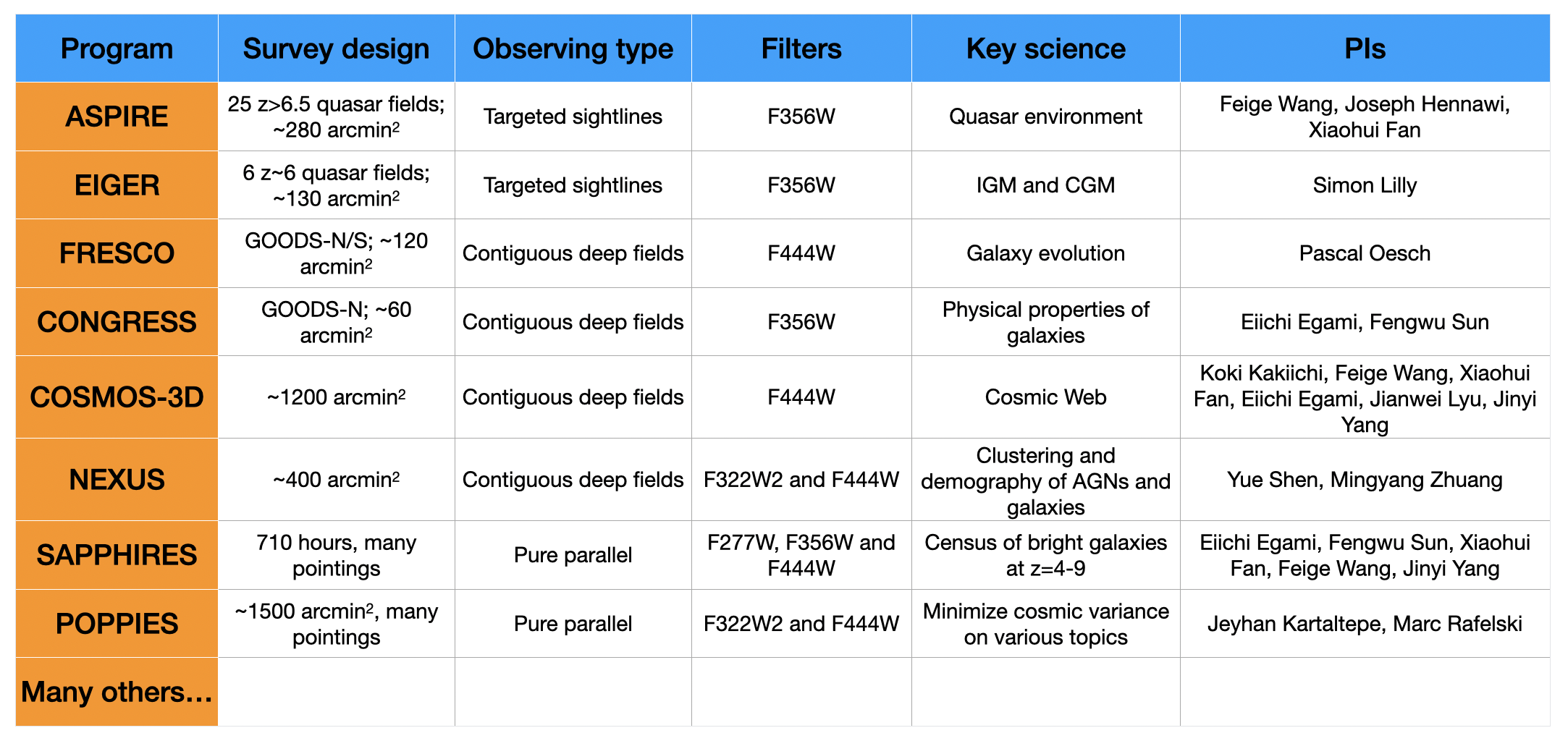}
    \caption{Figure from F. Wang showing the extragalactic NIRCam/WFSS program summaries.}
    \label{fig:nircam_wfss}
\end{figure}

{\textbf{Why NIRCam/WFSS?}}
Wide-field slitless spectroscopy provides a uniquely efficient way to construct spectroscopic samples in the early Universe because every source in the field is dispersed simultaneously, without requiring target pre-selection. This talk particularly focus on NIRCam/WFSS, which has opens a new window in early Universe studies given its unprecedented sensitivity in 3-5 $\mu$m. NIRCam/WFSS is particularly powerful for discovering rare populations, extreme emission-line galaxies, AGNs, and overdensities in the early Universe, including populations that were not known in advance. Its selection function is also fundamentally different from that of conventional targeted spectroscopy, being much closer to a line-flux-limited selection. JWST has explored three complementary survey strategies (\ref{fig:nircam_wfss}): targeted independent sightlines such as ASPIRE \citep{Yang2023, Wang2023} and EIGER \citep{Kashino2023, Mathee2023}, contiguous mappings such as FRESCO \citep{Oesch2023}, CONGRESS \citep{Egami2023}, NEXUS \citep{Shen2024,Zhuang2026}, and COSMOS-3D (Kakiichi in prep, \cite{Meyer2026}), and pure-parallel surveys such as SAPPHIRES \citep{Sun2025} and POPPIES \citep{Kartaltepe2024}. Together, these strategies explore the trade-off between depth, area, cosmic variance, and sensitivity to rare structures.

{\textbf{Connecting SMBHs, host galaxies and dark matter halos of luminous quasars.}}
A long-standing question is what dark-matter halos host the earliest luminous
quasars. ASPIRE was designed to address this statistically using a flux-limited
sample of 25 quasars at $z>6.5$, complemented by the deeper EIGER quasar
sightlines. With hundreds of spectroscopically confirmed galaxies, we can now
measure the quasar--galaxy cross-correlation rather than simply count companions.
The inferred correlation length is approximately $r_0^{QG}\simeq8.7\,h^{-1}\,\mathrm{cMpc}$, corresponding to a characteristic halo
mass of $\sim10^{12.3}\,M_\odot$. Remarkably, luminous quasars appear to occupy
roughly $10^{12}\,M_\odot$ halos over a large fraction of cosmic history \citep{Eilers2024, Wang2026, Huang2026}. This shifts the key question from whether early quasars reside in massive halos to why this particular halo mass appears to be especially favorable for rapid luminous black-hole growth. At $z\sim6$--7, such halos already host billion-solar-mass black holes, placing strong constraints on duty cycles and SMBH growth histories.

The eternal (20 years plus) question of how if and why SMBHs coevolve with their host galaxies. 
JWST also enables direct measurements of the stellar component of quasar host
galaxies \citep{Ding2023Natur,Yue2024}. The inferred stellar masses are broadly consistent with expectations from
stellar-to-halo mass relations and independent ALMA dynamical constraints, while
the luminous quasars are on average $\sim0.4$ dex above the local
$M_{\rm BH}$--$M_\star$ relation (J. Yang in prep.). Their morphologies are highly diverse, ranging
from compact systems to regular S\'ersic profiles, mergers, and extended systems,
suggesting that no single triggering mechanism dominates (J. Yang in prep.).
The puzzle then is the relation between the mass of the black hole and stellar masses are established in such a young universe and how Roman can answer this question. 

{\textbf{NIRCam/WFSS uncovered faint AGNs and little red dots.}}
NIRCam/WFSS has revealed a new population of faint broad-line AGNs and little red
dots \citep{Matthee2024ApJ,Lin2024ApJ}. These new populations have been extensively studied with all JWST observing modes. Recent COSMOS-3D constructed a large sample of such objects and found that spectroscopically selected broad-line
AGNs span a broader range of SEDs than classical photometric LRD definitions (O. Gilbert in prep.). Clustering measurements from FRESCO+CONGRESS \citep{Lin2026} and COSMOS-3D (F. Wang in prep.) indicate that BLAGNs, including LRDs,
have similar clustering to star-forming galaxies, in clear contrast to luminous quasars. In addition, \citep{Stone2026} found that LRDs have weaker variability compared with luminosity matched broad line AGNs.
These studies raise a key question: what is the connection, if any, between the abundant faint LRD population and the ordinary AGNs or luminous quasars?
Roman will help us to understand this question in a statistical manner by turning the study of few exciting objects to statistically significant surveys that beat cosmic variance. 


{\textbf{High-redshift galaxies: from discovery to population statistics.}}
One of the clearest lessons from NIRCam/WFSS is that extreme [O\,III] emission at
$z>5$ is common rather than exceptional, making NIRCam an extremely efficient
redshift survey instrument. These large spectroscopic samples now allow us to move
from source discovery to galaxy physics and population statistics. NIRCam/WFSS together with NIRSpec observations 
suggest a metallicity deficiency at high redshift \citep{li2025}, with evidence
that overdense environments influence the mass--metallicity relation, while
SAPPHIRES is beginning to reach extremely metal-poor systems \citep{hsiao2025}. FRESCO and
CONGRESS have produced hundreds of emission-line galaxies and robust luminosity
functions \citep{meyer2024}, while COSMOS-3D extends this work to much larger contiguous areas.
With hundreds of [O\,III] emitters at $6.7<z<9$, the measurements begin to constrain
luminosity functions, clustering, galaxy--halo connections, and cosmic variance
simultaneously (\citet{Meyer2026}; B. Jin et al. in prep; J. Huang et al. in prep.).

{\textbf{Protoclusters and cosmic reionization.}}
The same WFSS redshift surveys naturally map large-scale structure. ASPIRE and EIGER \citep{Wang2023,Eilers2024} demonstrated that quasar environments are extremely diverse: some fields show very few companions, while others contain dramatic redshift spikes. Spectroscopy turns these spikes into genuine three-dimensional structures \citep{Wang2026, Champagne2025ApJ}. FRESCO has revealed an
extreme overdensity at $z\sim5$ \citep{Helton2024}, while SAPPHIRES identified a remarkable overdensity
at $z=8.47$ in a pure-parallel field \citep{Fudamoto2025}. COSMOS-3D takes the next step with contiguous mapping, revealing a structure spanning $6.9<z<7.1$ that expands the entire footprint of COSMOS-3D (J. Champagne in prep.). This illustrates a natural progression from individual galaxies to redshift spikes, protoclusters, and ultimately the cosmic web. The same galaxy maps can also be connected directly to the IGM. During reionization, ASPIRE finds excess Ly$\alpha$ transmission around [O\,III] emitters \citep{Jin2024ApJ,Kakiichi2025arXiv}, while after reionization overdense regions instead tend to show suppressed transmission because of higher gas densities \citep{Kashino2026ApJ}. The sign reversal of the galaxy--IGM relation provides a
direct probe of the topology of reionization and ionizing photon escape.

{\textbf{From JWST to Roman.}}
JWST has demonstrated the key ingredients: blind emission-line selection, clustering measurements, protocluster discovery, and galaxy--IGM cross-correlations. Roman will extend the same experiment to much larger volumes. For AGNs and LRDs, Roman will transform individual discoveries into measurements of demographics, clustering, duty cycles, and cosmic evolution. For high-redshift galaxies, samples may approach $10^5$ objects, turning rare systems into statistically useful populations and enabling precision luminosity functions and clustering at $z>8$. For reionization and large-scale structure, Roman grism surveys will provide hundreds of thousands of emission-line redshifts and a large sample of Ly$\alpha$ emitters and protoclusters. In this sense, JWST provides depth, detailed spectroscopy, and resolved physics, while Roman provides area, statistics, and cosmological-scale topology.











 \subsection{Jasleen Matharu:Revealing how the Earliest Galaxies Evolved into Today’s Diverse Population with Slitless Spectroscopy \label{sec:JasleenM}}
 
\begin{figure}
    \centering
    \includegraphics[width=0.9\linewidth]{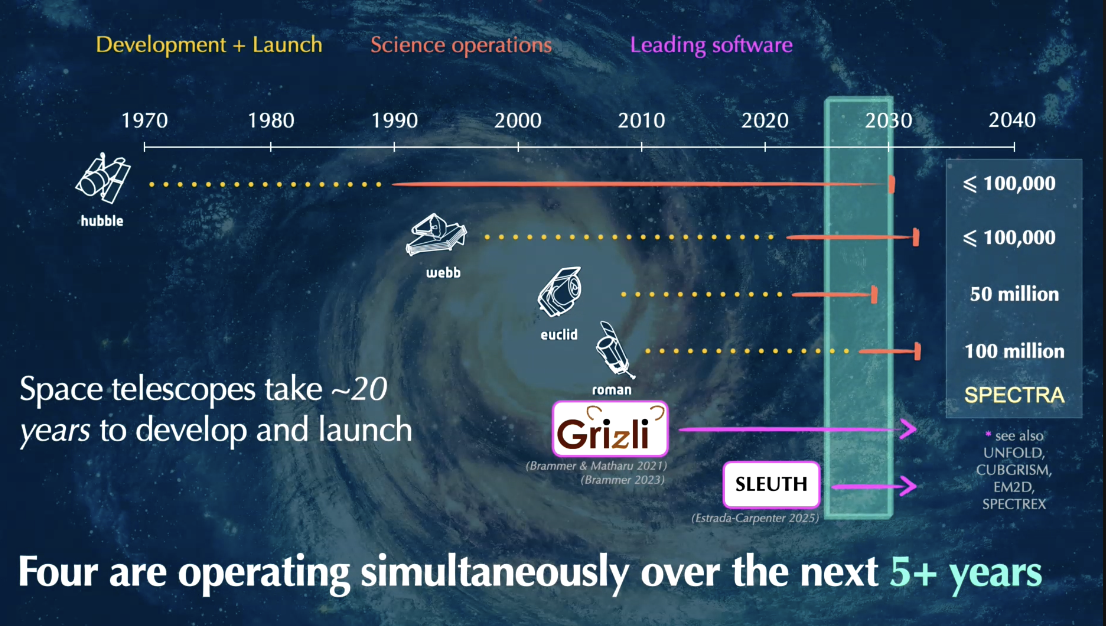}
    \caption{Figure from J. Matharu showing the evolution of WFSS instruments, survey capabilities, and tools available.}
    \label{fig:WFSStorrent}
\end{figure}

A torrent of WFSS data from multiple instruments (e.g., Figure \ref{fig:WFSStorrent} and facilities need to be leverage to answers diffcult questions. For example: Queching mass versus environment essential to work out galaxy evolution \citep[e.g.,][]{matharu2024, matharu2023}

\begin{figure}
    \centering
    \includegraphics[width=0.9\linewidth]{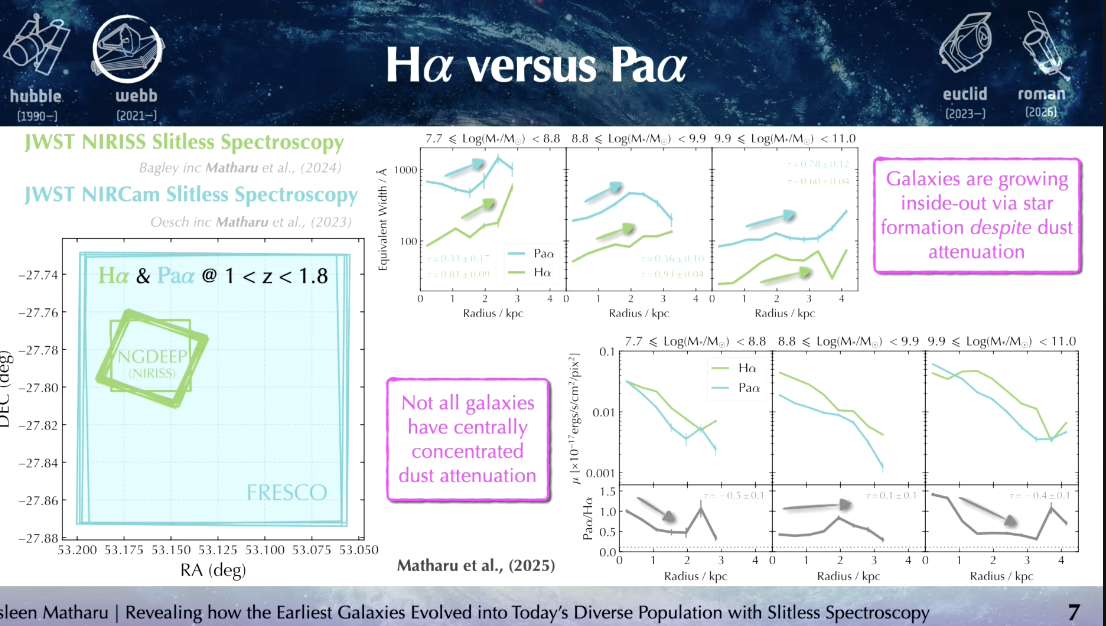}
    \caption{Examples of studies with NIRISS and NIRCam to study how star-formation rates change across galaxies all the way to the end of cosmic noon.}
    \label{fig:SFR_JM}
\end{figure}

\citet{matharu2026} uses multiple surveys and multiple techniques to work that galaxies are growing inside-out via star formation despite dust attenuation. Also not all galaxies have centrally concentrated dust attenuation (Figure \ref{fig:SFR_JM}. All this work on spatially resolved star formation from cosmic reionisation down to $z~0.5$ means we now have a pretty good understanding on spatially resolved star formation globally for star forming galaxies across cosmic time.

But there's more to WFSS than star-formation, e.g, active black holes, metallicities. There are AGN/SFR diagnostics that do not require deblending which means that we now have the capability to do spatailly resolved diagnostic diagrams with WFSS but may break down in the distant universe so more data and work are needed there. 

Roman and Euclid are poised to help, suggesting the future (of spatially resolved star-formation studies) is slitless!!

\subsection{Qiangqiao Zhou: Exploring the Ha Luminosity Functions at z 1.3 and z 2.0 in the CANUCS Lensing Fields with JWST NIRISS \label{sec:QZhou}}

\begin{figure}
      \centering
    \includegraphics[width=0.9\linewidth]{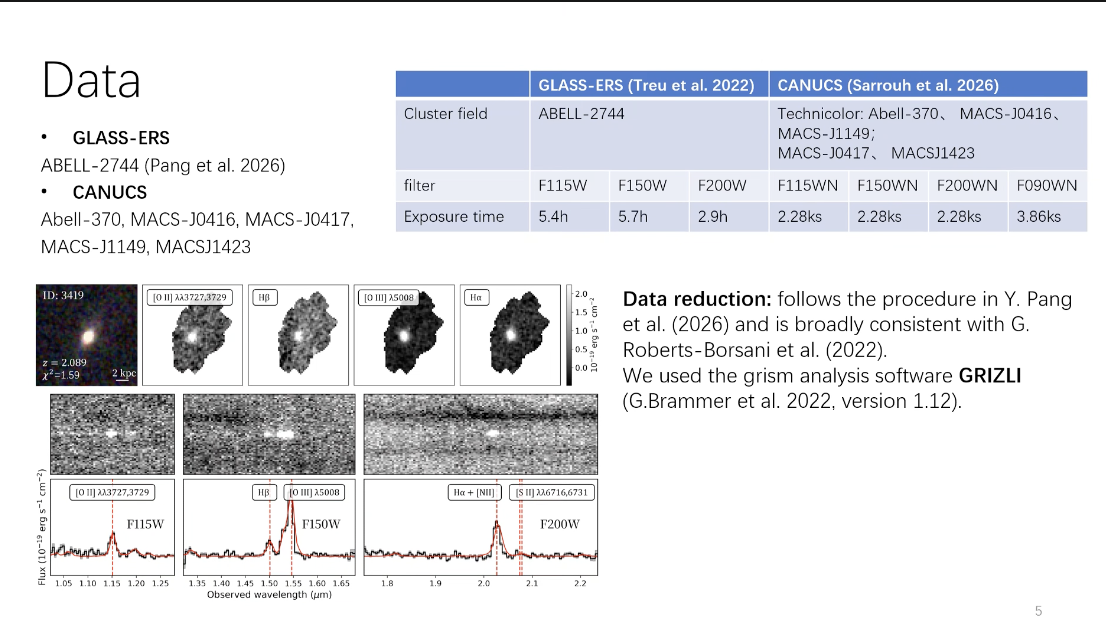}
    \caption{Example from Qiangqiao Zhou presentation on using lensing and WFSS.}
    \label{fig:canucs1}  
\end{figure}

We present H$\alpha$ luminosity function (LF) measurements at redshifts z $\sim 1.3$ and z $\sim 2.0$. We adopted JWST grism data in 5 lensing fields from CANUCS program: ABELL-370, MACS-J0416, MACS-J0417, MACS-J1149, MACS-J1423. This work extends to a much larger area the methodology first demonstrated on the GLASS-JWST NIRISS data by \citet{Pang2026}, who derived the H$\alpha$ LF at the same two redshifts from a sample of 99 emitters behind the Abell 2744 cluster. Based on emission lines spectroscopically identified in the F115W, F150W and F200W filters, we select 803 H$\alpha$ emitters from all the redshift bins and fields, an order of magnitude larger than that pilot sample. Through detailed effective volume and completeness analysis for each source, we construct the H$\alpha$ LF in two redshift bins. Thanks to the sensitivity of NIRISS WFSS and gravitational lensing magnification, our sample reaches intrinsic H$\alpha$ luminosities $\sim 10$ times deeper than previous grism surveys, down to L(H$\alpha $) 10$^{40.5}$ erg s$^{-1}$ at z$\sim 1.3$ and L(H$\alpha $) 10$^{40.9}$ erg s$^{-1}$ at z$\sim 2.0$ with completeness larger than 0.8, corresponding to star formation rates of 0.4 and 1.0 M$_\odot$, respectively. We robustly constrain the faint-end slope of the H$\alpha$ luminosity function after considering the cosmic variance, and find values consistent with, but substantially more precise than, those reported by \citet{Pang2026}. The emission-line samples presented here will enable further detailed studies of galaxy properties including metallicities, following the spatially resolved chemical-abundance analyses already demonstrated on lensed NIRISS targets \citep{wang2022glass}. We find a negligible contribution from bright active galactic nuclei in our sample. The methodology presented here can be readily applicable to other JWST slitless spectroscopic datasets and future wide-field slitless surveys, including those from Euclid, Roman, and the Chinese Space Station Telescope. WFSS is an efficient technique with which to measure H$\alpha$ equivalent widths, and luminosities with to study the evolution of star-formation \citep[e.g.,][]{FuCanucs2025, Sun2023, CoveloPaz2026, Pang2026}. Using WFSS on lensed sources is an efficient approach to measure the distributions of the faintest sources (Figure \ref{fig:canucs1}).

\subsection{Westley Brown:The Evolution of Galaxy Structure, Mass, and Environment Since the End of Cosmic Noon \label{sec:WBrown}}
The morphology-density relation i.e., the relation between galaxy morphology and local density in nearby galaxy clusters is being revisited with Euclid data. Specifically, \citep{cleland2026} has shown that the relationship between early-type galaxy fraction and local galaxy density exists across a general galaxy sample (not specific to clusters) up until $z\sim1$, although it appears to weaken with increasing redshift and shows some dependence on galaxy stellar mass.

\begin{figure}
      \centering
    \includegraphics[width=0.45\linewidth]{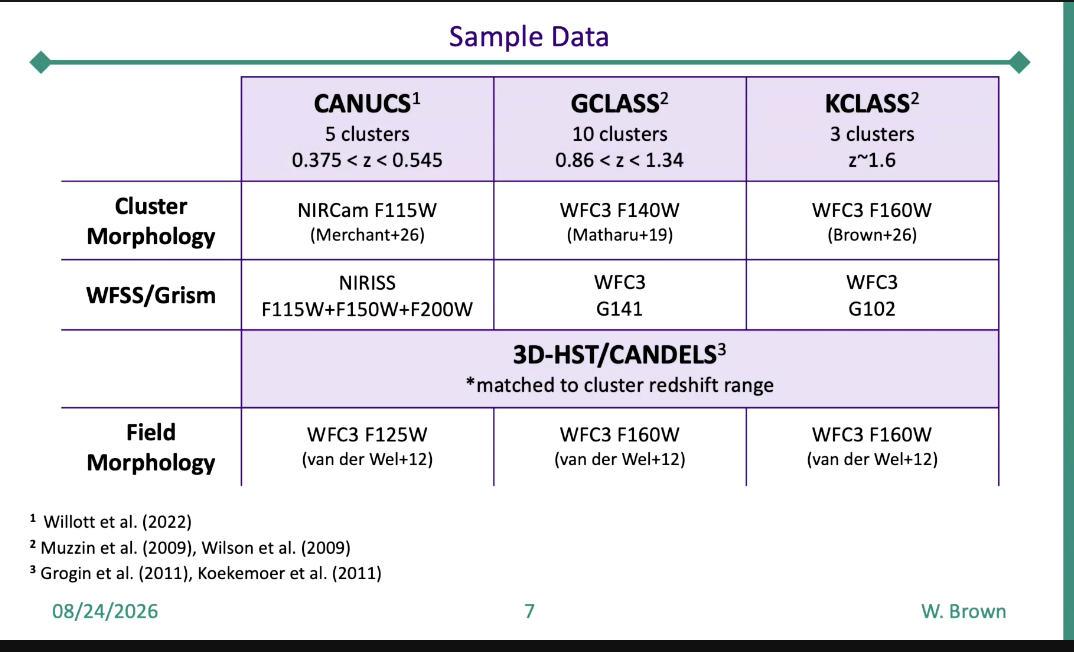}
    \includegraphics[width=0.45\linewidth]{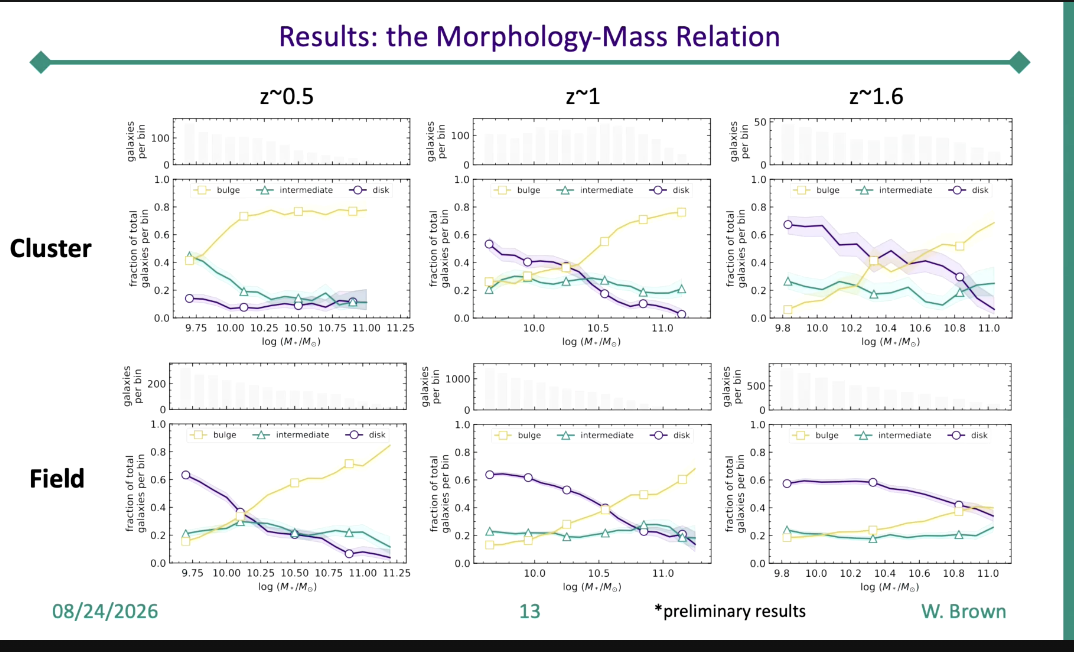}
    \caption{Figure from W. Brown showing the evolution of the morphology density relation from WFSS observations from multiple surveys.}
    \label{fig:morphdense}  
\end{figure}

WFSS are ideal to answer the questions: What is the relationship between galaxy structure, mass and environments, and how does this relationship evolve with redshift? We utilize cluster samples with WFSS at 3 different redshift ranges (CANUCS at $z\sim0.5$; SpARCS GCLASS at $z\sim1$; SpARCS KCLASS at $z\sim1.6$) and construct field comparison samples from 3D-HST/CANDELS (Figure \ref{fig:morphdense}). We study how galaxy structure/Sérsic index depend on local galaxy density and stellar mass in each sample. Both the morphology-density and morphology-mass relations (Figure \ref{fig:morphdense}) evolve with redshift, such that the cluster and field trends become more similar by $z\sim1.6$.

WFSS is our most efficient tool for accurate measurements of galaxy environments. With both imaging and WFSS, the Nancy Grace Roman telescope will provide a wealth of data and revolutionize our understanding of the morphology-density relation, large-scale galaxy environments, and how these relationships have evolved over time. The future is with Roman.

\subsection{Bingcheng Jin: COSMOS-3D: the Largest Census of [OIII] Emitters at z=7-9 and Implications for Emission-line Survey at
High-redshift \label{sec:Bjin}} 

Early JWST WFSS programs have hinted that our understanding of cosmic-dawn galaxy properties are potentially biased due to cosmic variance \citep[e.g.,][]{Meyer2026} and environmental effect \citep[e.g.,][]{Champagne2025ApJ, Zhu2025ApJ}. There are emerging efforts from the community including COSMOS-3D (\#GO 5893, PI: Koki Kakiichi) that target blank field with sufficient cosmic volume to mitigate the bias. COSMOS-3D's continuous coverage over a wide survey area ($\sim$1200 sq. arcmin.) allows us to compile hundreds of $z=6.8-9$ \OIII~emitters with a highly uniform selection function. 
Employing a model dependent framework based on FLAMINGO-10k simulation \citep[e.g.,][]{Pizzati2024MNRAS, Huang2026arXiv} to fit \OIII~LF simultaneously with clustering, B. Jin et. al. (in prep.) found that the uncertainties of LF increased by a factor of 1-1.5, as compared to Poisson counting error.


While deep grism surveys (e.g., EIGER, MAGNIF, ALT, FRESCO, SAPPHIRES, POPPIES)  characterize the faint end, COSMOS-3D delivers stringent constraints against cosmic variance at the bright ends ($\log\,(L_{ \rm{[O\,III]}}/{\rm erg\,s^{-1}})> 43.2$ and $M_{\rm UV}<-22\ \rm mag$). Jin et al. (in prep.) is building an unbiased [O\,III]/UV ratio as a function of galaxy properties ($M_{\rm UV}$), which enables direct constraints on the galaxy UV luminosity from the purely emission-line flux limited subset at redshifts up to 9. \citet{Topping2024MNRAS} and B. Jin et al. (in prep.) point to a population of metal-poor/exotic ISM condition or substantial ionizing photon escape (high $f_{\rm esc}$), that can be missed by grism line emitter search, as these faint sources exhibit steep UV slope and suppressed [O\,III]/nebular emission. .. 

\subsection{Anika Kumar and Jeyhan Kartaltepe: Constraining the High-Redshift Star-Forming Main Sequence with JWST Pure-Parallel WFSS \label{sec:Poppies}}

POPPIES (Public Observation Pure Parallel Infrared Emission-Line Survey) is a JWST cycle 3 pure parallel NIRCam WFSS program (PIs: J. Kartaltepe and M. Rafelski, JWST-GO-5398). This program has $\sim 400$ hours in $\sim110$ separate fields of observation. 
These data utilize multi-slot parallel opportunities to obtain WFSS in 1-3 filters spanning $ \sim 2 - 5 \mu$m and direct imaging over $ \sim 1 - 5 \mu$m, providing a wide area blind emission line survey \citep{Kartaltepe2024}. 

Because POPPIES obtains both spectroscopic and photometric data covering a large area on the sky all with varying depth and filter coverage, it is minimally impacted from cosmic variance. However, the large scale and pure-parallel nature of the survey also introduce several challenges, including non-uniform filter coverage and depth, variations in dither patterns between filters, instrumental calibration uncertainties, limited deep ancillary data outside of legacy fields, and the identification and interpretation of single-line emitters. Details about the POPPIES data reduction and cataloging will be presented by Vanderhoof et al. (in prep.) and Kumar et al. (in prep.).

\begin{figure}
    \centering
    \includegraphics[width=1.0\linewidth]{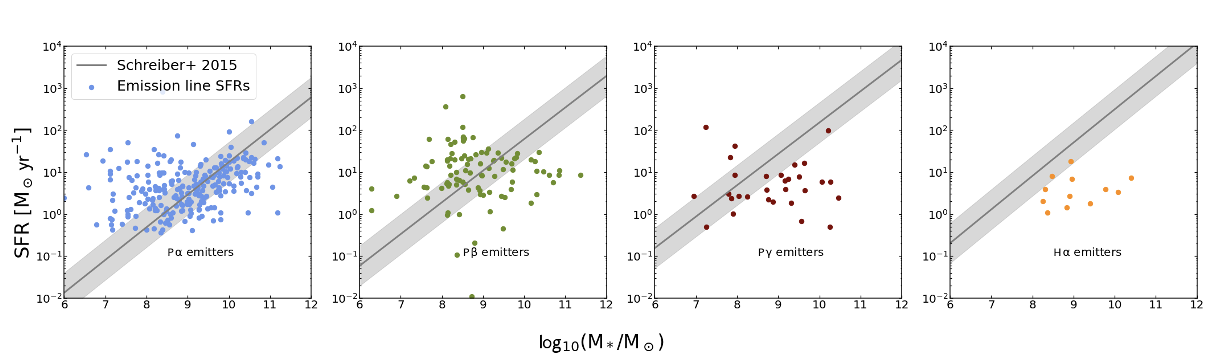}
    \caption{Preliminary results from Kumar et al. (in prep) from POPPIES fields overlapping with GOODS-N and CEERS, where galaxies are placed on the star-forming main sequence using star formation rates derived from multiple emission lines.}
    \label{poppies}
\end{figure}

Pure parallel NIRCam WFSS programs provide a unique and unbiased view of star formation across cosmic time. By using both the large volumes of photometric and spectroscopic data to derive and compare between SED SFRs and emission line SFRs, they can place meaningful constraints on relations like the star forming main sequence (JWST-AR-11320, PIs: A. Kumar and J. Kartaltepe). Figure \ref{poppies} shows preliminary results from POPPIES fields overlapping with GOODS-N and CEERS, where galaxies are placed on the star-forming main sequence using star formation rates derived from multiple emission lines (using relations from \cite{Nuefeld2024} and \cite{Reddy2023}) and stellar masses derived from SED fitting of the POPPIES photometric data (Kumar et al., in prep.).

\subsection{Vihang Mehta: Opening new avenues to study galaxy evolution with JWST/MIRI and Roman slitless spectroscopy \label{sec:vmehta}}
\begin{figure}
    \centering
    \includegraphics[width=0.9\linewidth]{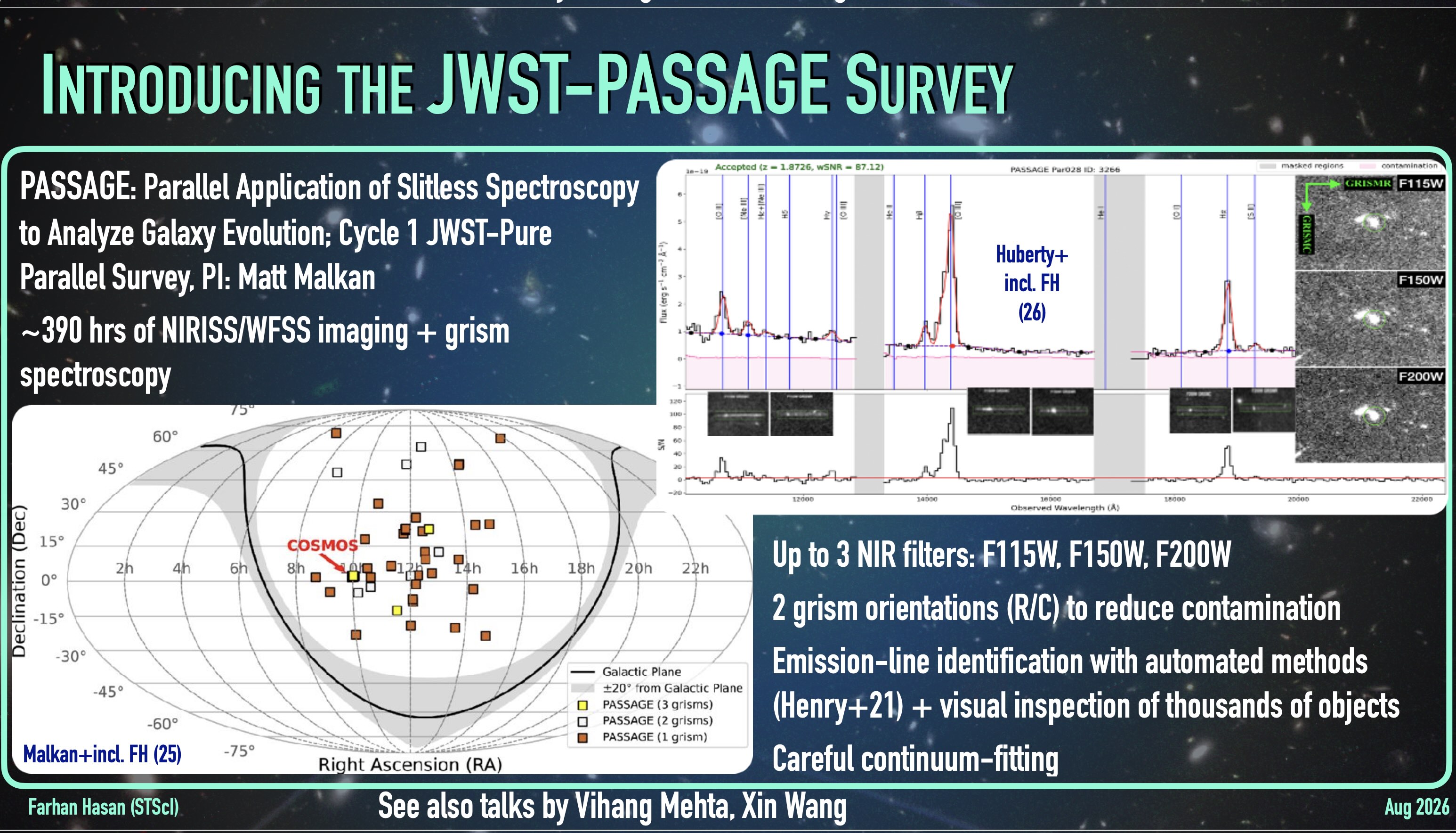}
    \caption{An overview of the PASSAGE survey.}
    \label{fig:passage}
\end{figure}

PASSAGE (Parallel Application of Slitless Spectroscopy to Analyze Galaxy Evolution) is a Cycle 1 JWST NIRISS pure-parallel program \citep[PI: M. Malkan, JWST-GO-1571,][]{malkan2025}. PASSAGE received a total allocation of $\sim591$ hours and comprises 63 fields totaling $\sim305$ arcmin$^2$. Each field was observed in up to 3 filters (F115W, F150W, and F200W) and up to 2 orientations (GRISMR and GRISMC). 15 PASSAGE parallel fields fall within the COSMOS footprint, 11 of which overlap with COSMOS-Web. The spectroscopic redshift catalog for PASSAGE in COSMOS is presented in \citet{huberty2026}, and the full survey has enabled studies of mass-metallicity, spatially resolved metallicity, protocluster environment analyses, constraining the star-forming main sequence, dust, high-z Lyman alpha emitters and Lyman break galaxies \citep[e.g.,][Nedkova et al. in prep, Hasan et al. in prep, Sattari et al. in prep, Acharyya et al. in prep]{runnholm2025,acharyya2026,sackih2026}. See Figure \ref{fig:passage} for an overview of the PASSAGE survey. 

\section{Cosmic noon} \label{sec:Day1PM}

\subsection{Xin Wang: Review on Tracing the Galaxy Baryon Cycle from the Epoch of Reionization to Cosmic Noon with HST and JWST WFSS \label{sec:XWang}}

\begin{figure}
    \centering
    \includegraphics[width=0.9\linewidth]{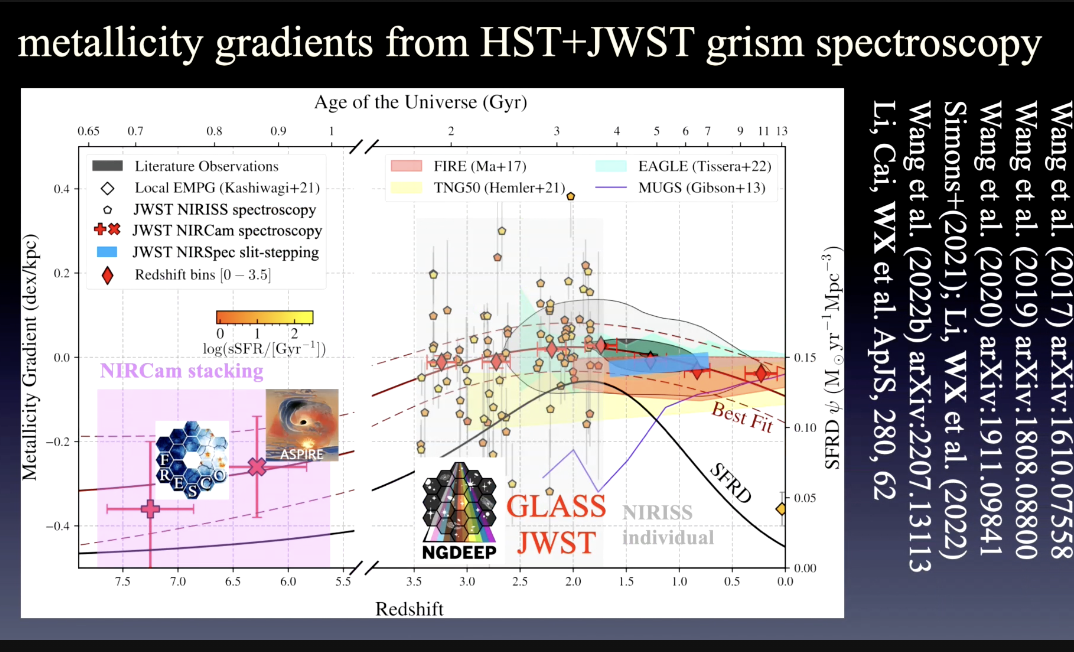}
    \caption{Figure from X. Wang's review talk giving an overview of the NIR grism studies to investigate galaxy evolution at cosmic noon.}
    \label{fig:metalgrad}
\end{figure}

{\textbf{Probing the Baryon Cycle: Stars and Gas}}

The baryon cycle can be probed with studies of gas-phase metallicity \citep[e.g.,][]{MaioRev19}. Oxygen is a useful tracer because it is produced by core-collapse supernovae (CC-SNe) and has the shortest formation timescale of any major metal, making it well-suited to probing the onset of star formation. The abundance of oxygen relative to hydrogen in the interstellar medium (ISM) is used as a proxy for gas-phase metallicity.

The particular strength of near-infrared WFSS in this context is that it delivers spatially resolved emission-line maps for every source in the field of view simultaneously. This converts metallicity from a single integrated number into a two-dimensional map, and does so for statistically meaningful samples rather than one galaxy at a time \citep{wang2017, wang2019, wang2020}. The radial gradient of that map is the observable that connects directly to the baryon cycle, since it records where gas is accreted, where metals are produced, and how efficiently feedback redistributes them.

{\textbf{GLASS Survey}}

The GLASS survey used 140 HST orbits with the G102 and G141 grisms to observe 10 galaxy clusters spanning z = 0.3--0.7. Combined with the magnification provided by the foreground clusters, these data yielded the first gas-phase metallicity maps of cosmic-noon galaxies at sub-kiloparsec resolution \citep{wang2017}, and subsequently a systematic census of such gradients in star-forming galaxies at cosmic noon \citep{wang2020}, complementary to the samples assembled from the CLEAR grism survey over a similar redshift range \citep{Simons2021}. A notable outcome of this program was the discovery of strongly inverted (positive) metallicity gradients in low-mass dwarf galaxies at z $\sim$ 2 \citep{wang2019}, a configuration that is difficult to produce without substantial inflows of metal-poor gas into galaxy centers or metal-enriched outflows carrying heavy elements to outskirts.


{\textbf{GLASS-JWST (with NGDEEP and CANUCS)}}

The HST experience \citep[e.g.,][]{wang2017, wang2020} helped leverage NIRISS spectroscopy \citep{treu2022}, an example of which is shown in Figure \ref{fig:metalgrad}. The first gas-phase metallicity map of a distant galaxy obtained with JWST came from these data: a gravitationally lensed galaxy at z = 3.06 behind A2744, resolved into $\sim$50 independent resolution elements despite a stellar mass of only $\sim 10^{8.6}$ M$_\odot$, and showing a strongly inverted gradient of $0.165 \pm 0.023$ dex kpc$^{-1}$ \citep{wang2022glass}. Reaching masses $\lesssim 10^{9}$ M$_\odot$ at these redshifts is possible only with JWST's combination of sensitivity and spatial resolution.

This program extends the work to the mass--metallicity relation (MZR), including the fundamental metallicity relation \citep[FMR, the joint dependence on stellar mass and star formation rate;][]{Mannucci2010, Henry21}. Lensing magnification behind A2744 first carried the MZR into the dwarf regime at $z = 1.8$--$3.4$, where the measured slope is consistent with that of more massive field galaxies \citep{He2024}; the same analysis quantified the systematic flux loss incurred when morphological broadening is ignored in grism spectra, an important guideline for Euclid, Roman, and the Chinese Space Station Telescope. Deep NIRISS slitless spectroscopy from NGDEEP then pushed the relation at $z = 1.1$--$3.4$ down to $10^{6.3}$ M$_\odot$ with 183 galaxies; the slope stays consistent over nearly four orders of magnitude in stellar mass, while no robust evidence emerges for an additional star formation rate dependence beyond the MZR at present depths \citep{He2026}. For reference, the field MZR over this redshift range is set by rest-optical spectroscopic surveys of more massive galaxies \citep[e.g.,][]{Sanders2021}. Using NIRISS WFSS, the H$\alpha$ luminosity function has been extended roughly one dex deeper at the faint end \citep{Pang2026}, reaching star formation rates of $\sim 0.4$--$1.0$ M$_\odot$ yr$^{-1}$ at z $\sim$ 1.3--2.0 and thereby providing the faint emission-line samples on which resolved metallicity work depends.

Stacked Balmer decrement measurements reveal a more extended, dust-corrected star-formation profile in massive galaxies, consistent with an inside-out growth pattern. Stacking H$\alpha$ and H$\beta$ maps for 79 lensed galaxies at $1.1 < z < 2.3$ yields radial profiles of nebular attenuation and dust-corrected star formation rate, with the most extended profiles found in massive systems at the low-redshift end \citep{Ren2026}; this extends the first resolved Balmer decrement measurements made with NIRISS \citep{matharu2023}. Metallicity gradients are informative in this context: negative gradients suggest inside-out growth, flat gradients may indicate mergers, and positive gradients suggest outside-in growth \citep[e.g.,][]{Ma2017, Hemler2021}.

These radial gradients help constrain how feedback regulates the mixing of gas flows. Steeper gradients (as a function of redshift) suggest less feedback, fewer outflows, and less mixing, while shallower gradients suggest more feedback, more outflows, and more mixing. Cosmological zoom-in simulations support this interpretation. In the FIRE-2 suite, positive gradients occur preferentially in galaxies with high specific star formation rate and weak rotational support, where feedback-driven galaxy-scale gas flows redistribute metals within the ISM \citep{XSun2025}; the predicted occurrence rate ($\sim$7\% for $0.4 < z < 3$, rising to $\sim$13\% at $1.5 < z < 3$) is broadly consistent with what the grism surveys find. Extending the same simulations into the epoch of reionization, the median gradient flattens from $\approx -0.15$ dex kpc$^{-1}$ at z $\sim$ 10 to $\approx -0.1$ dex kpc$^{-1}$ at z $\sim$ 6 with decreasing scatter, and correlates with stellar mass and with the strength of gas flows \citep{XSun2026}.

The mass--metallicity gradient relationship also evolves with redshift. At z = 0, there is a positive dependence on mass, consistent with inside-out growth. At cosmic noon, the relationship is instead negative with mass, consistent with feedback-regulated starburst activity. During reionization, star formation appears to be regulated in a more localized way. This full sequence has now been assembled observationally: synthesizing 455 spectroscopically confirmed galaxies at $1.7 \leq z \leq 9$ resolved on sub-kiloparsec scales by deep JWST NIRCam and NIRISS WFSS together with legacy measurements, gradients are found to be steeply negative ($\sim -0.4$ dex kpc$^{-1}$) at $z > 5$, to flatten to near zero around z $\approx$ 2 at the peak of the cosmic star formation rate density, and to become negative again toward z = 0 \citep{ZLi2025}. At z $\sim$ 1, slit-stepping JWST/NIRSpec observations from the MSA-3D survey confirm a tight relation between stellar mass and metallicity gradient with an intrinsic scatter of only 0.02 dex kpc$^{-1}$ \citep{Ju2025}, and further show that dynamically hotter disks (lower $v/\sigma$, shorter radial mixing timescales) have systematically flatter gradients, identifying turbulent radial mixing as a key regulator of chemical stratification \citep{Ju2026}.

{\textbf{Role of Environment}}

Looking at the MAMMOTH protoclusters at z = 2, the MZR appears shallower in overdense environments \citep{wang2022mammoth}. This result, first established from 36 member galaxies in the BOSS1244 protocluster core, has since been confirmed and strengthened using 63 members across three overdense structures at z = 2--3 \citep{YYang2026a}, and is consistent with the environmental metallicity offsets reported from rest-optical spectroscopy over a comparable redshift range \citep[e.g.,][]{Chartab2021}. In these environments, low-mass galaxies retain their metals more effectively, suggesting efficient recycling of feedback-driven winds. Cold-mode gas accretion appears to be more powerful in these regions, and this cold-mode accretion is subject to metal dilution driven by the cluster-scale dark matter halo; accordingly, massive protocluster members are found to be metal-poor relative to field counterparts of the same mass, while lower-mass systems show comparable or mildly enhanced metallicities \citep{YYang2026a}.

The same signature appears in the spatially resolved measurements. The first census of metallicity gradients in overdense environments at z $\gtrsim$ 2 found an unusually high fraction of flat and inverted gradients, together with an anticorrelation between gradient and global metallicity \citep{ZLi2022}. With the completed MAMMOTH-Grism sample, $\approx$69\% (29 of 42) of protocluster members at z $\approx$ 2.3 show positive gradients, a fraction significantly above that of field galaxies at matched mass and redshift, and these are preferentially the galaxies that are metal-deficient with respect to the field MZR \citep{YYang2026b}. Taken together, this is direct evidence that dense environments drive enhanced inflows of pristine gas into galaxy centers.

Deep Keck/MOSFIRE follow-up of these same fields adds a cautionary note for abundance work in overdensities: protocluster members at z $\sim$ 2 show significantly elevated [O\,{\sc i}]/H$\alpha$ ratios that require both H\,{\sc ii} regions and low-velocity ($v \sim 200$ km s$^{-1}$) shocks, plausibly environmentally driven by ram-pressure stripping or tidal interactions. Neglecting this shock component biases abundance measurements and can mimic nitrogen enrichment, offering a possible explanation for the long-standing puzzle of enhanced [N\,{\sc ii}]/H$\alpha$ ratios at z $\sim$ 2 \citep{HZhou2025}.

\subsection{Claire Bolda: Probing the low-mass end of the $z=0.5 - 3.5$ star-forming main sequence in the NGDEEP Survey \label{sec:cbolda}}

\begin{figure}
    \centering
    \includegraphics[width=0.9\linewidth]{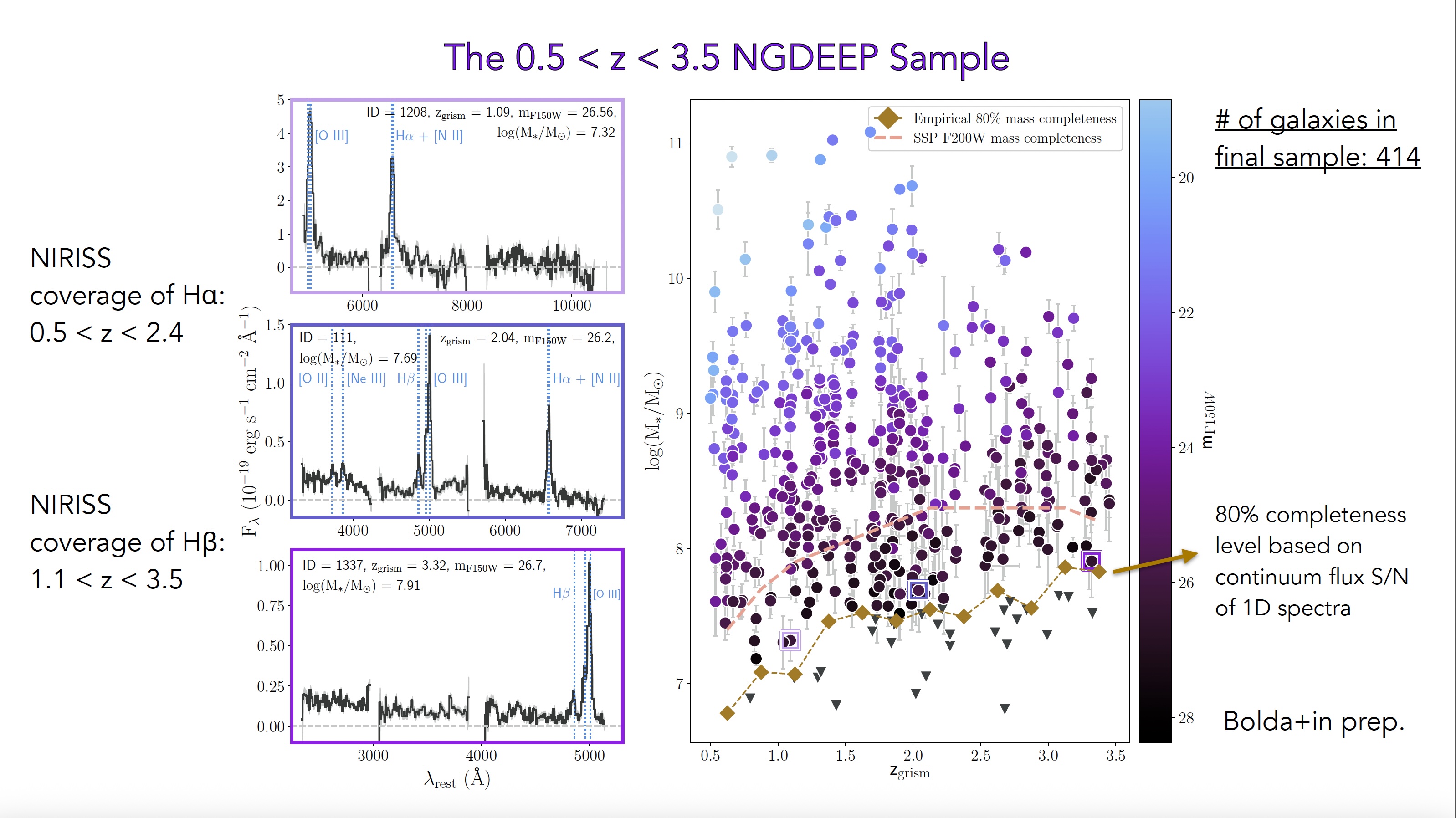}
    \caption{Figure from Bolda et al. in prep shows the stellar masses of the $0.5 < z < 3.5$, spectroscopically confirmed galaxy sample from the NGDEEP Survey. }
    \label{fig:ngdeepExample}
\end{figure}

Figure \ref{fig:ngdeepExample} from Bolda et al. in prep shows the stellar masses of the $0.5 < z < 3.5$, spectroscopically confirmed galaxy sample from the NGDEEP Survey \citep[NGDEEP; PID 2079; PIs: S. Finkelstein, C. Papovich, and N. Pirzkal;][]{Bagley_2024} as a function of redshift. This galaxy sample has an 80\% spectroscopic completeness level of $\mathrm{10^{6.8} M_{\odot}}$ at $z=0.5$ and $\mathrm{10^{7.8} M_{\odot}}$ at $z=3.5$. 

\begin{figure}
    \centering
    \includegraphics[width=0.9\linewidth]{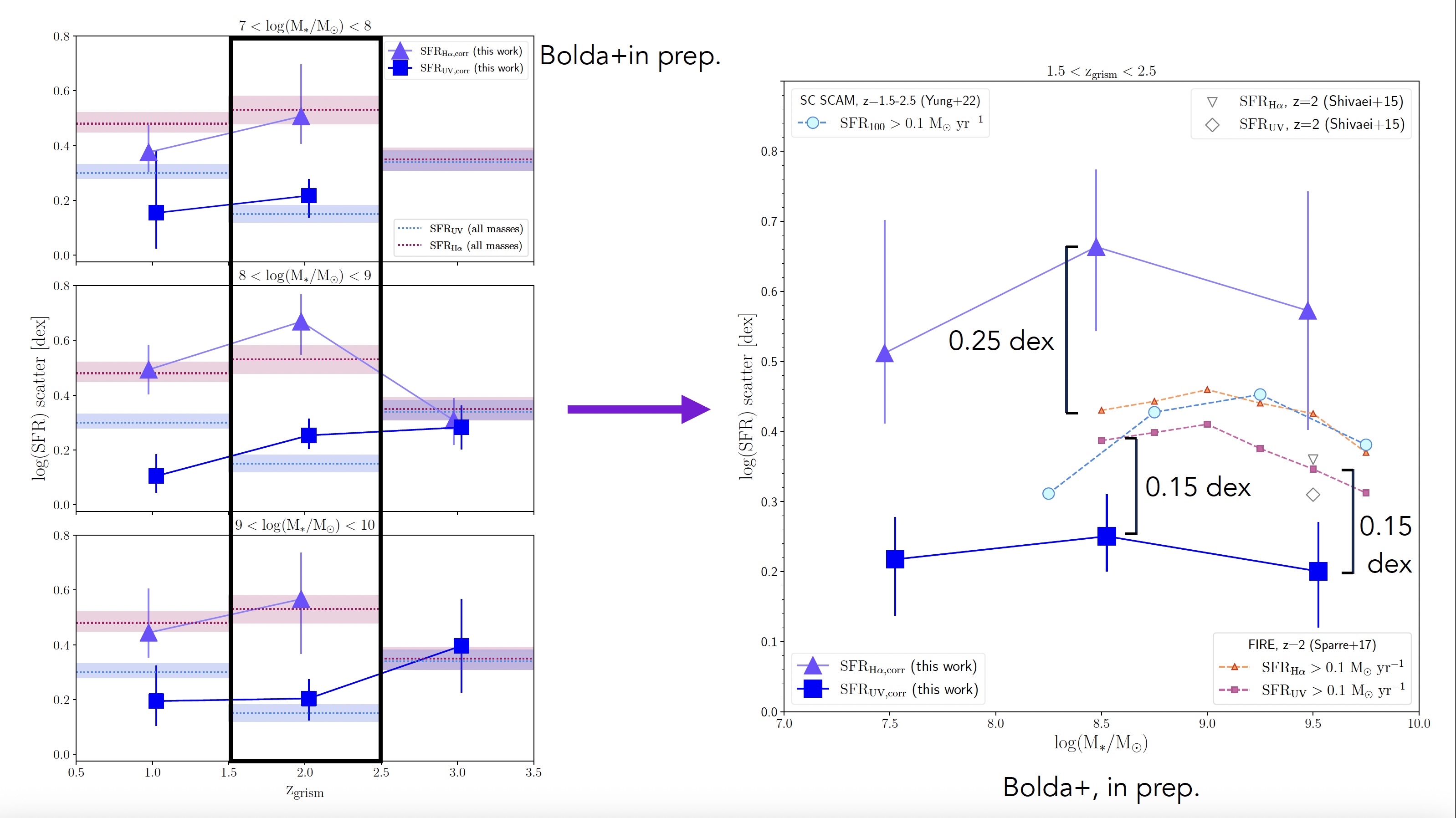}
    \caption{Figure from Bolda et al. in prep, the scatter in these SFRs, which trace star-formation on $\sim$10 Myr timescales, is compared as a function of stellar mass at $1.5 < z < 2.5$ to predictions from the FIRE simulations at $z=2$.}
    \label{fig:ngdeep_bursty_sfhs}
\end{figure}

For nearly 300 galaxies in this sample, Bolda et al. in prep determines \Ha\ SFRs from NGDEEP's 2D NIRISS spectra and corrects them for both dust and metallicity. In the right panel of Figure \ref{fig:ngdeep_bursty_sfhs} from Bolda et al. in prep, the scatter in these SFRs, which trace star-formation on $\sim$10 Myr timescales, is compared as a function of stellar mass at $1.5 < z < 2.5$ to predictions from the FIRE simulations at $z=2$ \citep{Sparre2017}. Additionally, the scatter in SED-derived SFRs averaged over 100 Myr, which probe star-formation on ~100 Myr timescales (Bolda et al. in prep refers to them as "UV" SFRs because they probe similar timescales to SFRs determined directly from the UV continuum), are shown as a function of stellar mass at $1.5 < z < 2.5$ in comparison to FIRE predictions at $z=2$ \citep{Sparre2017}. Bolda et al. in prep finds that their \Ha\ scatter measurements are notably higher than the FIRE predictions at low stellar masses, but agrees with the predictions at higher masses within its uncertainties. However, the UV scatter is lower than predictions at all stellar masses. Because the NGDEEP galaxy sample's \Ha\ SFR scatter is higher than the UV SFR scatter at all stellar massses for $z < 2.5$, these galaxies are experiencing variability on short, ~10 Myr timescales, but current feedback models cannot yet explain the intensity of the variability at low stellar masses.

\subsection{Farhan Hasan: Baryon cycling at the peak of galaxy formation with JWST/NIRISS WFSS \label{sec:fhasan}}

\begin{figure*}
    \includegraphics[width=0.325\linewidth]{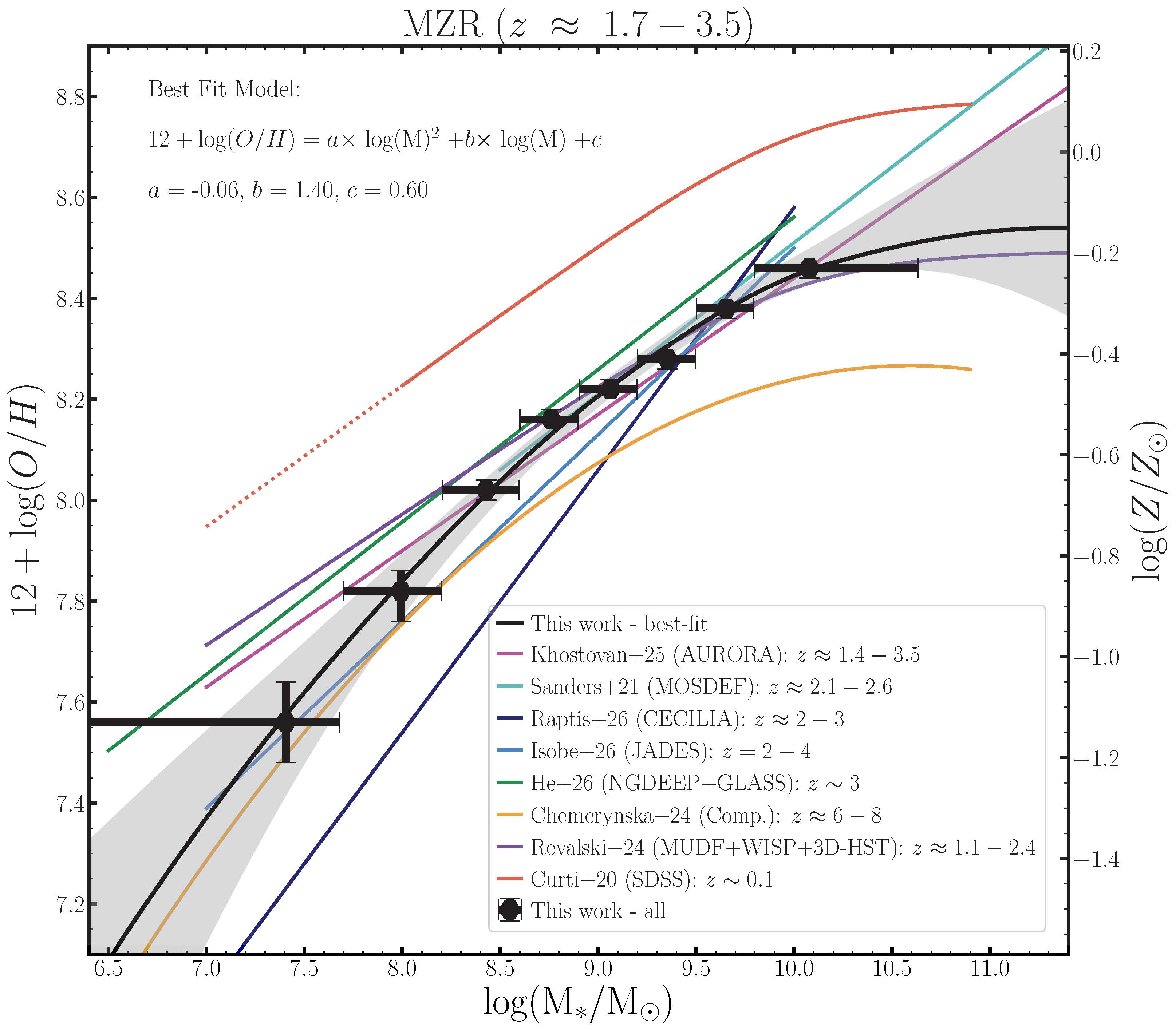}
    \includegraphics[width=0.325\linewidth]{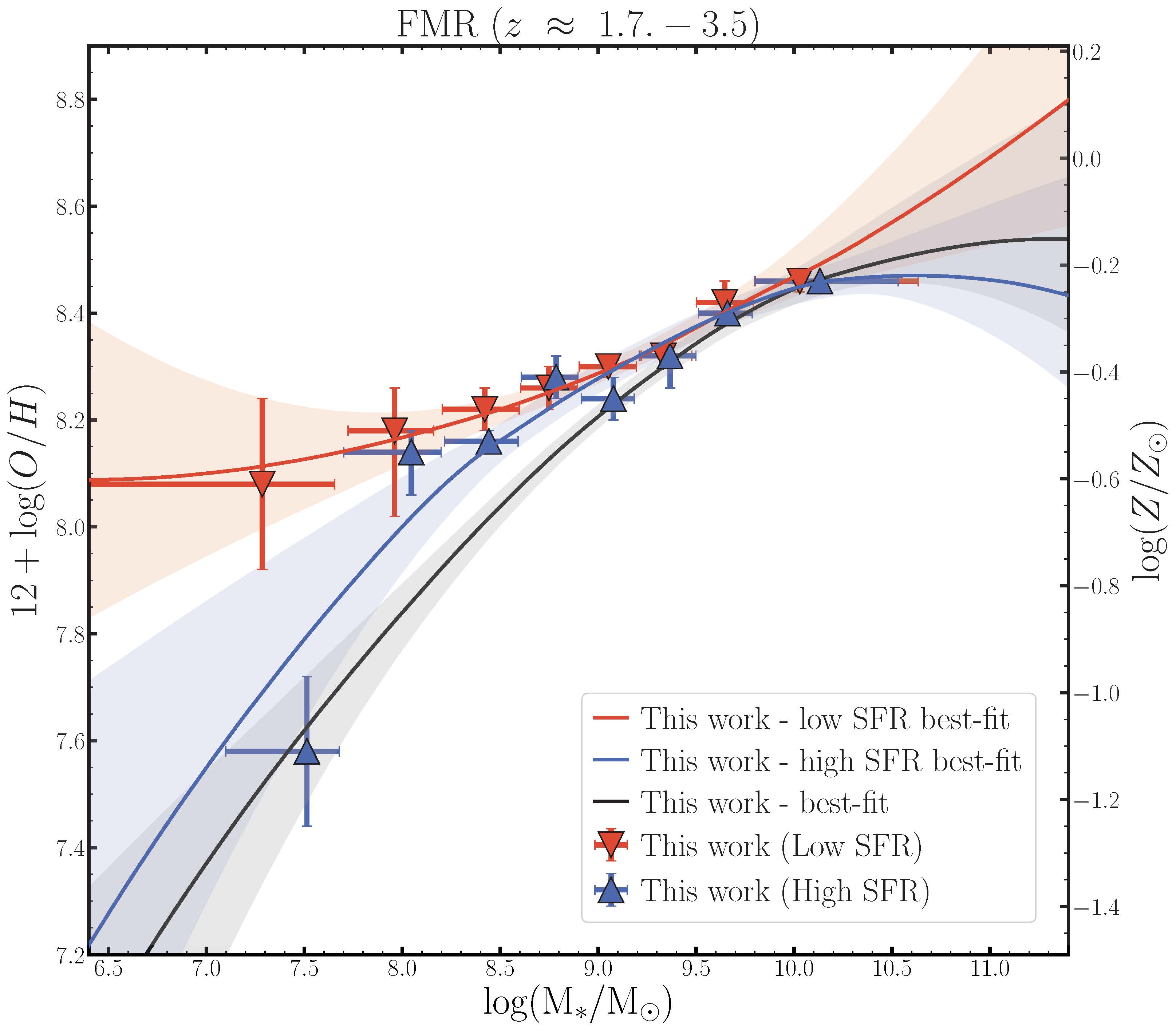}
    \includegraphics[width=0.325\linewidth]{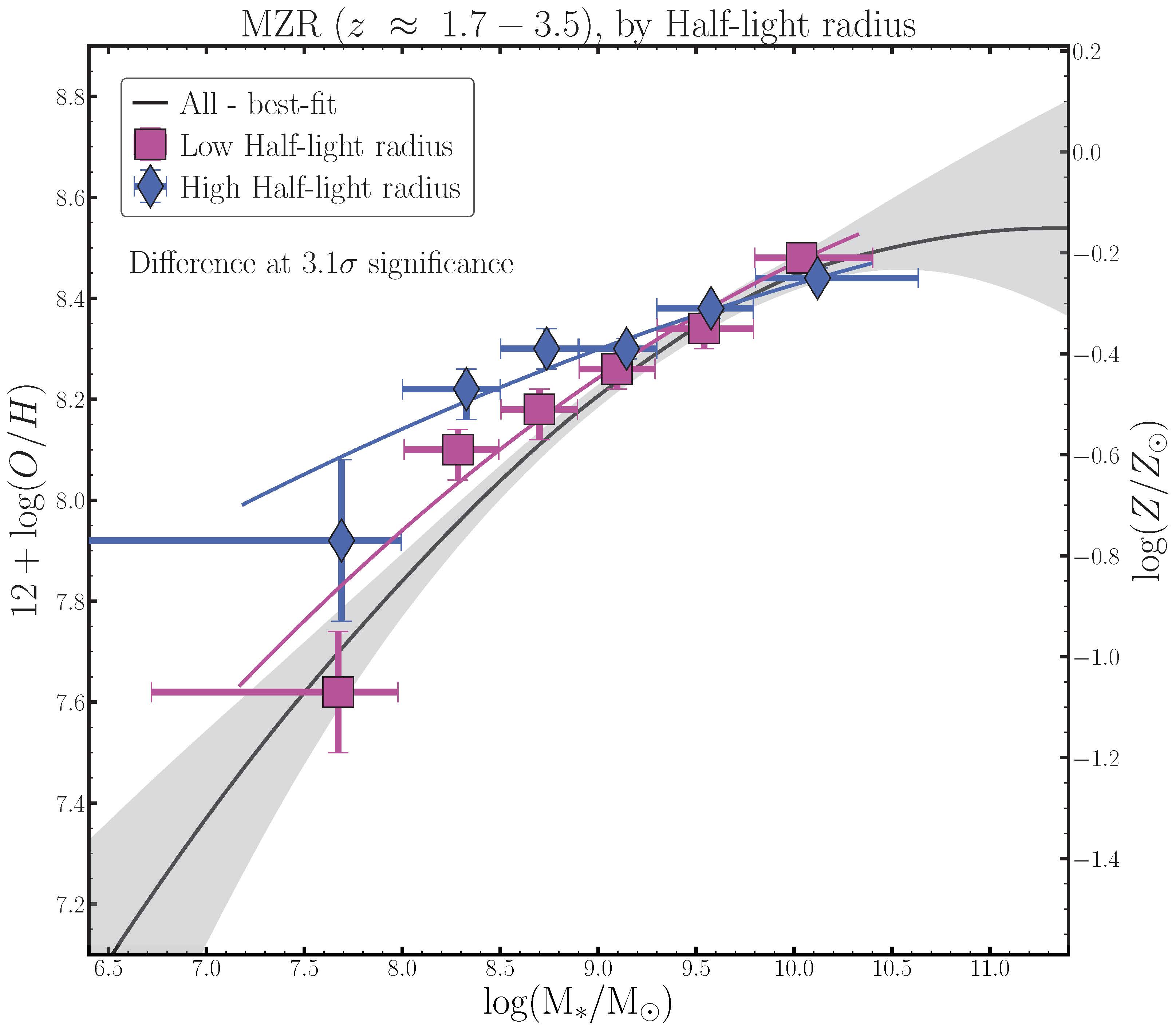}\\
    \caption{Results from PASSAGE: {\it Left}: Full MZR, compared to the literature;{\it Middle}: FMR;{\it Right}: MZR separated by size.}
    \label{fig:PASSAGE}
\end{figure*}

\begin{itemize}
\itemsep0em
\item PASSAGE survey: $\sim390$hrs of NIRISS WFSS + imaging in up to 3 NIR filters and 2 grism orientations \citep{malkan2025,huberty2026}.
The spectral analysis included automated line detection, visual inspection, and careful continuum fitting.

\item Gas-phase metallicities measured with multiple strong-line ratios ([OIII]/[OII], [OIII]/H$\beta$, etc.) and calibrations to ``direct'' metallicities measured with auroral lines from high-resolution JWST spectra. A sophisticated Bayesian approach was adopted to estimate probability distribution of metallicity by marginalizing over line ratios, dust, etc. \citep{Henry21,Revalski2024}.

\item M$_{\star}$, SFR, morphological properties estimated from ancillary imaging from COSMOS, COSMOS-Web, follow-up HST programs.
Due to low-resolution of NIRISS, M$_{\star}$-[OIII]/H$\beta$ diagnostic is used to separate AGN from star-forming galaxies ([OIII] doublet is deblended from H$\beta$).

\item Mass-metallicity relation (MZR) at $z\!\sim\!1.7-3.5$ is measured from stacked spectra (Hasan et al., in prep). The MZR is consistent with the literature at intermediate and high-masses, but a steep drop is seen at very low-masses ($<$10\% solar at $\log(M_{\star}/M_{\odot})\!<\!7.5$). 

\item Evidence for a fundamental metallicity relation (FMR) is also seen, i.e., low-SFR galaxies are metal-rich compared to high-SFR galaxies at fixed mass (Fig.~\ref{fig:PASSAGE}).

\item Statistically significant difference is seen in MZR of large (metal-rich) and small (metal-poor) galaxies, spiral (metal-rich) and non-spiral (metal-poor) galaxies. A small effect of line broadening is possible. 

\item Surveys such as POPPIES will enable higher-$z$ exploration with higher-resolution spectra, ultra-deep JWST spectroscopic surveys are essential for better constraining the MZR and FMR at the low-mass end, Roman and Euclid's deep WFSS surveys will be transformative in building large samples at intermediate and high-masses and statistically constraining the effect of large-scale environment on these scaling relations.

\end{itemize}

\subsection{Andreea Petric: Warm Molecular Gas in AGN Hosts from Wide Field Slitless Spectroscopy with MIRI \label{sec:apetric}}
The MIR 5-14 micron is the most recent wavelength regime to explored through wide field slitless spectroscopy \citep{petric2026, kendrew2026}. 

The MIRI instrument on JWST covers wavelengths from $\sim $ 5 to 28 $\mu m$ with imaging, coronagraphic imaging, low-resolution slit and (single-object) slitless spectroscopy (LRS), and medium-resolution integral field spectroscopy (MRS)~\citep{miri2015PASP, miri2023PASP}. The instrument has 3 Si:As Impurity Band Conduction (IBC) detectors, 1032 $\times$ 1024 px; 2 are dedicated to the MRS mode, and 1 is used for imaging, coronagraphic imaging, and low-resolution spectroscopy. The imaging detector has a number of subarrays, dedicated regions for coronagraphy using three 4-quadrant phase masks and a Lyot mask, and a fixed slit for the LRS slit mode. The LRS mode \citep{kendrew2015} uses a double prism mounted in the Imager filter wheel, which provides R$\sim$40-160 spectroscopy from $\sim$5 to 12$\sim \mu$m (the prism throughput extends to 14$\sim \mu$m, the transmission drops off steeply in the 10 -- 14$\sim \mu$m region).  

Starting in May 2024, a project has been ongoing to develop a Wide-Field Slitless Spectroscopy (WFSS) mode, using the LRS double prism in combination with the large Imager field of view to provide low-resolution dispersion of the full Imager field. This is the most significant new capability enabled for MIRI since the start of operations in 2022, and the only multiplexed spectroscopic capability in the mid-IR on JWST. It provides strong complementarity with similar WFSS modes on MIRI's near-IR counterparts on JWST, NIRCam and NIRISS. The mode was formally offered to the community for science observations from JWST Cycle 5 (July 2026).

This project aims to search for H$_2$ in AGN host galaxies using MIRI Wide Field Slitless Spectroscopy (WFSS). H$_2$ rotational lines serve as a tracer of shocked gas, making them a useful probe of AGN-driven shocks within their host galaxies \citep{petric2018, minsley2020, lam2019, u2022}.

\begin{figure}
    \centering
    \includegraphics[width=0.9\linewidth]{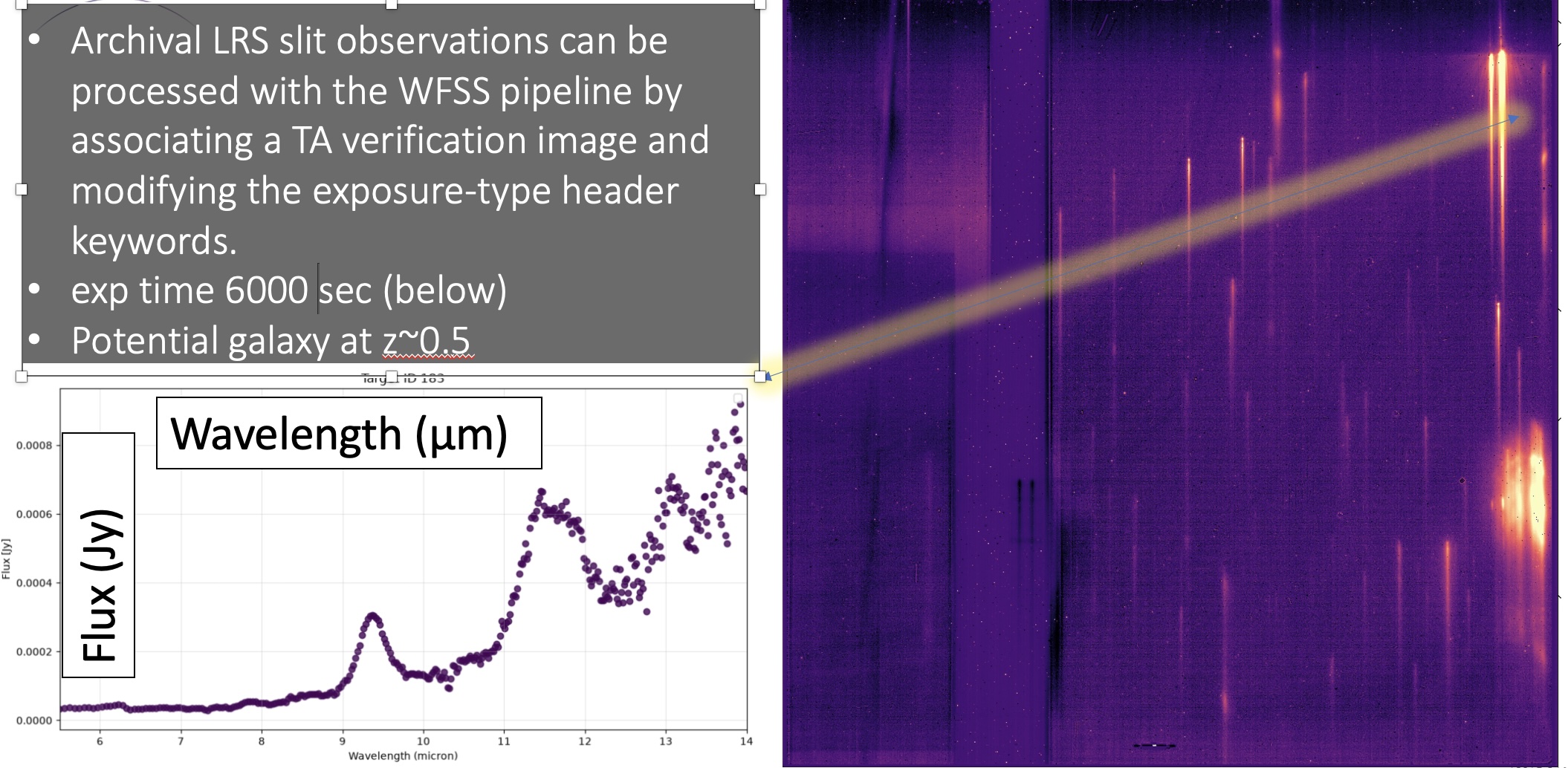}
    \caption{Figure showing the potential of MIRI WFSS observations to study the ISM of galaxies between now and cosmic noon. }
    \label{fig:ExampleAOP}
\end{figure}

The motivation for this work draws on several lines of evidence. Little Red Dots (LRDs) represent a population of faint AGN at high redshift, and mid-infrared color-color diagrams offer a way to identify obscured AGN more broadly. Notably, the redshift evolution of obscured AGN differs from that of unobscured AGN, which is where MIRI data becomes particularly valuable. Using MIR PAH lines to quantify the AGN contribution to the bolometric luminosity of galaxies is a demonstrated techniques \citep[e.g.,][]{petric2011}. Beyond energetics, MIR rotational H$_2$ lines can be used to track shock waves that compress molecular clouds and drive H$_2$ formation from ratios of warm H$_2$ relative to IR/PAH/[Ne II] emission. H$_2$ rotational excitation temperatures are found to be higher in AGN host galaxies. Furthermore, this warm H$_2$ excess correlates with [O I] and [O III] outflows, reinforcing the connection between shocked gas and AGN activity \citep{Hill2014, petric2018, lam2019, minsley2020, u2022, riff2020}. In terms of observational strategy, MIRI already has WFSS-compatible data available from Low Resolution Spectrometer (LRS) slit observations (Figure \ref{fig:ExampleAOP}).

\subsection{Xiangyu Jin: Charting the Late Stages of Reionization with JWST NIRCam Wide Field Slitless Spectroscopy \label{section:XJin}}





Cosmic reionization marks the transition of the intergalactic medium (IGM) from neutral to ionized. Two decades of effort have roughly pinned down its timeline, with a beginning at $z>10$ \cite[e.g.,][]{Witstok2025Natur}, a midpoint at $z\sim7-8$ \citep{planck2020}, and an end at $z<6$ \cite[e.g.,][]{Bosman2022MNRAS}. However, the topology of reionization and the dominant sources of reionization remain largely unconstrained. JWST NIRCam WFSS can probe both questions through efficient, unbiased redshift surveys of line emitters in quasar fields and large contiguous fields. JWST NIRSpec follow-up observations can further characterize the properties of ionizing sources. 

\textbf{Quasar fields.} The quasar Ly$\alpha$ forest traces the ionization state of the IGM: gas with a neutral fraction above $\sim0.1\%$ produces essentially zero transmission, while highly ionized regions are associated with high transmission. Pairing a NIRCam/WFSS redshift survey with the quasar sightline therefore connects IGM transmission directly to galaxies. ASPIRE (JWST Cycle 1 GO 2078; PI: F. Wang, X. Fan, J. Hennawi) targets 25 reionization-era quasars with a single F356W WFSS pointing in each field, and has identified $\sim500$ [OIII] emitters at $z\sim5-7$. \citet{Jin2024ApJ} split the Ly$\alpha$ forest in 14 fields into segments enclosed by spheres of a given ``influence radius'' around the [OIII] emitters and segments enclosed by none (i.e., regions ``away from [OIII] emitters"). Stacking these separately, they find the IGM effective optical depth ($\tau_{\rm eff}$) near the emitters is lower at $5\sigma$ significance, and reaches a given value at $\Delta z\sim0.1$ higher redshift. This indicates an ``inside-out" topology of reionization where reionization happened earlier near [OIII] emitters. 

\begin{figure}
    \centering
    \includegraphics[width=0.6\linewidth]{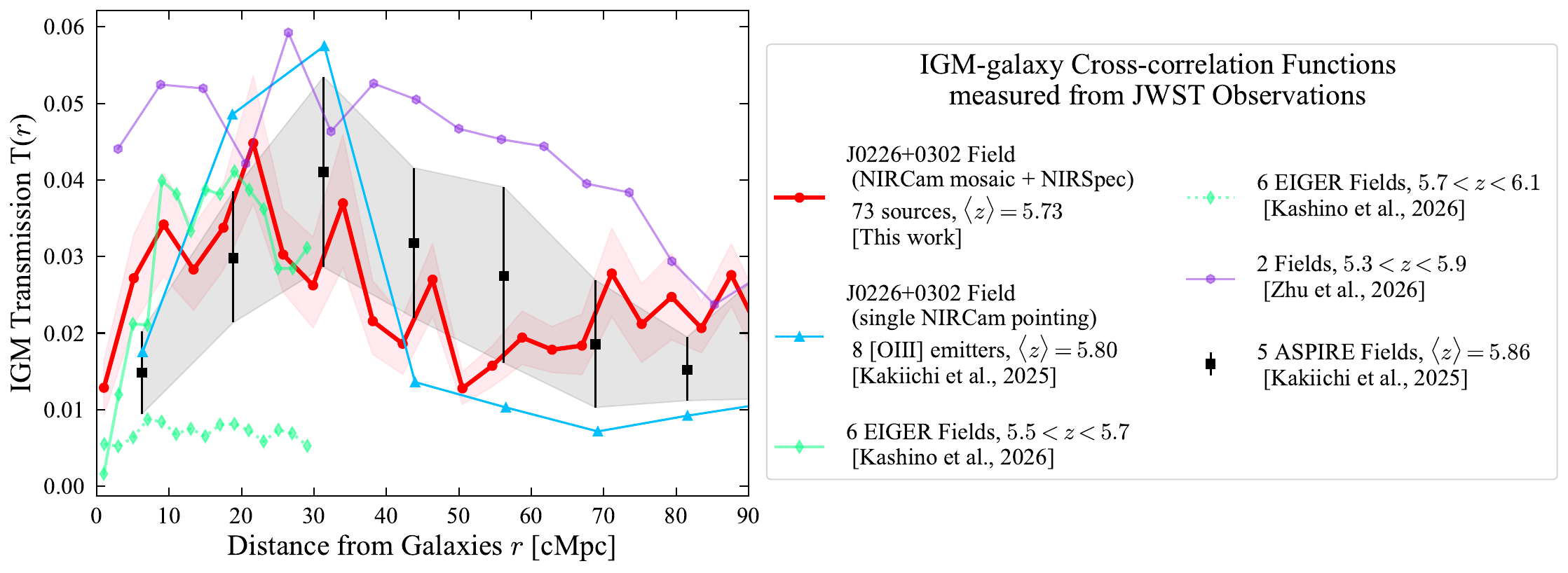}
    \includegraphics[width=0.35\linewidth]{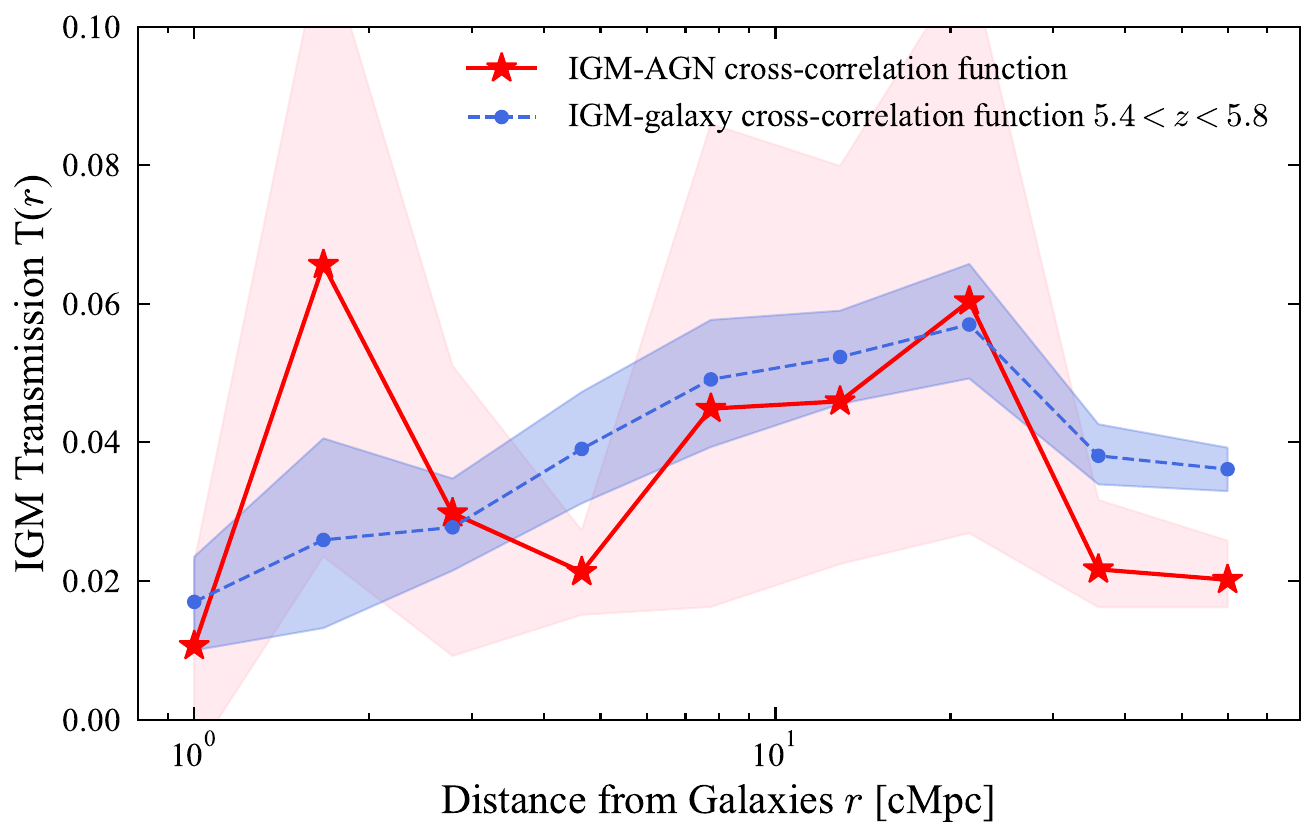}
    \caption{Left -- IGM-galaxy cross-correlation functions from JWST NIRCam WFSS surveys in quasar fields (black, blue -- \cite{Kakiichi2025arXiv}; red -- \citealt{Jin2026arXiv}; green -- \citealt{Kashino2026ApJ}; purple --\citealt{Zhu2026ApJ}); Right -- The IGM-AGN cross-correlation function (red) and the IGM-galaxy cross-correlation function (blue) in the J0226 quasar field. Both figures are from \citet{Jin2026arXiv}.}
    \label{fig:xj_figures}
\end{figure}

The IGM-galaxy cross-correlation function is a complementary statistic \citep{Kakiichi2018MNRAS}, measuring IGM transmission as a function of distance from galaxies. In five ASPIRE fields, \citet{Kakiichi2025arXiv} find transmission suppressed within 5 cMpc of the [OIII] emitters, as expected from the local gas overdensity, and a $2\sigma$ excess at 20-50 cMpc (Figure \ref{fig:xj_figures}--left). The observed [OIII] emitters cannot produce an excess even with a 100\% ionizing photon escape fraction, so faint, undetected galaxies clustered around them must contribute to the ionizing background. \citet{Jin2026arXiv} find a similar shape of the IGM-galaxy cross-correlation function in the J0226+0302 quasar field (JWST Cycle 2 GO 3325, PI: F. Wang \& J. Yang), with the excess IGM transmission shifted closer to galaxies ($\sim10-40$ cMpc). Comparison with EIGER \citep{Kashino2026ApJ} and GO 4092 \citep{Zhu2026ApJ} reveals substantial field-to-field scatter in the IGM-galaxy cross-correlation functions (see Figure \ref{fig:xj_figures}--left), consistent with the cosmic variance predicted by THESAN simulations \citep{Garaldi2025OJAp}. More than 50 independent quasar sightlines are needed to overcome the cosmic variance. 

NIRSpec follow-up also identifies the AGN population among [OIII] emitters. \citet{Jin2026arXiv} identify four broad-line AGN among 49 [OIII] emitters at $z<6.4$ with an AGN fraction of 8\%. The IGM-AGN cross-correlation shows a tentative excess transmission within 5 cMpc (Figure \ref{fig:xj_figures}--right), in contrast to the suppression of transmission seen around galaxies. An illustrative model requires a 50-100\% AGN ionizing photon escape fraction to reproduce such an excess transmission within 5 cMpc. Confirming this would require a larger AGN sample across independent quasar fields.

\textbf{Large contiguous fields.} COSMOS-3D covers 0.33 deg$^2$ in a single field with the F444W grism. Using COSMOS-3D data, X. Jin et al. (in prep.) identify more than 1000 H$\alpha$ emitters at $z\sim5-6.6$ and measure their ionizing photon production efficiency ($\xi_{\rm ion}$), which correlates significantly with UV magnitude. UV-faint galaxies produce ionizing photons more efficiently than UV-bright ones, pointing to reionization driven by faint galaxies.

\textbf{Outlook:} As the quasar Ly$\alpha$ forest absorption saturates at $z\sim6.5$, future observations from JWST and Roman will reveal the topology of reionization at $z>7$ through the galaxy Ly$\alpha$ emission. 

\subsection{Russell Ryan: slitlessutils demo/tutorial\label{sec:rryanD}}










While SLITELESUTILS was originally written for HST WFC3 and ACS it is highly versatile and its principles can be applied to most prism and grism observations. Examples are available, and the package can be installed via conda or pip.

Slitless spectroscopy saves the mapping between direct (imaging) x,y coordinates and dispersed (grism) x,y coordinates in a dispersion table, which encapsulates both the geometry of the field and instrument effects. Spectral extraction can be done in single or multi mode, with the latter relevant when multiple exposures are available. These tables enable downstream tasks like simulation and extraction.

An important step is syncing the WCS between the imaging and grism data. 

Cosmic rays are flagged by detecting sharp changes in the imaging data using a Laplacian operation, and sky background is subtracted using scaled templates. The latest reference files are packaged directly within slitlessutils.

Slitless can also reconcile astrometric offsets between imaging and grism data: it can either "undo" Gaia-based tweaks to the imaging to bring it back to grism-level alignment, or update the grism WFSS data to match the imaging shifts.

For grism simulations, a scene is created using the Pickles spectral library, with spectra associated to specific positions in the field and assigned magnitudes.

Spectral reconstruction uses a linear approach: each pixel's flux is the sum of contributions at different wavelengths from different (potentially overlapping) sources, and the input spectra are reconstructed by solving a large linear system. Using all orientations (position angles) simultaneously maximizes signal-to-noise and spectral resolution, with no trade-off between the two. Objects that are "linked" by spatial overlap should be grouped and solved together, while unlinked objects can be treated separately. Typically 5 orientations are sufficient, though up to 10 may be needed to fully decompose a complex object like a single galaxy. Segmentation can also be pushed further, from grouping by source ID down to treating every individual pixel as its own "source."

\section{Discussion Day 1}

{\textbf{Where will major paradigm shifts occur?}}

One area where a major paradigm shift is expected is in mass-metallicity relations. Metallicity remains challenging to measure, especially at high redshift, and more auroral-line-based direct measurements at high-z are needed to improve strong-line calibrations. Stellar metallicities are likewise very difficult to measure, and progress here would meaningfully advance the field.

Another shift is expected in time-domain science with Roman. A great deal of work has already gone into early-time supernovae, and balancing time allocation between ground-based and space-based observations is complicated — but Roman could make this considerably easier, opening up work across many supernova types.

Changing-look AGN represent a further opportunity, with Rubin's optical survey and WFSS data both expected to contribute, and Roman's prism also well-suited to identifying AGN. A Unified Transient Alert Monitor could be valuable in tying these efforts together.

Finally, there's a recurring theme of connecting object populations across cosmic time — identifying a given population as the likely precursor of another, then acquiring the area, redshift, and mass measurements needed to project that evolutionary link. One example raised was tracing protoclusters forward to systems like Coma, a present-day cluster merger.

{\textbf{What can we achive now and what will become possible? }}

Significant progress will come from synergies with other facilities, allowing complementary datasets and capabilities to be combined. Systematic errors are also expected to decrease substantially thanks to Roman's imaging and spectroscopy.

On the data processing side, using multiple orientations could enable MaNGA-level spatially resolved surveys with Roman at cosmic noon, bringing integral-field-like spectroscopic coverage to a much larger and more distant sample.

We also emphasized the value of open collaboration: sharing tools and datasets broadly so the community can make collective progress rather than working in isolation.

{\textbf{How should we approach for the next generation of capabilities?}}

For the next generation of capabilities, preparation for Roman should begin now including learning to work with ASDF files, since tools like DS9 won't be sufficient going forward.

Roman also needs community-driven benchmarks, as there is currently no unified approach to this. A key open question is how to simulate data in order to construct these benchmarks in the first place. Euclid offers a useful model here: since IPAC is involved in both missions, there's an opportunity to learn from Euclid's approach and apply those lessons to Roman.

\section{Transients and Roman}
\subsection{Jyotika Roychowdhury: Spectral Extraction of Type Ia Supernovae in OpenUniverse2024 with Roman WFI P127 Prism \label{sec:JRoychowdhury}}

\begin{figure}
    \centering
    \includegraphics[width=0.9\linewidth]{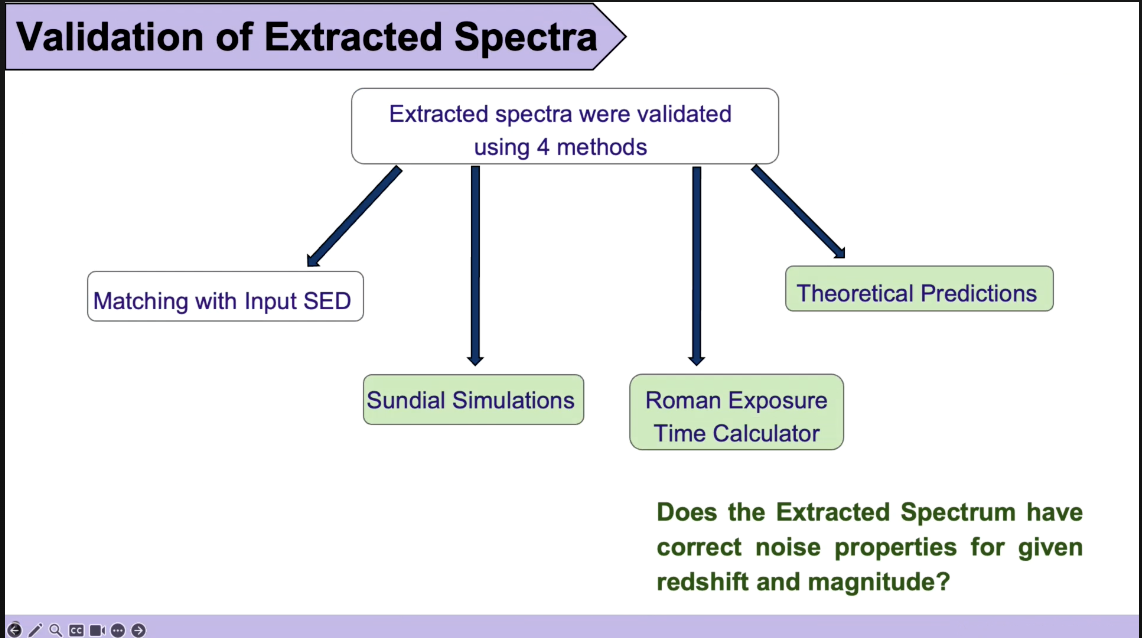}
    \caption{Validation of spectra extracted from OpenUniverse Images.}
    \label{fig:ValidSim}
\end{figure}

The Nancy Grace Roman Space Telescope Wide Field Instrument P127 Prism will provide low-resolution slitless spectroscopy that will play a key role in the classification and redshift determination of Type Ia supernovae (SNe Ia) for future cosmological studies. Preparing for these observations requires realistic simulations and validated analysis pipelines to understand the performance and limitations of Roman prism spectroscopy. In this work (Roychowdhury et al. (in prep.)), we use the OpenUniverse2024 simulation suite to generate prism dispersed images corresponding to the Wide and Deep tiers of Roman's High-Latitude Time-Domain Survey. We then apply a spectral extraction algorithm (developed by DerKacy et al. (in prep.)) to recover one-dimensional spectra of SNe Ia from the dispersed images. The extracted spectra are validated by comparing their fluxes, noise properties, and signal-to-noise ratios with the true input SNANA spectra, Sundial simulations, Roman Exposure Time Calculator and theoretical predictions \ref{fig:ValidSim}. This also allows us to connect image-level simulations with catalog-level simulations. We evaluate the extraction framework under two extreme scenarios: 1) isolated case containing only supernovae, and 2) contaminated case containing supernovae, galaxies, and stars without any host-galaxy subtraction. Supernova-only images give us clean isolated spectral traces of supernovae without any contamination so we can test and validate the extraction pipeline, before making things more complicated by including contamination. Together, these scenarios establish upper and lower bounds on the expected extraction performance.

 \begin{figure}
    \centering
    \includegraphics[width=0.9\linewidth]{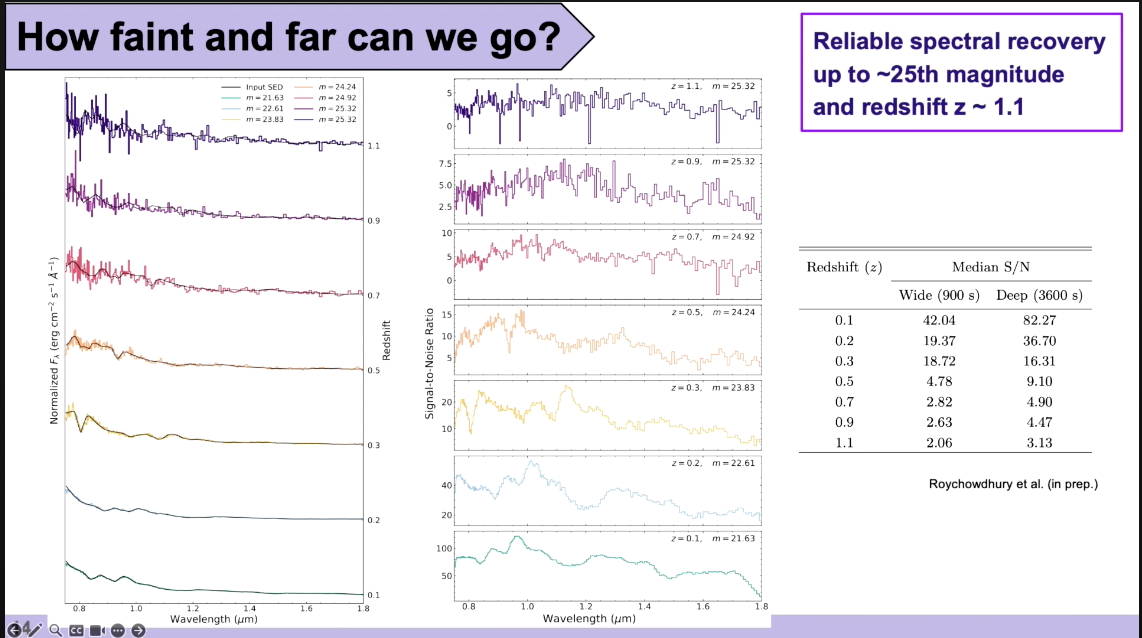}
    \caption{Performance of spectral extraction pipeline.}
    \label{fig:RomanIdeal}
\end{figure}

After host-galaxy subtraction, real Roman observations are expected to lie between these limits, closer to the supernova-only case. For these supernova-only images, the pipeline successfully recovers spectra for supernovae as faint as 25th magnitude (in F184 filter) at redshift z = 1.1, across five epochs spanning $-15$ to $+15$ days relative to peak brightness \ref{fig:RomanIdeal}. The recovered spectra are analyzed for supernova classification and redshift estimation, yielding high-confidence Type Ia classifications for 71\% (91\%) of the Wide (Deep) tier sample and redshift estimates with an NMAD of 0.0041 (0.0029) and 2.86\% (2.86\%) catastrophic $(\geq10\%)$ outliers, without any redshift prior. Contaminated simulations demonstrate that contamination from overlapping spectra of respective host galaxies and neighboring sources can significantly degrade spectral extraction, thereby highlighting the need for accurate host-galaxy subtraction see sec. \ref{sec:tastraatmadja} and sec. \ref{sec:MGriggio} This work thus establishes and validates a framework for simulating Roman prism images and extracting scientifically useful spectra, and motivates future developments in host-galaxy subtraction and Roman supernova spectroscopy.

\subsection{Sahana Kumar: WFSS for Supernovae and Explosive Transients in the Era of LSST \& Roman \label{sec:SKumar}}
Supernovae (SNe) are some of the most common and luminous transient phenomena and evolve on varying timescales. SNe are often categorized by their progenitors, sorting into two main groups: thermonuclear and core-collapse. Most SN types change most rapidly immediately after the explosion, evolving on timescales of $\sim$days to weeks. Several months later, most SNe reach the nebular phase when the ejecta has become optically thin and their time evolution has significantly slowed down, with spectral features evolving over the course of $\sim$months to years. WFSS can help increase the sample size of spectroscopic observations of different populations of supernovae, especially at nebular phases when observations are less time-sensitive. 

Current big questions in the field involve details of the progenitor star, the physics of the explosion mechanisms, and understanding the role SNe play in their local environments and host galaxies. This is an exciting time for time domain astronomy thanks to Rubin-LSST and other transient surveys (i.e. ZTF, ATLAS, etc.) providing a wealth of photometric observations. Spectroscopic follow-up of individual transients is becoming more challenging with the thousands of transient alerts generated each night, but WFSS can help. 

\begin{figure}
    \centering
    \includegraphics[width=0.9\linewidth]{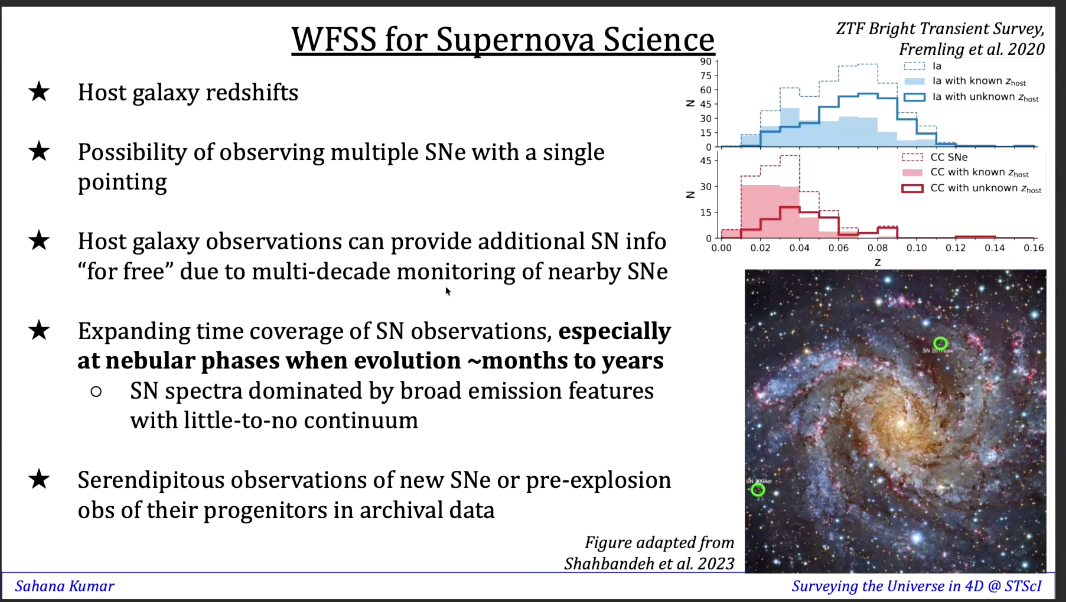}
    \caption{Figure from S. Kumar outlining the goals of WFSS observations for SNe science.}
    \label{fig:SNe4WFSS1}
\end{figure}

Spectroscopy is necessary for classification of supernovae, but also provides unique information on the kinematics, chemical composition, and geometry of the ejecta that cannot be obtained from photometry. In addition to observing the SNe themselves, WFSS can greatly benefit SN science by providing host galaxy redshifts and pre-explosion information on nearby stellar populations (Figure \ref{fig:SNe4WFSS1}). WFSS is beneficial for all types of supernovae and other explosive transients, but here are some specific examples of how WFSS can help address important questions in the field. 

\textbf{Type Ia Supernovae (SNe Ia)} - SNe Ia are well-known as standardizable candles for cosmology, but there are many questions on the details of their origins. For example, it is unknown if the progenitor white dwarfs much reach the theoretical Chandrasekhar mass limit of 1.44 $M_{\odot}$ or if they can explode at much lower masses. It is also unknown if all normal SNe Ia used for cosmology have the same explosion mechanism. 

Space-based WFSS provides the best opportunities to observe key spectroscopic features that may be able to resolve these important questions (Figure \ref{fig:SNe4WFSS1}). The detection of stable Nickel in SNe Ia can provide key insights for both observational and theoretical studies. In order to produce iron group elements (IGEs) such as Nickel, sufficiently high density conditions are required which are not predicted for all progenitor masses and explosion mechanisms. Nickel is the preferred IGE for this type of observation because the radioactive isotope $^{56}Ni$ (which powers the early-time luminosity) has a short half-life, so any Ni observed at later times must come from stable material. Ground based studies of nearby SN Ia populations reveal observed differences in late-time Ni spectral features, possibly indicating different physics involved in different sub-populations of SNe Ia \citep{Kumar2026a}.

Space-based WFSS is ideal for searching for stable Ni as atmospheric telluric features can obscure the most well-isolated Ni line in the NIR, and JWST is the only telescope that can reach additional Ni lines in the MIR. Recent observational efforts are building a legacy data set with JWST by observing a representative sample of SNe Ia that span different sub-populations and local environments (e.g. \citet{Kwok2026, DerKacy2024}). In addition to observing Ni lines, nebular phase observations of iron lines can provide important constraints on the progenitor mass and for modeling of energy in the ejecta and radiative transport (see \citet{Diamond2018, Kumar2023}). Iron lines are very broad and evolve slowly at nebular phases, and this method is suitable for lower resolution spectroscopic observations such as those with the Roman prism during the HLTDS. 


 \begin{figure}
    \centering
    \includegraphics[width=0.9\linewidth]{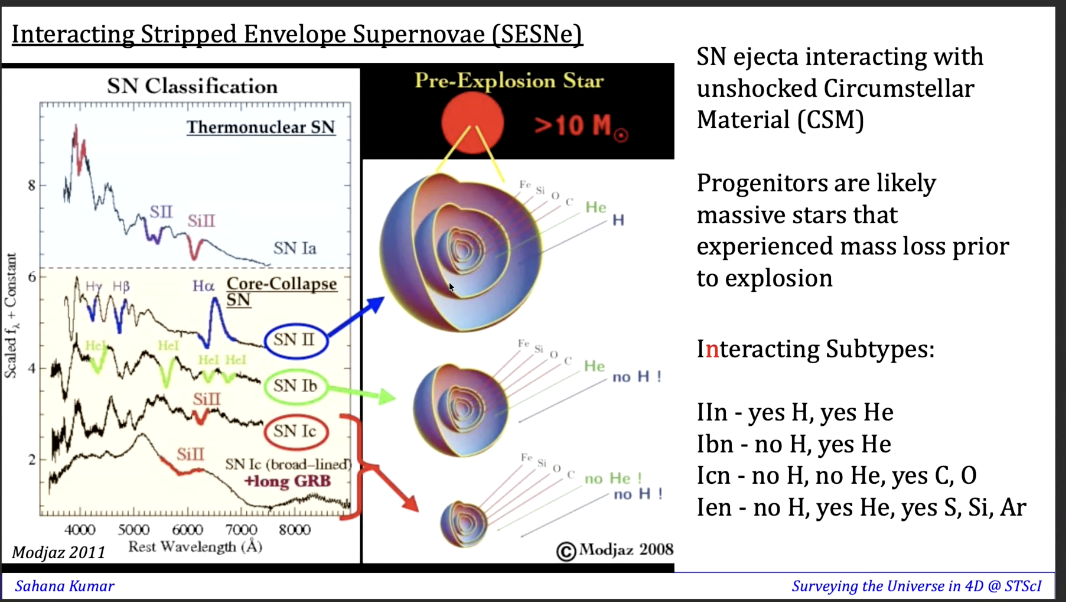}
    \caption{Figure from S. Kumar showing how in an important class of SNe the ejecta collides with nearby circumstellar material to produce spectral features revealing the composition of that surrounding gas. }
    \label{fig:SNe4WFSS2}
\end{figure}

\textbf{Interacting Stripped Envelope SNe} - Stripped Envelope supernovae (SESNe) involve the core-collapse of massive stars that have lost their outer layers prior to explosion. Interacting supernovae are observed subtypes where the SN ejecta interacts with nearby unshocked circumstellar medium (CSM) producing the characteristic narrow spectral features that reveal the chemical composition of the surrounding CSM. The different subtypes include interaction with H-rich, He-rich, and C-rich material (Figure \ref{fig:SNe4WFSS2}). 

The timescales for interaction between the SN ejecta and nearby CSM can be very large and is connected to the mass-loss history of the progenitor. Observations of interacting SNe have shown interaction can occur over the course of several days (i.e. \citet{Pellegrino2022}) or several years (i.e. \citet{Baer-Way2025}). WFSS is well-suited for the slower-evolving interacting SNe due to the large window of opportunity for observations. Furthermore, the high resolution of some WFSS instruments is necessary to resolve the narrow spectral features (line widths $\leq \sim 2000~km/s$) that arise from interaction and to distinguish possible different velocity components. 

Spectroscopic observations of interacting SNe are of interest not only to the supernova community, but also massive stars and the late stages of stellar evolution. By studying the CSM interaction over time, we can re-construct the most recent mass loss history of the progenitor. Furthermore, the CSM masses derived from observations of interacting SNe indicate mass-loss that is too high for stellar winds leading to growing interest in other possible mass-loss mechanims, including binary interaction.

\textbf{Calcium Strong Transients (CaSTs)} - The past few decades of rapid photometry and wide field transient searches have revealed new populations of explosive transients. One new category is Calcium Strong Transients (CaSTs), characterised by weak oxygen emission and strong calcium emission at later times. Due to their identifying features not emerging until $~1+$ month post explosion, CaSTs are often classified as peculiar SESN or peculiar SNe Ia shortly after discovery. 

The small but growing sample of known CaSTs reveals interesting diversity and spectral featueres that resemble both thermonuclear and core-collapse events \citep{Kumar2026b}. Future WFSS observations can provide crucial information on the local environments of CaSTs and possible clues on the progenitor stars. Some CaSTs are found in the outskits of their host galaxies, possibly indicating older progenitors associated with thermonuclear explosions, whereas others are found closer to galactic centers (see \citet{JacobsonGalan2021}). Roman WFSS observations during the HLTDS and HLWAS are of particular interest because Roman has the wavelength coverage and sensitivity to possibly reveal higher-redshift CaSTs or identify previously unknown or misidentified CaSTs.  


\subsection{Russell Ryan: From Dispersed Images to Spectral Time Series: The Roman SNIa Spectroscopy Pipeline \label{sec:rryanT}}
\begin{figure}
    \centering
    \includegraphics[width=0.9\linewidth]{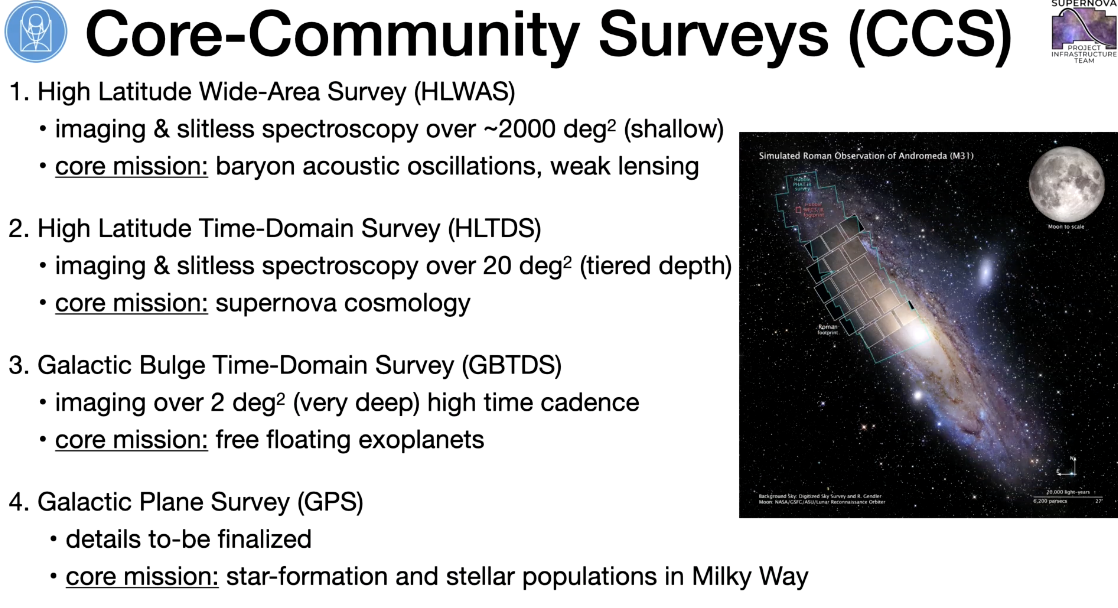}
    \includegraphics[width=0.9\linewidth]{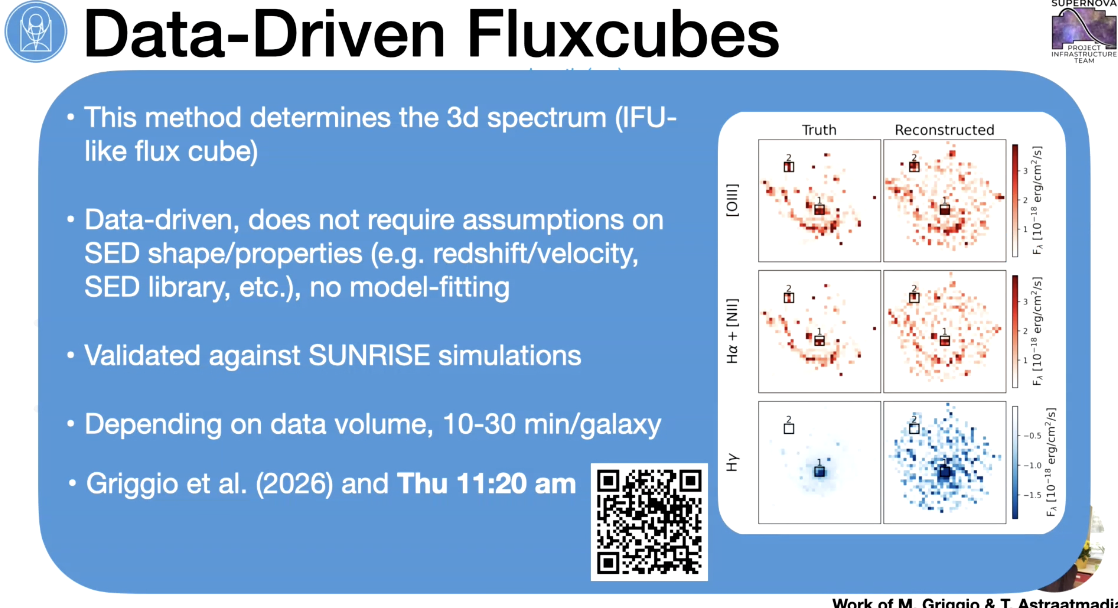}
    \caption{List of the he Roman core community surveys (top) and schematic of pipeline by the HLTD team (bottom).}
    \label{fig:romancc}
\end{figure}

Roman's core community surveys paradigm change on how we do science (Figure \ref{fig:romancc}). The High Latitude Time Domain group developed a specialized pipeline for prism data primarily to decontaminate SNe spectra from their hosts. This code will be made public soon. 

The wide-field slitless spectroscopy pipeline, including the data and software products derived were produced by the Supernova Project Infrastructure Team for the P120 (prism) on the Roman Space Telescope. The observations will be predominately in the southern High-Latitude Time-Domain Survey, which will have deep and wide tiers that cover $\sim 0.6 \deg^2$ and $\sim 4.6 \deg^2$, respectively.  The deep tier will be in the Euclid-Deep Field-South and have 72 position angles, cadenced over 1-2 years and reaching a total depth of ~100 nJy(3$\sigma$).  The primary design ethos of the pipeline is to produce a spectral time-series for SN Ia, whose central challenge is mitigating the contamination from the host galaxies.  For this, the team has developed multiple strategies based on differing assumptions, data richness, and observational timescales.  These data and software products are entirely public and, like all Roman data, have no proprietary period. Although the pipeline is highly tailored to transient and point sources, it has some components of particular interest to general extragalactic research.


\subsection{Ivelina Momcheva: Scaling Slitless Spectroscopy from HST and JWST to Euclid and Roman \label{sec:IMomcheva}}

Wide-field slitless spectroscopy has entered a transformative era with JWST, Euclid, and the upcoming Roman Space Telescope. While these facilities enable unprecedented spectroscopic surveys across cosmic time, extracting robust physical information from slitless data remains challenging. 
Recent results were presented from the OutThere pure parallel program, one of the largest JWST/NIRISS datasets, which uses JWST slitless spectroscopy to study distant galaxy populations and develop new analysis strategies for these data. I will focus in particular on new calibration frameworks derived through a self-calibration approach using the first few years of data, enabling improved treatment of contamination and spectral extraction.

New numerical approaches are being developed to address slitless spectroscopy challenges, including inverse-modeling approaches based on large-scale linear matrix minimization, as well as emerging machine learning methods for deriving physical properties from joint spectroscopic and imaging datasets. Together, these approaches aim to improve the robustness and scalability of inference for increasingly large and heterogeneous datasets.

We can use from HST and JWST slitless spectroscopy operational lessons to inform future wide-area spectroscopic surveys with Euclid and Roman, where many of the same calibration, extraction, and inference challenges will arise at dramatically larger scales.

Faster numerical methods are key to conquering the avalanche of data and tailored approaches to certain fields of view can be powerful as is strategic use of machine learning tools to overcome scarce data challenges.

\subsection{Zihao Zuo: A Neural Spectral Deblender for Wide-Field Slitless Spectroscopy \label{sec:ZZuo}}

Presentation address this WFSS deblending through learned image restoration: a computer-vision model transforms a blended cutout into the target's isolated trace, trained on a realistic MIRAGE + grizli simulation comprising approximately 3,900 scenes. In simulation, the model achieves 0.95–0.99 shape fidelity, preserves emission lines, and recovers continuum even under 10–25× neighbor contrast, for both cross-dispersion and dispersion-direction blends. The key limitation is separation rather than contrast: blends that fall on the same row and within sub-pixel offsets remain unresolved.

\subsection{Mainak Singha: Roman Grism Forecasts for the Intermediate-Luminosity $z=7.5-10$ Quasar Population  \label{sec:MSingha}}

Recent JWST observations have increased the known population of faint high-redshift AGN and motivated renewed interest in their possible contribution to hydrogen reionization. Singha et al. 2026a considered a scenario in which faint AGN contribute to the ionizing photon budget while remaining consistent with existing constraints on the ionization state of the intergalactic medium. In the models presented in the talk, AGN can account for approximately 31--75\% of the ionizing photons at $z>5$, depending in part on the assumed escape fraction from galaxies. The current observational uncertainty remains substantial because the faint AGN population is primarily constrained from relatively small JWST survey areas.

Roman offers a complementary approach through its combination of survey area and slitless spectroscopy. The WFI grism covers approximately $1.0$--$1.93\,\mu{\rm m}$, which places the Ly$\alpha$ break within the observed band for high-redshift quasars. \citet{MS2026b} simulated Roman observations of quasars over $z=7.5$--10 using synthetic QSO spectra, IGM attenuation, wavelength-dependent PSF convolution, slitless dispersion, detector and photon noise, and optimal extraction of the resulting one-dimensional spectra. Three different assumptions for the intrinsic Ly$\alpha$ equivalent-width distribution were considered in order to bracket one of the main astrophysical uncertainties in the forecast.

The simulations suggest that Ly$\alpha$-break detections should be relatively complete for brighter sources and become increasingly dependent on the assumed Ly$\alpha$ equivalent-width distribution toward the survey limit. At $J_{\rm AB}\leq23.5$, the completeness is approximately $0.95$--$1.00$ across the simulated redshift range. At $J_{\rm AB}=24.5$, completeness ranges from approximately $0.32$--$0.59$, depending on redshift and the adopted equivalent-width prior. At $J_{\rm AB}=25$, it ranges from roughly $0.08$--$0.32$. The simulations therefore indicate a relatively robust reach to about $J_{\rm AB}\simeq24.5$, with some sensitivity extending toward $J_{\rm AB}\simeq25$ depending on the underlying QSO population. At $z\simeq8$, this corresponds approximately to $M_{1450}\simeq-22.5$ to $-22.0$, an intermediate-luminosity regime between classical wide-field quasar samples and many of the faint AGN identified with JWST \citep{MS2026b}.

The study also examined contamination from M/L/T dwarfs and passive red galaxies, which can overlap with high-redshift quasars in broad-band color space. The extracted spectra were compared using a spectral angle mapper (SAM), which measures similarity in the overall spectral shape rather than absolute normalization. In the simulated sample, the QSO recovery rate was at least 98.2\% across the three adopted Ly$\alpha$ priors. No contaminants crossed the adopted classification boundary among the 8,000 brown-dwarf and 11,360 red-galaxy noisy realizations considered, corresponding to 95\% upper limits on the false-positive rate of $3.7\times10^{-4}$ and $2.6\times10^{-4}$, respectively. These values should be interpreted as performance within the assumptions of the simulations rather than as expected contamination rates for the final Roman survey \citep{MS2026b}.

An important distinction is between detecting the Ly$\alpha$ break and classifying a candidate once a spectrum is available. The former becomes difficult toward $J_{\rm AB}\sim25$ and depends appreciably on the assumed Ly$\alpha$ distribution, whereas the spectral-shape classification remains comparatively stable in the simulations. For candidates already identified through photometric selection, adding the simulated grism classification gives a posterior selection efficiency of approximately 99.8\% at $J=25$ and 96\% at $J=25.5$. The main uncertainty in forecasting the number of Roman-detected faint quasars is therefore the underlying high-redshift QSO population and its Ly$\alpha$ properties, rather than the simulated QSO--contaminant separation itself \citep{MS2026b}.

\subsection{Fabian Hervas-Peters: Beating Shape Noise with Slitless Spectroscopy: Kinematic Lensing with Roman \label{sec:FHervas}}

\begin{figure}
    \centering
    \includegraphics[width=0.9\linewidth]{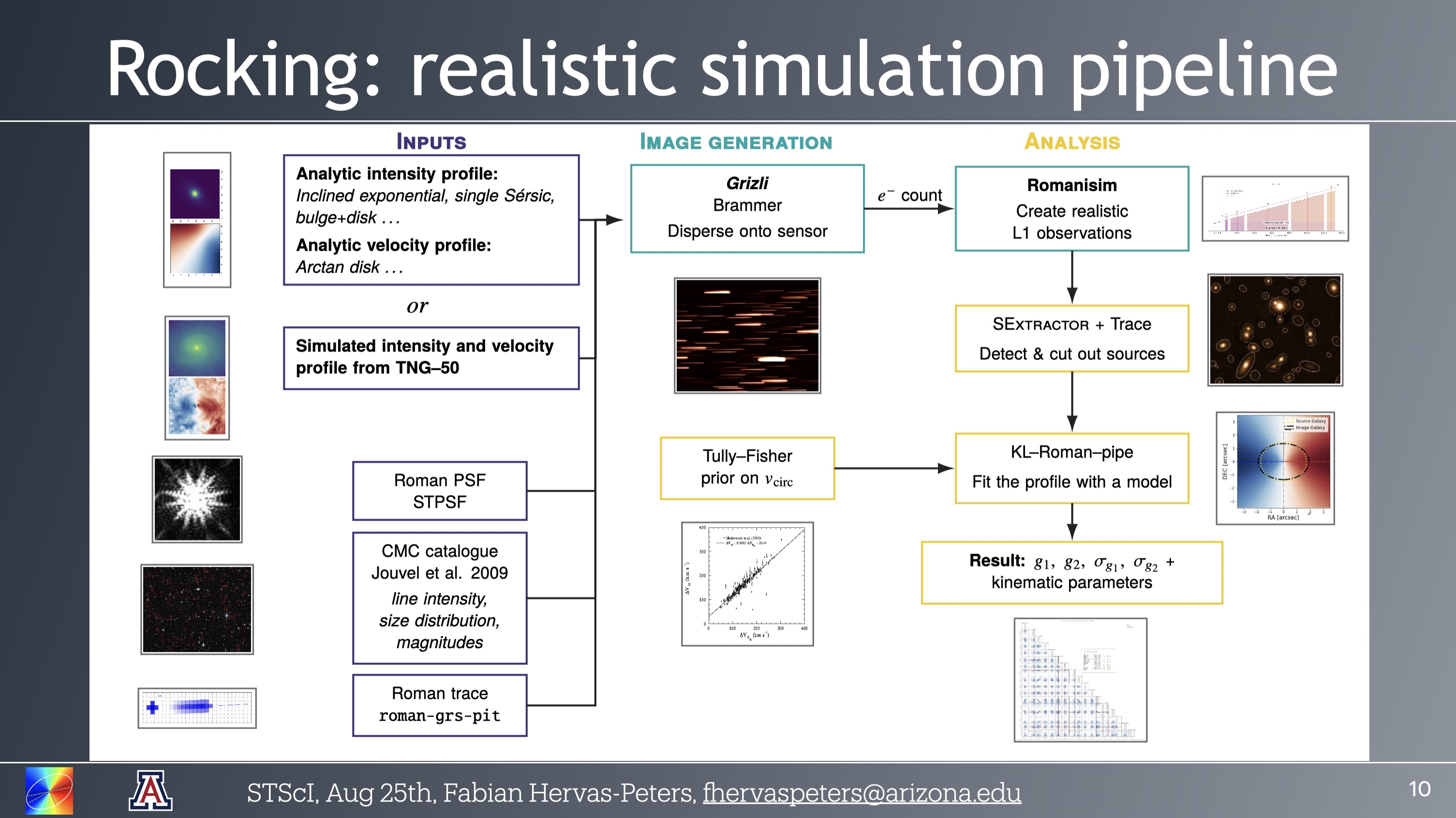}
    \caption{Description of the \textsc{Rocking} pipeline producing simulated grism dispersed images from the Roman HLWAS survey with resolved kinematics.}
    \label{fig:rocking}
\end{figure}

Gravitational lensing has established itself as a primary tool to uncover the dark matter distribution of the Universe. The traditional approach relies on averaging over multiple galaxy ellipticity estimates, such that the average ellipticity $\epsilon$ is an estimator of the gravitational shear $\gamma$, i.e.$\, \langle \epsilon \rangle \approx \gamma$. In kinematic lensing, the goal is to measure the shear components per galaxy by leveraging the information coming from the distinct imprints of the shear transformation on the intensity and velocity profiles respectively \cite{Huff_KL}. The kinematic information can be obtained via the Doppler shift of an emission line using an IFU observation, slit spectra \cite{Pranjal_KL}, or slitless spectroscopy. For the Roman space mission, the goal is to leverage the H$\alpha$ sample, using 2D emission-line maps to obtain the kinematics by jointly forward modeling the grism and direct image. This sample, found at intermediate redshifts, i.e. $z\in[0.5,2]$, fulfills the tradeoff between being sufficiently resolved by \textit{Roman} to measure kinematics, while being at a far enough distance for the large-scale structure to cause detectable lensing. This will produce improved cosmological constraints compared to the traditional weak lensing approach \cite{Xu_KL, Huang_Kl}. The fully differentiable implementation of the forward modeling framework is publicly available in the \textsc{kl-roman-pipe} library. To further validate \textsc{Rocking}, simulations based on an oversampled grizli \cite{grizli} dispersed images, later passed as electron counts to \texttt{romanisim} to include realistic noise and detector effects, allow for validation of the pipeline. The pipeline is shown in details in \ref{fig:rocking}. These simulated images help to test model performance, and will be used to identify potential sources of shear bias. In the medium tier of the Roman HLWAS, four roll angles are planned, with two exposures each \cite{WangEtAl2022}. To optimally extract kinematic information while omitting contamination, two roll angles separated by $180^\circ$ are the optimal configuration. Since the Doppler effect is not symmetric under a sign flip, if the velocity and dispersion axes are aligned, the galaxy is maximally extended; if one axis is flipped, the galaxy becomes maximally contracted. Some open questions are under investigation regarding the optimal model, the emission-line map construction, and the morphological correspondence between emission-line and total flux.

\subsection{Tri Astraatmadja: From data cube to electrons: A more realistic simulation of Roman spectroscopy \label{sec:tastraatmadja}}

\begin{figure}
    \centering
    \includegraphics[width=\linewidth]{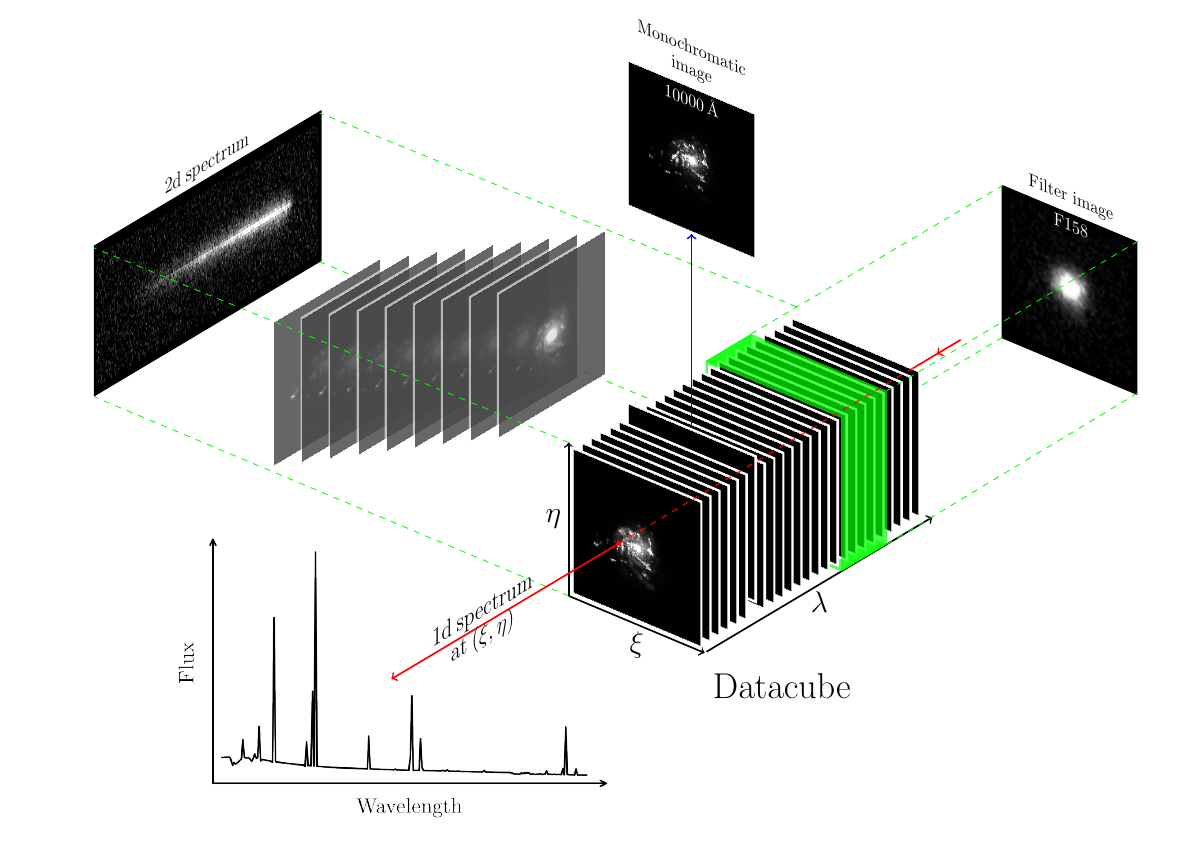}
    \caption{An illustration of a datacube and its various products through projections. A datacube is a flux mapping of a scene of the sky in two spatial axes $(\xi,\eta)$ and the wavelength axis $\lambda$. If we slice the datacube at a constant wavelength, each slice is a monochromatic image at wavelength $\lambda$ (blue arrow). If we look through a particular spatial coordinate $(\xi,\eta)$ along the wavelength axis (red arrow), we will obtain a 1d spectrum. In imaging mode, we convolve the datacube with the corresponding PSF and throughput of a particular filter, and integrate it along the wavelength range of the filter (highlighted as the green box), obtaining a filter image as recorded by the detector. In slitless spectroscopy mode, the prism is put in the optical path rather than the filter. The datacube is convolved with the PSF and the prism passband then integrated along a path describing the trace of the prism, to obtain a 2d spectral image. Note that the horizontal axis of the 2d spectral image is actually not an exact mapping of the wavelength axis, but rather it is a mapping of both the spatial axis $\xi$ and spectral axis $\lambda$ (and in some cases also $\eta$) through the trace. Illustration adapted from \cite{pon16}.}
    \label{fig:datacube}
\end{figure}

\begin{figure}
    \centering
    \includegraphics[width=\linewidth]{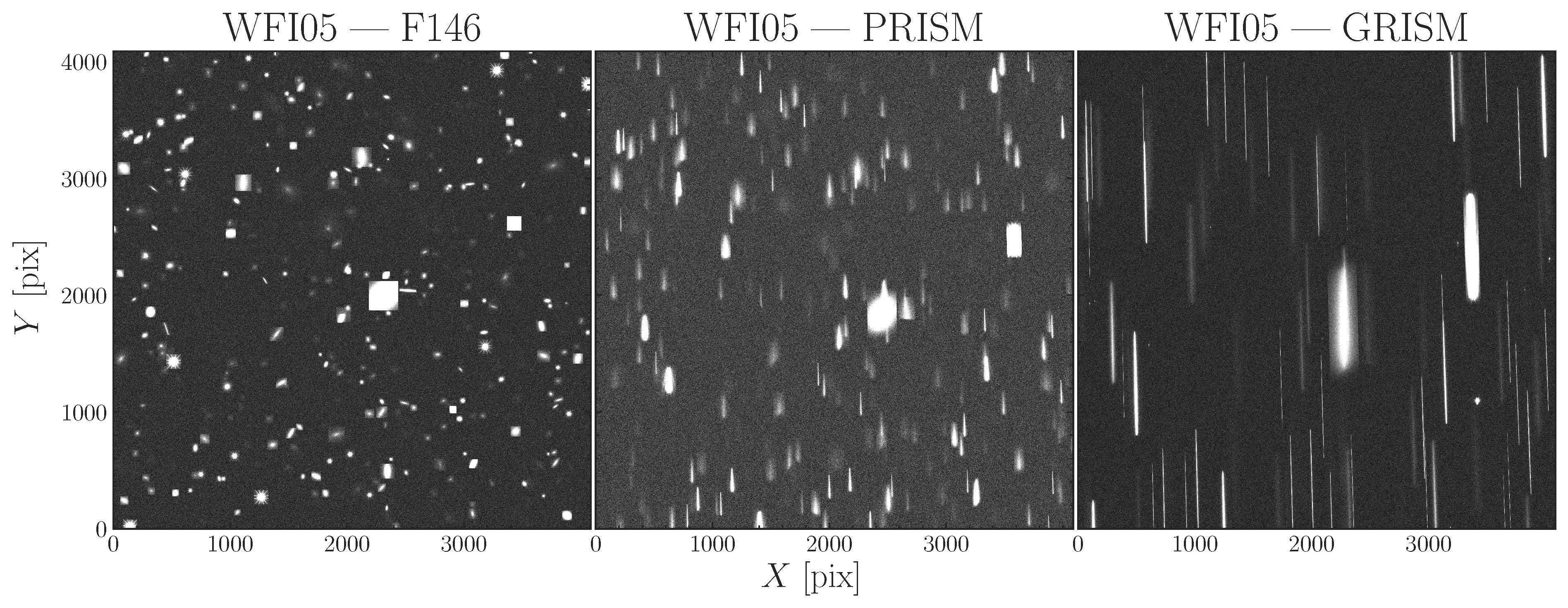}
    \caption{An example of simulated \textit{Roman} images in three filters: F146, PRISM, and GRISM, for the WFI05 detector. Stars are generated using the \textit{Gaia} catalogue, while galaxies are generated using the COSMOS catalogue. Coordinate transforms between celestial and detector coordinates for a given pointing of the telescope is facilitated by \texttt{pysiaf} \citep{sal19}.}
    \label{fig:ilia}
\end{figure}

The initial motivation to create a new kind of spectroscopic simulation came from the need to take 3d data cubes (2d spatial, 1d spectral) of scenes as input, then see how the corresponding 2d spectroscopic images would look like as observed by a given instrument, e.g. \textit{Roman}. Another motivation is to develop the algorithm that inverse the process: Given multiple spectroscopic images of the same scene but with varying roll angles, can we recover the underlying data cubes? 

Figure~\ref{fig:datacube} shows how a data cube can be projected into data space according to the observing mode. In imaging mode, the data cube is convolved with the PSF and throughput of a particular filter, then integrated along the wavelength range of the filter. In slitless spectroscopy mode, the integration is performed along a (slightly) curved path called the trace, in addition to convolution with the relevant point-spread-function (PSF) and throughput.

Without going into too much details, we can see that a 2d spectral image $\textbf{s}$ can be formed by projecting 3d datacube $\textbf{d}$ onto the detector plane by means of a projection matrix $\textbf{H}$, i.e.
\begin{equation}
    \textbf{s} = \textbf{H}\textbf{d}.
\end{equation}
The projection matrix $\textbf{H}$ encapsulates all the operations needed for the projection, among others PSF convolution; interpolation of datacube wavelength grid into pixel wavelength grid; and the rotating, shifting, and sampling of the datacube slices. The projection matrix $\textbf{H}$ is essentially a forward model, and hence it is possible to reverse the process to recover the underlying data cube \citep[e.g.][]{ast26, griggio26}.

The ingredients to simulate the detectors are field and detector dependent. The PSF, for example, are simulated using \texttt{stpsf} \citep{per11,per12,per14} for all filters, detectors, and 9 cardinal points on the detector. These are then used as lookup tables for interpolations. The throughput is taken from the latest definition from the \texttt{Roman Space Telescope Technical Information}\footnote{\href{https://github.com/spacetelescope/roman-technical-information}{https://github.com/spacetelescope/roman-technical-information}}.

While mapping between sky coordinates in the datacube onto the detector coordinates is facilitated by \texttt{pysiaf}\footnote{\href{https://roman-docs.stsci.edu/simulation-tools/additional-simulation-tools/pysiaf-for-roman}{https://roman-docs.stsci.edu/simulation-tools/additional-simulation-tools/pysiaf-for-roman}} \citep{sal19}, the Drizzle algorithm \citep{fru02} is used to redistribute the fluxes onto the focal plane of the detector. 

Suppose we have a catalog of objects in a given field of view (FOV) of a given Wide Field Instrument (WFI) detector, we can simulate the full $4088\times 4088$ image by repeating the process of mapping the beam of lights from spatiospectral coordinates $(\alpha,\delta,\lambda)$ in the data cubes onto the corresponding detector coordinates $(k,m)$, for all sources in the FOV. The results of such simulations are shown in Figure~\ref{fig:ilia}.

In addition to noisy image (with the noise generated using Poisson distribution), the simulator also produces noise-free image as part of its output. User can use this noise-free image to generate noise from their own noise model. Output of the simulator is in ASDF format, which is the format we expect from real \textit{Roman} data. The simulator is available publicly\footnote{\href{https://gitlab.com/astraatmadja/Ilia}{https://gitlab.com/astraatmadja/Ilia}}. Plans for future updates includes more noise model (e.g. $1/f$ noise) and possible extensions to other instruments.

\subsection{Sabnam Shrestha: Identifying Optically Obscured Infrared Apparent Dual Nuclei with Euclid and Roman \label{sec:SShreshta}}

\begin{figure}
    \centering
    \includegraphics[width=0.9\linewidth]{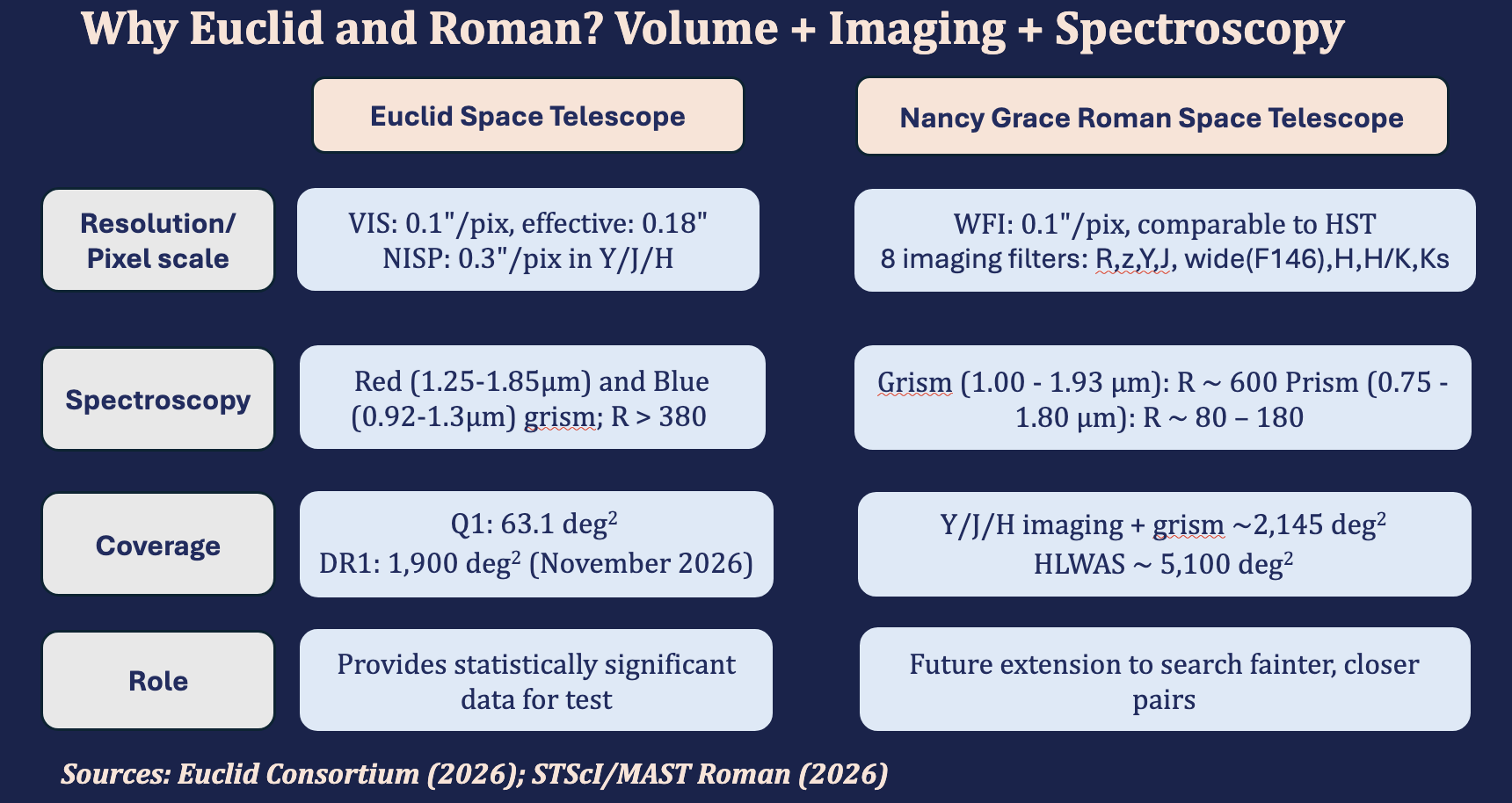}
    \caption{Euclid and Roman imaging, spectroscopy, and survey coverage relevant for searching dual nuclei candidates.}
    \label{fig:dualagn1}
    \includegraphics[width=0.9\linewidth]{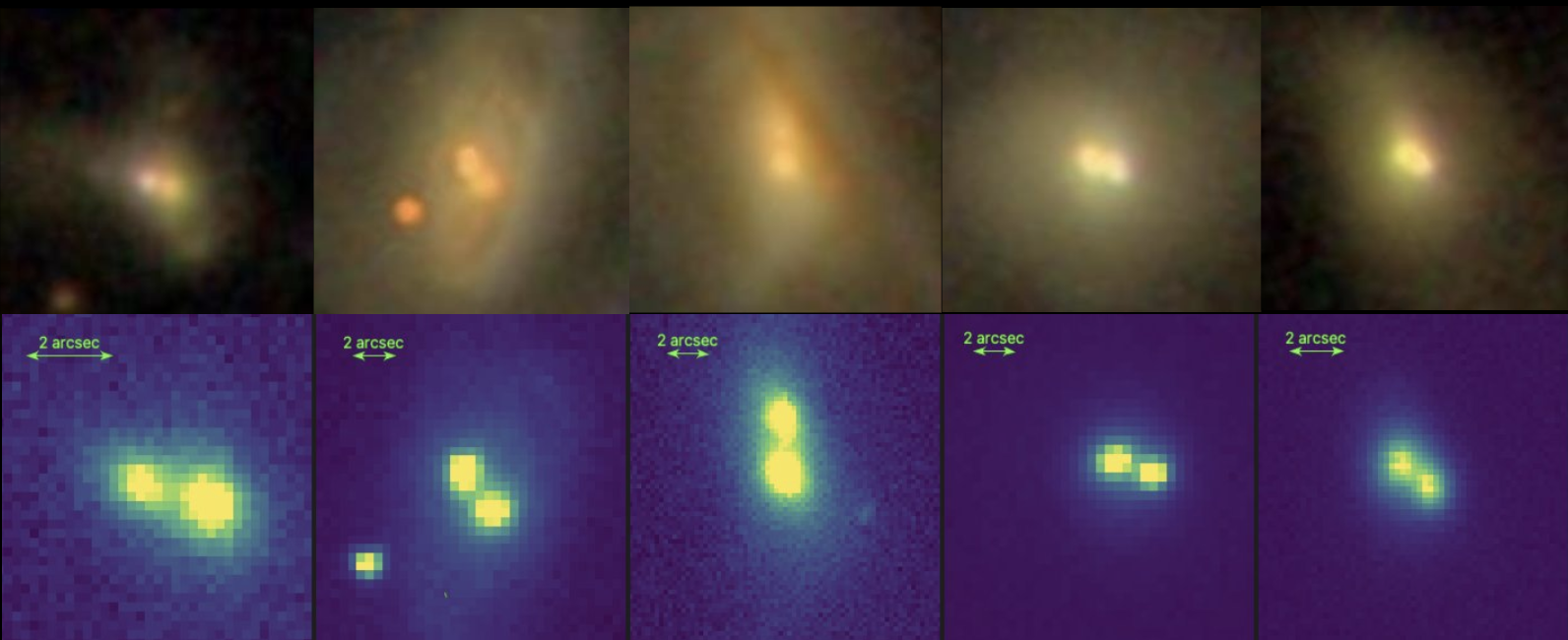}
    \caption{25" SDSS optical cutouts of merger systems (top) and corresponding UKIRT J-band images (bottom).}
    \label{fig:dualagn2}
\end{figure}

\noindent We aim to identify close dual nuclei and dual AGN with projected separations $<$30,kpc, an important population for understanding supermassive black hole growth, merger timescales, and galaxy evolution. Cosmological simulations predict a wide range of dual AGN abundances with number densities spanning several orders of magnitude depending on the subgrid physics and black hole dynamics adopted \citep{ps2025}.  
This motivates a large volume observational test. Euclid and Roman will provide the sky coverage, wide-field imaging, and slitless spectroscopy needed to build and characterize this population at the volumes needed to test the competing models (Fig.~\ref{fig:dualagn1}).\vspace{0.2cm}
\\
Within this search, we will also investigate optically obscured, infrared apparent dual nuclei. These are systems that appear as a single or blended nucleus in optical surveys but reveal hidden compact nuclear components in the near-infrared (Fig.~\ref{fig:dualagn2}). This population is motivated by existing HST imaging, where multi-wavelength comparisons reveal nuclear structure invisible at optical wavelengths but clearly resolved in the near-infrared. We will use source segmentation maps across optical and near-infrared bands to identify candidates where a single optical detection resolves into two infrared components, and follow up with grism spectroscopy to confirm their AGN nature. This work aims to build a complete close pair catalog within 30\,kpc and test dual AGN predictions.

\subsection{Isak Wold: Mapping the Ionization State of the IGM: Large-Volume Lyman-Alpha Surveys in the Roman Era \label{IWold}}

\begin{figure}
    \centering
    \includegraphics[width=0.9\linewidth]{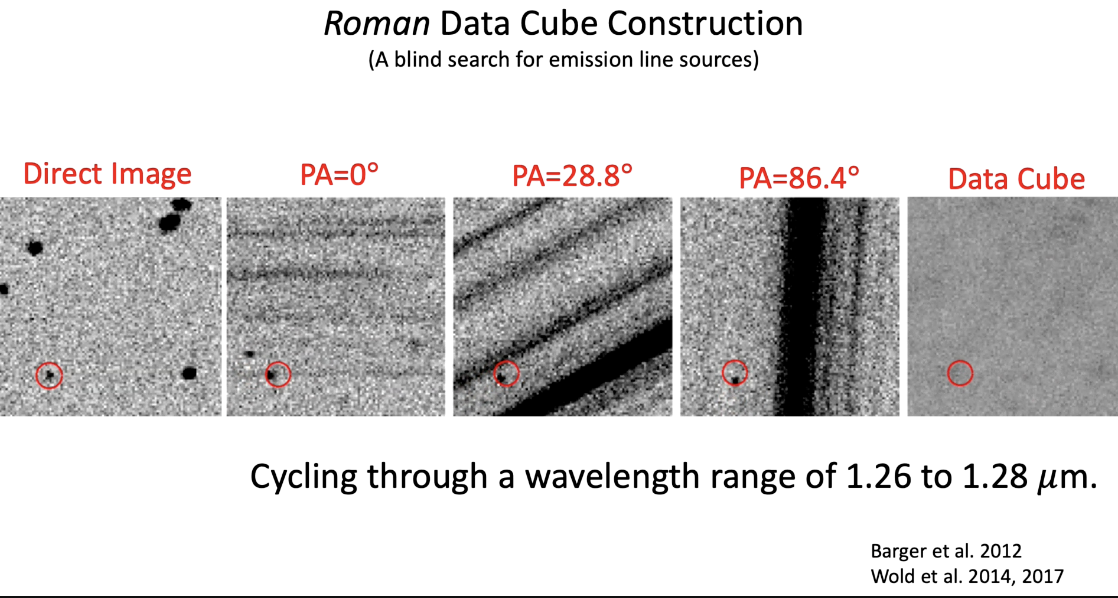}
    \includegraphics[width=0.9\linewidth]{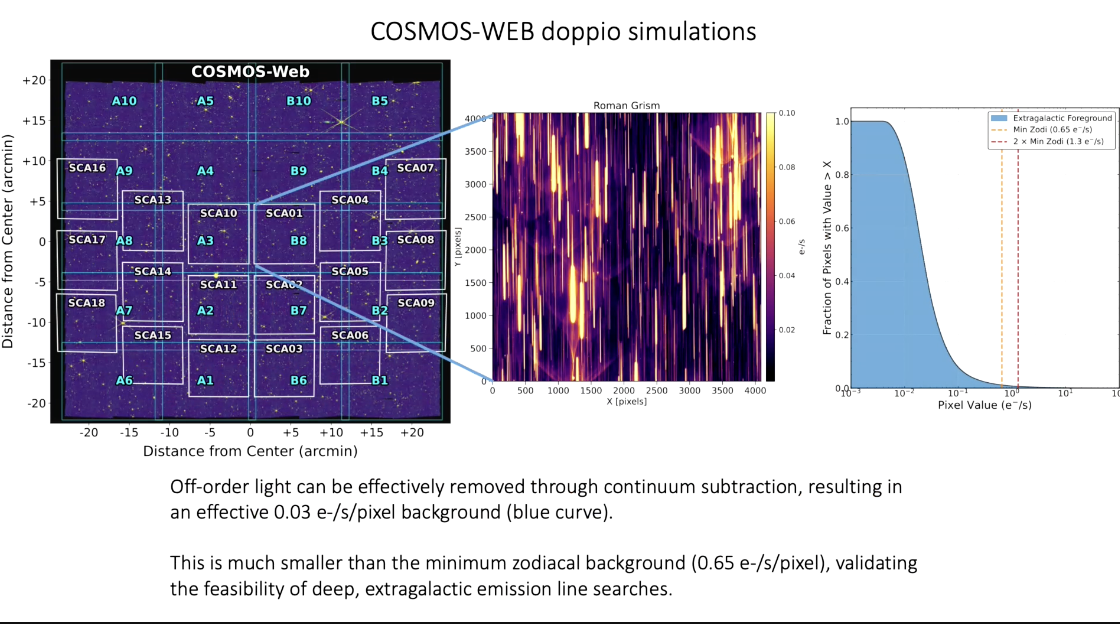}\\
    \caption{Figure from I. Wold showing that Roman will detect Ly$\alpha$ galaxies at Cosmic dawn. }
    \label{fig:IWfig1}
\end{figure}

Roman's ability to obtain deep NIR spectra over a wide FOV will allow us to measure the evolution of they Lyman alpha luminosity function at $z >7$. This will provide ionization measurements of the IGM at cosmic dawn. Doppio can model the impact of off-order spectra given an observational strategy. Working on making Doppio publically available (Figure \ref{fig:IWfig1}). 

\subsection{Sangeeta Malhotra: GRACE: The Grism Reionization and Cosmic Evolution Survey \label{sec:graceR}}

\begin{figure}
    \centering
 \includegraphics[width=0.9\linewidth]{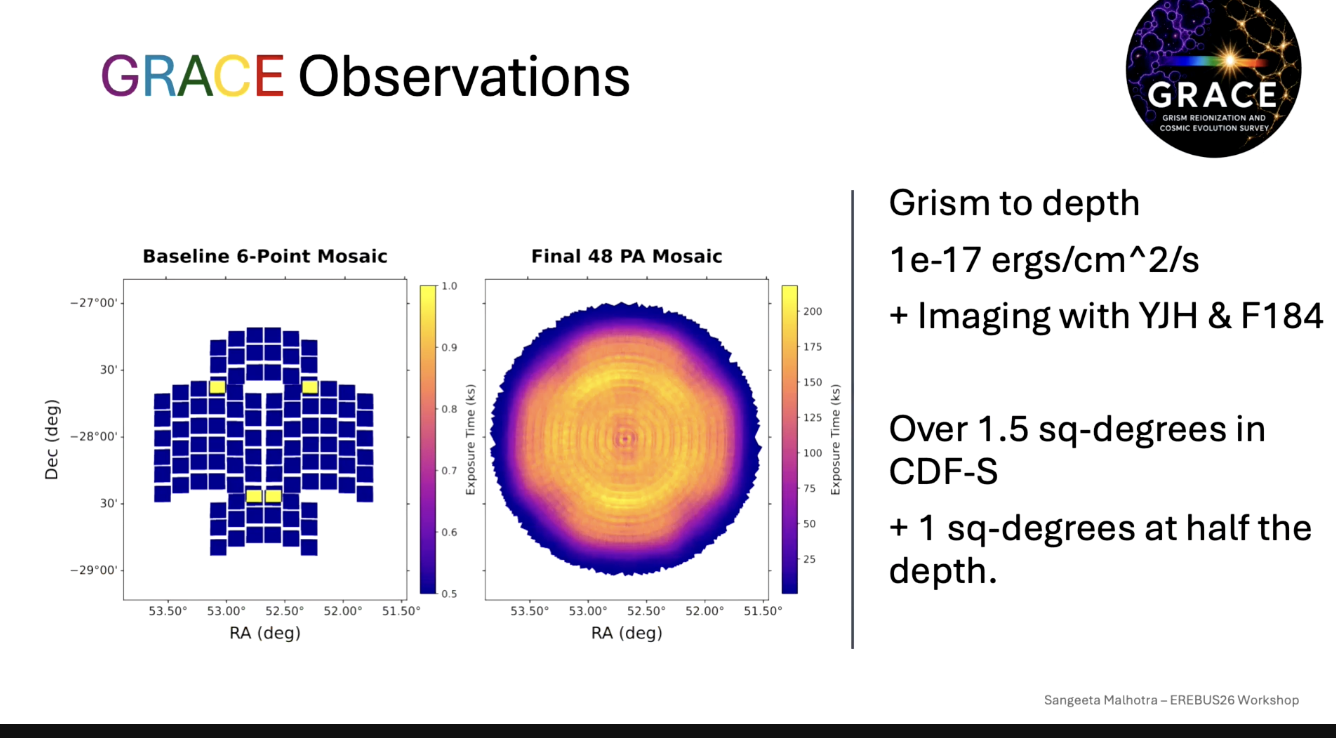}
    \caption{Progress from HST to Roman in studying reionization and cosmic evolution.}
    \label{fig:Grace}
\end{figure}

\begin{figure}
    \centering
 \includegraphics[width=0.9\linewidth]{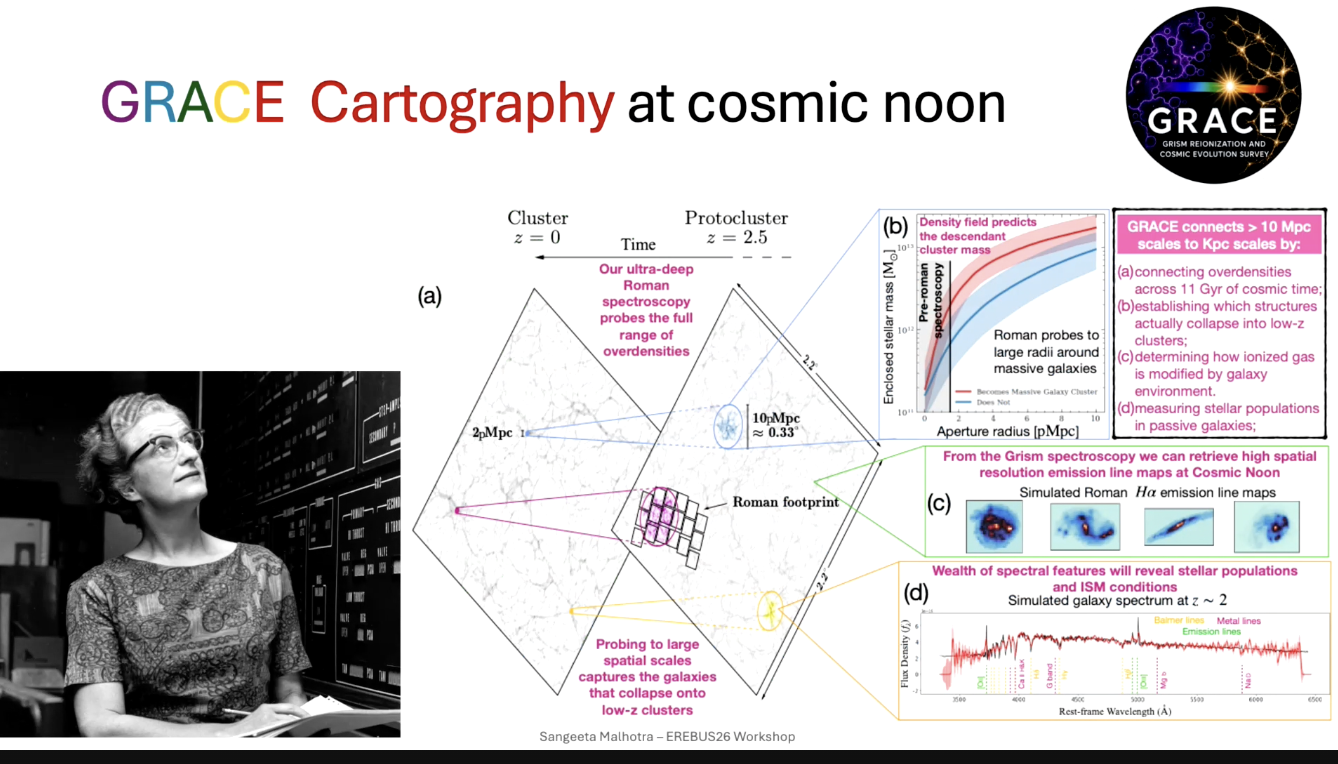}
    \caption{Figure from S. Malhotra presenting the ultra-deep Roman Spectroscopy Cycle 1 program GRACE: The Grism Reionization and Cosmic Evolution Survey, }
    \label{fig:Grace2}
\end{figure}

The combination of depth, survey area, and spectral resolution is unique and essential for a range of science cases. It enables us to reach typical, (L$^*$) galaxies during the epoch of reionization, including detecting Ly($\alpha$) at (z$\sim$ 7), while also probing typical galaxies at cosmic noon with sufficient source densities for statistical studies. The data will enable measurements of kinematic lensing, as well as studies of the metallicity and properties of dwarf stars, including Y, L, and T dwarfs. For galaxies, the combination of spectroscopy and depth will allow measurements of star-formation rates, metallicities, and stellar populations in both emission-line and passive systems, reaching down to representative rather than only the brightest galaxies. Finally, the survey will provide a powerful view of the evolution of galaxies and the role of their environments, connecting galaxy properties and their evolution to the larger-scale structures in which they reside.
From GRAPES \citep{malhotra2005} to GRACE (Figure \ref{fig:Grace} Malhotra 2026) "You have to do crazy things sometimes". 
The combination of depth, area, and spectral resolution will allow us to get galaxy properties and evolution with environment. Grace allows precision mapping of cosmic web \citep{wold2022, wold2024}(Figure \ref{fig:Grace2}). 

\subsection{Mason Footh: Dual AGN in the era of Euclid and Roman: preliminary results from past surveys and promise
for the future \label{sec:mfooth}}

Tight scaling relations like the M-$\sigma$ relation suggest a link between the growth of supermassive black holes (SMBH) and their host galaxy. In a galaxy merger, SMBH can grow by merging with the other SMBH or by accretion (i.e an active galactic nucleus or AGN). AGN triggering is predicted in mergers, with accretion rates peaking at pair separations of < 10 kpc in the later stages of the merger. One expects dual AGN (dAGN) to be relatively common, but detections have been rare and controversial. Studies disagree on dAGN fractions by orders of magnitude. Our preliminary results leveraging optical spectroscopy from the Sloan Digital Sky Survey and multiwavelength imaging from a variety of telescopes, accounting for merger stage, show a high dual AGN excess in late-stage mergers relative to a matched control dataset. These excesses ($\sim$3) are not well constrained due to small sample sizes, despite being one of the largest extant samples of late-stage galaxy mergers ($\sim$1,300 pairs). This is due to the patchwork nature of our dataset leveraging a broad variety of telescope and wavelengths. The low resolution of SDSS imaging restricts us to a redshift range of 0.02 to 0.06, further hampering the search for AGN in galaxy mergers due to the low merger rate in the nearby universe.  
\par The Roman high latitude wide area survey will give us the resolution ($\sim$.1”) and spectra necessary to identify late-stage merging galaxy kinematic pairs out to a z of around 2-3, and the volume($\sim$2400 deg$^2$) necessary to have a much larger sample size. This is where most simulations suggest the merger rate should peak. Alongside that, the so called 'OHNO' BPT diagnostic diagram will allow us to identify AGN. Targeted observations from other telescopes like JWST or serendipitous archival observations from telescopes like Chandra will be able to probe the accretion of the SMBH and constrain the dAGN fraction at lower redshift far beyond our current capabilities. This will also allow us to constrain merger rates overall, and via low-z observations, potentially the SMBH dynamical friction timescale which is wildly unconstrained as well.

\subsection{Day 2: Mainak Singha: Roman spectroscopy data challenge demo \label{sec:msinghaD}}
Materials for the third \textit{Roman}-SPQR data challenge can be found \href{https://github.com/sangeetak20/RomanSpectraDataChallenge/tree/main/DataChallenge3}{here}.

\subsubsection{Prism data analysis demonstration}
Using the simulator discussed in Section~\ref{sec:tastraatmadja}, a scene containing stars and galaxies have been simulated in F146 and prism filters for the WFI05, and in four roll angles. The images for one roll angle is shown in Figure~\ref{fig:ilia}. The notebook demonstrates a method to extract a (uncalibrated) 1d spectrum and finding the trace of the spectrum by centroiding along columns for each row.

\subsection{Day 2: Discussion Session}


\textbf{Era of Euclid and Roman: How can statistical galaxy
studies and large-scale structure science advance?}

\begin{itemize}
\item We will discover more subpopulations of galaxies.
\item We may obtain a more complete fossil record of galaxy evolution.
\item We may discover new populations and phenomena, such as dark stars, pair-instability events, and Pop III supernovae.
\item Large-scale structure studies will expand to include connections along the cosmic web and connections with voids.
\item The definition of environment may change.
\item There will be more connections between galaxy evolution and cosmology.
\item This may require advances in cosmological simulations, which will need both large volumes and high resolution.
\item Simulations will need to improve to enable better comparisons with observations in the large-data era.
\item The galaxy simulation and large-scale structure modeling communities are somewhat siloed from the observational community.
\item Observers want libraries of spectra across all wavelengths, but simulation teams may not be able to provide these products, and vice versa.
\item Environment is a key aspect. HST/JWST fields that cover one Ly$\alpha$ bubble may become important. These fields could help identify the physical mechanisms driving galaxy evolution.
\end{itemize}

\textbf{What are the key priorities, expected challenges, and
connections to cosmology?}

\begin{itemize}
\item Science outcomes depend on having science-ready spectra.
\item Producing these products is a major undertaking.
\item Funding this work and providing career advancement for those who do it remain challenges.
\item A major legacy to the community will be the availability of these products.
\item We need good mechanisms for developing treasury and archival programs.
\item The issue is not necessarily access to calibration programs, but having the right data available.
\item Photometry may be needed to provide the calibration required for specific science topics.
\item Roman data will be public, and much of the Cycle 1 support was focused on analysis.
\item Appreciation for those developing and maintaining pipelines is growing, but perhaps not fast enough.
\end{itemize}

\textbf{How do these surveys complement detailed, targeted
observations with HST and JWST?}

\begin{itemize}
\item Multiepoch surveys can be used to monitor transients and variable AGN, enabling DDT proposals.
\item Roman can provide large samples of LRDs and other similar populations, while MIRI can extend their SEDs into the mid-IR.
\item HST is well suited for deep UV imaging and can identify many emission-line galaxies that may be Ly$\alpha$ emitters or LyC leakers.
\item JWST IFU observations can provide spatially resolved systems for studying dynamics and evolution.
\item Roman and Euclid may identify only one emission line, while JWST can detect additional lines.
\item The complementarity also works in the other direction: HST and JWST can provide information that complements Roman.
\item X-ray data are also needed.
\end{itemize}

\textbf{What are synergies with upcoming Rubin data?}

\begin{itemize}
\item Yes, there are strong synergies, particularly for transients.
\item Rubin provides temporal sampling over large areas.
\item Roman does not perform as well at $z<0.5$.
\item To cover the full temporal parameter space, we need observations from all of the major telescopes.
\end{itemize}

\textbf{Key insights:}

\begin{itemize}
\item Interest centers on environment studies and a successful launch. This suggests that readiness is as important as the science goals.
\item Joint imaging and spectroscopy provide a major opportunity, followed by larger sample sizes and improved statistics for rare objects.
\item Calibration, data volume, and data reduction are major concerns.
\item Pipelines and computing support may therefore need to be priorities early in the survey era.
\end{itemize}

\section{JWST pipeline}
\subsection{Jo Taylor: JWST/WFSS Pipeline and Data Products Review \label{sec:JTaylor}}
Three instruments have WFSS capability on JWST: NIRISS (Near Infrared Imager and Slitless Spectrograph, 0.8 - 2.2 microns), NIRCam (Near Infrared Camera, 2.4 - 5.0 microns), and MIRI (Mid Infrared Instrument, 5.0 - 14 microns). NIRISS and NIRCam allow for perpendicularly dispersed light, while MIRI can achieve the same result by changing the position angle.
Unified model for NIRISS, NIRCam, and MIRI.  Resource include JDox, PipelineReadTheDocs; JWebbinars, JWSTPipelineNotebooks, JWSTDataAnalysisTool (JDAT) Notebooks, help desk. 

\begin{figure}[h!]
    \centering
    \includegraphics[width=0.9\linewidth]{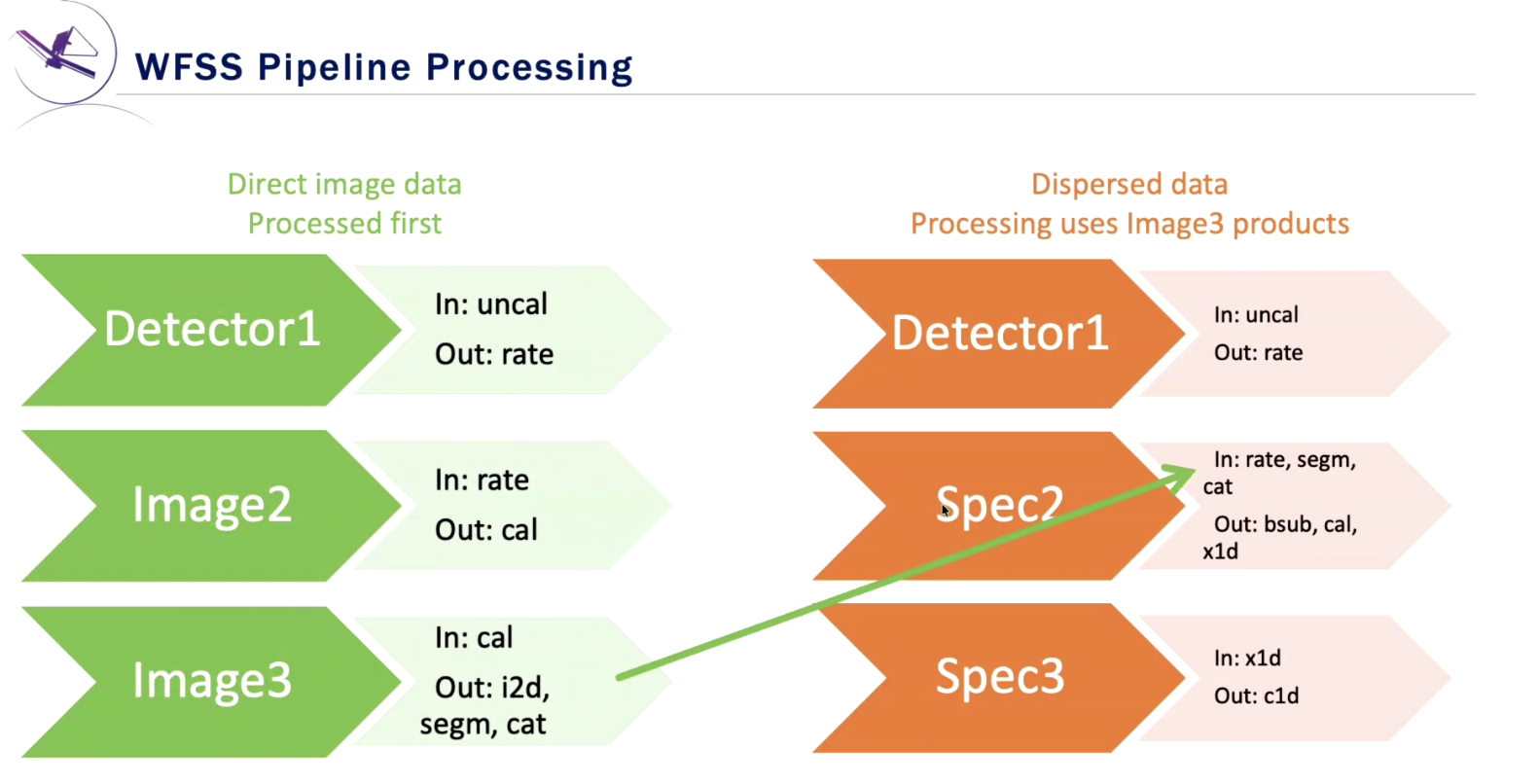}
    \caption{Figure from J. Taylor presentation showing a schematic representation of the main steps of the pipeline.}
    \label{fig:wfssPipe}
\end{figure}

Figure \ref{fig:wfssPipe} shows a schematic representation of the main steps of the pipeline. Direct imaging data are processed first and result in an i2d file (combined and resampled dithered images), source catalog, and segmentation map. The default parameters used to create source catalogs and segmentation maps ingested into MAST (Mikulski Archive for Space Telescopes) are a best attempt to create a usable estimate of the scene. Users are encouraged to re-run the Image3Pipeline to create custom catalogs tailored to their science needs. Dispersed data are then processed and require providing the Image3 outputs as input (to know which sources to extract).
Detailed information about each step, including the code, are public and available through the STScI webpages: \href{https://jwst-pipeline.readthedocs.io/en/latest/index.html}{ReadTheDocs} and \href{https://jwst-docs.stsci.edu/jwst-science-calibration-pipeline#gsc.tab=0}{JDox}.

Recent pipeline upgrades include bug fixes for contamination correction, source catalog creation with deblend=True, extraction of a subset of sources, allowing a user-defined background mask, a rehaul of cal, x1d, and c1d file structure, creation of a new intermediate file (bsub) by default, and improved background template scaling method tied to the latest CRDS background templates \citep[][]{Noirot2025}.

Pipeline improvements expected to be released in February 2027 include: applying contamination correction in default MAST products, new iterative polynomial contamination modeling method (similar to what grizli uses), new "multi-band" contamination modeling method, new contamination output products, improved grism WCS, and moving flux calibration to after 1D extraction.

Possible pipeline improvements that may be in the Februrary 2027 release include: contamination modeling for saturated sources, utilizing multiple orients in contamination, and improved background correction in crowded fields. 

If JWST/WFSS users have any questions about the pipeline or products, want to report issues, or request any changes, the best way to advocate for change is to contact the \href{https://stsci.service-now.com/jwst}{STScI Help Desk}.

\section{Galaxies, AGN, LRDs and Quenching}

\subsection{Roman Fernandez Aranda: Through the PRISMS: a survey of more than 80 hours of slitless JWST MIRI WFSS \label{sec:RFernandez}}

\begin{figure}
    \centering
    \includegraphics[width=0.99\linewidth]{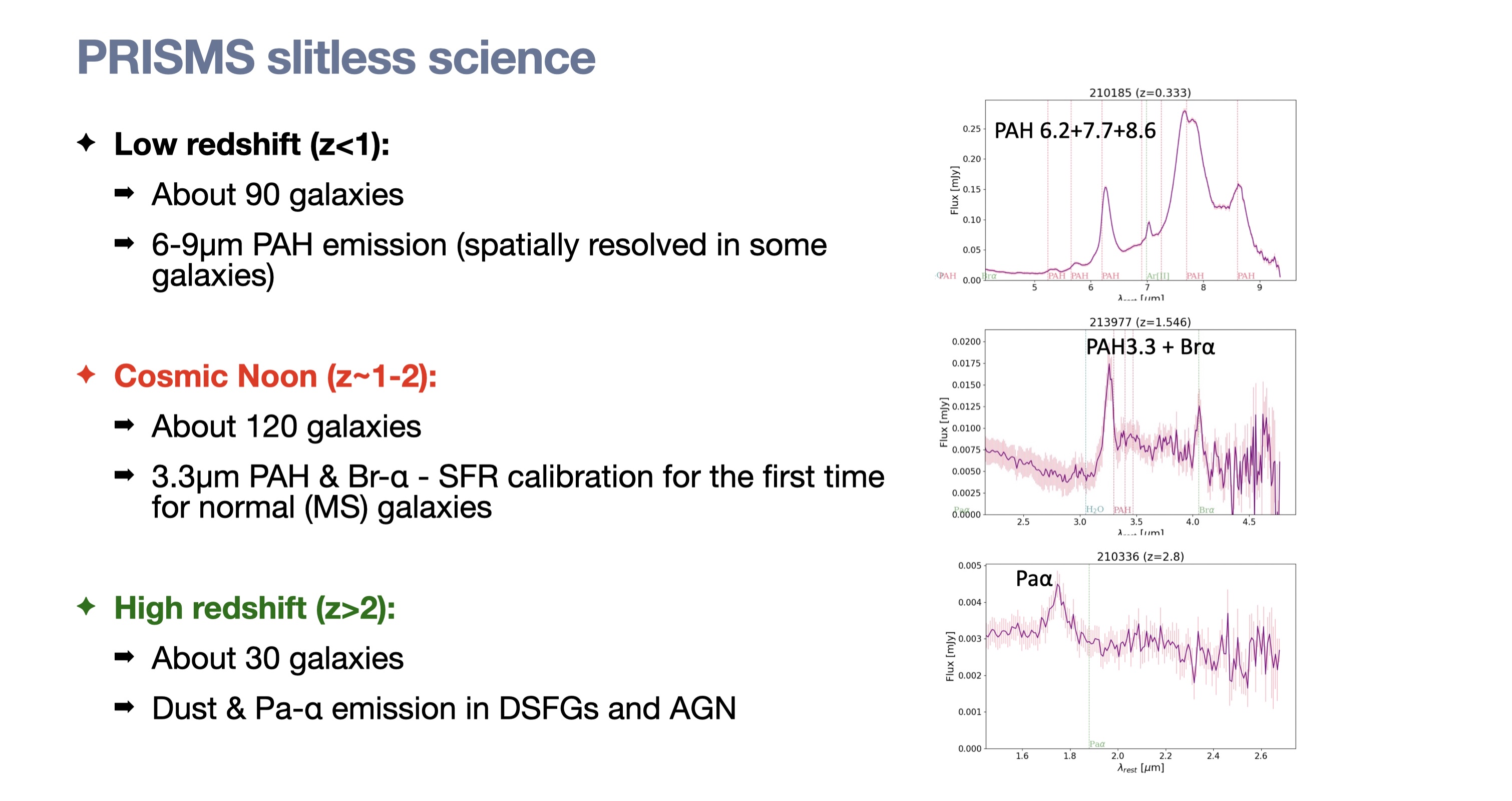}
    \caption{Science cases at different redshifts for the PRISMS MIRI WFSS survey}
    \label{fig:RFA_PRISMS}
\end{figure}

JWST MIRI's low-resolution spectrograph (LRS) provides parallel slitless spectroscopy from 5-14 $\mu$m in the large field of view of the MIRI imager, making it well suited for a variety of science cases. However, until Cycle 5 slitless spectroscopy has not been offered as a prime mode for MIRI, so it remains unexplored. We recently obtained more than 80 hours of deep (1$\sigma$=0.4-0.7 $\mu$Jy) LRS data with WFSS spectra alongside in several cosmological fields, as part of the PRImordial galaxy Survey with MIRI Spectroscopy (PRISMS), a JWST Cycle 4 program (PID 8051). PRISMS WFSS data has rich scientific potential, with hundreds of galaxies detected, on three redshift bins (Figure \ref{fig:RFA_PRISMS}): (i) low redshift (z$<$1), where we detect PAH emission features at 6-9$\mu$m, in some galaxies spatially resolved; (ii) Cosmic Noon (1$<$z$<$2), when the 3.3$\mu$m PAH feature remains scarcely explored and we detect it alongside Br-$\alpha$, providing a PAH and SFR calibration for the first time for normal (MS) galaxies; and (iii) high-redshift (z$>$2), with hot dust and Pa-$\alpha$ emission observed in dusty-star forming galaxies and AGN. These results highlight the unique potential of PRISMS and MIRI WFSS to probe dust, star formation, and nuclear activity across cosmic time.

\subsection{Sarah Kendrew: MIRI Wide-Field Slitless Spectroscopy of Cosmic Noon Galaxies in the Hubble Ultra-Deep Field \label{sec:skendrew}}

\begin{figure}
    \centering
    \includegraphics[width=0.9\linewidth]{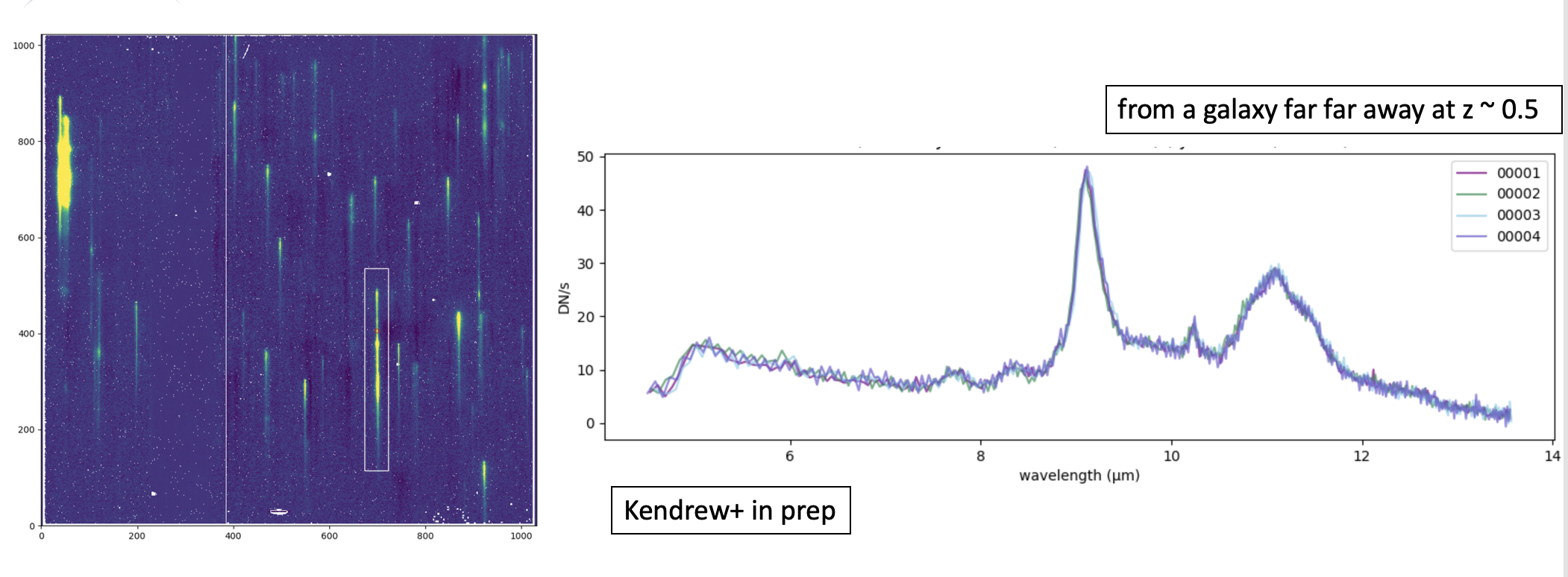}
    \caption{MIRI GTO data on HUDF showing the impressive potential for a blind search of PAHs with MIRI WFSS, specifically the 3.3 $\mu$m PAH.}
    \label{fig:SKfig1}
\end{figure}

 This presentation showed the first results from a JWST/MIRI Wide-Field Slitless Spectroscopy (WFSS) survey of the Hubble Ultra-Deep Field (Fig.~\ref{fig:SKfig1}), demonstrating the capabilities of this new observing mode and its calibration \citep{kendrew2026}. Spectra for 47 galaxies with secure spectroscopic redshifts were used to study the 3.3 $\mu$m PAH feature across to redshifts spanning the peak epoch of cosmic star formation. The measured PAH luminosities are consistent with established correlations with total infrared luminosity and SED-derived star-formation rates, despite the current calibration uncertainties. These results demonstrate the potential of MIRI WFSS to efficiently characterize dust-obscured star formation and galaxy evolution during Cosmic Noon.

\subsection{Preethi Nair: Constraining AGN fractions in merging galaxies and the implication on supermassive black hole
occupation fractions \label{sec:pnair} }


\begin{figure}[h!]
    \centering
    \includegraphics[width=0.9\linewidth]{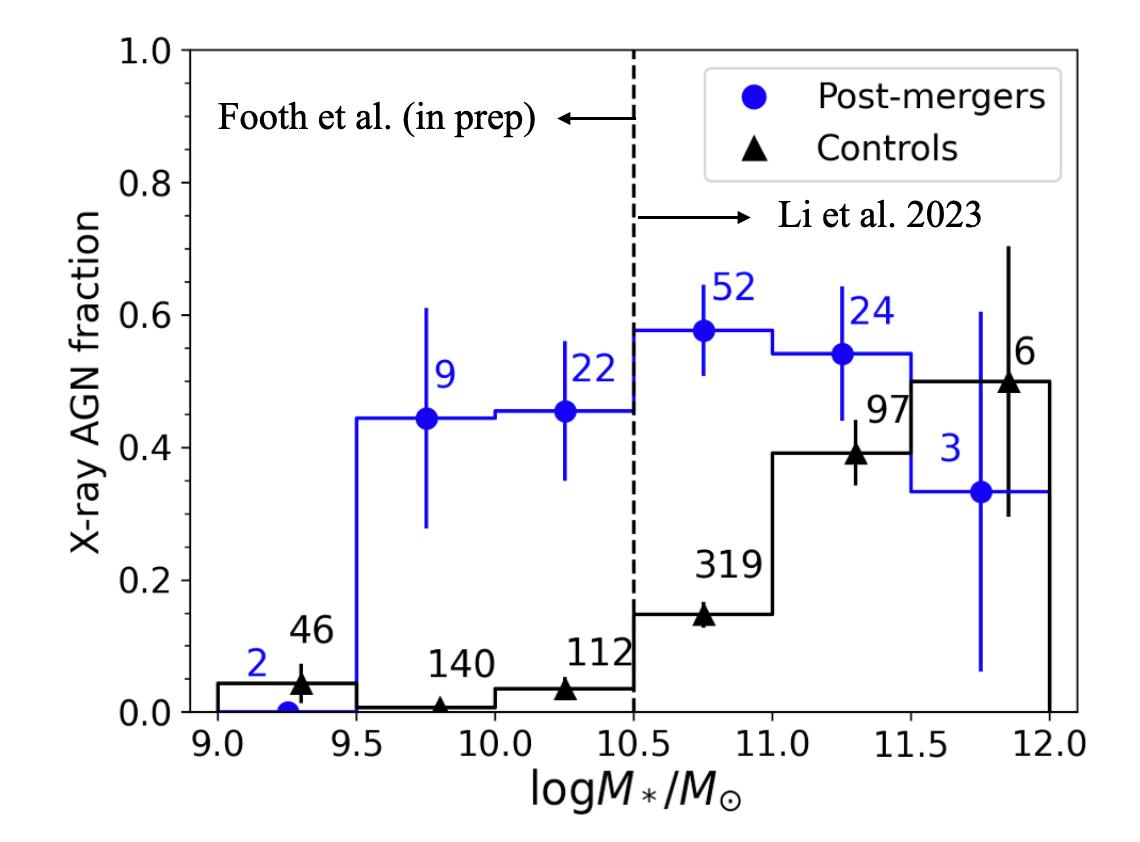}
    \caption{X-ray AGN fraction as a function of stellar mass for post-merger remnant galaxies}
    \label{fig:PMXlogM}
\end{figure}

Post-mergers are rare objects probing the end coalesced stage of galaxy mergers. They have high optical AGN fractions, high post-starburst fractions, and span the range in the color-mass space suggesting they are good probes to characterize supermassive black hole accretion, growth, coevolution and the quenching of galaxies \citep{li2023,2023A&A...674A..24C}. As post-mergers are shocked systems, their optical AGN natures are uncertain. In a previous work, we showed that the X-ray AGN fractions in massive post-mergers ($>10^{10.5} M_{\odot}$) are at least $\sim$55$\%$ and 2--3 times higher than a mass and redshift matched control sample with a greater than 6.5 sigma significance \citep{2023ApJ...944..168L}. This suggests that post-mergers could be a good probe of SMBH occupation fraction. In two currently ongoing follow-up works, we have extended this analysis down to 10$^{8}$ solar masses. We find that X-ray AGN fractions are nearly constant down to 10$^{9.5} M_{\odot}$ and then drops strongly, favoring a high mass seed hole model. See Figure~\ref{fig:PMXlogM}. (The figure does not show the sample from our ongoing Chandra work targeting galaxies with mass less than 10$^{9.5} M_{\odot}$.) Roman and Euclid will allow the extension of this work to higher redshifts by enabling the identification of post-mergers.

\subsection{Annie Giman: NIR Spectroscopy of Optically Luminous Type 1 and 2 AGN \label{sec:agiman}}
NIR spectroscopy of type 1 and type 2, optically luminous QSOs, with multi-wavelength coverage. The type 1 QSOs are from the Palomar-Green sample and the type2 selected to match the [OIII] luminosity and redshift distribution of the PG QSOs. Giman et al. in prep shows NIR spectroscopy in JHK with Gemini GNIRS showing potential differences in the properties of H$_2$ flux as measured in ro-vibrational lines. The spectra are then median combined following the methods of \citep{Glikman2006} to produce NIR templates for the Type 1 and Type 2 QSOs. 

Wide field slitless spectroscopy (WFSS) with Roman and Euclid will provide statistically significant samples, and robust composite templates will be of particular importance for AGN identification and classification. Investigating these relationships in the nearby universe is key for grounding studies at higher redshifts and ultimately understanding the role of co-evolution and obscuration across cosmic time as a function of mass and environment.


\subsection{Zhaoran Liu: Little Red Dots in the Grism Era: Census and Black Hole Mass Constraints from Spectroscopic Variability \label{sec:Zliu}}
LRDs are an abundant population of compact, red sources found predominantly at $z \approx 4$--$9$ \citep[e.g.,][]{Matthee2024ApJ, Kocevski25, deGraaff26}. They exhibit broad Balmer emission and point-like morphology in the rest-frame optical, seemingly indicative of AGN activity. Yet they are unlike any well-studied AGN population. They are remarkably X-ray faint \citep{Yue24, Sacchi25}, while virial mass estimates based on local calibrations \citep[e.g.,][]{GreeneHo05} imply ``overmassive'' black holes with masses comparable to those of their host galaxies, pushing the implied black hole mass density at $z\sim5$ to uncomfortable levels \citep[e.g.,][]{Inayoshi24}.

Variability is a key missing piece of the LRD puzzle. AGN are variable on virtually every observable timescale and across the electromagnetic spectrum, making variability a powerful and independent diagnostic of the nature of LRDs. The JWST TWINKLE Survey (JWST Cycle~4, PID~7404; PIs: R. P. Naidu, J. Matthee, and J. Chisholm) is a NIRCam slitless spectroscopy program specifically designed to systematically probe LRD variability. Using TWINKLE observations of 27 LRDs, \citet{liu26_twinkle} found no detectable variability over observed-frame baselines of 150--238 days, providing direct evidence that LRDs differ fundamentally from typical AGN (see also \citealt{Stone26}).

Indeed, there are growing arguments that the broad lines of LRDs arise from non-virial processes such as radiative transfer in dense gas \citep[e.g.,][]{Naidu25, Rusakov2026, Sun26_decomposition}, and that these systems host lower-mass black holes accreting at or above the Eddington limit \citep[e.g.,][]{King24, Lupi24, Umeda25, Greene26, Liu26, Naidu26_ge_lrd}. If so, virial estimates are not applicable to this class of objects, and a larger sample is needed to characterize their variability and to trace their evolution across cosmic history --- particularly at lower redshift \citep{Kapoor26_eiger}.

Roman will be key to probing this evolution at $z = 0$--$2$. The HLWAS will deliver a low-redshift census: its slitless spectroscopy traces LRDs across this entire range, while the HLTDS will sample their light curves over rest-frame years, placing direct constraints on their nature and central engine.

\subsection{Olivier Gilbert: JWST Constraints on Stellar Mass - Black Hole Mass Relation at $5<z<6.5$ \label{sec:ogilbert}}

With the help of NIRCam imaging and slitless spectroscopy, the COSMOS-3D JWST survey uncovered the largest uniformly selected spectroscopic sample of broad-line AGN (BLAGN) at $z=5-6.5$. These BLAGN have a significant overlap with the mysterious "little red dots" (LRD) discovered with JWST. Both BLAGN and LRD seem to be dominated in the rest-frame UV by their host galaxy, and their rest-frame optical by an AGN. The $M_\mathrm{BH}-M_\star$ relation shows no strong correlation, and, if virial estimations hold \citep{Juodzbalis2026, Scholtz2026}, black holes are overmassive in the early universe (Gilbert, in prep). There are also preliminary hints that the central engines of these BLAGN and LRD could be different than in local AGN, possibly having a dense gas envelope, as suggested by \citet{Rusakov2026}. This shows promising results for the use of upcoming surveys (e.g., Roman) to select large samples using a combination of slitless spectroscopy for line detections and photometry.

\subsection{Swetha Sankar: When Jets Don’t Quench: Near-Infrared H2 in Star Forming Low-Excitation Radio Galaxies \label{sec:ssankar}}

\citet{Sankar2026} show Gemini/GNIRS near-infrared spectroscopic observations of eight low-redshift ($z<0.1$) blue low-excitation radio galaxies (BLERGs)---a rare subset ($\sim2.5\%$ of low-excitation radio galaxies; LERGs). These star-forming BLERGs exhibit significant warm H$_2$ emission traced via ro-vibrational transitions at $T \sim 2,000$--$4,000$~K. BLERGs span a broad range of mass-normalized warm H$_2$ luminosities ($L_{\rm H_2}/M_\star$), comparable to radio-emitting early-type galaxies, yet without a clear positive dependence on radio power. Instead, the strongest H$_2$ emission preferentially occurs in morphologically disturbed and advanced-merger systems, while compact radio sources ($\lesssim 20$~kpc) remain plausible sites of localized jet--ISM interaction. 

Together, these results suggest that merger-driven processes---including tidal shocks, gas inflows, and disturbed interstellar medium conditions---are the dominant drivers of warm molecular gas excitation in BLERGs, although localized jet-driven heating may contribute in individual systems. The compact radio morphologies, gas-rich hosts, and rarity of BLERGs are consistent with a short-lived evolutionary phase in which radio AGN activity coexists with an interaction-driven, molecular-rich interstellar medium prior to the onset of large-scale maintenance-mode feedback. Spatially resolved spectroscopy and higher-resolution radio imaging will be essential to disentangle the relative roles of mergers and jets in regulating the molecular gas of jet-mode AGN.

\subsection{Gaël Noirot: The color evolution and quenching timescales of comic noon galaxies through the green valley  \label{sec:gnoirot}}

\citet{Noirot2022a} presents a photometric and spectroscopic study of the color evolution and quenching timescales of comic noon galaxies through the green valley.

\begin{itemize}
\item Multi-band CANDELS photometry and deep archival HST grism spectroscopy ($>10$ orbits of G102 and G141) were used to derive the rest-frame NUVrK colors and delayed-tau SFHs of a sample of ~250 $>10^8~M_\odot$ galaxies at cosmic noon ($z=1.0-1.8$).

\item The color-age relation of these galaxies is fitted which allows to estimate how their rest-frame colors change with time at these redshifts. This is used to track galaxies as they move along different quenching channels.

\item It is found that only fast quenchers ($\tau < 0.5$~Gyrs) got enough time to fully quench and reach the red sequence at these redshifts.

\item It is measured that it takes about 1 Gyr for such galaxies to cross the green valley at these redshifts (Figure~\ref{fig:greenvalley1}), at an instantaneous rate of $\sim 0.8$mag/Gyr at the bottom at the green valley (independent of green valley boundary definitions).

\begin{figure}[ht!]
    \centering
    \includegraphics[width=0.9\linewidth]{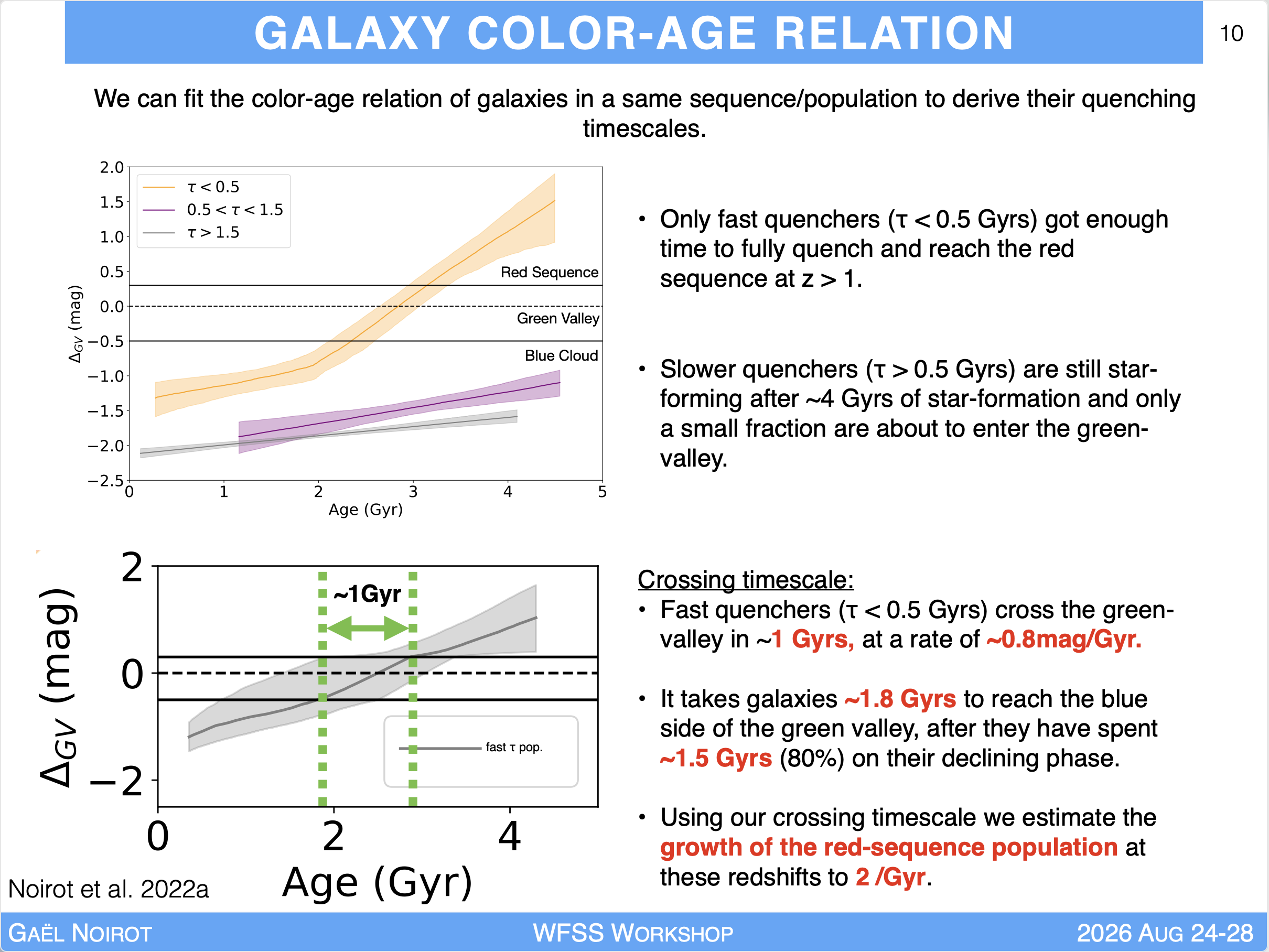}
    \caption{This slide shows the fit to the color-age relation for three quenching pathways: fast (orange), intermediate (purple), and slow (grey) quenchers (top panel). The bottom panel is a fit to the fast population only, incorporating several fitting assumptions. We find a green-valley crossing timescale of $\sim 1$~Gyr, at a rate of $\sim 0.8$mag/Gyr at the bottom at the green valley independent of green valley boundary definitions.}
    \label{fig:greenvalley1}
\end{figure}
\newpage

\item The crossing timescale and the green valley and red sequence number densities are used to estimate a growth of the red-sequence population of a factor of 2 per Gyr at these redshifts, in excellent agreement with the growth of the quiescent SMF at the redshifts.

\item The H$\alpha$ vs stellar continuum size evolution as galaxies move along their quenching track is derived based on the grism data alone. Only a very tentative trend of smaller H$\alpha$ sizes compared to the continuum as a function of time since onset of quenching is found. This perhaps indicate outside-in quenching \citep[][Figure~\ref{fig:greenvalley2}]{Noirot2022b}.

\begin{figure}[ht!]
    \centering
    \includegraphics[width=0.9\linewidth]{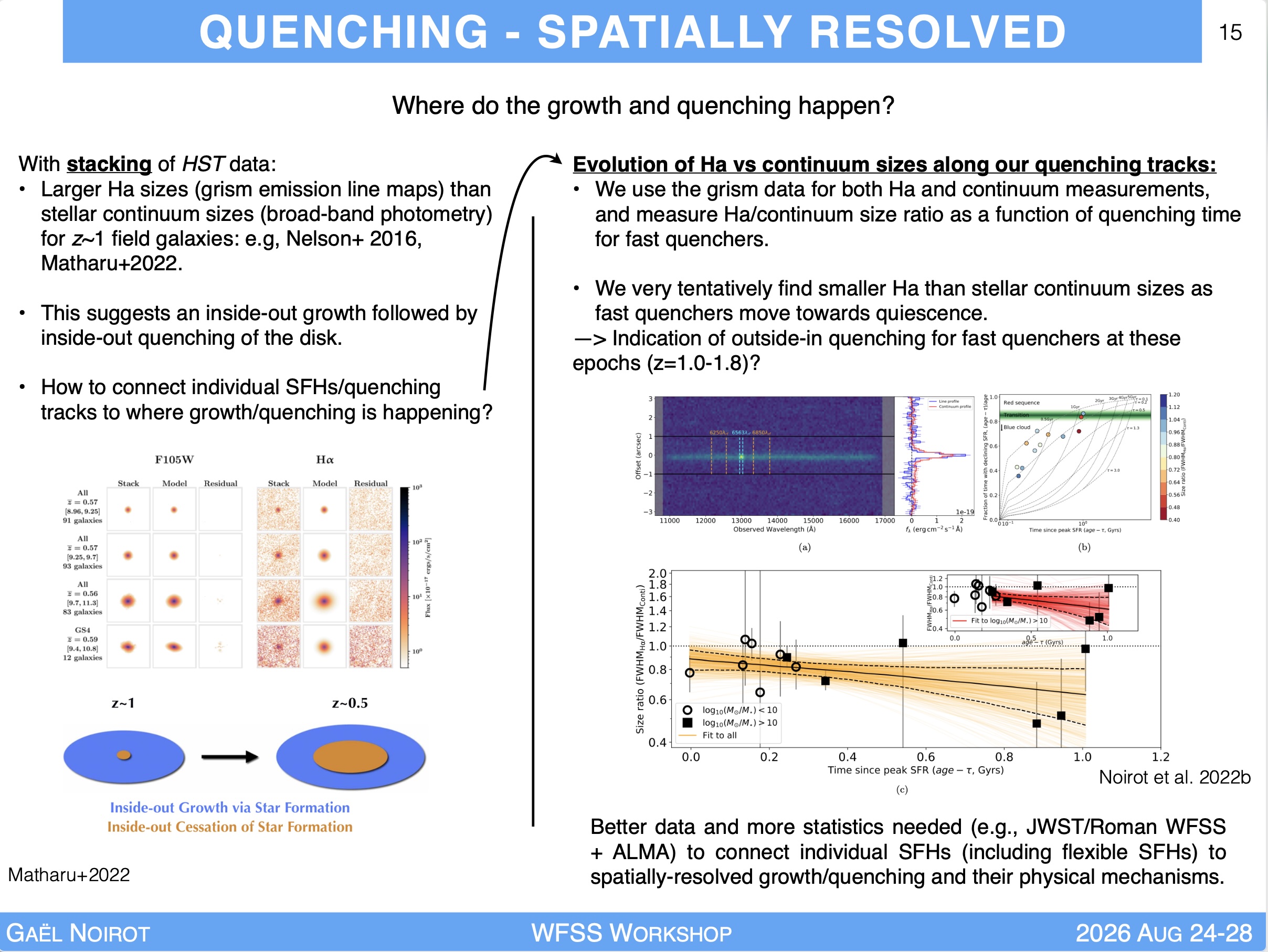}
    \caption{This slide shows the evolution of the H$\alpha$ vs continuum sizes along the fast quenching track for a sub-sample of galaxies, as measured directly form the HST grism data. A tentative trend of smaller H$\alpha$ than continuum sizes as quenching progresses is found, but more statistic is needed to conclude.}
    \label{fig:greenvalley2}
\end{figure}
\item Better data and more statistics are needed (e.g., JWST/Roman WFSS + ALMA) to connect individual SFHs to spatially-resolved growth/quenching and their physical mechanisms.
\end{itemize}
\newpage

\subsection{Rachel Plesha Demo \label{sec:rpleshaD}}

This demo goes through an example of running the jwst pipeline on NIRISS WFSS program in a Jupyter notebook. The demo is run using the absolutely "bleeding edge" version of the pipeline as STScI is actively working to improve particularly the contamination step and the output products. The demo is broken down into four sections plus a bonus section:
\begin{enumerate}
    \item The notebook starts by downloading data directly from MAST using astroquery. Sometimes this can time out with WFSS data, so to get around that, we need to submit the requests in batches of five.
    \item The next section is about running the spec2 pipeline.
    \begin{itemize}
        \item The pipeline runs on association files. For the spec2 pipeline, that file contains the science array, the direct image, the source catalog, and the segementation map.
        \item The pipeline takes a dictionary of different pipeline steps which then in turn have a dictionary of their individual parameters.
        \begin{itemize}
            \item In this example, we modify some of the default parameters for the extract\_2d product. There was some discussion about the different ways to limit which products are saved and it was brought up that limiting to a specific magnitude limit might be more beneficial than just limiting to the Nth brightest source (100 by default due to file size limits).
            \item The example also turns on the wfss\_contam step, which is currently skipped by default, but is planned to be turned on by default in the next pipeline build.
        \end{itemize}
        \item There is an example of running spec2 with the contamination correction on and with it turned off. This is for visualization purposes later in the demo, but should not be necessary with the next pipeline release.
    \end{itemize}
    \item The demo next shows file content and visualization for each of the spec2 output products.
    \begin{itemize}
        \item bsub (background subtracted, full-frame grism data)
        \item simul (Full-Frame Simulations used for Contamination)
        \item extract\_2d (2D spectral cutouts); not a default output product, but it is helpful to look in the demo.
        \item cal (contamination subtracted 2D spectrum); actively being modified from the current jwst release to include contamination information
        \item x1d (extracted 1D spectrum); actively being modified to include contamination information
    \end{itemize}
    \item The last main section of the demo highlights some of the custom ways to modify the pipeline to enhance your specific scene.
    \begin{enumerate}
        \item Provide a custom background mask.
        \begin{itemize}
            \item create a mask that is the same shape as the image. We show an example by creating a segementation map and giving that as input.
            \item feed the saved masked back into the "bkg\_subtract" step using the "wfss\_mask" parameter and setting the "wfss\_maxiter" step to zero.
        \end{itemize}
        \item Extract specific sources by providing a list of RA and Dec coordinates and/or a list of source IDs (as defined by the source catalog).
        \item Customize WFSS contamination by using:
        \begin{enumerate}
            \item Iterative polynomial fitting (similar to the grizli approach).
            \item Multi-band imaging cube
        \end{enumerate}
        \begin{itemize}
            \item Contamination via iterative polynomial fitting and multi-band image cube are still works in progress, but a preview is included in the demo.
            \item Contamination models are intended to be extendable/customizable depending on what level of contamination the scene needs.
            \item Example residuals of both models as well as the current default flat spectrum model are shown.
        \end{itemize}
    \end{enumerate}
    \item There is some additional content shown to highlight some more areas of the jwst pipeline:
    \begin{enumerate}
        \item customize the image3 pipeline to reflect parameters that best define your field. This is essential to getting good results in the spec2 pipeline as it expects well-defined sources and segementation map. It may be best to run this twice with different parameters for point sources and extended sources as they are treated differently.
        \item run the spec3 pipeline to combine the 1D spectra. Examples are shown of these data products.
        \item A final snippet is shown to visualize the extraction cutouts on the full-frame image by using columns in the x1d file.
    \end{enumerate}
\end{enumerate}

\subsection{Andreea Petric Demo \label{sec:apetricD}}
Petric showed a MIRI WFSS demo to be released before February 2027 as a James Webb Space Telescope Data Analysis Tool Notebooks (JDAT) notebook. 

\subsection{Ned Molter: WFSS JWST Products and Pipeline Feedback Discussion \label{sec:PFeed}}

\textbf{Do you use default Stage 2 or 3 products?}
\begin{itemize}
    \item By a show of hands, most people in the room indicated they do not currently use the STScI WFSS pipeline after Detector1.
\end{itemize}

\textbf{What doesn’t work well with current STScI WFSS products?}
\begin{itemize}
   \item To compare different WFSS pipelines (e.g. jwst pipeline vs grizli) and having standards for folks to feel confident in the calibrations and algorithms.
   \item Some people have had problems with extract1d, especially with different behavior across modes. More flexibility with extract1d would be nice, i.e. more algorithms for extraction.

\end{itemize}

\textbf{What calibration are you doing outside the pipeline and why?}
\begin{itemize}
    \item Many folks are using grizli, because it was giving reasonable results first
    \begin{itemize}
        \item grizli provides resolved 2D emission line maps
        \item However, grizli does not support all science use cases, e.g. absorption spectra presents different challenges.
    \end{itemize}
    \begin{itemize}
        \item People use grizli/own tools because they’re used to it and need to be convinced it’s easy to switch.
        \item STScI needs to clearly demonstrate accuracy of pipeline and algorithms, i.e. show benchmarks.
    \end{itemize}
    \item NIRCam Calibration areas of improvement:
    \begin{itemize}
        \item order 2 wavelength calibration for science purposes. \textit{Note: there was wavelength calibration data obtained in cycles 4 and 5}
        \item background correction
        \item trace
        \item sensitivity curves
        \item Follow-up question: when was the last time you used STScI pipeline and reference files and found calibration was insufficient?
    \end{itemize}
\end{itemize}

\textbf{Would product changes, or new products, make analysis easier?}
\begin{itemize}
    \item There was discussion about how best to support early career researchers in both funding and future career prospects when they develop their own pipelines. This led to a broader question of if we should expect early career researchers to write their own pipelines and/or do their own calibration in their limited time?
    \item STScI is in a more unique position that they will be able to support the code unlike other code bases maintained by only a few people.
    \item  Additional pipeline/analysis tools that allow people to take advantage of having 2D spectra, e.g. forward modeling of 2D spectra and convolving template set with 2D image
\end{itemize}

\textbf{How easily have you been able to get technical support?}
\begin{itemize}
    \item Sometimes when people send a help desk question they are just sent JDox links and question is not sufficiently answered.
    \item Some people prefer cold calling email, but this requires you know someone on the inside.
    \item Might help if STScI folks are more embedded in community programs. This is already happening to some degree.
    \item Are there existing slack channels or forums where people are having discussion about WFSS? 
    \begin{itemize}
        \item No one in the room currently has any channels like this.
        \item ST has JWST office hours every other week, a representative from each instrument team is always present. This compliments the help desk by offering in-person support.
    \end{itemize}
    \item The ETC youtube tutorials are very useful, but they should be linked in JDox.
\end{itemize}
\textbf{What is the best way for users to learn about the pipeline?}
\begin{itemize}
    \item Demos you can work through. This in part already exists, although we would be open to areas we  can expand on:
    \begin{itemize}
        \item \href{https://spacetelescope.github.io/jwst-pipeline-notebooks/notebooks/NIRISS/WFSS/JWPipeNB-niriss-wfss.html}{https://spacetelescope.github.io/jwst-pipeline-notebooks/notebooks/NIRISS/WFSS/JWPipeNB-niriss-wfss.html}
        \item \href{https://spacetelescope.github.io/jdat_notebooks/}{https://spacetelescope.github.io/jdat\_notebooks/}
        \item Caveat about tutorials: technical explanations are sometimes divorced from tutorials themselves (say on jdox). These need to be included together.
    \end{itemize}
    \item Hands-on workshops, like this one. However, the discussions at this conference has been useful, especially hearing all the different fields present their issues. However, we didn’t have this conversation until Cycle 6. We may need to meet with community more often.
    \item People are including AI agents with code releases. This may be something to look into including from STScI.
\end{itemize}    

\textbf{What other JWST/WFSS pipelines or calibrations exist?}
\begin{itemize}
    \item Helen Fraser: OU WFSS pipeline (public in 2027)
    \item grizli (includes custom trace calibration)
    \item NGDEEP custom trace and wavelength calibration
\end{itemize}


\textbf{Misc Feedback}
\begin{itemize}
    \item  For both STScI and research groups: The code and methodology used to calibrate data/create reference files should be published/shared publicly.
\end{itemize}

\section{Spatially Resolved Spectroscopy}
\subsection{Helen Fraser: Developing a publicly accessible code for WFSS in crowded and confused fields. \label{sec:HFraser}}
JWST can be exploited to study the nearby ISM (interstellar medium) with exquisite spectral detail to the spatially resolved chemistry and properties of the star- and planet- forming material.

Ice tracks the largest reservoir of molecular material. It plays a pivotal role in star formation feedback processes, especially as it acts as a sink, removing coolants such as CO from the gas phase. 
Almost all CNO budget of ISM. So we want to know how much, what type, and where the ice is. 

Though not the original extragalactic use envisaged for WFSS modes with JWST, NIRCAM (and soon MIRI WFSS) are proving invaluable tools to study molecular clouds and star forming regions; the added multiplex advantage of slitless spectroscopy means data can be obtained at a fraction of the telescope time and overheads compared to surveying the same objects one pointing at a time.

\begin{figure}
    \centering
    \includegraphics[width=0.9\linewidth]{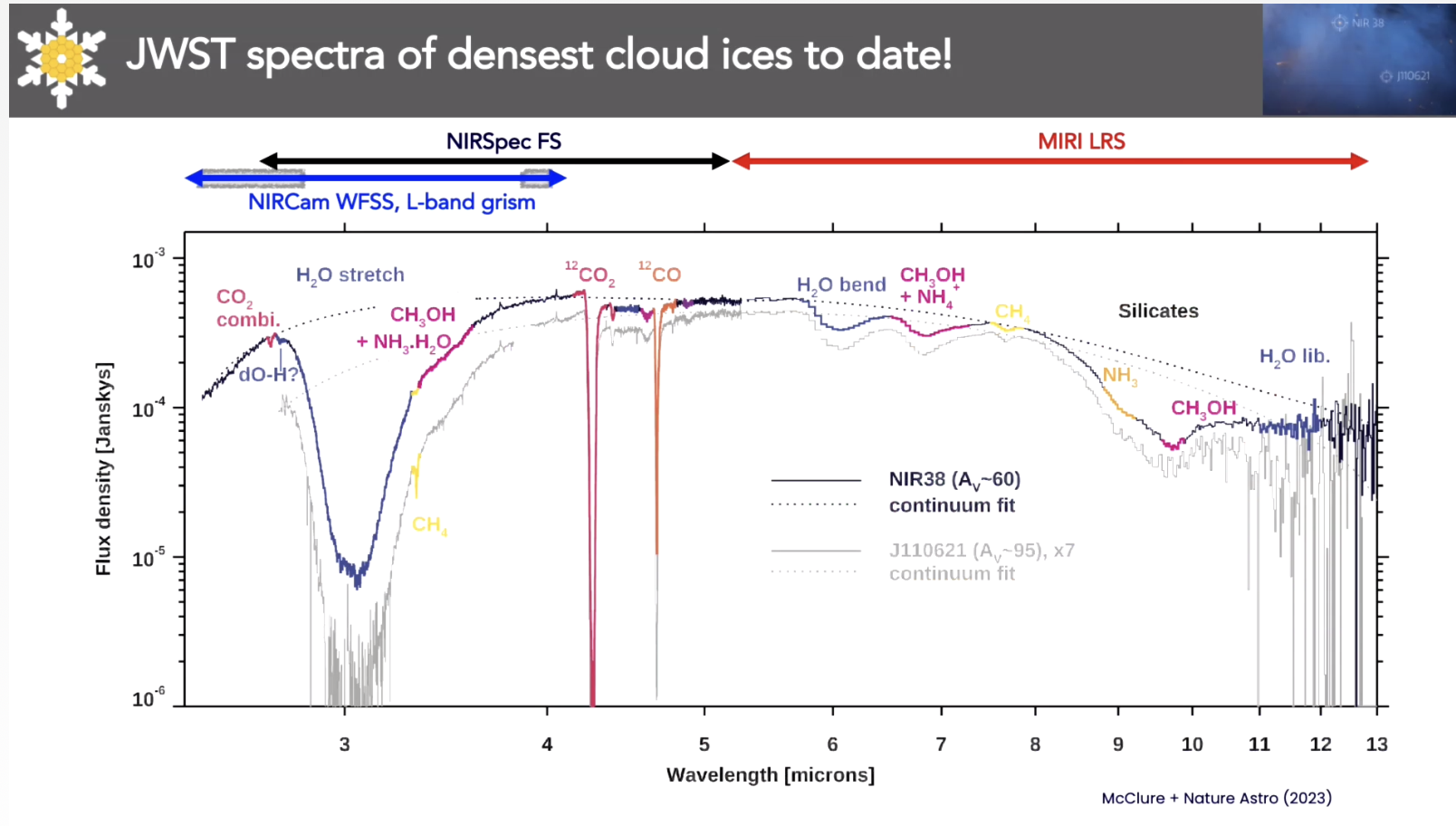}
    \includegraphics[width=0.9\linewidth]{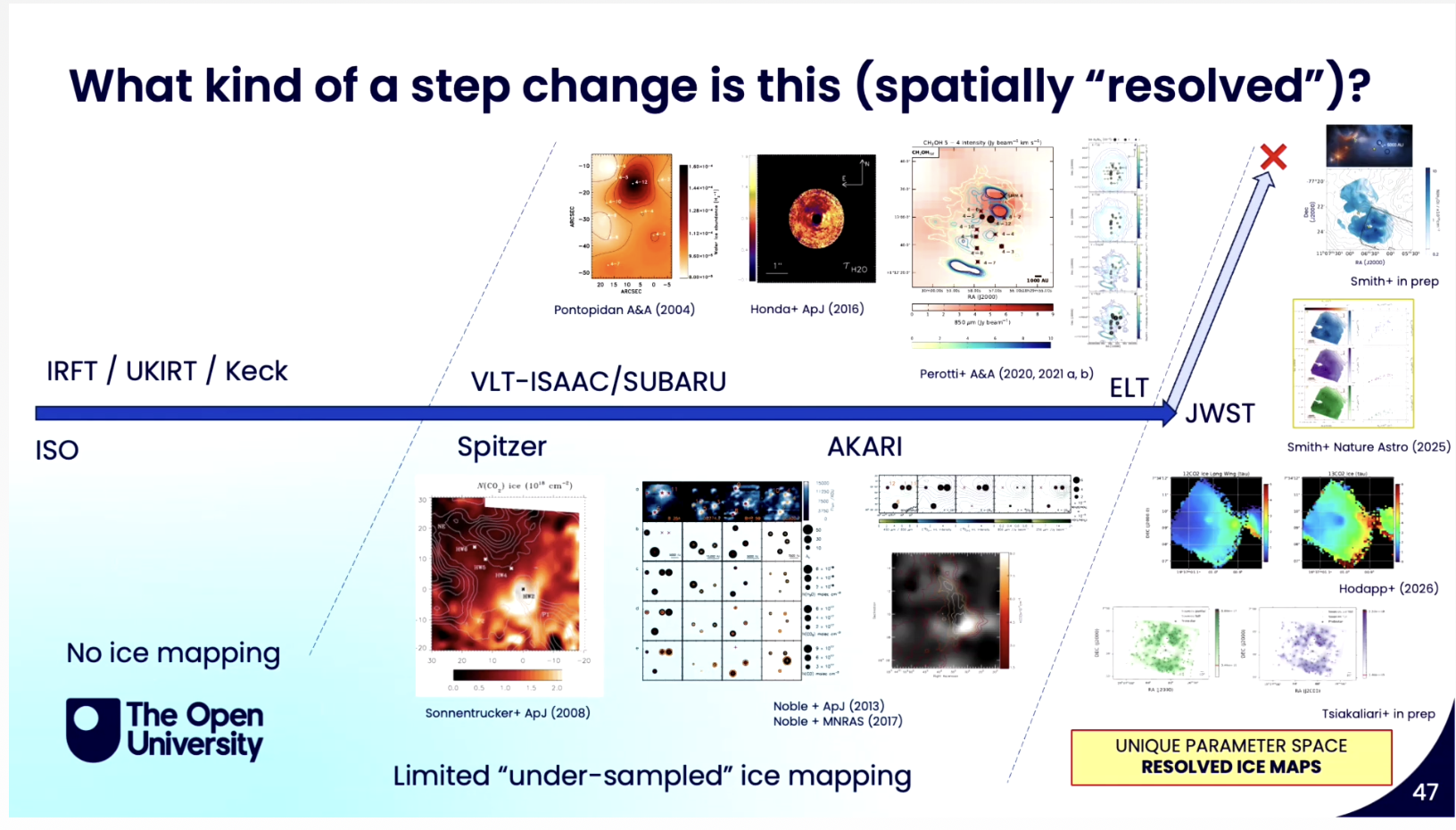}\\
    \caption{Top: example of ice spectra. Bottom: Figure representing the overall strategy of spatially resolved observations of ice.}
    \label{fig:IceF1}
\end{figure}

A significant challenge for WFSS observations on ice is that it is only seen in absorption (Figure \ref{fig:IceF1}) and the presentation showed the development of an accessible code for WFSS in crowded and confused fields which successfully recovers ice absorption.

  By building through WFSS data reduction in the IceAge ERS and NIRCAM GT programmes, it has been a journey learn where official JWST pipelines work and fail when trying to extract spectral data from confused and crowded nearby fields. However it can be demonstrated that NIRCAM and NIRSPEC spectra of the same objects are entirely commensurate - with some careful data reduction (not official pipelines). Through a series of molecular clouds at different evolutionary stages including B335 and Cha I (so far) it has been possible to build on knowledge gained from ice mapping with Spitzer and AKARI (plots of solid material abundance in a molecular cloud  as a function of spatial distribution). With the very first results from SPHERE-X available now too, it is possible to compare JWST AKARI and Sphere-X ice maps - with NIRCAM there is a demonstrable resolution advantage - spatially and spectrally. 

\subsection{Vicente Estrada-Carpenter: The Curious Case of the Missing UV Clumps \label{VEstradaC}}

\begin{figure}
    \centering
    \includegraphics[width=0.9\linewidth]{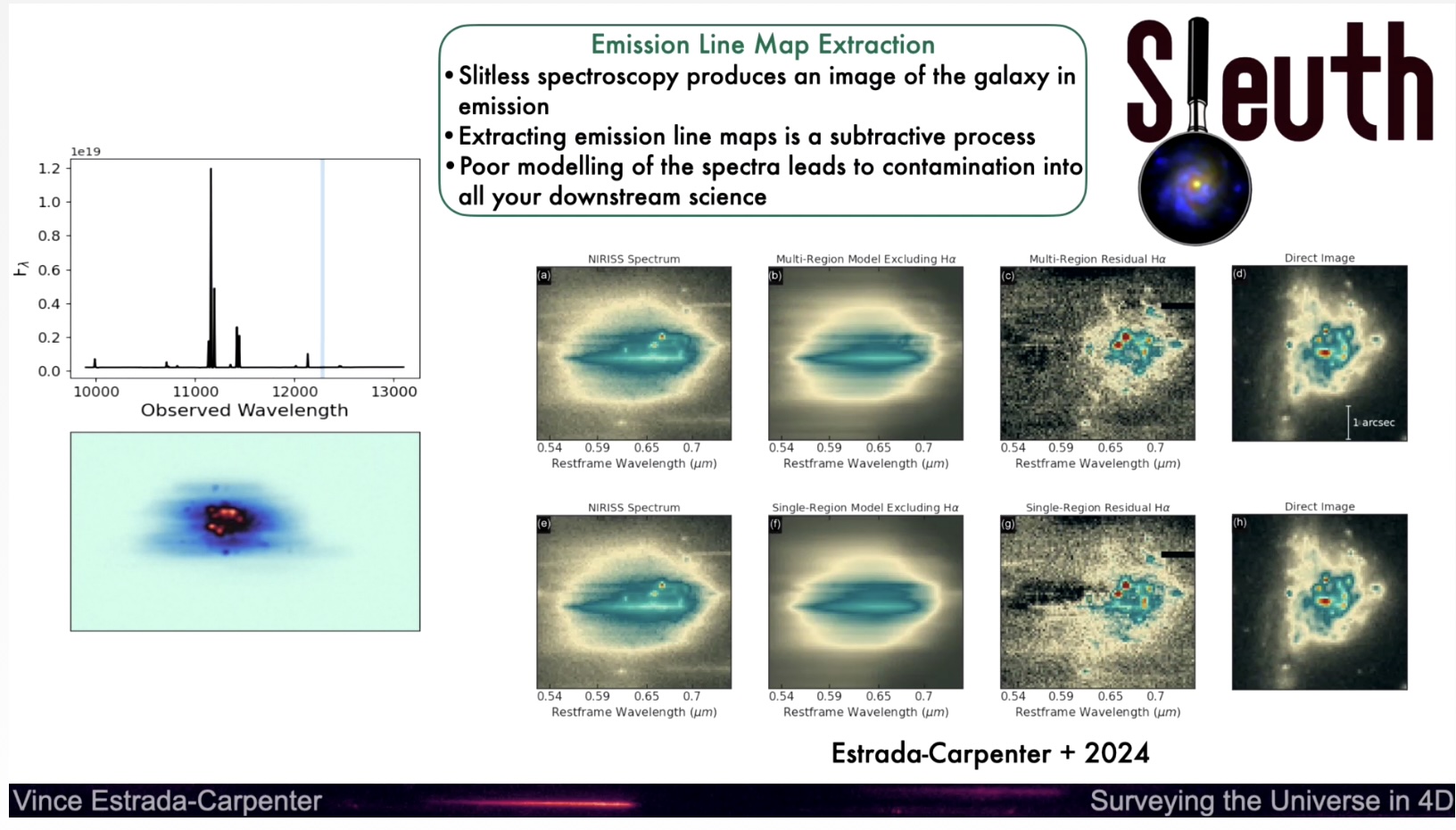}
    \caption{Example of how Sleuth's multiregional approach to modeling properly subtracts the galaxy continuum, leaving behind an accurate high-fidelity map of the galaxy's emission.}
    \label{fig:sleuth}
\end{figure}

Star-forming clumps are intense compact regions of star formation. They act as building blocks of galaxies, adding large amounts of mass and accounting for a sizeable amount of the host galaxy's star formation. They are difficult to study due to their small sizes and contamination from their host galaxy, making spatially resolved spectroscopy the best option to study them. 

Sleuth is a tool to perform emission line mapping for such complicated sources \citep{estrada2024, estrada2025}. Sleuth does this by properly modeling the spatially resolved spectra (Figure \ref{fig:sleuth}), producing high-fidelity emission-line maps. By using these maps, it has been found that star-forming clumps have diluted metallicities; combined with their elevated star formation rates, this likely indicates that they are being fuelled by inflowing pristine gas \citep{estrada2025}. Evidence has also been seen that these clumps form in short bursts of star formation and rapidly quench.

\subsection{Jeremy Favaro: The CANUCS NIRISS WFSS Catalog of 1900 galaxies at $z = 0.2 - 9.2$ and Metallicities at Cosmic Noon \label{sec:JFavaro}}
\begin{figure}
    \centering
    \includegraphics[width=0.9\linewidth]{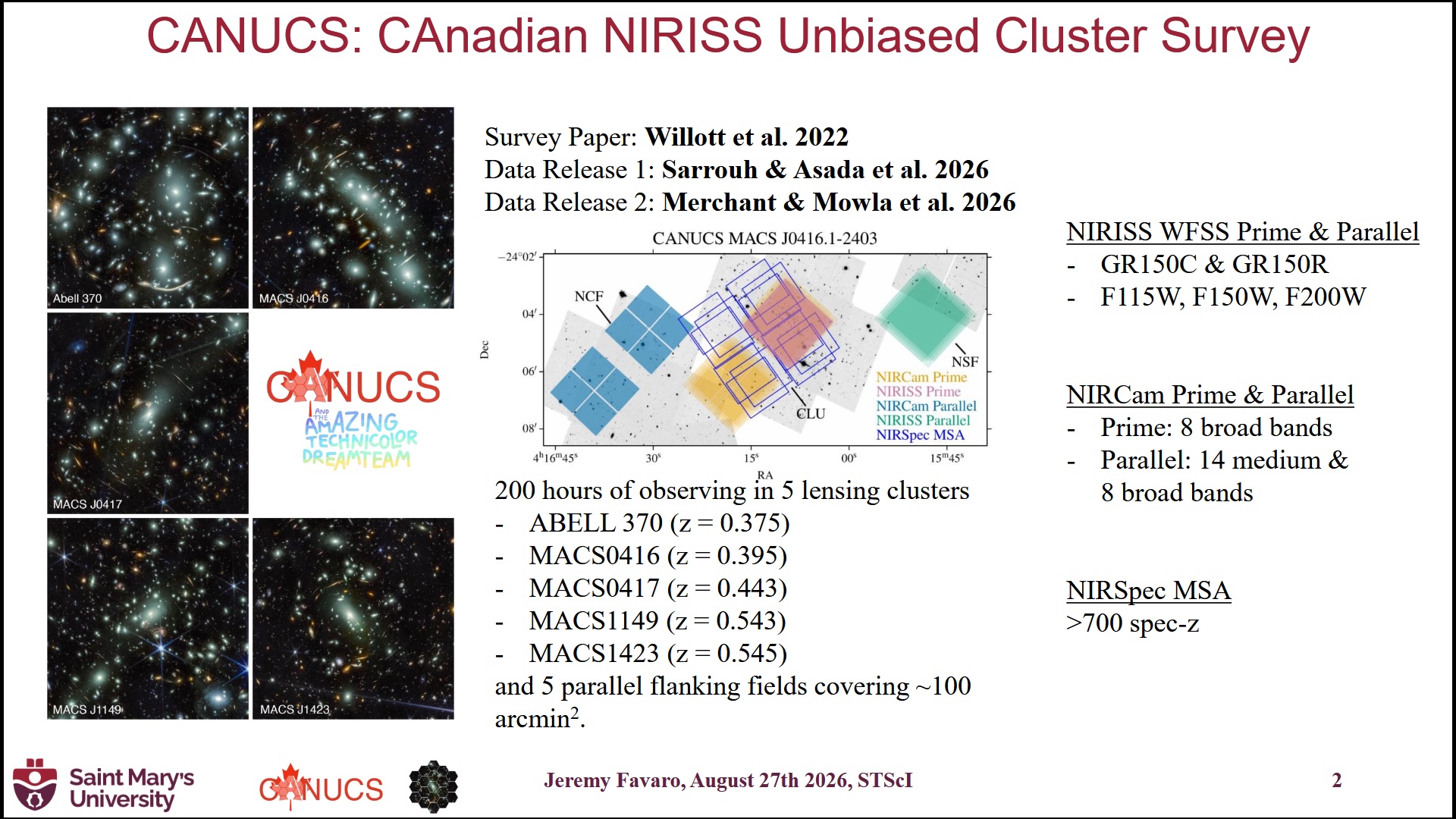}
    \caption{CANUCS survey overview.
    To the left, the cores of the five lensing clusters are shown in false RGB color.
    In the center, the NIRISS, NIRCam, and NIRSpec coverage of the MACS0416 Prime and Parallel fields are shown.}
    \label{fig:canucs3}
\end{figure}

\begin{figure}
    \centering
    \includegraphics[width=0.9\linewidth]{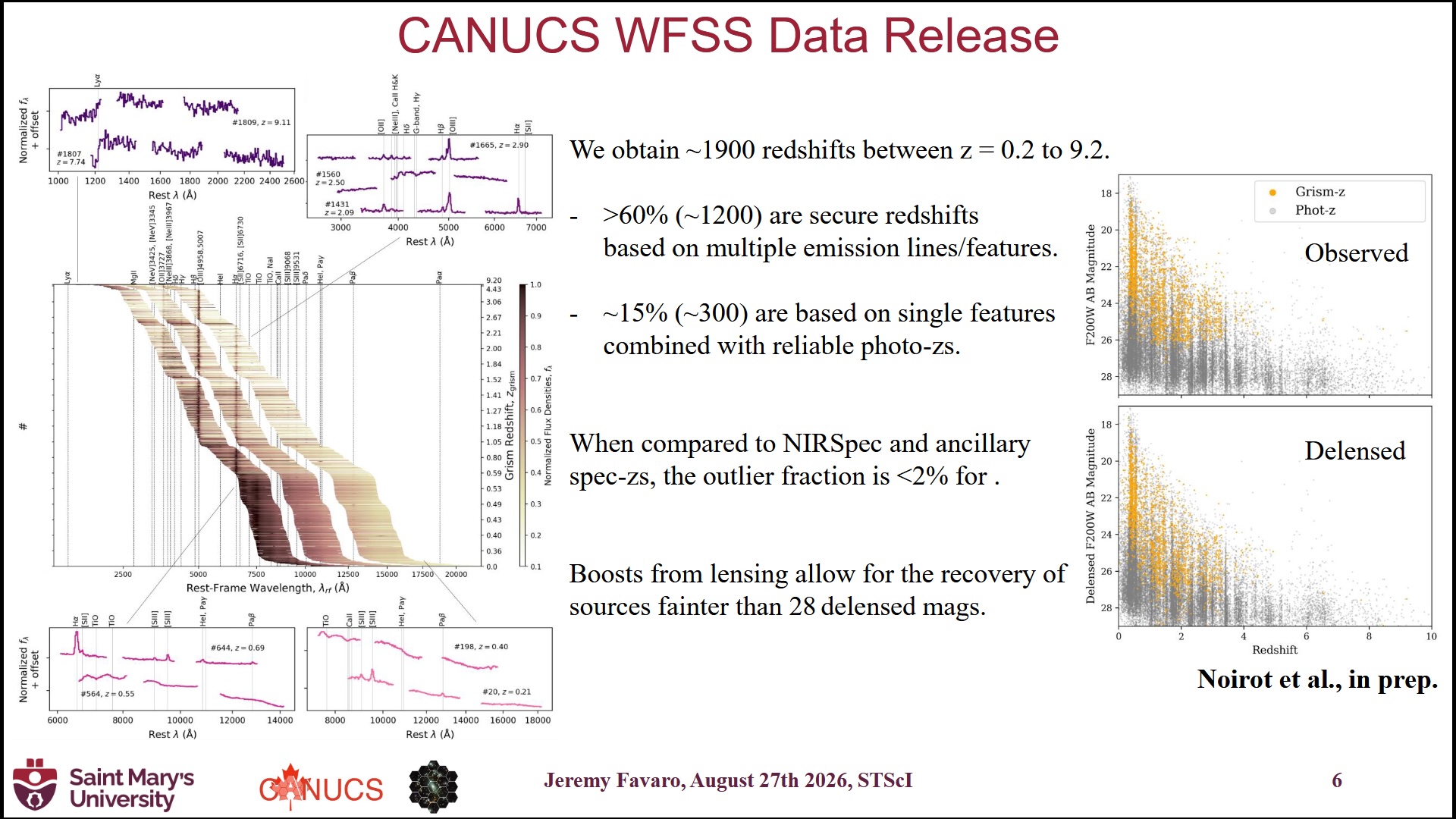}
    \caption{Upcoming CANUCS NIRISS WFSS data release overview.
    The vertical stack of WFSS spectra and sample 1D spectra highlight the variety of spectral features, from strong emission lines to TiO absorption, captured in the catalog.
    Observed and delensed F200W magnitudes are compared to highlight the boost in detection from strong lensing.}
    \label{fig:canucs4}
\end{figure}

The Canadian NIRISS Unbiased Cluster Survey (CANUCS, \citealt{Willott2022}; Data Release 1, \citealt{Sarrouh2026}) is a large program that targets 5 strongly lensing cluster fields (Figure \ref{fig:canucs3}) with an upcoming NIRISS data release that contains 1900 redshifts between $z = 0.2$ to $9.2$, of which $\sim \!\! 1500$ are based on multiple spectral features or single features plus a robust photometric redshift (Figure \ref{fig:canucs4}). 
Foreground contaminants (BCGs and large ellipticals) are carefully modeled and removed with in a spatially resolved manner with Sleuth \citep{estrada2024} and redshift fits are vetted by at least two inspectors.
The NIRISS data release will provide an ideal catalog for statistical studies of cosmic noon galaxies, studies of $z \sim 0.4$ elliptical cluster galaxies, cluster lens modeling, and more. 

Leveraging the CANUCS dataset enables the spatially resolved study of metallicities and the baryon cycle at Cosmic Noon. 
Spatially resolved modeling with Sleuth enables the extraction of emission line fluxes from high quality line maps.
Global emission line fluxes are measured by summing pixel fluxes in the corresponding line map, excluding the outermost low SN region of the source. 
This spatially resolved approach reveals lower [OIII] line fluxes and higher metallicities than predicted by single-region modeling. 
These findings mirror the effect of outshining in stellar mass measurements, which causes spatially unresolved mass measurements to underestimate stellar masses when compared to spatially resolved SED fitting.

\subsection{Eleni Tsiakaliari: Exploiting JWST NIRCam WFSS to study extended objects in crowded fields: The example of Ice
Mapping in B335. \label{sec:ETsiakaliari}}

\begin{figure}
    \centering
    \includegraphics[width=0.9\linewidth]{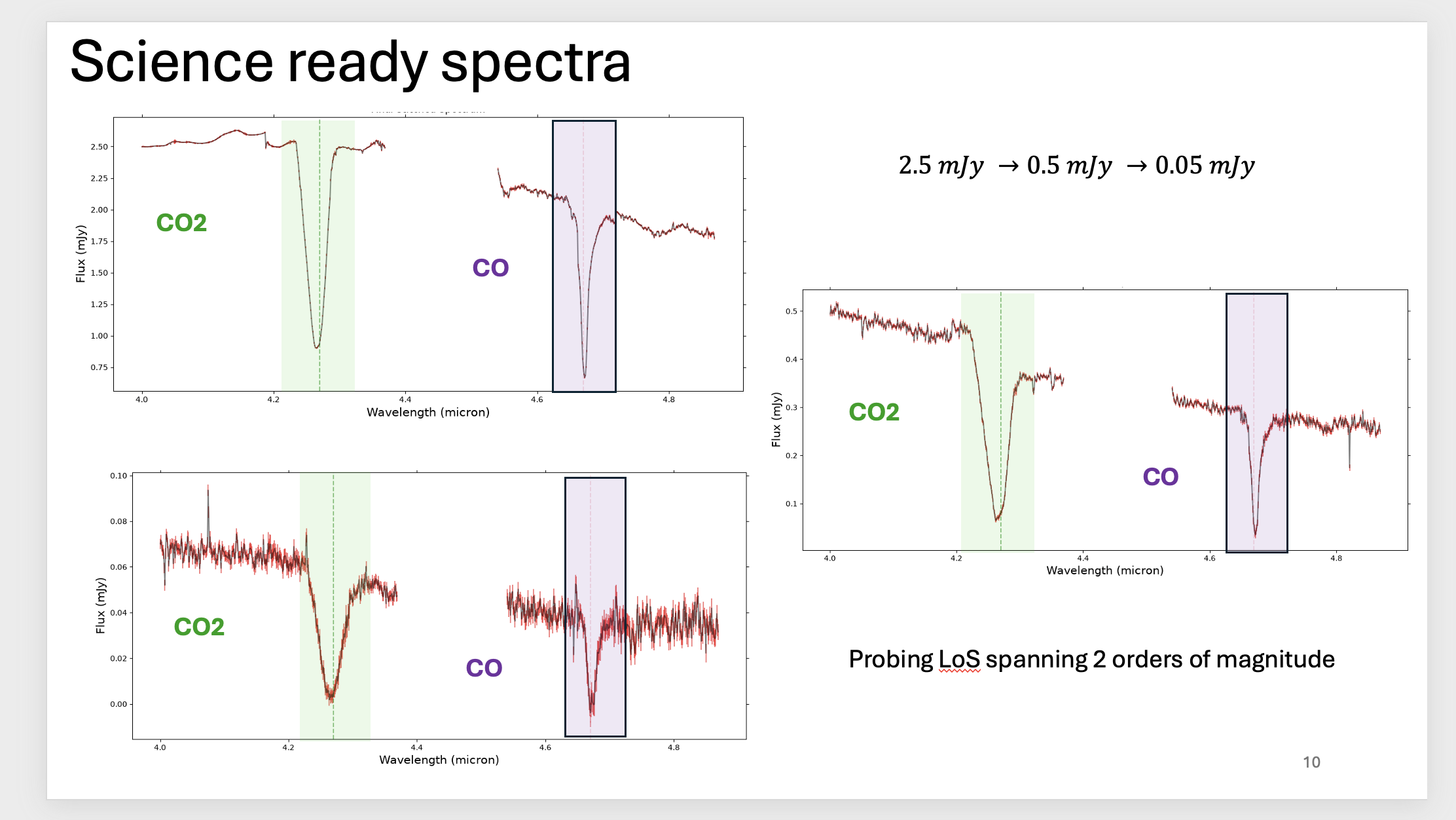}
    \caption{Preliminary CO2 ice map of B335 - Tsiakaliari et al (in prep)}
    \label{fig:specs}
\end{figure}

NIRCam WFSS has enabled the spatial ice mapping of entire molecular clouds, such as B335 (Fig.~\ref{fig:specs}). We can probe lines of sight spanning the entire cloud and we can start discussing the interaction the parent envelope with the embedded protostar in B335 by studying ice in absorption. We do this by using WFSS to look at stars \textit{behind} the cloud and then look at there the ice absorption features are against the background star continua. One of the problems we face is that we do not know anything about the population of the stars behind the cloud. This makes it rather difficult to have an accurate representation of the background star continuum for each of the lines of sight probed. Traditionally, ice spectroscopists have fitted a polynomial of nth degree to obtain what the baseline would look like. However, this is somewhat handwavy and results in loss of the error asymmetry when converting from flux space to optical depth space. We instead use a gaussian process model, which we train on a number of fixed conditioning points, to obtain a more accurate representation of the continuum. We implement bootstrapping to go from flux space to optical depth space and eventually to column densities. By using these two techniques we ensure the error propagation is done correctly and the errors on the column density are asymmetric (as they should be). The major ice species discussed in this work were CO2 and CO. With all these tools in our toolkit, we can now zoom back and see the bigger picture of YSO-envelope interactions, something we were not able to do before. (My PhD is funded partly by STFC and partly by the Open University UK). 

To summarise: \newline
We step away from using polynomial fits to get the background star continuum - instead we use bootstrapping. \newline
The baseline is now the obtained via a Gaussian Process model trained on a set of fixed conditioning points, set in flat parts of the spectrum, to the left and right of the ice feature.\newline
 By using bootstrapping we can now preserve the asymmetric nature of the flux errors and correctly propagate it all the way to column densities. \newline 
 Novel treatment for partial spectra, recovering the full shape by fitting a Gaussian-Lorentzian Sum function to the partial spectrum allows us to recover spectra with an ice feature but which have suffered from background over-subtraction. \newline

\subsection{Massimo Griggio: Spatially resolved spectroscopy from slitless observations: non-parametric, multi-orientation 3D fluxcube reconstruction \label{sec:MGriggio}}

\begin{figure*}
    \centering
    \includegraphics[width=.9\textwidth]{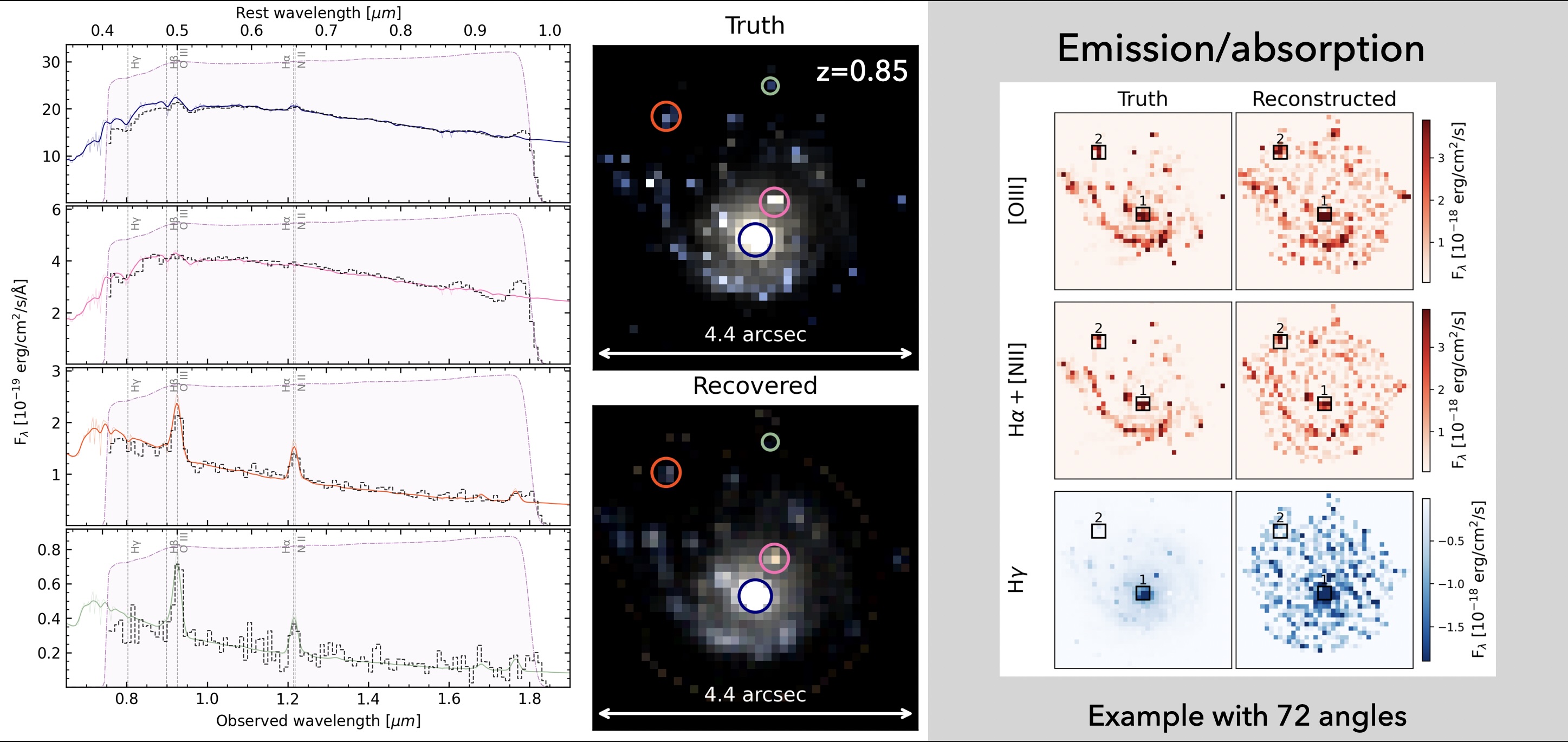}
    \caption{Example of flux cube reconstruction from \cite{griggio26}.}
    \label{fig:fluxcube}
\end{figure*}

Slitless spectroscopy provides an efficient, unbiased way to collect wide-field astronomical spectra, but extracting spatially resolved spectroscopy from it is hindered by severe spatial-spectral overlap. Without a physical slit, light from adjacent sources and different subregions within the same extended galaxy overlap on the detector, creating self-contamination that varies with the dispersion angle. Standard analysis methods struggle to isolate embedded features, such as faint transients or localized star-forming regions, without relying on restrictive spatial assumptions or oversimplified models.

To solve this, we developed a data-driven linear reconstruction technique that converts multi-orientation slitless data into a full three-dimensional spatial-spectral flux cube, effectively providing low-resolution integral field unit (IFU) capabilities \citep{griggio26}. By leveraging observations taken across multiple dispersion angles and dithers, the method maps how every spatial and spectral bin projects onto detector pixels across all exposures. The framework incorporates instrument pointing, dispersion curves, transmission, and point-spread function (PSF), and then iteratively solves the resulting regularized linear system to reconstruct the scene's spatial and spectral flux distribution.

Validated on simulated Roman simulated data \ref{fig:fluxcube}, this approach proves that spatially resolved spectral information can be recovered purely from slitless data. Crucially, the algorithm is entirely non-parametric and data-driven: it operates on a pixel-by-pixel basis without using galaxy spectral templates, structural component models, or prior redshift assumptions. The reconstructed flux cubes directly enable accurate redshift determination, recovery of spatially resolved emission and absorption features, and host-light subtraction from transient events without introducing spectral bias.

\subsection{Aryana Haghjoo: From JWST to Roman: A Physics-Informed AI Super-Resolution Framework for Grism Spectroscopy \label{sec:AHaghjoo}}
The information a galaxy spectrum carries is set by its resolution, but the
deep surveys that provide statistical samples mostly observe at low resolution. At low spectral resolution the diagnostic complexes stay blended, putting the ratios that measure dust, metallicity and ionization out of reach for the largest samples. The question is, whether a machine learning model trained on paired low- and high-resolution spectra can recover the lost resolution and reconstruct the fine detail.

The pilot uses JWST/NIRSpec spectra from \textit{JWST Advanced Deep Extragalactic Survey (JADES)}, where we are provided with the spectra of 2,858 galaxies with both the
prism ($R\sim100$) and the medium gratings ($R\sim1000$) resolution. A three-stage model super-resolves the spectrum, infers a
redshift from it, and refines the lines with attention across emission-line
tokens. It successfully separates [O\,\textsc{iii}]\,$\lambda\lambda4959,5007$ where the input prism spectrum shows a single bump \citep{haghjoo2026a}. Benchmarked against seven classical deconvolution techniques, the distinctive gain of the machine learning technique is line shape, a FWHM bias below $1$\,nm against $11$--$35$\,nm, paid for with a higher false-detection rate \citep{haghjoo2026b}.

\begin{figure}
    \centering
    \includegraphics[width=0.95\linewidth]{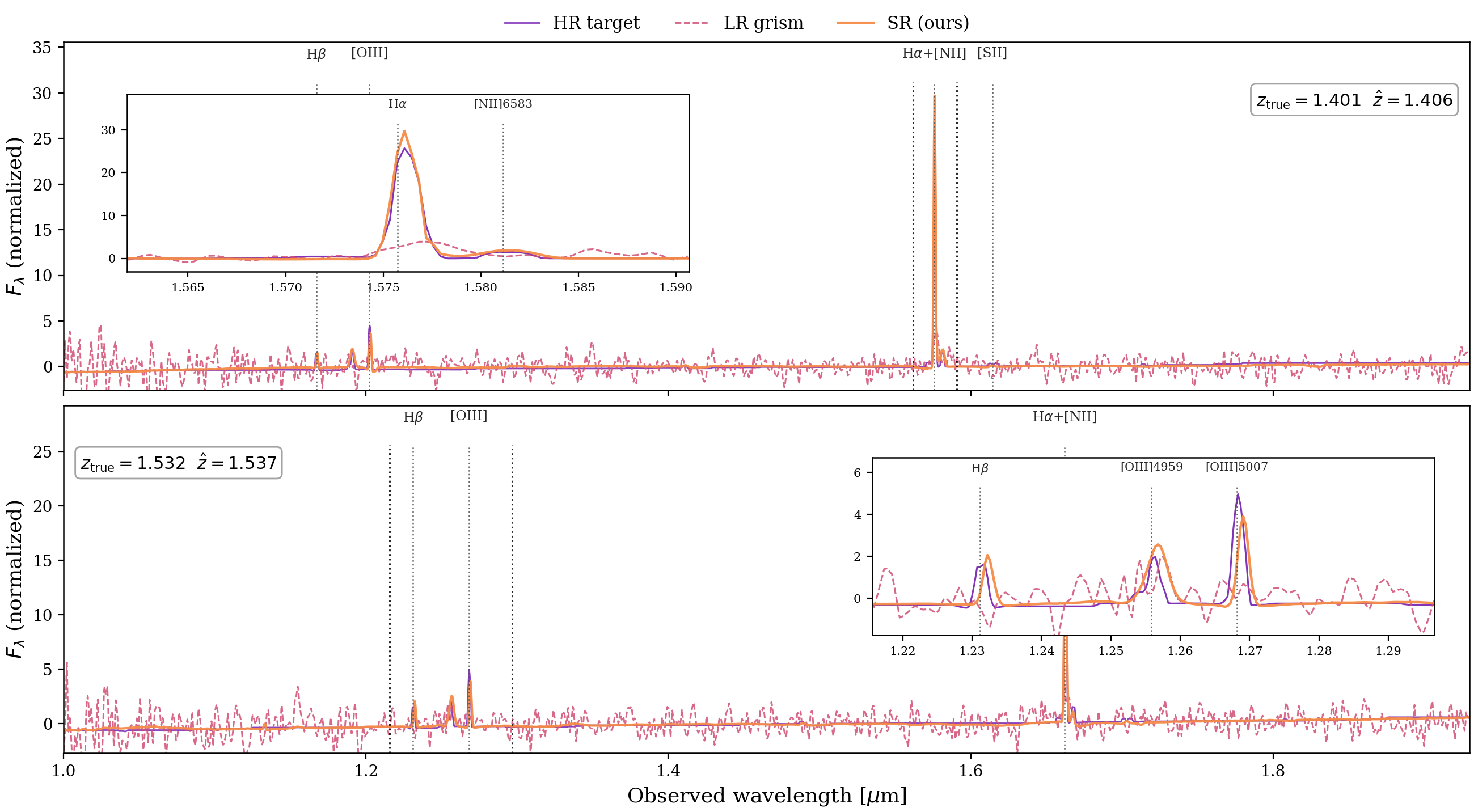}
    \caption{Two simulated Roman grism galaxies from the held-out test set:
    the extracted grism spectrum (dashed), the model SED target (dark), and the
    super-resolved output (orange), each normalized per row. The insets zoom on
    the complexes the grism blends, \Ha+\NII\ at $z=1.401$ (top) and
    \OIIIab\ with \Hb\ at $z=1.532$ (bottom). Redshifts printed in each panel
    are the true value and the pipeline estimate. Note that the emission lines such as H$\alpha$, H$\beta$, and [O\,\textsc{iii}]\,$\lambda\lambda4959,5007$, are present in the super-resolved spectra but wash out in the HLSS spectra \citep{haghjoo2026c}.}
    \label{fig:AHroman}
\end{figure}

Extending the same three-stage machine learning pipeline to 36,404 simulated Roman \textit{High Latitude Spectroscopic Survey (HLSS)} spectra ($R\approx461$, $1$--$1.93\,\mu$m), a lone in-band emission line leaves the redshift alias-degenerate. Conditioning the redshift head on the three Roman Medium imaging bands (\textit{Y106, J129, H158}) solves the issue and takes the catastrophic-outlier rate from $26\%$ to $5.1\%$. Consecutively, the model is able to recover the emission lines such as H$\alpha$, H$\beta$, and [O\,\textsc{iii}]\,$\lambda\lambda4959,5007$, which were present in the underlying SED but completely wash out in the simulated realistic spectra (Figure \ref{fig:AHroman}, \cite{haghjoo2026c}).

\section{High z large scale structure / clusters}

\subsection{Yuming Fu: Mapping quasars in the Universe with Euclid spectroscopy: lessons learned and synergy with Roman \label{sec:YFu}}
Euclid Q1 spectroscopic quasar identification strategy: matching more than 9200 pre-selected QSO candidates from Gaia/WISE to Euclid spectroscopy. Gaia QSOs have better SNR in general, and higher success rate. Combining Euclid NIR colors with WISE can find highly red QSOs efficiently. Wide field depth for secure redshift determination is around $H_{\rm E}=21.3$ \citep{Fu_2026_Euclid_Q1_quasar}. 

\begin{figure}
    \centering
    \includegraphics[width=1\linewidth]{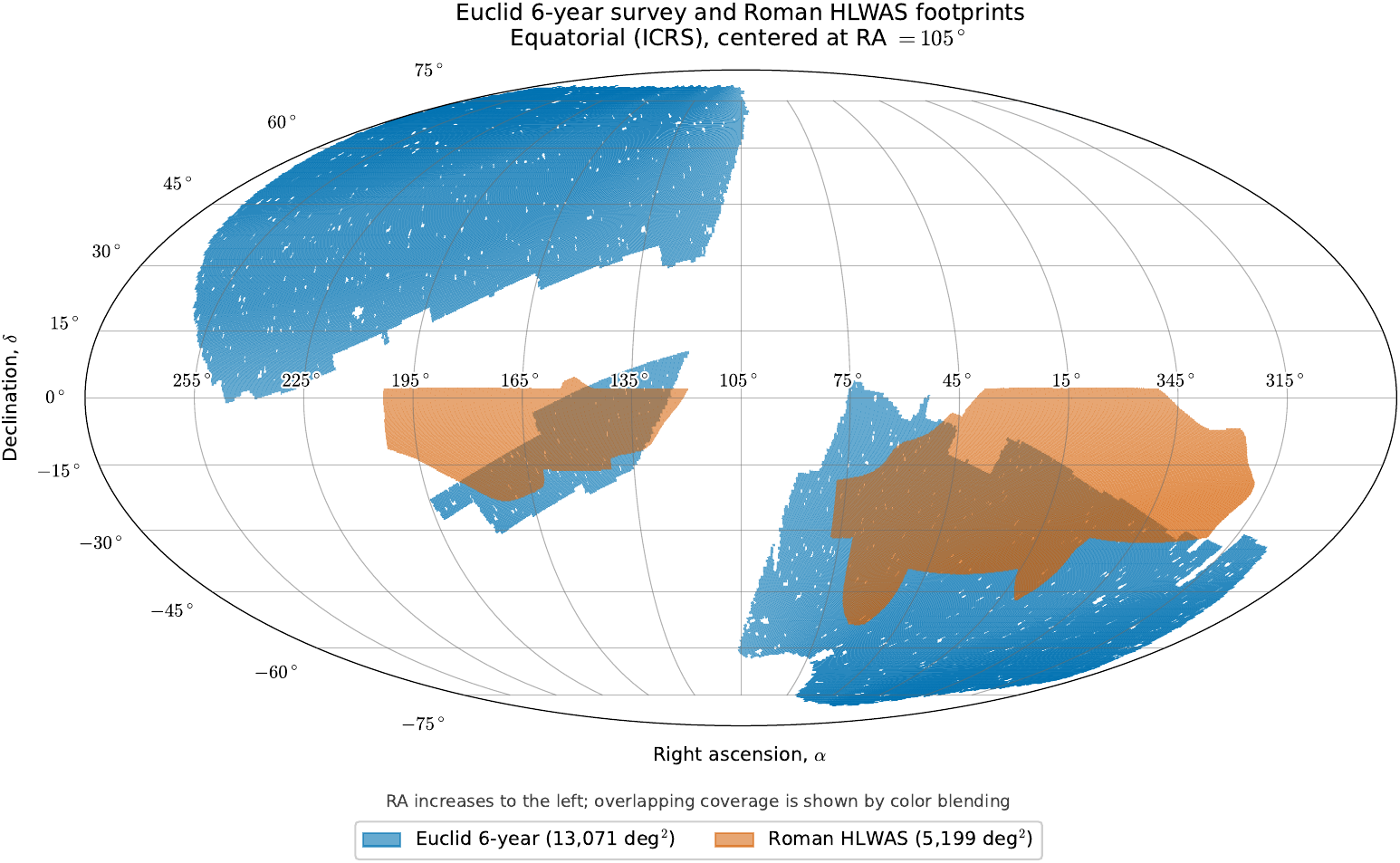}
    \caption{The planned Euclid survey footprint in 6 years of the nominal survey and the Roman HLWAS footprint (Y. Fu).}
    \label{fig:roman-euclid-sky-coverage}
\end{figure}

Euclid DR1 spans around 2000 $\deg^2$ (Figure \ref{fig:roman-euclid-sky-coverage}), and a new strategy is adopted for the large number of spectra sources:
\begin{itemize}
    \item Magnitude cut: spectra sources with $H_{\rm E}<21.5$ for wide fields and $H_{\rm E}<22.7$ for deep fields. 
    \item Euclid native photometric pre-selection of quasar candidates using machine learning.
    \item Hybrid redshift estimation combining machine learning and template matching, slitless spectroscopy and photometry. Human-in-the-loop visual inspection to further improve the measurements.
\end{itemize}

\begin{figure}
    \centering
    \includegraphics[width=1\linewidth]{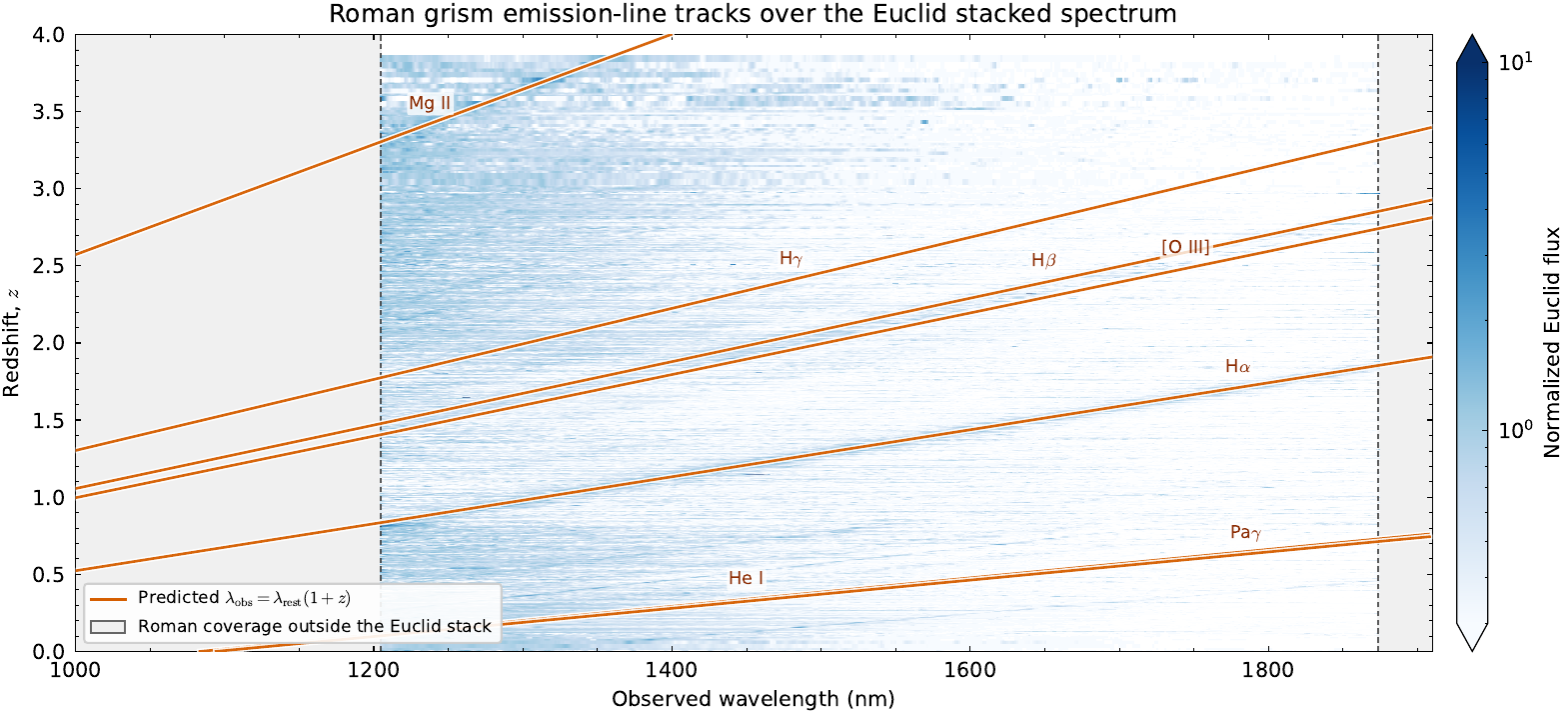}
    \caption{Major quasar emission lines within the Roman Grism wavelength range overlaid on the Euclid Q1 quasar emission line map \citep{Fu_2026_Euclid_Q1_quasar}.}
    \label{fig:roman-euclid-line-map}
\end{figure}

Redshift degeneracies still exist with Euclid red grism (RGS) alone. Roman extends the wavelength coverage to bluer wavelengths, enabling more robust redshift determination, and calibration for the Euclid measurements (Figure \ref{fig:roman-euclid-line-map}).

\subsection{Jiamu Huang: Probing Quasar and Galaxy Dark Matter Halo Masses at z>6 with Clustering Analysis \label{sec:JHuang}}

Clustering measurements connect high-redshift quasars and galaxies to their underlying dark matter halos, constraining their environments and duty cycles. Using JWST/NIRCam WFSS from the ASPIRE survey, we measure the quasar--[O\,III] cross-correlation and [O\,III] emitter auto-correlation across 25 quasar fields at $\langle z\rangle \simeq 6.6$. We interpret these measurements using halo populations from the FLAMINGO simulation, with mock catalogs accounting for the survey selection and cosmic variance.

\begin{figure}
    \centering
    \includegraphics[width=0.99\linewidth]{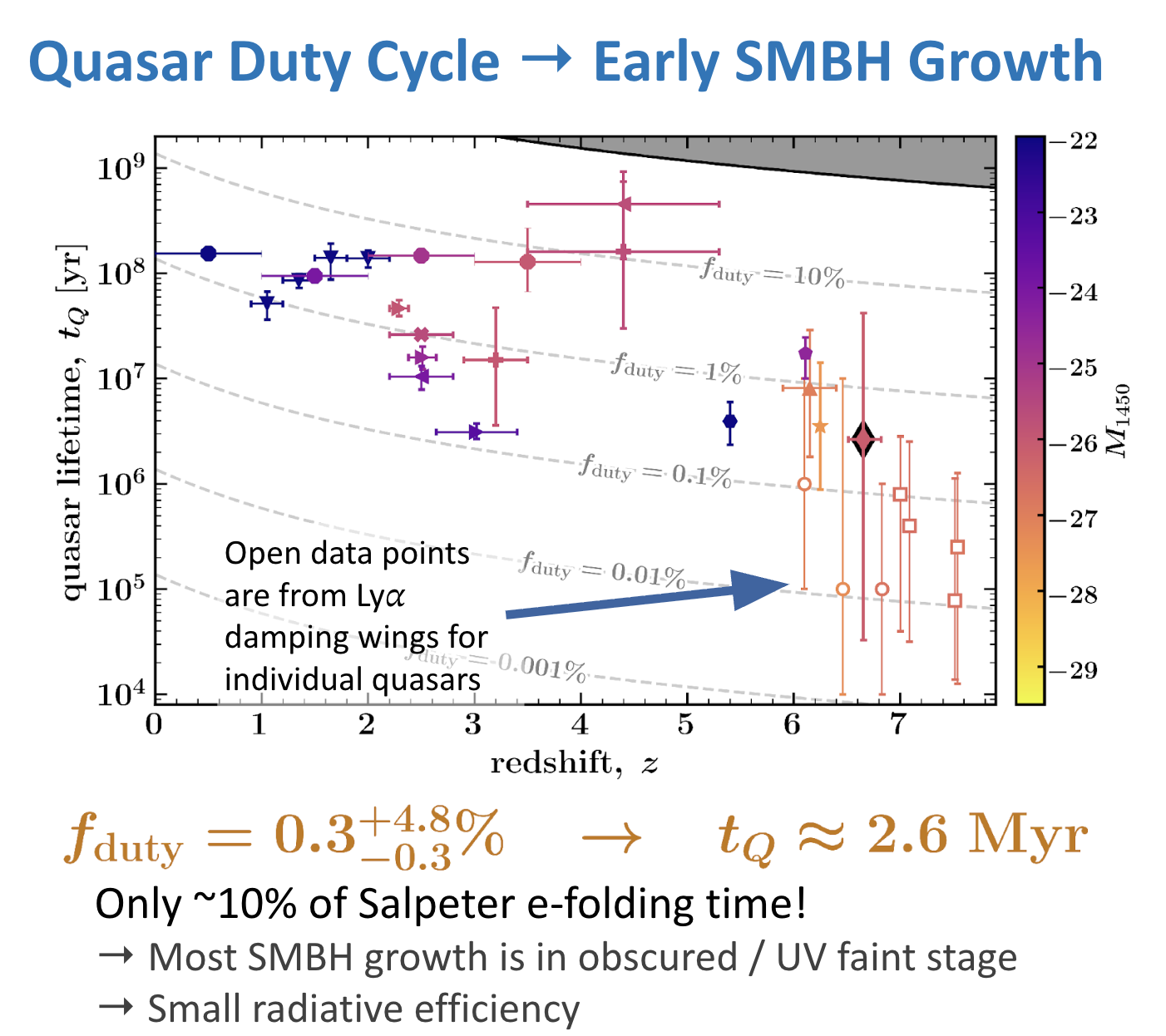}
    \caption{Evolution of quasar UV-bright lifetimes and duty cycles, including constraints from the ASPIRE clustering analysis.}
    \label{fig:JHfig_qso}
\end{figure}

Our analysis indicates minimum host halo masses of approximately $10^{12.1}\,M_\odot$ for quasars and $10^{10.6}\,M_\odot$ for [O\,III] emitters. Combining the inferred abundance of quasar host halos with the observed quasar number density yields a sub-percent central estimate for the quasar duty cycle, albeit with substantial uncertainties. This corresponds to a UV-bright lifetime of order a few million years, suggesting that the observed luminous phase is brief compared with the time required to assemble the central supermassive black holes. Figure~\ref{fig:JHfig_qso} places these constraints in the context of quasar lifetime and duty cycle evolution. Cosmic variance is a major source of uncertainty, highlighting the importance of simulation-based covariance estimates for interpreting clustering in small JWST fields.

\begin{figure}
    \centering
    \includegraphics[width=0.99\linewidth]{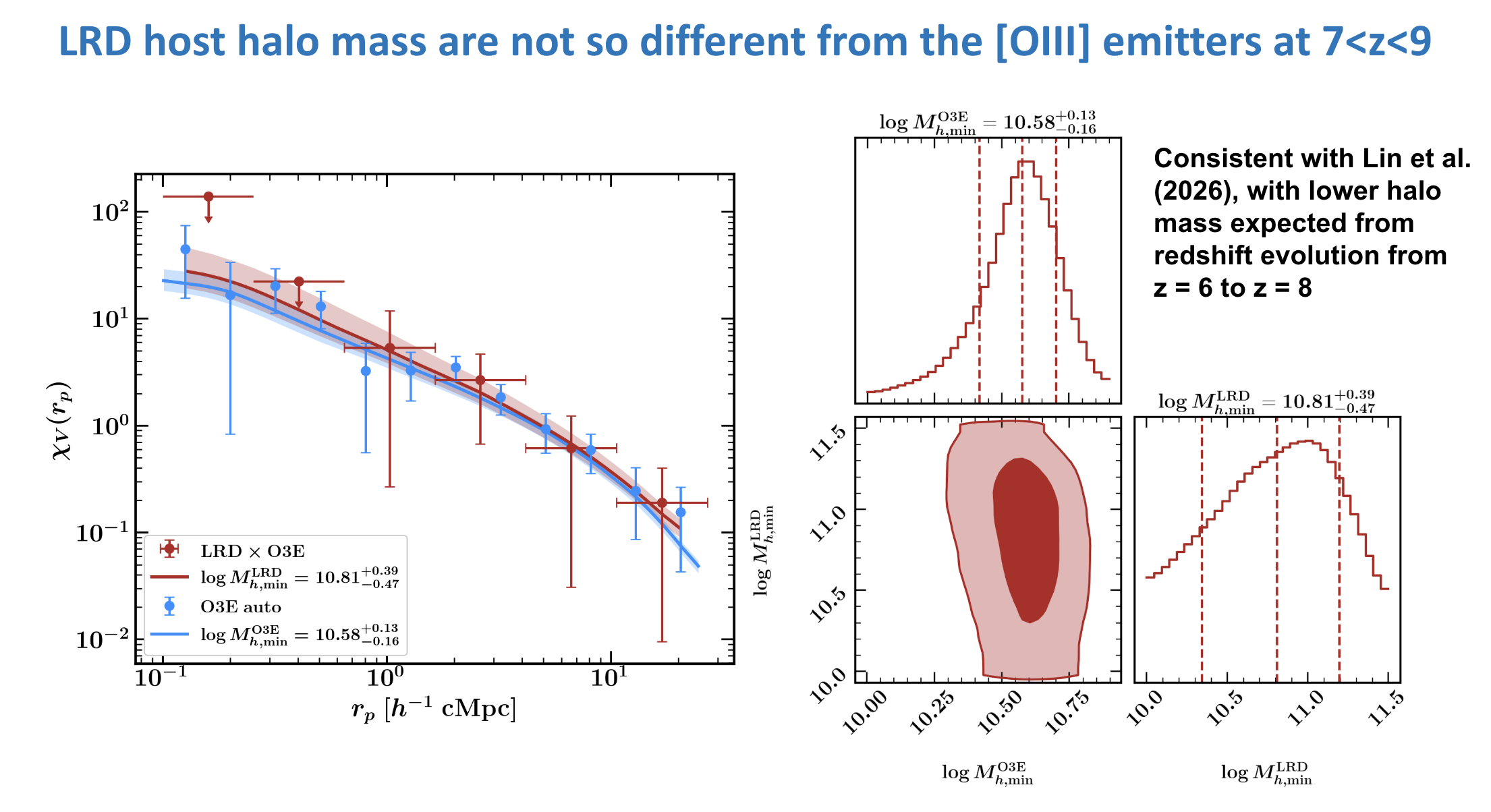}
    \caption{Auto-correlation function of [O\,III] emitters and their cross-correlation with spectroscopically confirmed little red dots at $z=7$--9 in COSMOS-3D.}
    \label{fig:JHfig_o3_lrd}
\end{figure}

We also apply this approach to [O\,III] emitters and spectroscopically confirmed little red dots (LRDs) using COSMOS-3D WFSS. Enabled by the survey's sample of [O\,III] emitters, we present the first spectroscopic-redshift-based measurement of the LRD--[O\,III] cross-correlation at $z=7$--9. Combining this measurement with the [O\,III] emitter auto-correlation (Figure~\ref{fig:JHfig_o3_lrd}), we infer a minimum LRD host halo mass of $\log_{10}(M_{\mathrm{h,min}}^{\mathrm{LRD}}/M_\odot)\simeq10.8$. LRDs therefore occupy halos of similar mass to those hosting star-forming galaxies at the same redshifts and are substantially less strongly clustered than the luminous quasars in ASPIRE, whose inferred minimum host halo masses are approximately $10^{12}\,M_\odot$.

\subsection{Marc Rafelski: JAGGER, NIRCam WFSS of Galaxies and the CGM at the End of Reionization \label{sec:MRafelski}}

\begin{figure}[htbp]
\centering
\includegraphics[width=0.8\linewidth]{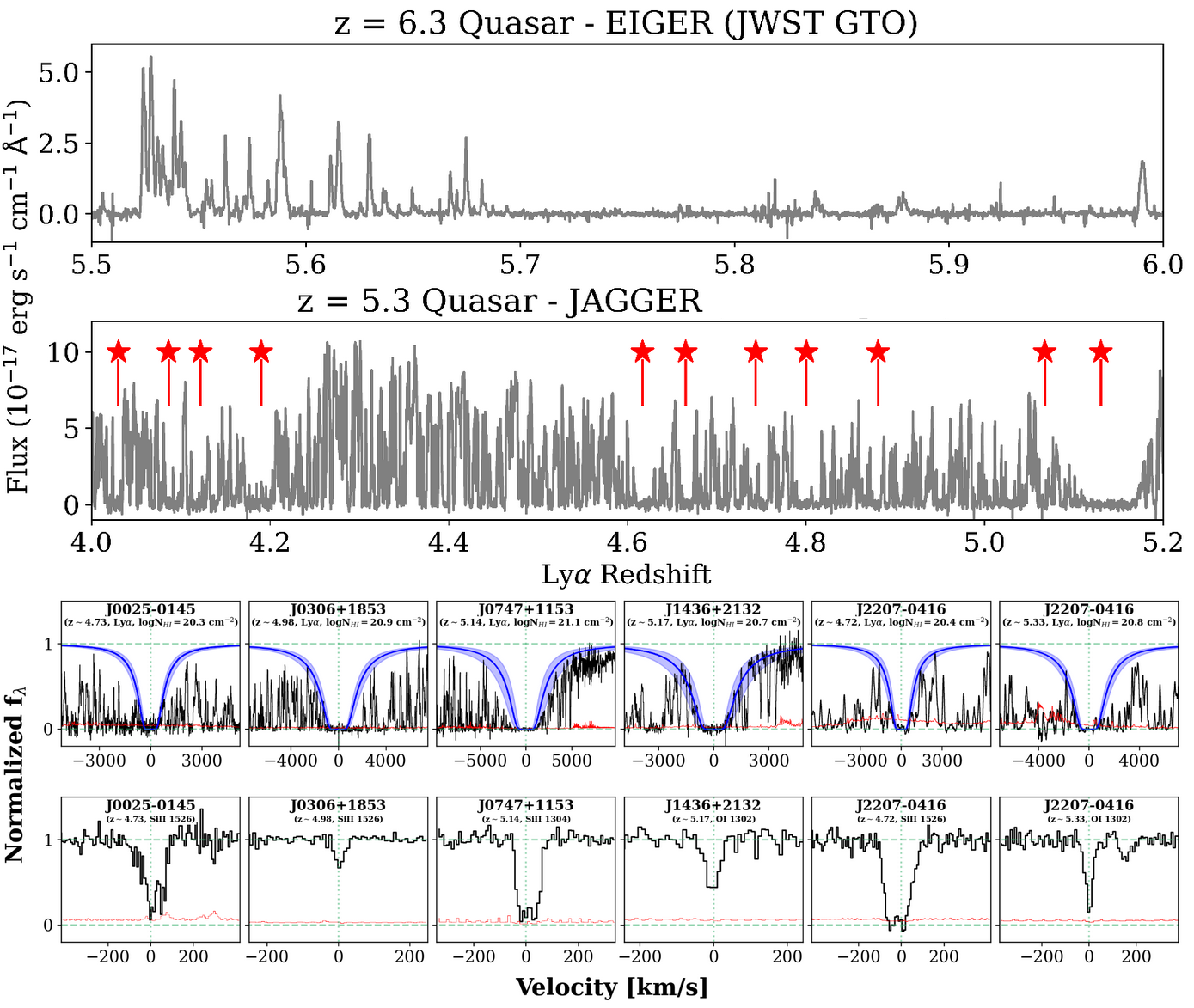}
\vspace{-5pt}
\caption{X-shooter spectra in the \lya\ forest of a $z \approx 6.3$ quasar
from the JWST EIGER program (top) and a $z \approx 5.3$ JAGGER quasar
(bottom), with strong H\,{\sc i} absorbers marked by stars. At
$z \gtrsim 5.5$ a thick jungle of \lya\ forest lines prevents the
measurement of individual absorption systems, so the higher-redshift WFSS
quasar programs cannot constrain $N$(H\,{\sc i}) or metallicity. The
redshift interval targeted by JAGGER ($z \approx 4$--$5.5$, bottom) is the
last window in which individual H\,{\sc i} absorption lines can be
measured, enabling $N$(H\,{\sc i}) determinations and thereby also
metallicities.}
\label{fig:jagger_forest}
\end{figure}

\begin{figure*}[htbp]
\vspace{-10pt}
\centering
\includegraphics[width=0.325\linewidth]{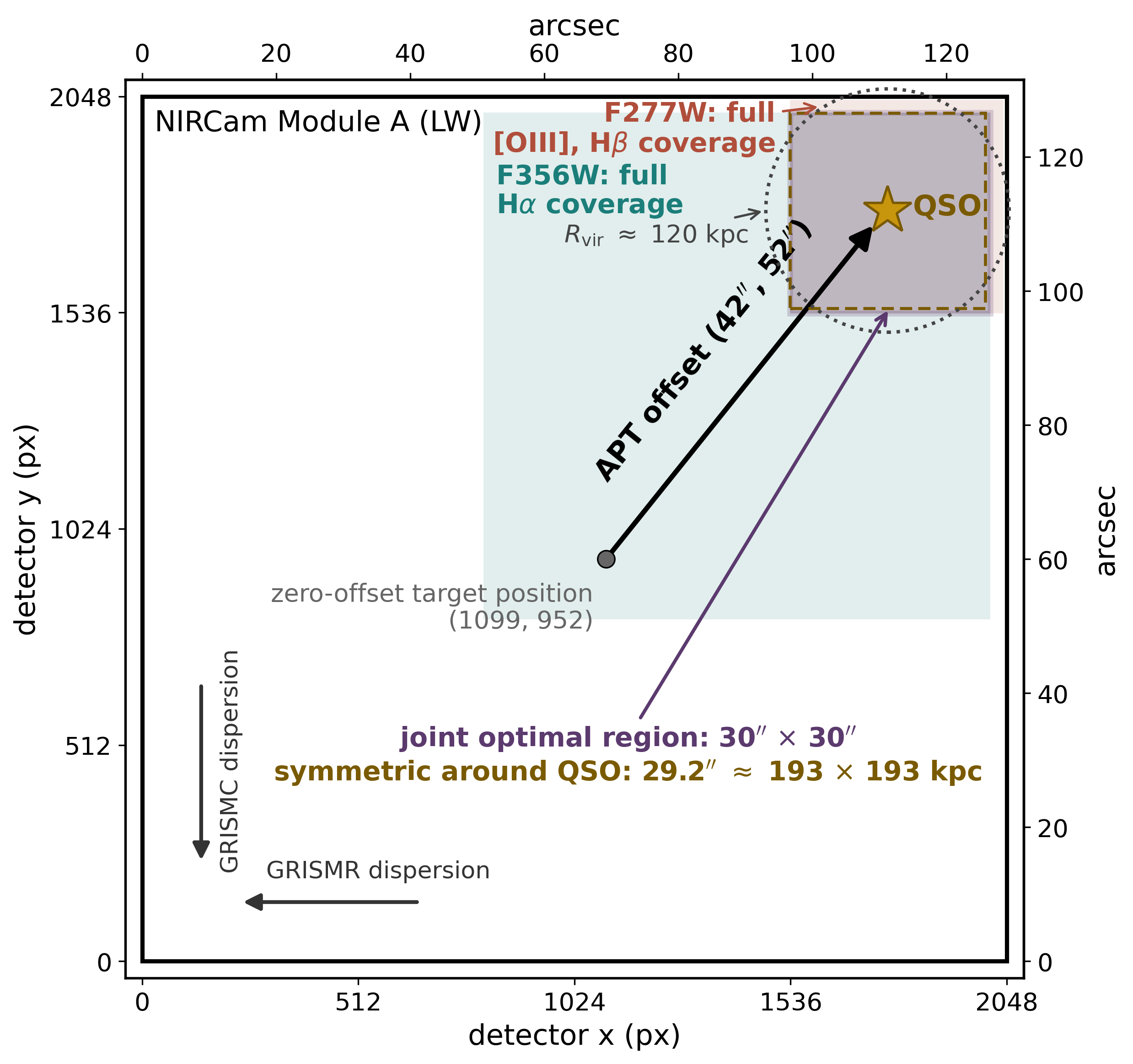}
\includegraphics[width=0.325\linewidth]{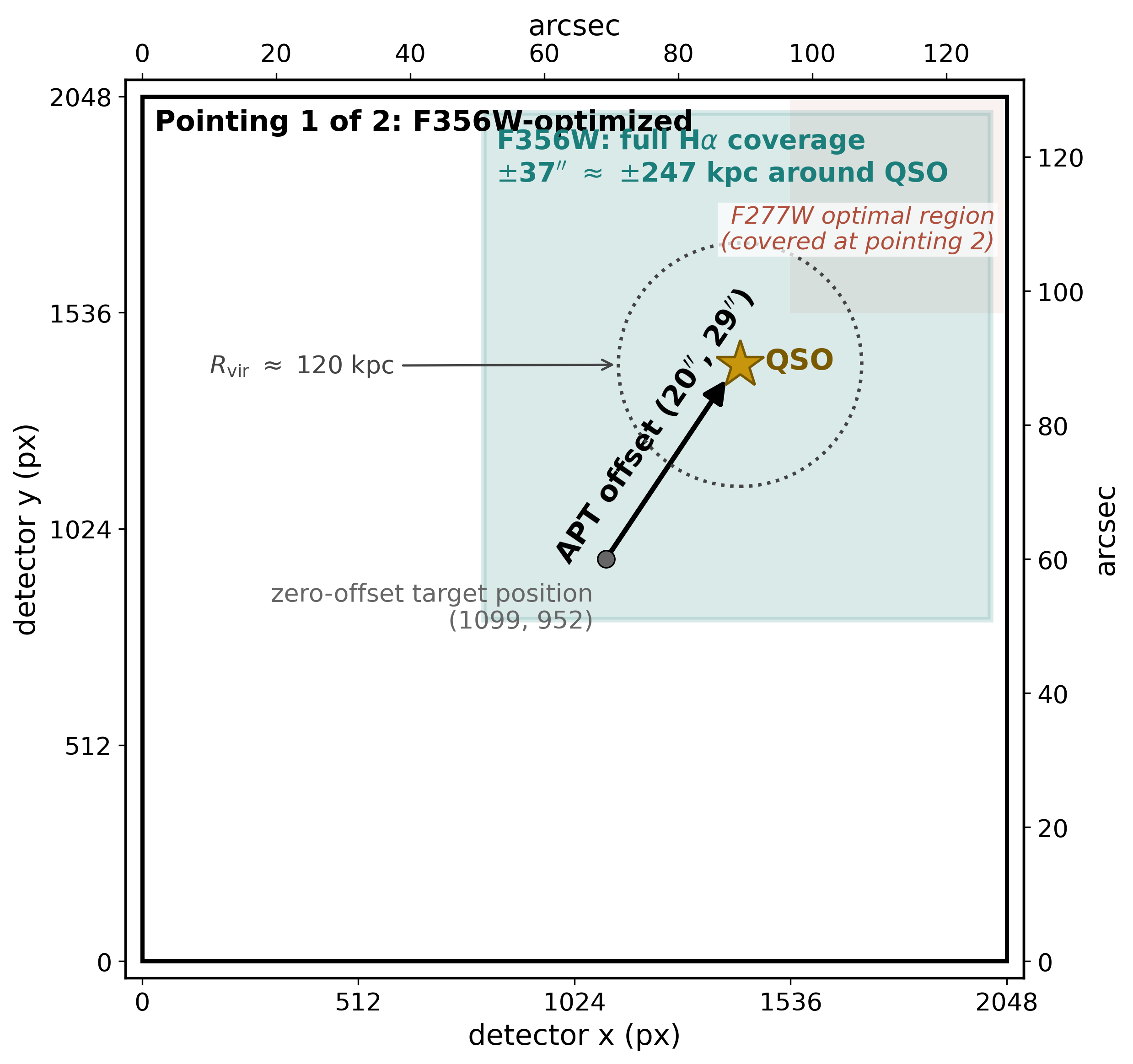}
\includegraphics[width=0.325\linewidth]{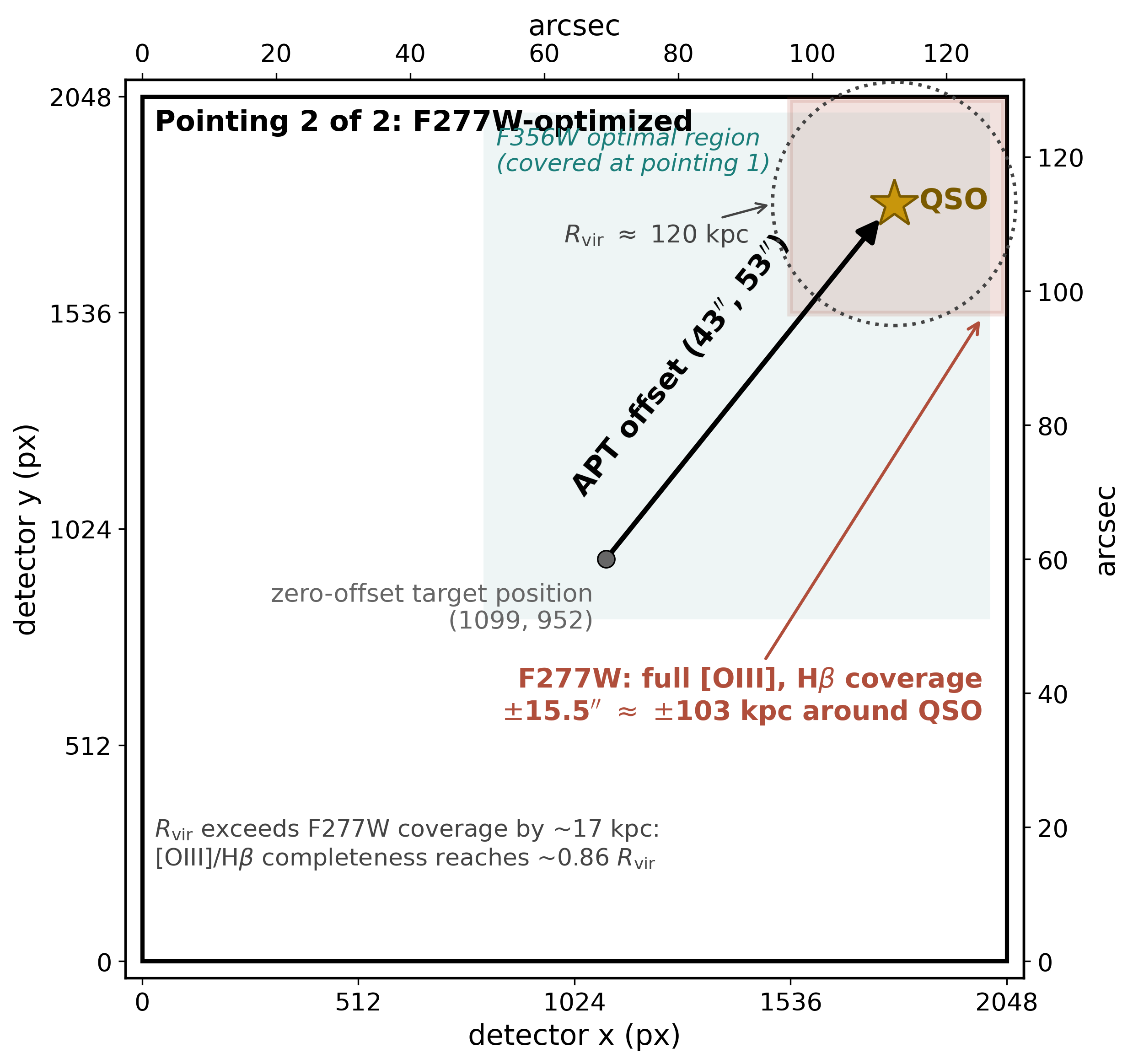}\\
\caption{JAGGER target placement on the NIRCam Module A LW detector.
{\it Left}: with both filters in a single observation, one APT offset serves
both, and full wavelength coverage in both filters is limited to the
$30'' \times 30''$ joint region; the coverage symmetric around the QSO
($29''$, $\approx 193$ kpc) does not fully enclose the virial radius
(dotted circle; $R_{\rm vir} \approx 120$ kpc for a $10^{13}\,M_\odot$ halo).
{\it Middle}: splitting into two observations lets the F356W-optimized
pointing center the QSO in its full-coverage region, with \Ha\ complete to
$\pm 37'' \approx \pm 247$ kpc in all directions.
{\it Right}: the F277W-optimized pointing reaches
$\pm 15.5'' \approx \pm 103$ kpc ($\approx 0.86\,R_{\rm vir}$),
the full-bandpass detector-edge limit.}
\label{fig:jagger_placement}
\end{figure*}

\begin{itemize}
\itemsep0em

\item JAGGER (JWST Cycle 5 GO 10096; PIs Dutta, Rafelski, Fumagalli) targets six $z \approx 5$--$5.5$ QSO fields from the Qz5 survey, whose archival X-shooter and HIRES spectra contain 44 metallicity-unbiased strong H\,{\sc i} absorbers at $z = 4$--$5.5$, with column densities from \citet{oyarzun2025} and metallicities from \citet{wisz2026}. This is the last redshift window in which individual absorption systems can be resolved within the \lya\ forest (Figure~\ref{fig:jagger_forest}), so both $N$(H\,{\sc i}) and metallicity remain measurable; the higher-redshift WFSS quasar programs (EIGER, ASPIRE) can not make those measurements.

\item JAGGER observes each field with NIRCam WFSS in F277W and F356W using both GRISMR and GRISMC. F356W covers \Ha\ and F277W covers \Hb\ and \OIIIab\ at $z \approx 4$--$5.5$, providing dust-robust SFRs, metallicities, and systemic redshifts. The two filters and two orthogonal dispersion directions also jointly secure emission-line and source identification.

\item The program is motivated by the limitations of ground-based IFU surveys of quasar fields (MAGG, MUSEQuBES, KAGG), which are structurally limited to \lya\ beyond $z \approx 1.5$. New Keck/KCWI results (KAGG; Oyarz\'un et al., submitted) find \lya\ counterparts for only 7 of 36 metallicity-unbiased DLAs, none at $z > 4$, while ALMA-confirmed [C\,{\sc ii}]-emitting hosts in the same fields are \lya-dark. Our working hypothesis is that the tracer, rather than the depth, is the limitation, although the data cannot yet rule out other explanations.

\item JAGGER will identify the galaxies associated with the 44 absorbers and measure their systemic redshifts, dust-robust SFRs, gas-phase metallicities (\NII/\Ha, \OIII/\Hb), and stellar masses, sizes, and morphologies from six-band NIRCam imaging. Correlating absorber $N$(H\,{\sc i}) and metallicity with galaxy properties, impact parameter, and environment directly tests feedback and accretion models at the end of reionization.

\item A placement lesson emerged from the program design: when both filters share a single observation, one APT offset must serve both filters, and the region with full wavelength coverage in both filters is limited to $\approx 30'' \times 30''$; even at the corrected offset, the coverage symmetric around the QSO spans only $\approx 29''$ ($\approx 193$ kpc at $z = 4.5$) and does not fully enclose the virial radius ($R_{\rm vir} \approx 120$ kpc for a $10^{13}\,M_\odot$ halo; most associated galaxies occupy lower-mass halos with smaller $R_{\rm vir}$; Figure~\ref{fig:jagger_placement}, left).

\item Splitting each field into two observations, so that each filter receives its own offset, greatly improves the coverage (Figure~\ref{fig:jagger_placement}, middle and right): \Ha\ complete to $\pm$247 kpc ($\approx 2\,R_{\rm vir}$) and \Hb\,$+$\,\OIIIab\ complete to $\pm$103 kpc ($\approx 0.86\,R_{\rm vir}$, the detector-edge limit) in all directions around the QSO, at a $\sim$10\% S/N cost from the added overheads. Relaxing the F277W blue-end requirement (using only the wavelengths needed for \Hb\,$+$\,\OIIIab\ at the survey redshifts) recovers full coverage to $R_{\rm vir}$ even for the most massive halos, while the smaller virial radii of typical lower-mass galaxies are covered without relaxation. Sources outside these regions also retain full wavelength coverage in a single grism orientation over a much larger area.

\end{itemize}

 \subsection{Scott Tompkins: The Extragalactic Background Light and Cosmic Spectral Energy Distribution in the Era of Large Galaxy Surveys \label{sec:STompkins}} 

 The extragalactic background light (EBL) is composed of all light originating from outside of our Galaxy from $\gamma$-rays to radio waves. It is typically broken down into different wavelength windows such as the cosmic radio background (CRB), cosmic infrared background (CIB), and cosmic optical background (COB). The integration of galaxy number counts continues to be a reliable method for measuring the EBL as it is insensitive to foreground contamination from our Galaxy and Solar System. Historically, cosmic variance has been the largest systematic uncertainty in measurements of the integrated galaxy light (IGL). The need to combine data from different observatories also introduces uncertainty as differences in magnitude zero-points must also be considered. 
 
\begin{figure*}
     \centering
     \includegraphics[width=0.9\textwidth]{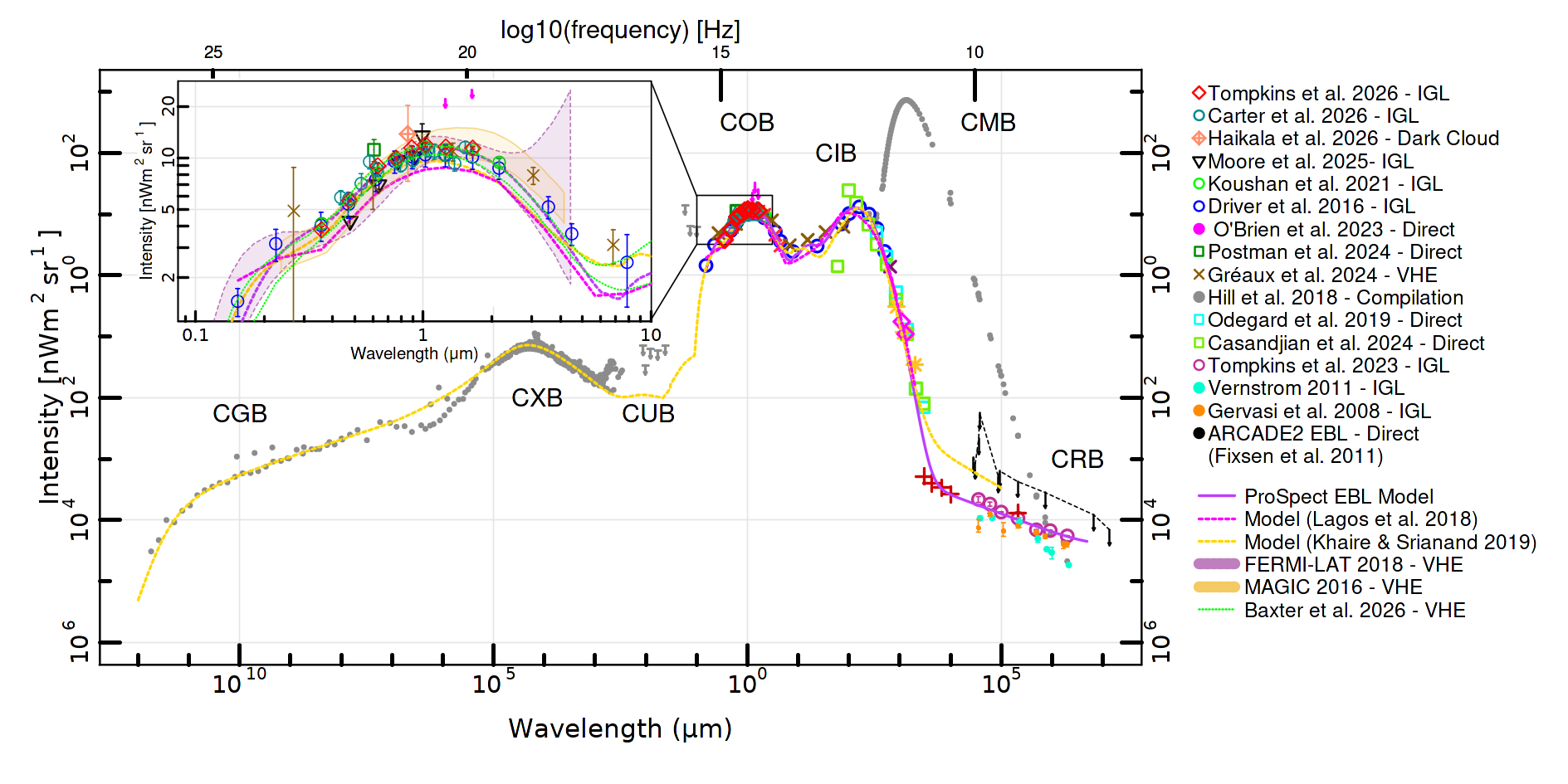}
     \caption{A compilation of EBL measurements and models from $\gamma$-rays to radio wavelengths. Measurements are labeled as either IGL, direct, dark-cloud, or indirect very high energy (VHE) measurements. Upper limits are displayed as downward facing arrows. An incomplete set of phenomenological and semi-analytical models are displayed as well. Measurements presented as a part of this conference are from \cite{Tompkins2026} (red diamonds). The new model discussed during the conference, \textsc{ProSpect EBL} (solid purple line) will be introduced in Tompkins et al. (in prep). The inset panel (top left) shows the status of COB and CIB measurements.}
     \label{fig:fig_Tompkins}
 \end{figure*}

 In the era of wide-area deep galaxy surveys from the Nancy Grace Roman Space Telescope and Euclid Space Telescope, a single observatory will provide deep imaging across large areas. These facilities have the potential to revolutionize measurements of the IGL. As the sum of all galaxy light, the IGL should be directly predictable from models of galaxy formation and evolution. Comparing model predictions of the IGL to observations will allow us to test models and begin to refine inputs to models. See Figure \ref{fig:fig_Tompkins} for a compilation of EBL measurements and models spanning from $\gamma$-rays to radio wavelengths. The inset panel (top left) highlights the cosmic optical background (0.1-8$\mu$m). This wavelength window is observable by one or more the space telescopes discussed during this conference.

 Future WFSS data from Roman and archived data from HST and JWST will provide accurate redshifts for tens of millions of galaxies. This will allow us to measure the IGL within specific redshift windows and measure its evolution with time, the cosmic spectral energy distribution (CSED). Measurements of the CSED are also directly predictable. Finally, it should be noted that none of the measurements of the IGL presented during this talk required dedicated observing time. Measurements of the IGL and CSED can be performed as ancillary science from future and archival data, highlighting that archiving and maintenance of legacy data sets continues to produce science after the original proposals and survey goals have been completed.

 \subsection{Nicolas Gomez Cruz: Assessing pipeline-dependent systematics in JWST/NIRSpec spectra of high-redshift emission-line galaxies  \label{sec:NGomez}}

After the observation of a possible dynamical dark energy component reported by the  Dark Energy Spectroscopic Instrument (DESI) \citep{DESI:2024mwx}, a validation from other distance estimators seem necessary. HII galaxies, through the $L-\sigma$ relation, provide perhaps the most reliable distance estimator  in the redshift range $2<z<8$ \citep{Chavez:2024twa}. However, the rather large scattering in the $L-\sigma$ relation  requires a large sample, and a careful treatment of systematics. One source of systematics comes already from the reduction of the raw data into calibrated scientific data products. For example,  explicit comparison of the calibrated spectra of two different groups reveals that sometimes the reduced 1D spectra are quite different. 

\begin{figure*}
     \centering
     \includegraphics[width=0.99\textwidth]{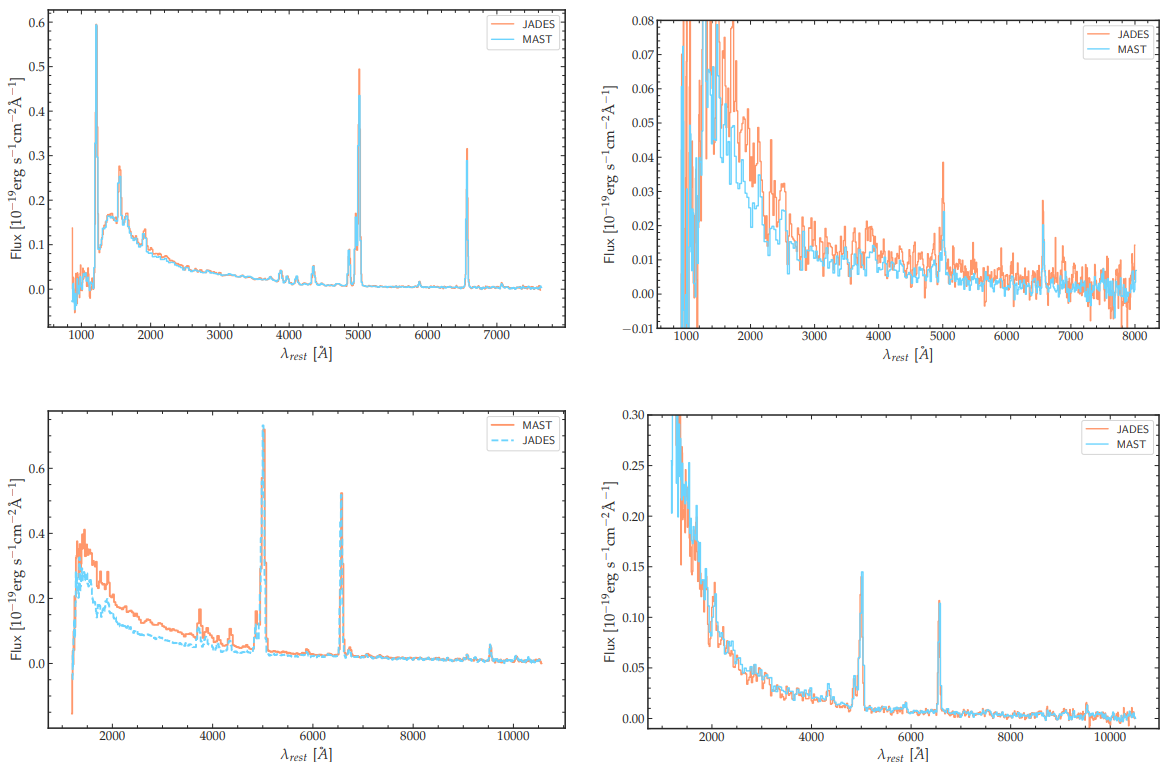}
     \caption{Comparison of the spectrum of four HII galaxies in the GOODS-South Deep Field taken by JWST/NIRSpec. The difference in the spectra comes from two  reduction pipelines applied to the same raw data: one from the standard calibration pipeline, version 2.0.1 (MAST), the other a customized version of the JWST Advanced Deep Extragalactic Survey (JADES) group, from DR4. Each spectrum represent 3 distinct visual dissimilarities, and one similarity (bottom right) between them. Taken from Gómez-Cruz et al. (work in prep).}
     \label{fig:NGomez}
 \end{figure*}

In \textbf{Figure \ref{fig:NGomez}}, we can observe two reduced spectra of the same raw data: one from the standard calibration pipeline (version 2.0.1) and the other from the Data Release 4 of the JWST Advanced Deep Extragalactic Survey (JADES) group. The figure displays four examples of HII galaxies (HIIGs) in the GOODS-South field. The four spectra represent four categories in which the dissimilarity can be classified: brighter lines but same noise and continuum (top left), brighter lines and higher noise (top right), different continuum (bottom left) and \textit{circa } coincident spectra (bottom right); upon comparison of one pipeline's product with respect to the other.

In order to assess the systematics introduced by the calibration process we must estimate derived differences in physical properties that result from the use of one pipeline or the other. It is found in the context of emission line galaxies at high redshift (and between JADES and MAST spectra), that the third case (spectrum in bottom left) may affect the equivalent width of bright emission lines, whereas the first case (top left) affects flux and flux ratio. However, the resulting discrepancy impacts on individual cases (outliers) in more and less consistent measurements on a large sample, mean quantities also being generally in accordance. Further and more exhaustive analysis is still required to establish a clear discrepancy that  affects, say, measurement of metallicities, SFR relations, or in constraining  cosmological parameters like the Hubble constant, the density parameter of matter today $\Omega_{m0}h^2$, or, notably in the case of a dynamical dark energy, the $w_0$ and $w_a$ from the  CPL parametrization. 

In principle, since JADES pipeline is built on top of MAST's, and targeted to deep sky objects, our report on the potential discrepancy on physical data will help users of MAST's products re-calibrate their analysis of high-redshift objects with the accuracy and of JADES' products without having to wait for the latter to be released.

 \subsection{Ivan Kramarenko: Allegro, A Next-Generation Tool for WFSS Data Reduction and Analysis \label{sec:IKramarenko}}
 Allegro is a new WFSS data reduction and analysis tool focused on speed, robustness, and emission-line science. It applies a median filter to the pre-processed grism images to isolate emission lines and runs Source-Extractor on the 2D spectral cutouts extracted from these images to detect the lines. Allegro is especially useful for fast searches of bright emission-line sources (e.g., LRDs) and population-level analyses that involve sophisticated completeness modeling (e.g., forward-modeling of the H$\alpha$ luminosity function in a lensed field).

\begin{figure}
    \centering
    \includegraphics[width=0.99\linewidth]{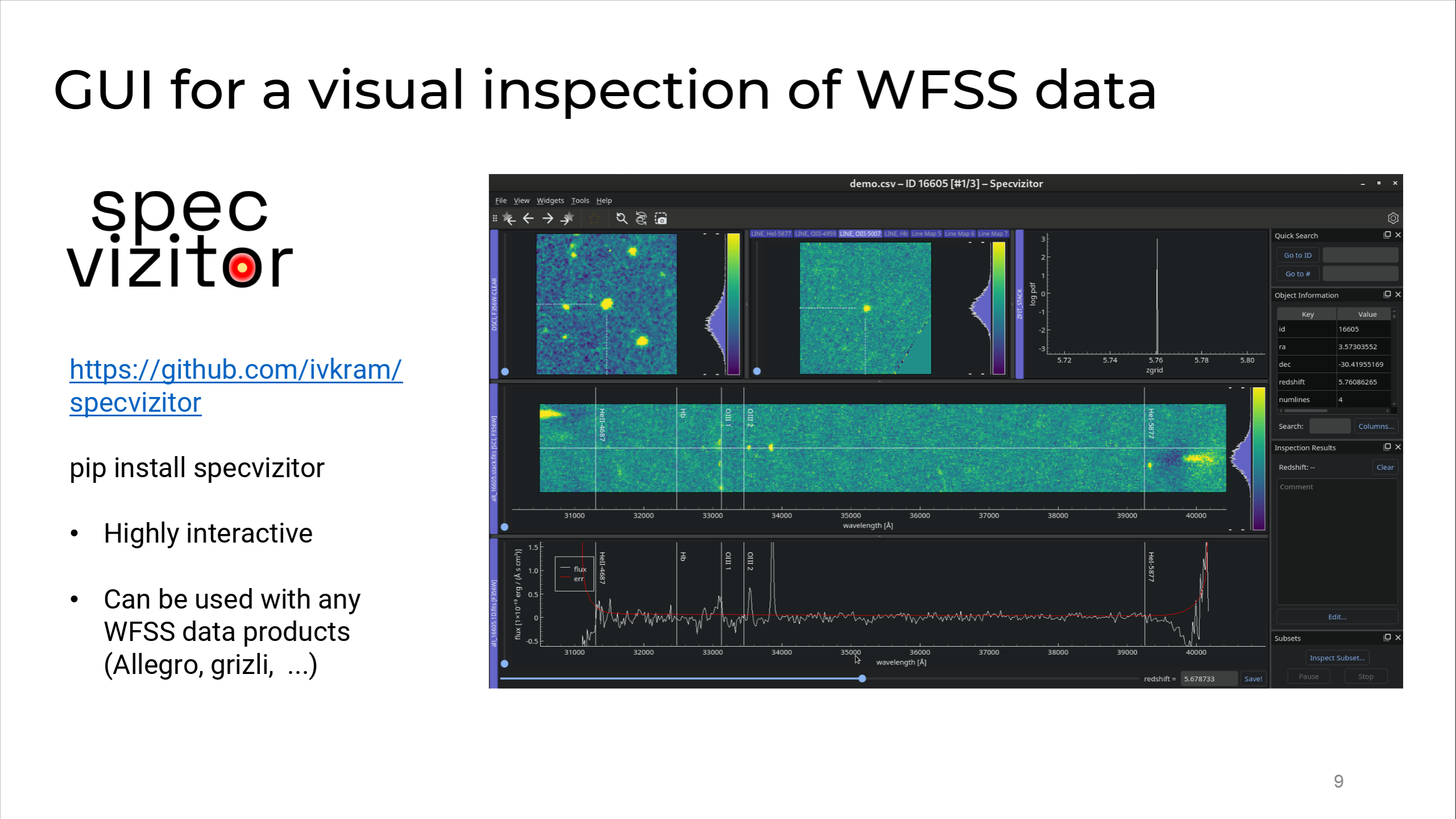}
    \caption{Figure from I. Kramarenko showing: Specvizitor (\href{https://github.com/ivkram/specvizitor}{https://github.com/ivkram/specvizitor}) is a Python GUI application for a visual inspection of WFSS data.}
    \label{fig:IKfig1}
\end{figure}

 Specvizitor (\href{https://github.com/ivkram/specvizitor}{https://github.com/ivkram/specvizitor}) is a Python GUI application for a visual inspection of WFSS data (Figure~\ref{fig:IKfig1}). It is highly interactive and can be used with any WFSS data products (including the grizli data products) thanks to customizable widgets.

 \subsection{Mohamed H. Abdullah: Connecting Galaxy Evolution and Cosmology with Spectroscopic Cluster Benchmarks in the Roman and
Euclid Era \label{sec:MAbdullah}}

The presentation summarizes the construction of the GalWCat spectroscopic galaxy-cluster catalog \citep{Abdullah18,Abdullah20a} and highlights several of its scientific applications. It focuses on the environmental effects on galaxies in clusters, including studies of the galaxy size--mass relation \citep{Abdullah26c}, three-dimensional galaxy classification \citep{Abdullah26a}, and the dependence of star-formation activity on environment \citep{Abdullah26b}.
It also presents cosmological applications, including cluster-abundance constraints \citep{Abdullah20b}, richness–mass calibration \citep{Abdullah23}, and measurements of the cluster correlation function and BAO signal \citep{Abdullah24}. Finally, it outlines plans to extend the cluster-detection methodology to DESI and, in the future, to Euclid and Roman data.

 \begin{figure}
    \centering
    \includegraphics[width=1\linewidth]{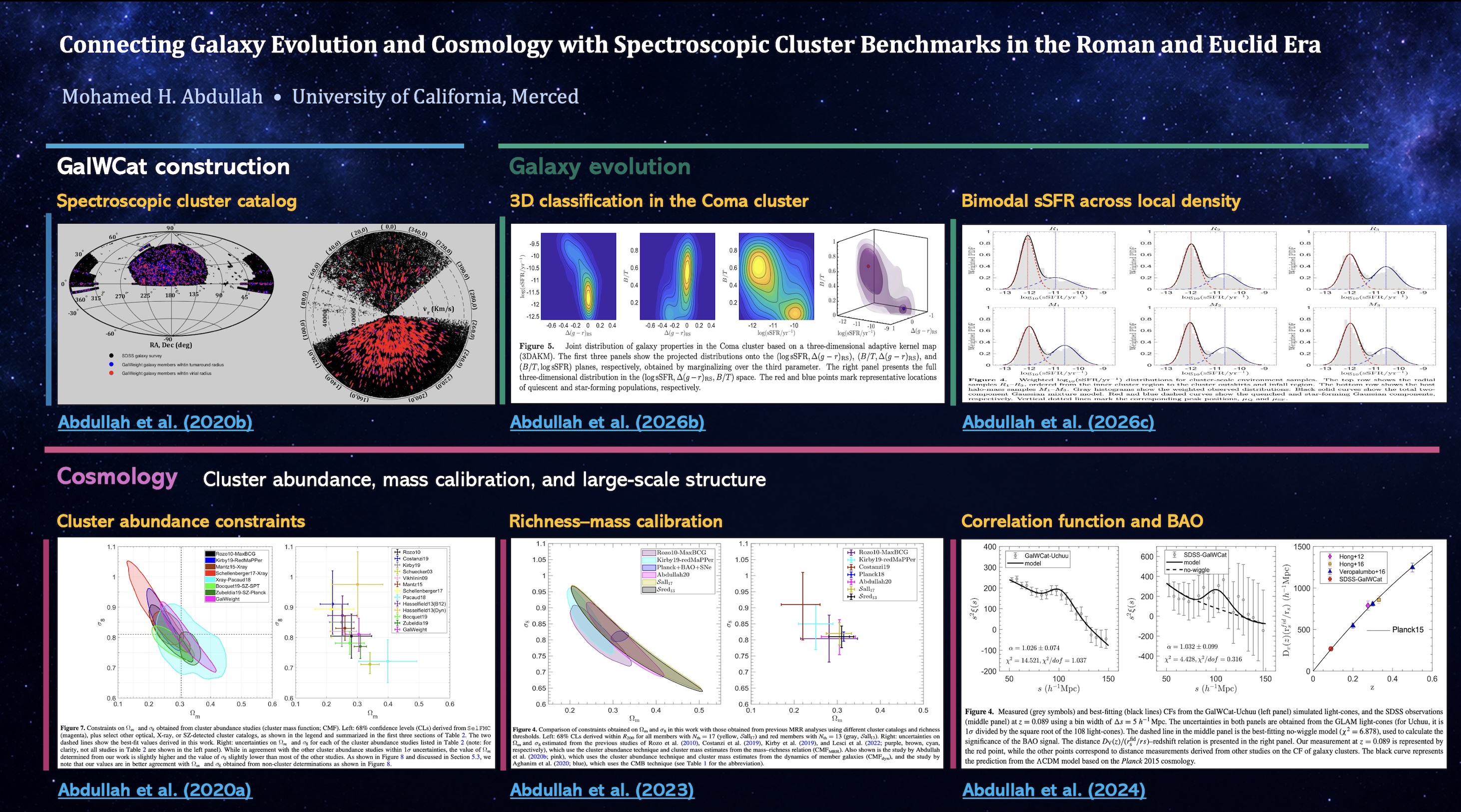}
    \caption{Connecting Galaxy Evolution and Cosmology with Spectroscopic
Cluster Benchmarks in the Roman and Euclid Era.}
    \label{fig:Abdullah}
\end{figure}

\section{Demo Sleuth \label{sec:VinceDemo} }
\begin{figure}
    \centering
    \includegraphics[width=1\linewidth]{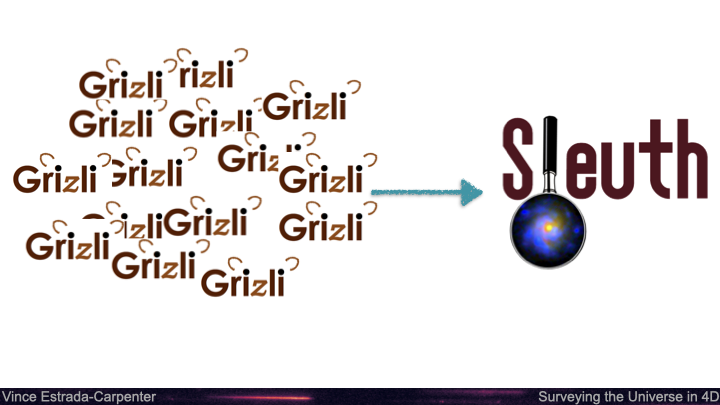}
    \caption{Sleuth utilizes techniques developed by Grizli \citep{grizli}, but applies them in a spatially resolved manner, producing more accurate, high-fidelity emission line maps.}
    \label{fig:sleuth1}
\end{figure}

Sleuth (https://github.com/Vince-ec/sleuth) is a Python package for forward-modeling and extracting spectra from JWST Wide Field Slitless Spectroscopy data. Given imaging, catalogs, segmentation maps, and grism exposures for a field, Sleuth builds per-object 2D forward models (continuum + emission line models), fits them to the grism data, and extracts 2D emission line maps. Its spatially resolved approach (Figure \ref{fig:sleuth1}) overcomes limitations of current modeling methods (Figure~\ref{fig:Abdullah}). 

\section{Discussion Session Day 4: How do we take advantage of WFSS data? What do we need to take advantage of WFSS data? \label{sec:DiscD4}}

Different teams write custom pipelines - how do we appropriately give recognition for this work, acknowledging that considerable effort is needed to generate science-ready products which can be a barrier for entering the field, while also minimizing independent groups reinventing the analysis wheel?
\begin{itemize}
\item How do we assess whether reduction pipelines are providing science data products that are sensible? One possibility, especially for pipelines focused on enabling spatially resolved spectroscopy from WFSS, is to benchmark against NIRSpec IFU data of the same source.
\item How do we benchmark reduction codes, recognizing that observatories have different requirements and capabilities? Some pipelines designed for specific science cases and it can be helpful to broaden the WFSS community by making it clear which pipelines are better suited for which science cases.
\item Pipeline developers promote their own codes, and all are designed for different purposes. As a community, we have not identified a unified benchmark for evaluating pipelines. Simulations can be a starting point but they are not representative of real data that often have noise sources and properties that are not perfectly replicated in simulations.
\item A lot of pipelines already exist. Could we utilize AI agents to evaluate pipelines or to optimally reduce WFSS data based on specific science goals?
\item More focused workshops (like this one!) would be advantageous to facilitate discussion and collaboration across the community, both for pipeline development and to advertise scientific results enabled by WFSS.
\item Early career researchers are incentivized to create their own codes to burnish their reputations and CVs. This reward structure promotes "reinventing the analysis wheel," and there should perhaps be greater recognition for those who do the work to test the existing pipelines against each other - such a paper could be highly cited.
\item The use of web-based science platforms, such as the Roman Research Nexus \citep{Desjardins26} \footnote{\url{https://roman.science.stsci.edu}}, is also able to help with taking advantage of WFSS data in those use cases where the data are too big or difficult to download or fit on personal computers and local compute clusters.
In fact, the Nexus platform can help with the petabyte-scale data deluge coming from Roman by making its cloud-based data sets accessible to public users through a Jupyter Notebooks interface on the web. This interface can provide software environments composed of a dedicated suite of analysis and visualization code, packages, and libraries, such as those presented in this document, to facilitate remote analysis or processing of the data.

\end{itemize}

\section{Posters}
\subsection{Natalia Alvarez-Ibanez: Population-level Fe II emission in Little Red Dots (LRDs) observed with JWST/NIRSpec}\label{sec:NAlvarez}
\begin{figure}
    \centering
    \includegraphics[width=0.5\linewidth]{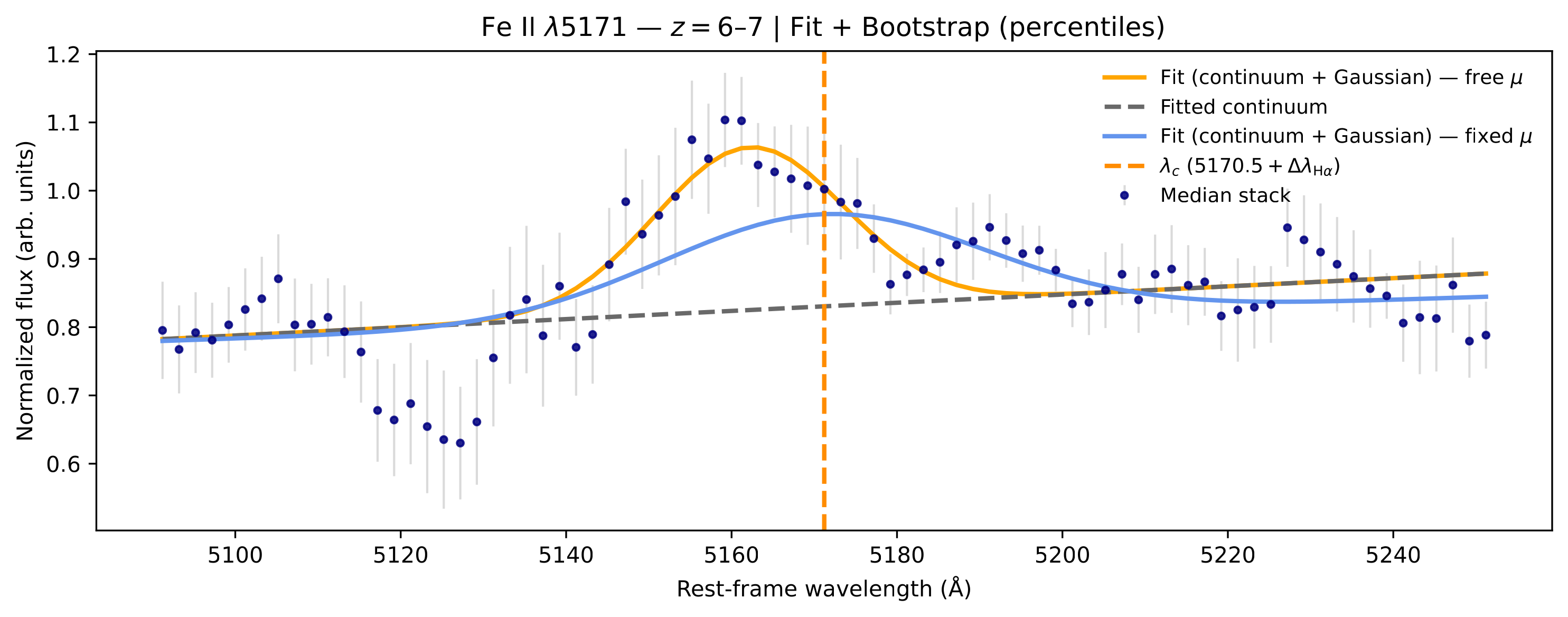}
    \caption{Population-level Fe II emission in Little Red Dots observed with JWST/NIRSpec: Stacked spectra shows an excess complatible with Fe II emission line at $z\sim6-7$.}
    \label{fig:NAIfig}
\end{figure}

A population-level search for rest-frame optical Fe~II emission was conducted using reduced JWST/NIRSpec PRISM spectra from version 4.4 of the DAWN JWST Archive. The parent sample was selected from the DJA/NIRSpec LRD catalog presented by de Graaff et al. (2025). The final analysis included 105 spectra. Median and inverse-variance-weighted mean spectral stacks were constructed in redshift bins to investigate weak Fe~II emission not detected at comparable significance in individual spectra. The central diagnostic presented in the poster is reproduced in Figure~\ref{fig:NAIfig}. The strongest integrated excess around Fe~II 5171~\AA{} is found in the median stack at $6<z<7$, comprising 13 LRDs, and remains positive in 99\% of the galaxy-level bootstrap realizations constructed using median stacking.

Nevertheless, the broad profile of the excess, the approximately 9~\AA{} blueshift obtained when the centroid is allowed to vary, and deviations of comparable significance in adjacent control windows indicate a complex local spectral structure, making the isolated identification of Fe~II $\lambda5171$ difficult. If Fe II contributes to the excess, it may suggest the presence of a dense gas associated with the nuclear environment. However, the present stack cannot distinguish this contribution from blended emission, gas kinematics, or radiative-transfer effects.

Although the analysis is based on NIRSpec/PRISM spectroscopy, Fe~II $\lambda5171$ falls at 3.62--4.14~$\mu\mathrm{m}$ over \(6\leq z<7\), within the combined F356W and F444W coverage of NIRCam/WFSS. Deep WFSS surveys could therefore extend this search to larger samples.

\subsection{Kasra Mokhtarpour: Exploring Gaia Data Release 3 for Changing State AGN Candidates \label{sec:KMokhtarpour}}
Changing state active galactic nuclei (CSAGNs) are rare systems in which the optical/UV spectrum changes as the accretion flow of the super massive black hole (SMBH) changes\citep{2023NatAs...7.1282R}. These objects are especially useful because they enable us to study the accretion physics on human timescales. However, finding such rare sources is really difficult as large time domain surveys contain numerous variable AGNs, while only a very tiny fraction could be selected as changing state candidates. In my work, I searched for CSAGN candidates in Gaia DR3 using Gaia's catalog of variable AGN\citep{2023A&A...674A..24C}. The goal of this research is to use Gaia photometry as a first-pass filter, narrowing a very large parent sample to a smaller set of sources whose variability is consistent with a change-of-state transition during Gaia DR3 observing baseline. The next step is conducting a spectroscopic follow-up on the current photometrically selected CSAGN candidates. This work is motivated by the coming era of large spectroscopic and time-domain surveys, where methods that connect photometric variability to efficient spectroscopic follow-up will be needed to identify the rare changing accretion-state AGNs.

\subsection{Zhiwei Pan: NEXUS: A Demographic Study of AGN and Little Red Dots with JWST Wide-Field Slitless Spectroscopy \label{sec:zpan}}

\begin{figure}
    \centering
    \includegraphics[width=1\linewidth]{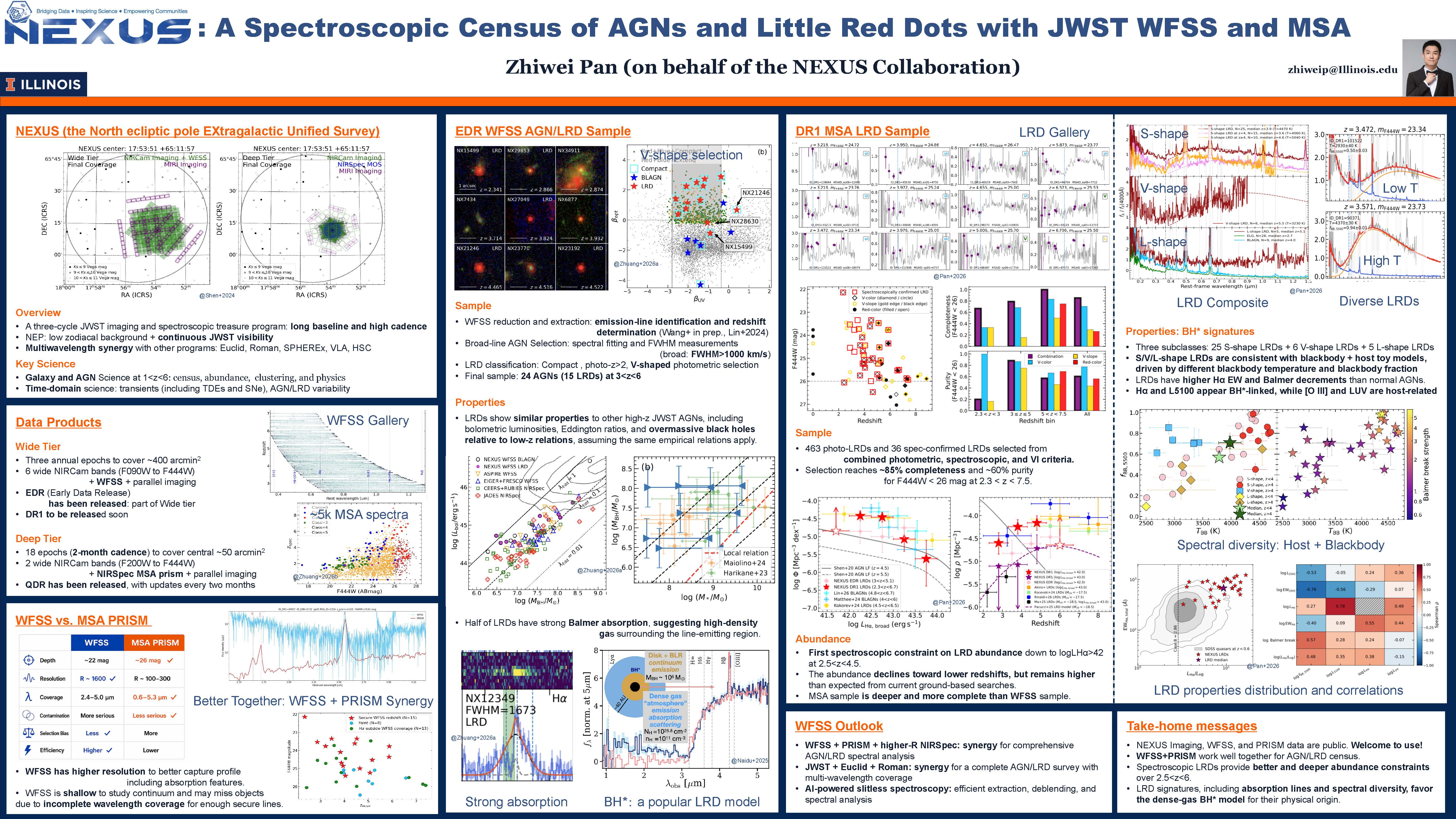}
    \caption{NEXUS combines JWST WFSS and NIRSpec/MSA spectroscopy to build a complementary census of AGNs and LRDs, constrain their abundance at \(2.5<z<6\), and reveal diverse spectral signatures consistent with the dense-gas scenario.}
    \label{fig:ZWPposter}
\end{figure}

As shown in Figure \ref{fig:ZWPposter}, NEXUS presents a spectroscopic census of AGNs and LRDs using complementary JWST WFSS and NIRSpec/MSA observations. The samples include 24 WFSS-selected broad-line AGNs and 36 MSA-confirmed LRDs, providing deeper constraints on LRD abundance and revealing diverse S-, V-, and L-shaped spectra. Their strong Balmer absorption, large \Ha{} equivalent widths, and spectral diversity support models involving a compact blackbody-like source surrounded by dense gas and mixed with host-galaxy emission.

\subsection{Sandra Jaison: Off-Nuclear AGNs in MaNGA with Megacubes: Spatially Resolved Identification \label{sec:Sjaison}}
\textit{Off-Nuclear AGN Signatures in MaNGA-Spatially Resolved Selection Using {\sc megacubes}}: Jaison et al. (in prep.) identify 201 off-nuclear AGN ionization candidates in MaNGA {\sc megacubes} using spatially resolved BPT, WHAN, and WHAD diagnostics. Of these, 99 are validated by 3D BPT \citep{ji2020} as exhibiting true offset AGN-like activity. While limited to the local universe, this low-redshift catalog provides a benchmark for high-redshift studies. Follow-up JWST (NIRSpec/MIRI) observations of these candidates will help constrain their physical nature, whereas forthcoming WFSS surveys will yield the large, unbiased samples needed to characterize offset AGN populations statistically.


\subsection{Gaël Noirot: Star Formation Activity in Confirmed Clusters and Proto-Clusters at $1.4 < z < 2.8$ with HST Slitless Grism Spectroscopy\label{sec:GNoirot}}

The presentation focused on results from confirmed clusters at $1.4<z<2.8$ with {\it HST} slitless grism spectroscopy.

\begin{itemize}
\item CARLA \citep[Clusters Around Radio-Loud AGN;][]{Wylezalek2013, Wylezalek2014} is a Spitzer program that identified 420 high-redshift cluster candidates around Radio-Loud AGN across the entire extragalactic sky. Follow-up HST G141 grism spectroscopy of the 20 densest candidates confirmed 16 clusters at $z=1.4-2.8$, doubling the census of known $z>1.5$ clusters at the time \citep[][]{Noirot2016, Noirot2018}.

\item The sample consists of a diversity of clusters: e.g., SF-dominated clusters \citep[][]{Noirot2018}, centrally-concentrated quiescent-dominated clusters \citep[][]{Cooke2015, Cooke2016}, assembling clusters with large number of both red-sequence and SF members \citep[][]{Noirot2016, Noirot2018}, a $z=2$ cluster with nascent intra-cluster light \citep[][]{Noirot2018, Werner2023}, etc. 

\item From the HST emission line fluxes, it is found that on average massive cluster members are located below their SF main-sequence, while at the same time, the cluster centers are the main sites of star-formation. This indicates the accelerated evolution of massive members, while the clusters are still actively forming \citep[][]{Noirot2018}.

\item The HST imaging also reveals that the morphology-density and passive-density relations are already established in the clusters \citep[][]{Mei2023}, and that cluster members have larger mass-normalized sizes than in the field, suggesting different growth mechanisms for the field (small sizes followed by rapid growth) and cluster galaxies (larger sizes followed by slow growth) at these early epochs \citep[][]{Afanasiev2023}.

\item The sample probes the critical epoch of cluster assembly, ideal to study the mechanisms responsible for galaxy growth and quenching in these environments at the time where they are happening.

\end{itemize}

\subsection{Guillermo Romero Cruz: Spatially-Resolved Stellar Populations in the Inner Halo of M82 reveals Intermediate star formation \label{sec:GRomero}}

\textit{Evidence for recently-formed stars in the inner halo of M82 using AGB stars in the JWST images}: AGB stars were identified from JWST NIRCam photometry to trace recent formation of stars in M82. The age of the stellar population increases with height above the disk and this indicates ongoing, extended star formation in the disk.

\subsection{Arshia Akhtarkavan: Forward-Modeling Galaxy Kinematics and Morphology with JWST NIRCam Grisms \label{sec:aakhtarkavan}}

\textit{Forward-Modeling Galaxy Kinematics and Morphology with JWST NIRCam Grisms}: SAPPHIRES survey with NIRCAM/WFSS to recover intrinsic galaxy kinematics at redshifts $z\sim1-1.5$. The DINGO python package was used to forward model kinematics with dual grism spectra and obtain resolved Tully-Fisher relationship.

\subsection{Maya Seagraves: 50,000 Spectra and Counting: Preparing for a Roman Spectral Atlas \label{sec:MSeagraves}}

The Wide Field Camera 3 Infrared Channel (WFC3/IR) on the Hubble Space Telescope enables both direct imaging and slitless spectroscopy, in which a grism disperses the light from every source in the field. 
This team developed a Python-based pipeline to extract more than 50,000 stellar spectra from 84 HST fields spanning six filter-grism combinations, providing a framework that can be adapted to the Wide Field Instrument (WFI) on the Roman Space Telescope, which will produce vastly larger spectroscopic datasets. Using this pipeline we successfully derived effective temperature, surface gravity, and metallicity for 411 stars, demonstrating that large-scale, data-driven stellar characterization is feasible with slitless infrared spectroscopy. We will further improve the extraction by masking unwanted grating orders and optimizing metallicity determination, using the resulting spectral library as a testbed for the large-volume slitless spectroscopy expected from Roman (Seagraves et al. in prep).


\subsection{Yuxuan Pang: The Structure and Evolution of LRDs: Insights from JWST NIRSpec Medium and High Resolution Spectroscopy at z~4 \label{sec:YPang}}

\textit{The Structure and Evolution of LRDs: Insights from JWST NIRSpec Medium and High Resolution Spectroscopy at $z\sim4$}. This poster presents a study of the structure and evolution of Little Red Dots (LRDs) using JWST NIRSpec medium- and high-resolution spectroscopy at z$\sim$4. By decomposing the broad and narrow components of Balmer emission lines, we investigate the connections between line emission and the UV/optical continua. We find that the broad H$\alpha$ luminosity strongly correlates with the optical continuum—but not the UV—indicating a common AGN origin for both. In contrast, the [O III] line strength correlates with the UV continuum rather than the optical. We estimate central black hole masses of $10^6-10^8M_{\odot}$, with the black holes accreting at high Eddington ratios. Assuming a constant mass accretion rate within a slim-disk framework, we infer growth timescales of $10^5-10^7yr$, suggesting that LRDs may evolve into NLS1 galaxies. Upper-limit constraints reveal weaker optical Fe II emission in LRDs compared to typical AGN. To simultaneously match the inferred broad-line region size and the observed luminosity, we propose a "Clumpy Envelope" model, in which the optical emission arises from extended, clumpy gas with a characteristic radius of tens of light-days. The diversity in the observed optical continuum shapes can be explained by radial temperature gradients and self-absorption effects within this structure. 

\subsection{Faezeh Manesh: Constraining Star Formation Burstiness in Lensed Dwarfs at z ~= ~0.5~–~2.3 with HST + GLASS-JWST/NIRISS \label{sec:FManesh}}

\textit{Burstiness in Dwarf Galaxies at $0.5 < z < 2.3$}: Bursty star formation has been observed, studied, and modeled extensively in low-mass galaxies in the local Universe. With the advent of JWST, we can now investigate this phenomenon at higher redshifts. Using JWST/NIRISS grism spectroscopy from GLASS \citep{Treu_2022, Watson_2025, watson2026spatiallyresolvednebularstellarreddening}, together with rest-frame UV imaging from HST \citep{Alavi_2016} and multiwavelength photometry from UNCOVER and MegaScience \citep{suess2024mediumbandsmegascience}, we study more than 150 galaxies to constrain the degree of star-formation burstiness during cosmic noon. By taking advantage of the strong lensing provided by the Abell 2744 cluster, we extend this analysis to dwarf galaxies with stellar masses as low as $M_\star \sim 10^7\,M_\odot$. We combine these data to derive SED-based physical properties, including stellar masses and star-formation rates. We compare H$\alpha$ emission from the NIRISS grism spectra with rest-frame UV emission from HST imaging, which trace star formation on timescales of approximately 10 and 100 Myr, respectively. The H$\alpha$-to-UV ratio provides tentative evidence that dwarf galaxies become increasingly bursty from the local Universe to $z>0.5$.

\subsection{Peter Gwartney: The Role of Bars in Galaxy Evolution: Prospects for Euclid and Roman \label{sec:Pgwartney}}
\textit{The Role of Bars in Galaxy Evolution: Prospects for Euclid and Roman}: 113,696 SDSS galaxies with $9 < log\frac{M_{\star}}{M _{\odot}} < 12$ and $0.02 < z < 0.06$ were visually inspected for bars. When compared to ML classification, visual classification yields a higher bar fraction for high mass galaxies, undisturbed close pairs, cluster galaxies, and in dense environments. ML classification of bars will be necessary in the era of Roman and Euclid. However, the choice of bar selection technique can have a significant effect on studies of bar-derived properties.

\section{WFSS in the Age of ML and IFUs}

\subsection{Adrian Bayer: Optimally Surveying the Universe in 4D: Machine Learning and Field-Level Inference for Wide-Field
Slitless Spectroscopy \label{sec:ABayer}}

\begin{figure*}
\centering
\includegraphics[width=\textwidth]{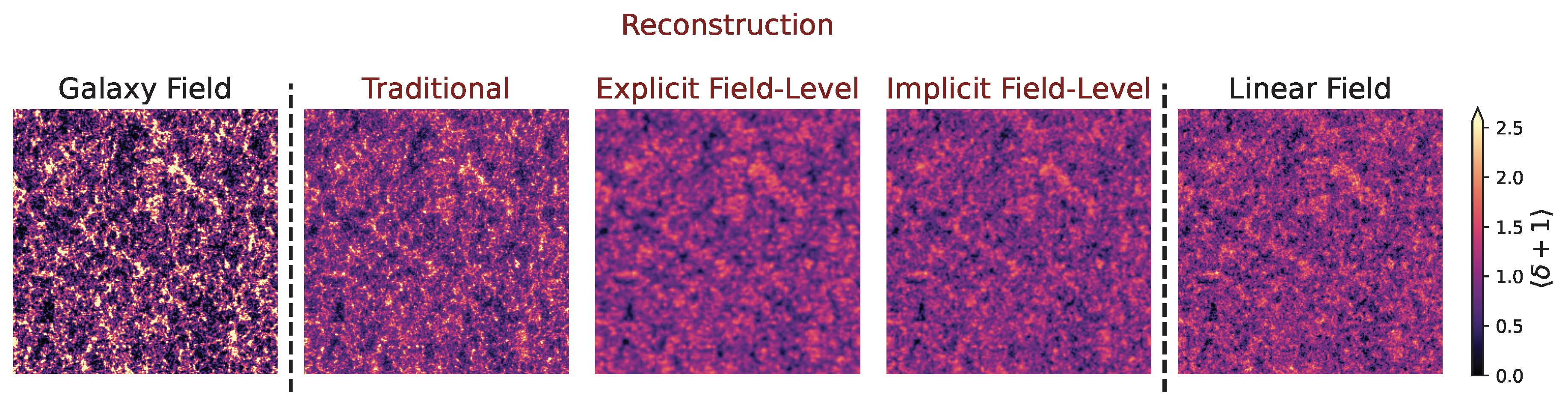}
\caption{\textbf{Illustration of field-level reconstruction methods.}
Starting from the observed galaxy field (left), traditional BAO reconstruction, explicit field-level inference, and implicit field-level inference are used to reconstruct the underlying linear density field, shown for comparison at right. The agreement with the linear field improves from traditional reconstruction to explicit and then implicit field-level inference, illustrating how field-level approaches can recover additional spatial information. Each panel shows a two-dimensional projection of thickness $125,{\rm Mpc}/h$ perpendicular to the line of sight. Reproduced from \citet{BayerEtAl2026BAO}.}
\label{fig:ABfieldlevel}
\end{figure*}

Statistics beyond the two-point function can recover cosmological information hidden by nonlinear evolution, including information about neutrino mass contained in halos, voids, and their spatial relationships \citep{BayerEtAl2021}. Field-level inference extends this idea by modeling the observed field itself rather than first compressing it into a small set of summary statistics. Figure~\ref{fig:ABfieldlevel} illustrates these methods in a concrete BAO reconstruction example, comparing traditional reconstruction with explicit and implicit field-level reconstructions of the underlying linear density field. Explicit approaches define a likelihood and can jointly infer the initial density modes, cosmological parameters, astrophysical parameters, and systematic parameters through optimization or sampling \citep{SeljakEtAl2017,SchmidtEtAl2020,MCLMC,Dong:2026wpn}. Implicit approaches instead use simulations and machine learning to learn the connection between observations and the quantities of interest. They do not require a tractable likelihood and can accommodate complicated survey effects, provided that these effects are accurately represented in the forward simulations.

A central challenge is ensuring that the additional information extracted by these methods is robust to theoretical and observational systematics. In DESI, for example, fiber assignment preferentially removes close angular pairs and imprints a density-dependent selection on the observed galaxy field \citep{BianchiEtAl2025}. Such effects must be included in the forward model, marginalized over, or removed through validated scale cuts. Similar validation is required for machine-learning analyses when changing simulation codes, numerical resolution, or astrophysical assumptions \citep{BayerEtAl2025Robustness}. Interpretability is equally important for identifying which physical structures and scales drive an inference, and for determining whether a model has learned genuine cosmological information rather than simulation-specific artifacts \citep{NuCNN,arnab}. Transfer learning may improve training efficiency by adapting models trained on large sets of approximate simulations using smaller sets of more realistic simulations or observations \citep{Thiele:2025mlo,KrishnarajEtAl2026}.

Although BAO provides a useful controlled demonstration that can be compared directly with mature reconstruction and two-point analyses \citep{ParkerEtAl2025,BayerEtAl2026BAO}, the broader goal is to use the same framework to extract nonlinear information and jointly model multiple observables, cosmological parameters, astrophysical processes, and survey systematics.

Roman and Euclid offer complementary opportunities for extending field-level inference in this direction. Both combine slitless spectroscopy with imaging and weak-lensing measurements over common areas. Euclid provides a very large survey volume and access to the largest cosmological scales, while Roman will survey a smaller area more deeply and with a higher source density \citep{EuclidCollaborationMellier2025,WangEtAl2022}. Field-level analyses could use a common latent matter field to model the three-dimensional spectroscopic galaxy distribution, photometric clustering, and weak-lensing shear simultaneously. This would enable joint constraints on cosmology and astrophysics, including galaxy bias, the galaxy--halo connection, emission-line properties, and the dependence of galaxy populations on their environments.

Beyond parameter constraints, field-level inference can produce constrained reconstructions---and, when the posterior is sampled, ensembles of constrained realizations---of the latent initial and evolved density fields, with well-measured modes anchored by the data and poorly constrained modes varying consistently with the model. These spatially resolved reconstructions can be evolved forward or used to predict other tracers, providing a natural basis for joint multi-probe analyses of the 4D Universe.

Cross-correlations are central to this opportunity. Correlating spectroscopic galaxies with photometric galaxies and weak lensing can add information while helping to calibrate photometric-redshift distributions, galaxy bias, and survey selection effects \citep{JohnstonEtAl2024}. Correlations with CMB lensing and the thermal Sunyaev--Zel'dovich signal provide further information and valuable consistency tests \citep{EiflerEtAl2024Roman}. At the same time, slitless spectroscopy introduces scene-dependent systematics: overlapping spectra, source morphology, detector effects, extraction, and redshift failures can make the selection function of the final catalogue depend on the surrounding scene \citep{GabrielpillaiEtAl2024,EuclidCollaborationPassalacqua2026}. A longer-term goal is therefore to propagate information and uncertainties through the full chain from detector pixels to spectra, catalogues, and cosmological inference. Correlated simulations such as HalfDome can provide a foundation for developing and validating these joint Roman, Euclid, Rubin, and CMB analyses \citep{BayerEtAl2025HalfDome}.

\subsection{Jerry J.-Y. Zhang: ROSSINI: the ROtational Slitless Spectrograph for INtegral field spectroscopy and Imager \label{sec:JZhang}}

\begin{figure}
    \centering \includegraphics[width=0.8\linewidth]{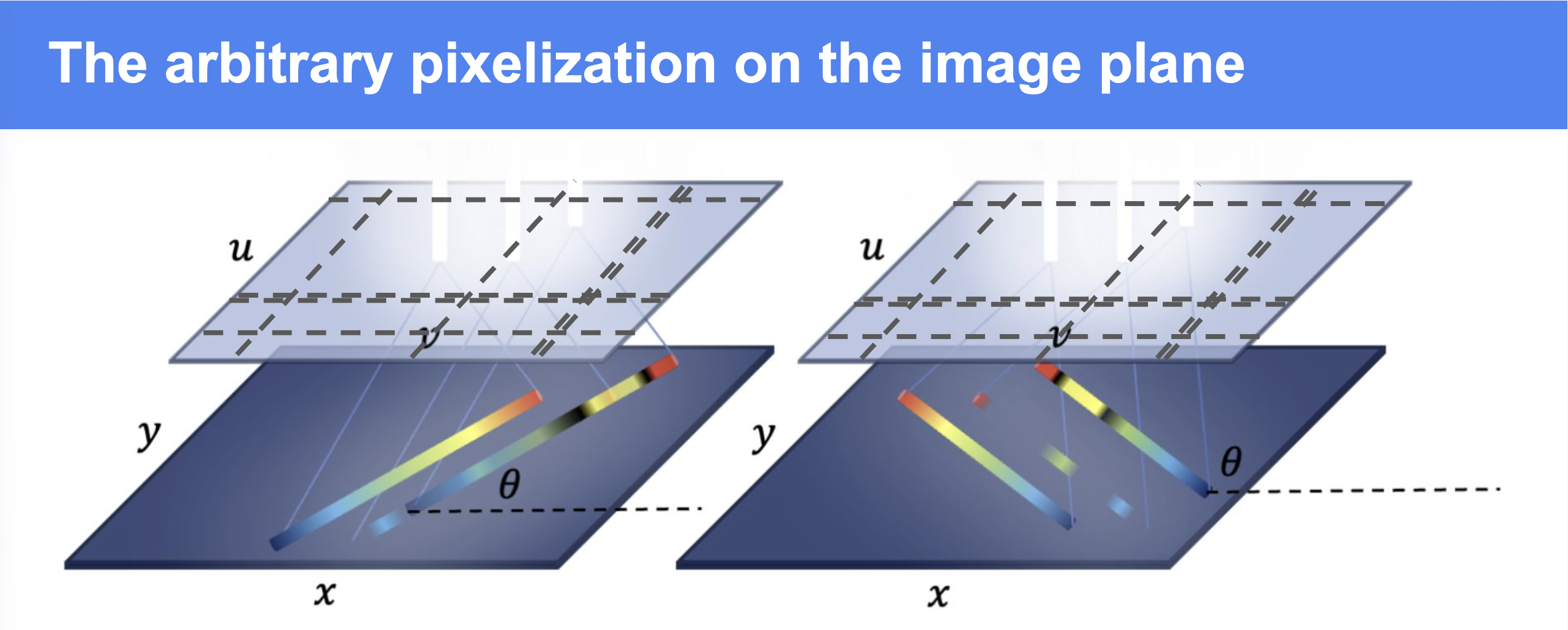}
    \caption{Reconstructing a datacube is just an arbitrary representation on the continuous reality on the focal plane ($u,v$). Thus, we can choose an arbitrary pixelization on the plane, which can even allow pixels with irregular shapes, to sample “regions of interest” at higher resolution.}
    \label{fig:RossiniFig}
\end{figure}

Among the astronomical spectroscopic techniques, Integral Field Spectroscopy (IFS) is regarded as one of the most versatile and powerful, but it is limited by small fields of view, complex instrument designs, and extremely high costs. We present ROSSINI: the ROtational Slitless Spectrograph for INtegral field spectroscopy and Imager, a spectrograph design aimed at performing IFS without integral field units (IFUs) but in a slitless fashion instead. The device relies on generating an arbitrary number of independent detector images of the same field by rotating the whole telescope and/or the dispersion direction of the optical element (prism, grating, or grism) and on a postprocessing tomographic reconstruction algorithm yielding a full IFS datacube, with arbitrary angular and spectral resolution defined by the user (Figure \ref{fig:RossiniFig}). By combining the very high efficiency of slitless spectroscopy with the high information content of IFS, ROSSINI is the ideal instrument to perform IFS in wide fields. 

\begin{figure}
    \centering \includegraphics[width=0.8\linewidth]{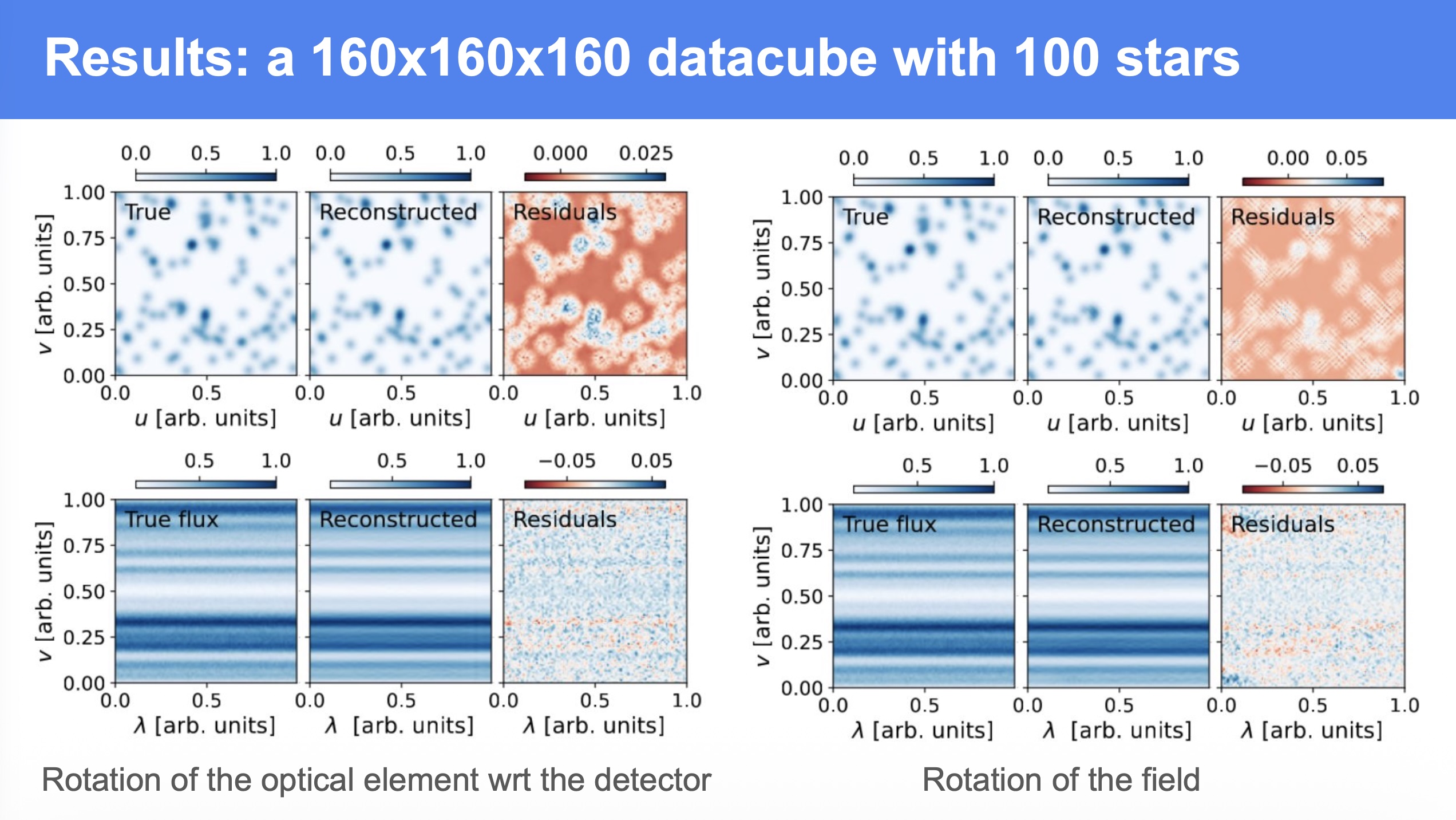}
    \caption{An iterative tomographic reconstruction for a small test datacube shows a satisfactory recovery.}
    \label{fig:RossiniFig2}
\end{figure}

We first develop the mathematical formulation of the problem and derive the solution as a linear matrix inversion. Then, we provide an interpretation of ROSSINI as a particular application of tomography and leverage this to propose a practical numerical solution based on iterative reconstruction. We test this novel conception through a series of numerical experiments: we first generate toy datacubes on the sky, then simulate the spectrograph and the corresponding detector images, and finally apply tomographic reconstruction, trying to recover the input datacube (Figure \ref{fig:RossiniFig2}). 

Conceptually, the rotational slitless spectrograph can handle datacubes with an arbitrary pixelization. The number of needed rotations can be easily computed from the relative pixelization of the datacube and the pixel number of the detector. From the numerical experiments, ROSSINI turns out to be able to reconstruct the benchmark datacubes with percent accuracy and in just a few hundred iterations, using negligible computational resources. We conclude that this novel conception is a promising way forward for future spectroscopic facilities, as it allows us to perform wide-field IFS in an efficient and cheap fashion by relying mostly on existing slitless spectrograph technologies.

\subsection{Mason Huberty: The Capability of Using JWST/NIRSpec as a Slitless Spectrograph \label{sec:MHuberty}}
Some science cases require slitless observations at $1-2\rm \mu m$ at $\rm R\gtrsim1000$. Unfortunately, none of the existing WFSS instruments (JWST/NIRISS/NIRCam/MIRI, Euclid NISP, and Roman WFI) achieve this wavelength/resolution combination. 
We overcome this obstacle with JWST/NIRSpec Multi-Object Coronagraphy for Slitless Spectrosocopy (MOCSS, PIs: Hayes and Scarlata): by opening essentially all of the microshutters on the NIRSpec microshutter assembly (MSA), NIRSpec can be effectively turned into a slitless spectrograph. Simultaneously, MSA shutters associated with known, bright sources that are not relevant for one's science case are closed. This prevents non-relevant bright spectral traces from contaminating those from faint sources. Huberty et al. (in prep.), demonstrates that this method achieves the intended WFSS observations and securely constrains spectroscopic redshifts for 22 sources, 18 of which are first time spectroscopic measurements. The 4 sources with existing spectroscopic redshifts from JADES match those from measured in MOCSS. This method has also allowed for the identification of an unusual orphan single-line with no counterpart in broadband imaging. To summarize, this method holds great promise for constraining some of the faintest lines and sources in the Universe.

\subsection{Mitchell Revalski: Slitless Meets IFU: Low-Mass Galaxy Evolution in the MUSE Ultra Deep Field \label{sec:Mrevalski}}

The MUSE Ultra Deep Field (MUDF) is a unique region of the sky that hosts two quasars at $z \approx$~3.22 that are physically separated by only 500 kpc. The light from these quasars allows us to connect the properties of galaxies observed in emission with their surrounding gas viewed in absorption along two converging sight lines. This field has been observed for 142 hours with VLT MUSE, and with 90 orbits of HST WFC3/IR G141 grism spectroscopy, making it the deepest single grism field ever observed with Hubble \citep{Revalski2023, Revalski2024}.

\begin{figure}
     \centering
     \includegraphics[width=0.99\linewidth]{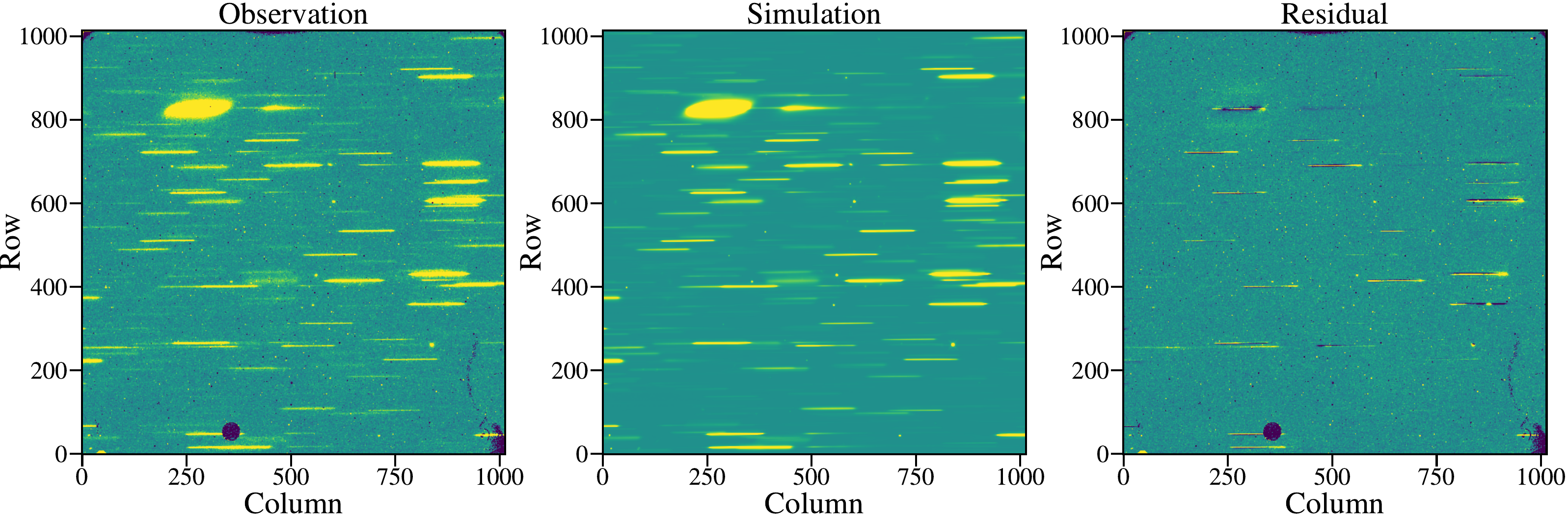}
     \includegraphics[width=0.75\linewidth]{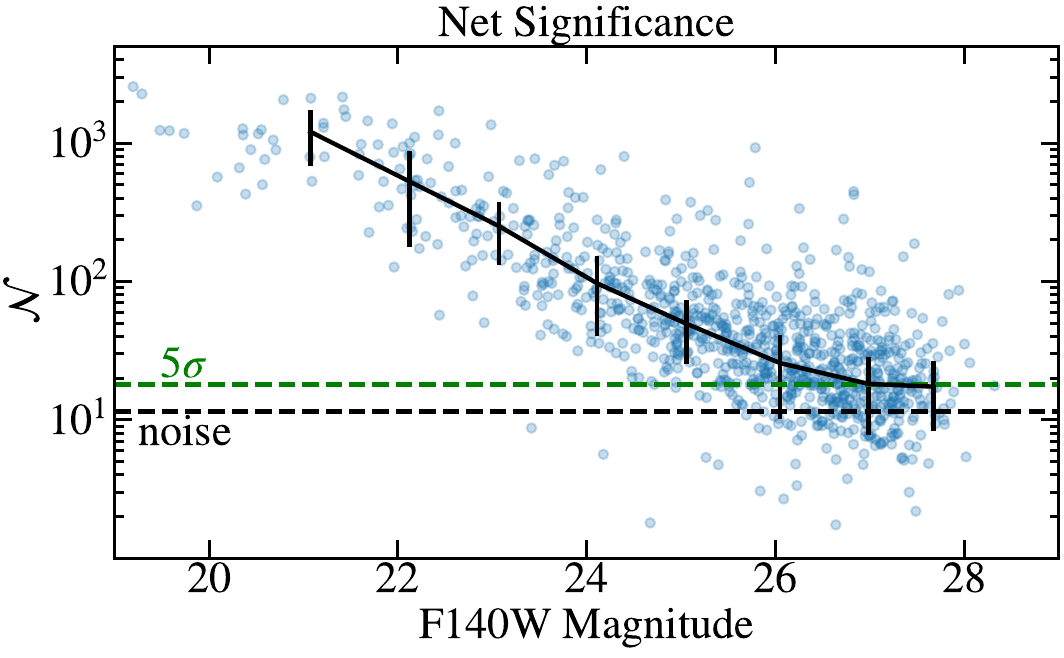}
     \vspace{-0.5em}
     \caption{Figures 6 and 8 from \cite{Revalski2024}, showing the observed and modeled spectral image for one position angle (upper panels) and the net significance ($\mathcal{N}$) for the extracted spectra (lower panel). The net significance captures the information content of a spectrum and has a higher value for sources with emission lines, a bright continuum, or a combination. In this case, the HST WFC3/IR G141 spectra of the MUDF reach a 5$\sigma$ continuum detection for sources as faint as \textit{m}$_\mathrm{AB} \approx$ 27.}
    \label{fig:MRfig}
 \end{figure}

A key metric in quantifying the quality and depth of grism spectra is the Net Significance ($\mathcal{N}$), or the maximum cumulative S/N as defined in \cite{Pirzkal2017}. Specifically, the net significance is calculated by dividing each flux value by its uncertainty, sorting the original flux and error arrays in order of decreasing S/N, and then computing the cumulative S/N by successively adding more bins until a maximum value and turnover is reached. This metric captures the information content of a spectrum whether it has strong emission lines, a bright continuum, or a combination of these elements. A specific value for the net significance can be related to e.g. a 5$\sigma$ detection of the continuum, using the equations in \cite{Pirzkal2004}. In this way, the net significance provides a useful metric for comparing the information content of WFSS across different surveys, instruments, and telescopes (Figure \ref{fig:MRfig}).

\subsection{Clive Binu: Revolutionizing Extragalactic Spectroscopy: Leveraging Transformer-Based Models for Automated
Redshift Prediction \label{sec:cbinu}}

We demonstrate the effectiveness of the Spectroscopic Pre-Trained Transformer (SpecPT) by adapting this deep learning model to analyze challenging Hubble Space Telescope (HST) WFC3 grism data through transfer learning. Initially pre-trained on ground-based spectra from the Dark Energy Spectroscopic Instrument (DESI), we fine-tune SpecPT using spectra from the 3D-HST+AGHAST surveys, effectively bridging the gap between ground and space-based spectroscopic analysis. Our results show marked improvement on the noisier, low-resolution WFC3 grism data, with the redshift prediction's Normalized Median Absolute Deviation (NMAD) improving from 0.2095 to 0.0039 and the catastrophic outlier fraction ($\eta$) reduced from 47.97\% to 10.4\%. We further combine photometric flux data with the predicted spectroscopic redshifts to break degeneracies arising from single emission line spectra, reducing $\eta$ further to 2.3\%. The fine-tuned model also yields superior spectral reconstructions with less noise and more accurately aligned features. The resulting model establishes a robust, scalable framework directly applicable to future grism-based surveys from observatories like the James Webb Space Telescope (JWST), Euclid, and Roman, paving the way for automated, high-precision analysis of large-scale spectroscopic data and accelerating discoveries in galaxy evolution and cosmic structure formation.

\section{DEMO CASTOR - Gaël Noirot \label{sec:GNoirotD}}
This hands-on demo presented the CASTOR Grism Simulator and Exposure Time Calculator (Noirot et al., in prep).
\begin{itemize}
\item CASTOR\footnote{\url{https://www.castormission.org}} \citep{Cote2025} is a planned Canadian-led Wide-Field (0.25 $\deg^2$) space telescope for the UV, blue and optical regimes ($\lambda=1500-5500~\rm \AA$), with the field of view of Roman (Figure~\ref{fig:castor1}) and the resolution of Hubble in the UV ($0\farcs{15}$ FWHM).

\begin{figure}
    \centering
    \includegraphics[width=0.9\linewidth]{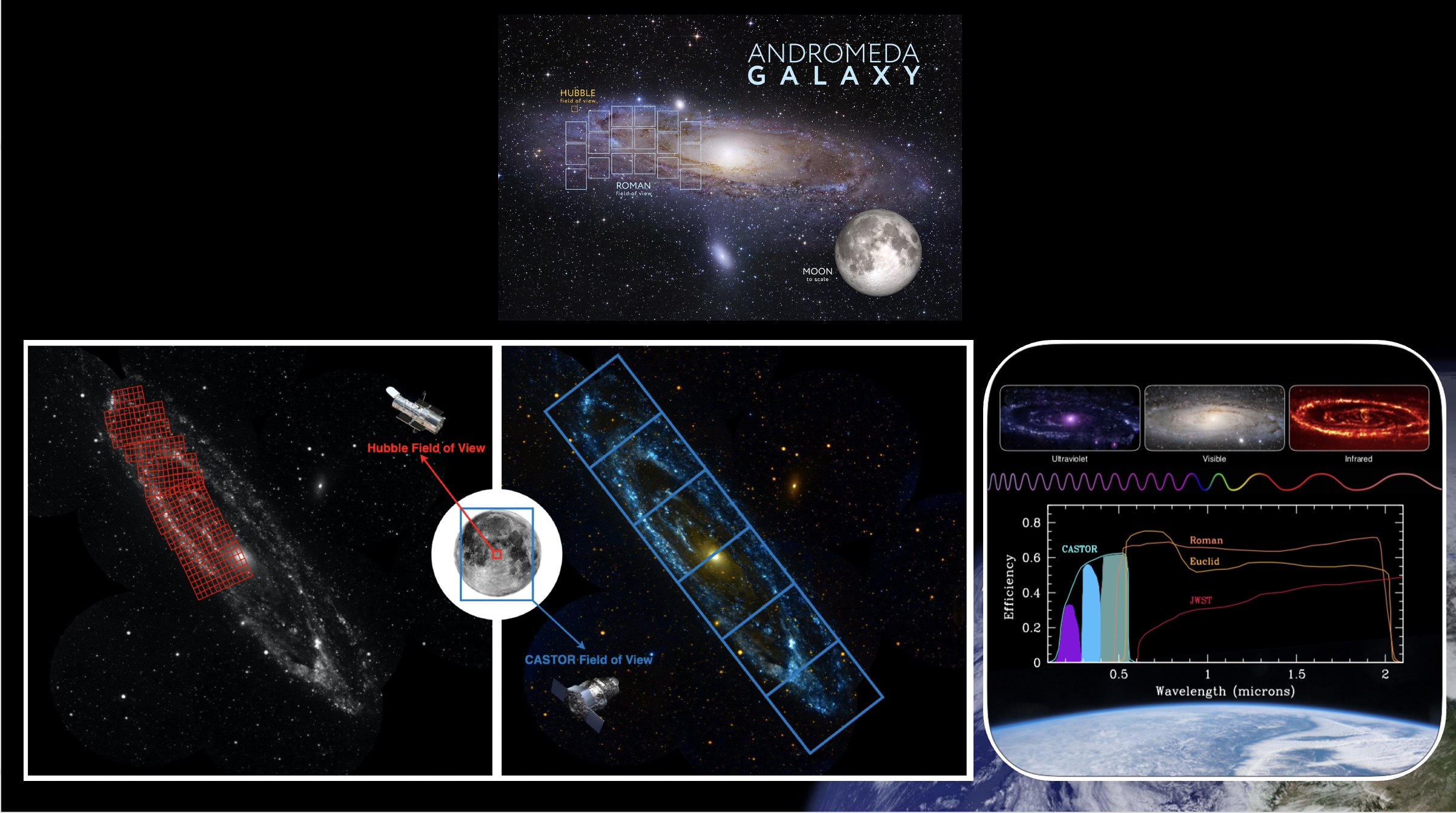}
    \caption{This slide shows a comparison between the HST (left), Roman (top), and CASTOR (middle) field of views. Roman and CASTOR have similar field of views, roughly covering the entire full moon. CASTOR is also very complimentary to other missions such as Roman, Euclid and JWST as it probes the UV to optical regimes not available to these other telescopes (right).}
    \label{fig:castor1}
\end{figure}

\item It provides simultaneous photometry in the UV ($\lambda=1500-3000~\rm \AA$), u ($\lambda=3000-4000~\rm \AA$) and g ($\lambda=4000-5500~\rm \AA$) bands, reaching mag $27$~AB in $\sim 600~\sec$, and simultaneous Wide-Field Slitless Spectroscopy in the UV and u bands (R~$\sim$~350), over the entire field of view. It also possesses an ultra-high precision photometer and UV Multi-Object Spectrograph (R~$\sim$~2000); see Figure~\ref{fig:castor2}.

\begin{figure}
    \centering
    \includegraphics[width=0.9\linewidth]{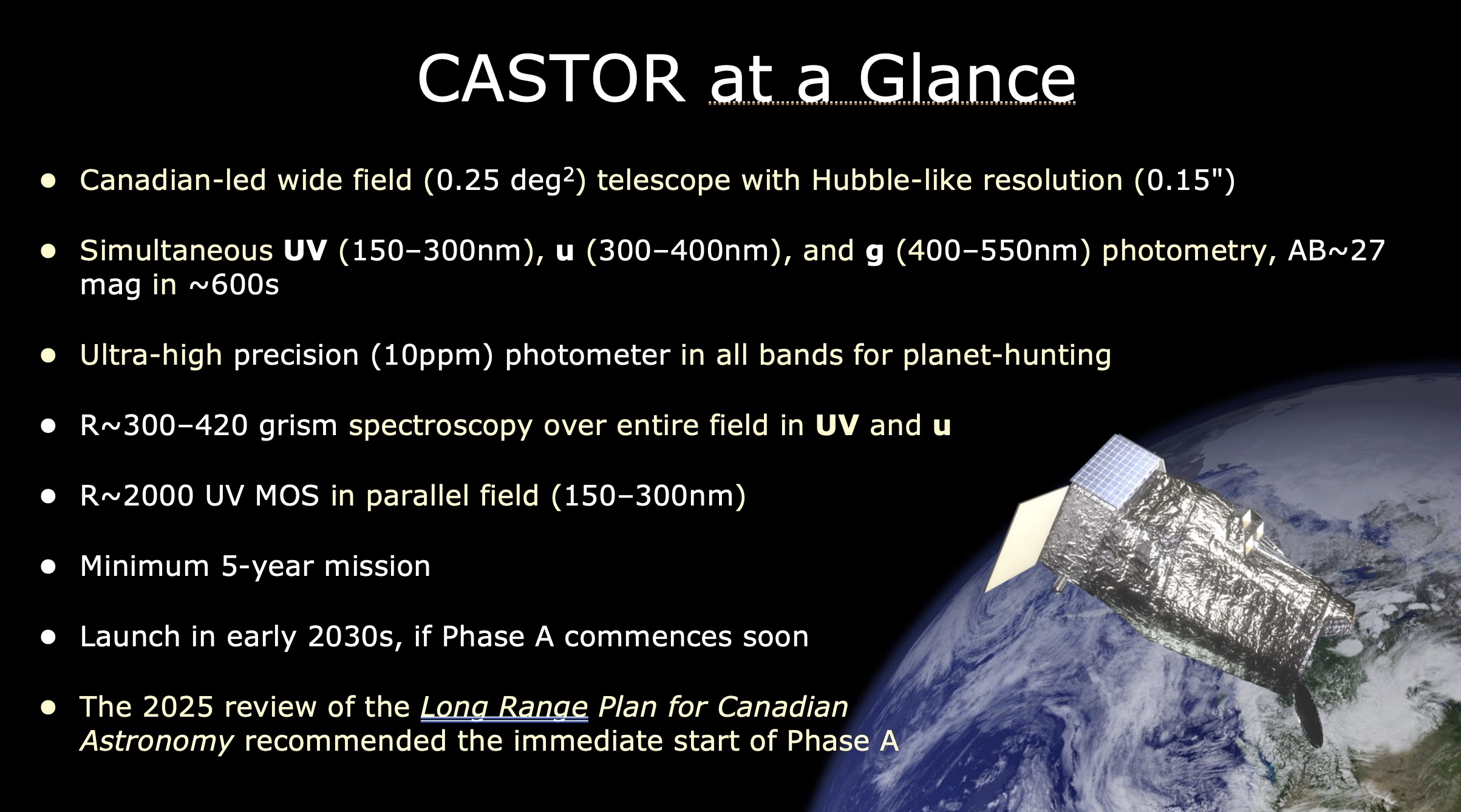}
    \caption{This slide highlights the main CASTOR specifications.}
    \label{fig:castor2}
\end{figure}

\item The ``CASTOR Grism ETC and Simulator'' is a tool built to simulate CASTOR grism data, first built as an Exposure Time Calculator (ETC) and further developed to simulate full scenes, perform forward-modeling template fitting (e.g., redshift/SED fitting), spatially-resolved simulations, etc.

\item The code is currently hosted at \url{https://github.com/gnoir0t/ETC_grism}. The demo notebook hosted on the Google Colab platform used for the WFSS workshop can be accessed at \url{https://tinyurl.com/castor4d}.

\item The demo notebook starts with simulating a single source and obtaining the 2D SNR map (per pixel) and 1D SNR profile of the dispersed grism data in both the UV and u bands for a given exposure time. The tool offers flexibility in the choice of sky background parameters as well as dark current and red noise. Users can provide their own spectral templates or directly use FSPS within the tool. Users can also segment their direct image in multiple regions to perform spatially-resolved simulations.

\item The demo then shows how to redshift-fit a simulated observed spectrum. The functionality properly accounts for morphological broadening by forward-modelling the input template to the grism dispersed frame at each step of the redshift range explored during fitting. The demo also shows how to loop over realizations to estimate the robustness of the fitting.

\item The notebook then introduces the full scene simulation capability, which allows to simulate entire fields and trace overlap. It also shows how to build scenes with foreground and background objects overlapping in the direct imaging (e.g., chance alignments, mergers, etc), which are then properly dispersed to the grism scene (Figure~\ref{fig:castor3}). This capability allows to add an extra layer of complexity in spatially-resolved simulations, where multiple stellar populations can sit on top of one another even within the same direct-image pixel. This functionality is often not available even in other spatially-resolved tools which assume that each direct image region/pixel has a unique stellar population.

\begin{figure}
    \centering
    \includegraphics[width=0.9\linewidth]{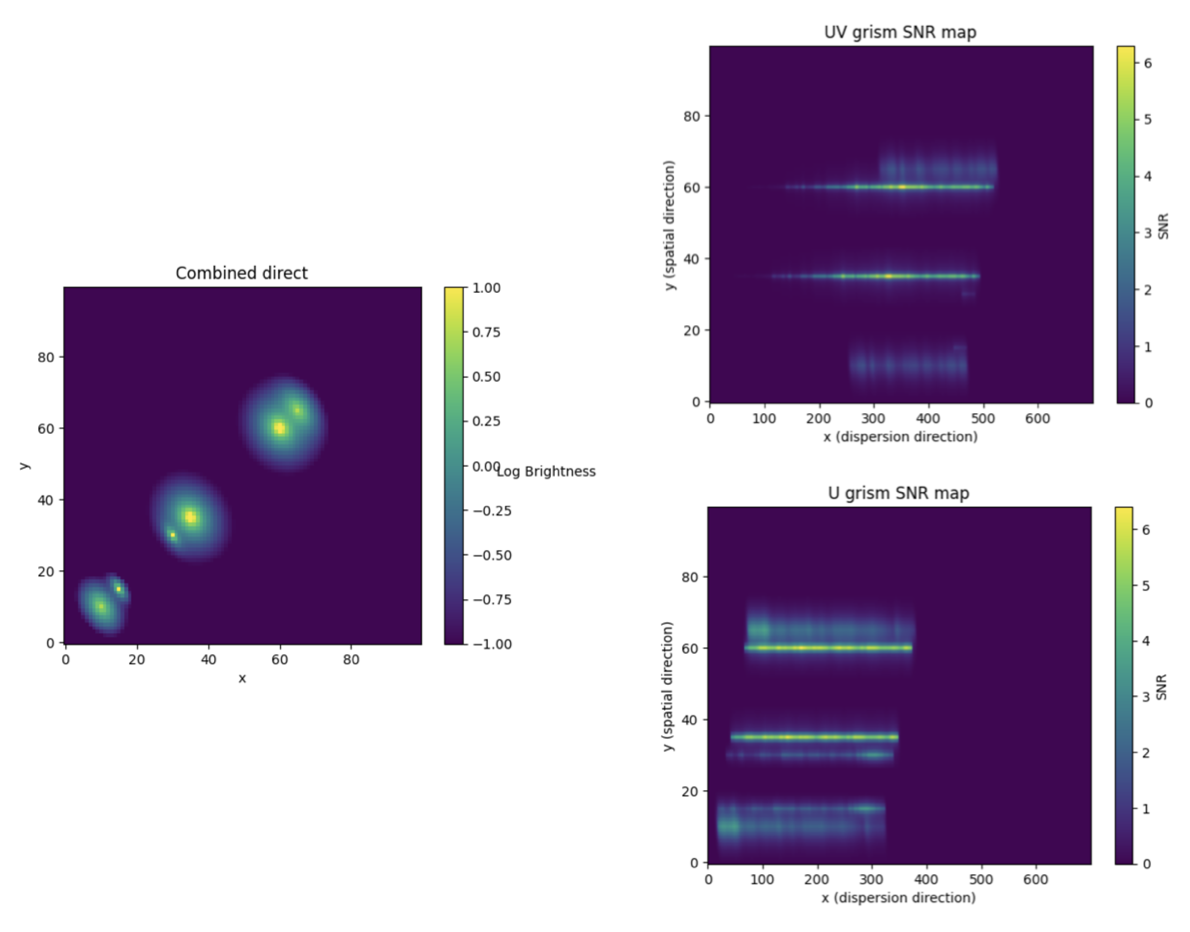}
    \caption{Simple example of a full scene simulation using the multi-layer capability of the CASTOR grism tool. The left panel shows the result of generating a foreground and a background scene where sources with different SEDs can overlap in the direct image plane. The right panels show the resulting 2D SNR map of the dispersed grism spectra in the UV (top) and u (bottom) bands for the specific observing strategy used during the tutorial.}
    \label{fig:castor3}
\end{figure}

\item The demo finally shows how to perform SED template fitting to recover physical parameters.

\end{itemize}





\begin{acknowledgments}
GN acknowledges support by the Canadian Space Agency under a contract with NRC Herzberg Astronomy and Astrophysics. VEC acknowledges support from the Beus Center for Cosmic Foundations.

\end{acknowledgments}

\begin{contribution}



\end{contribution}

%
\facilities{HST (ACS, STIS, WFC3), JWST (MIRI, NIRCam, NIRISS, NIRSpec), Euclid, Roman, CASTOR}

\software{astropy \citep{2013A&A...558A..33A,2018AJ....156..123A,2022ApJ...935..167A},  
          Cloudy \citep{2013RMxAA..49..137F}, 
          Source Extractor \citep{1996A&AS..117..393B}
          }


\bibliography{references,PASPsample701}{}
\bibliographystyle{aasjournalv7}

\end{document}